\documentclass[11pt, a4paper]{scrartcl}
\usepackage[a4paper, bindingoffset=6mm,marginratio={4:6,5:7}]{geometry}
\usepackage[T1]{fontenc}
\usepackage{multicol}
\usepackage[british]{babel}
\usepackage{stmaryrd,enumerate,bbm,latexsym}
\usepackage{amsthm,amsmath}
\usepackage[numbers]{natbib}
\usepackage[utf8]{inputenc}
\usepackage[bitstream-charter, cal=cmcal]{mathdesign}
\usepackage{hyperref}

\numberwithin{equation}{section}
\theoremstyle{plain}
\newtheorem{foo}{Foo}[section]
\newtheorem{theorem}[foo]{Theorem}
\newtheorem{lemma}[foo]{Lemma}
\newtheorem{corollary}[foo]{Corollary}
\newtheorem{proposition}[foo]{Proposition}
\theoremstyle{remark}
\newtheorem{remark}[foo]{Remark}
\setkomafont{section}{\normalfont\Large\bfseries}
\setkomafont{subsection}{\normalfont\large\bfseries}
\setkomafont{title}{\normalfont\large\bfseries}
\setkomafont{paragraph}{\normalfont\large\bfseries}

\newcommand{\Iota}{\mathrm{I}}
\newcommand{\TeXmacs}{T\kern-.1667em\lower.5ex\hbox{E}\kern-.125emX\kern-.1em\lower.5ex\hbox{\textsc{m\kern-.05ema\kern-.125emc\kern-.05ems}}}
\newcommand{\assign}{:=}
\newcommand{\backassign}{=:}
\newcommand{\mathD}{\mathrm{D}}
\newcommand{\mathd}{\mathrm{d}}
\newcommand{\mathe}{\mathrm{e}}
\newcommand{\nin}{\not\in}
\newcommand{\nobracket}{}
\newcommand{\noplus}{}
\newcommand{\nosymbol}{}

\newcommand{\tmdummy}{$\mbox{}$}

\newcommand{\tmop}[1]{\ensuremath{\operatorname{#1}}}
\newcommand{\tmscript}[1]{\text{\scriptsize{$#1$}}}

\newcommand{\tmtextbf}[1]{\text{{\bfseries{#1}}}}
\newcommand{\tmtextit}[1]{\text{{\itshape{#1}}}}
\newcommand{\subjclass}[2][]{%
  \medskip\noindent\textbf{MSC (#1).} #2\par}
\newcommand{\keywords}[1]{%
  \medskip\noindent\textbf{Keywords.} #1\par}
\newenvironment{enumeratealpha}{\begin{enumerate}[a{\textup{)}}] }{\end{enumerate}}

\title{The FBSDE approach to sine--Gordon up to $6 \pi$}
\author{
Massimiliano~Gubinelli\thanks{Mathematical Institute, University of Oxford,\\\noindent
\texttt{\{massimiliano.gubinelli, sarah-jean.meyer\}@maths.ox.ac.uk}}
\and
Sarah-Jean Meyer\footnotemark[1]
}
\date{}

\begin{document}
\maketitle

\begin{abstract}
	We develop a stochastic analysis of the sine-Gordon Euclidean quantum field
	$(\cos (\beta \varphi))_2$ on the full space up to the second threshold,
	i.e. for $\beta^2 < 6 \pi$. The basis of our method is a forward-backward
	stochastic differential equation (FBSDE) for a decomposition $(X_t)_{t
				\geqslant 0}$ of the interacting Euclidean field $X_{\infty}$ along a scale
	parameter $t \geqslant 0$. This FBSDE describes the optimiser of the
	stochastic control representation of the Euclidean QFT introduced by
	Barashkov and one of the authors. We show that the FBSDE provides a
	description of the interacting field without cut-offs and that it can be
	used effectively to study the sine-Gordon measure to obtain results about
	large deviations, sub-gaussian tails, decay of correlations for local
	observables, singularity with respect to the free field,
	Osterwalder--Schrader axioms and other properties.
\end{abstract}


\begingroup
\setcounter{tocdepth}{2}
\setlength{\columnsep}{1.2em}
\raggedcolumns
\begin{multicols}{2}
	\tableofcontents
\end{multicols}
\endgroup

\section{Introduction}
The aim of this paper is to provide a rigorous \tmtextit{description} of the
two-dimensional sine-Gordon Euclidean quantum field theory (EQFT) on the full
space in the regime $\beta^2 < 6 \pi$. The sine-Gordon EQFT is formally given
by the Gibbs measure
\begin{equation}
	\text{``}\nu_{\tmop{SG}} (\mathd \varphi) = \Xi^{- 1} \exp (- V_{\tmop{SG}}
	(\varphi)) \mu (\mathd \varphi) \text{{\phantom{.}}''}, \quad \varphi \in
	\mathcal{S}' (\mathbb{R}^2) \label{eq:int-Gibbs-ms},
\end{equation}
where $\mu$ is a massive Gaussian free field on the space of Schwartz
distributions $\mathcal{S}' (\mathbb{R}^2)$, the constant $\Xi$ is a
normalisation to make $\nu_{\tmop{SG}}$ a probability measure, and
$V_{\tmop{SG}}$ corresponds to the cosine interaction, formally defined as
\[ V_{\tmop{SG}} (\varphi) \assign \lambda \int_{\mathbb{R}^2} \cos (\beta
	\varphi (x)) \mathd x. \]
The sine-Gordon model is a prototypical example of a non-Gaussian EQFT and of
particular interest as both a theory with infinitely many phase transitions as
$\beta^2$ varies between $0$ and $8 \pi$ and more generally as a test-bed for
non-polynomial interactions.

The approach we take here is based on a scale dependent interpolation
$(G_t)_{t \in [0, \infty]}$ of the covariance $G_{\infty} = (\Delta - m^2)^{-
			1}$ of the Gaussian free field. This allows us to interpret the Gaussian free
field as the terminal value $W_{\infty}$ of a Brownian martingale $(W_t)_{t
	\in [0, \infty)}$ defined by
\begin{equation}
	W_t \assign \int_0^t \dot{G}_s^{1 / 2} \mathd B_s, \qquad t \geqslant 0,
	\label{eq:GFF}
\end{equation}
where $B = (B_t)_{t \geqslant 0}$ is a cylindrical Brownian motion on $L^2
	(\mathbb{R}^2)$. From this point of view, we can produce a scale dependent
stochastic dynamics $(X_t)_{t \in [0, \infty]}$ for the target measure
\eqref{eq:int-Gibbs-ms} which provides a pathwise
scale-by-scale coupling $(X_t, W_t)_{t \in [0, \infty]}$.
\begin{equation}
	\mathd X_t = - \dot{G}_t \mathbb{E}_t [\mathD V_{\tmop{SG}} (X_{\infty})]
	\mathd t + \dot{G}_t^{1 / 2} \mathd B_t, \qquad t \geqslant 0.
	\label{eq:fbsde-intro}
\end{equation}
Here, $\mathD V_{\tmop{SG}} (\varphi) = - \lambda \beta \sin (\beta \varphi)$ is
formally the functional derivative of the interaction potential
$V_{\tmop{SG}}$, and we write $\dot{G}_t \assign \partial_t G_t$ for the scale
derivative of $G$. The conditional expectation $\mathbb E_t$ is taken with respect to the
augmentation of the filtration generated by the Brownian motion $W$ and can
be interpreted as averaging out the small scales $s^{-1}\leqslant t^{-1}$.

In this paper, we show that, once properly renormalised, the
FBSDE~\eqref{eq:fbsde-intro} provides an effective stochastic quantisation
equation for \eqref{eq:int-Gibbs-ms} for any size of the coupling constant $\bar{\lambda}:=|\lambda|\in \mathbb{R}$. This allows to construct the measure
\eqref{eq:int-Gibbs-ms} without cut-offs from a straightforward analysis of the
equation and only basic estimates of the convolution $\dot{G}$ (see
Theorem~\ref{thm:int-wp}). Moreover, we can efficiently transport properties
from the Gaussian free field to the sine-Gordon EQFT via
\eqref{eq:fbsde-intro}. In particular, we obtain
\begin{enumeratealpha}
	\item an explicit description of the infinite volume measure via a
	variational principle (Theorem~\ref{thm:infinite-vol-var});

	\item a proof of the mutual singularity of the Gaussian free field and the
	finite volume sine-Gordon measure for $\beta^2 \geqslant 4 \pi$ (Theorem~\ref{thm:singularity});

	\item a simple proof for the exponential decay of correlations of general
	local observables (Theorem~\ref{thm:decay-of-correlations});

	\item an analysis of the semi-classical limit $\hbar \rightarrow
		0$~(Theorem~\ref{thm:LDP});

	\item a full verification of the Osterwalder--Schrader axioms and a proof of
	non-Gaussianity (Theorem~\ref{thm:OS-axioms});
\end{enumeratealpha}
In order to give a rigorous meaning
to~\eqref{eq:fbsde-intro} we start, as usual, from a well-defined
approximation of the sine-Gordon measure given by
\begin{equation}
	\nu^{\rho, T} (\mathd \varphi) = \Xi_{\rho, T}^{- 1} \exp (- V^{\rho, T}
	(\varphi)) \mu^T (\mathd \varphi) \label{eq:int-approx-measures},
\end{equation}
where $\rho$ is an infrared cut-off and $\mu^T$ denotes the law of an
approximation $W_T$ to the massive Gaussian free field~$W_{\infty}$ as
in~\eqref{eq:GFF}. In Section~\ref{sec:CP} we will show that in this
regularised setting, the FBSDE \eqref{eq:fbsde-intro} produces the correct
measure, that is the solution $(X_t)_t$ to the FBSDE
\begin{equation}
	\mathd X^{\rho, T}_t = - \dot{G}_t \mathbb{E}_t [\mathD V^{\rho, T} (X^{\rho, T}_T)] \mathd t +
	\mathd W_t, \qquad t \in [0, T] \label{eq:approx-fbsde}
\end{equation}
has terminal law $\tmop{Law} (X_T^{\rho, T}) = \nu^{\rho, T}$. As a byproduct, we show
that it is associated with the solution to the stochastic optimal control
problem
\begin{equation}
	- \log \mathbb{E} [\mathe^{- V^{\rho, T} (W_T)}] = \inf_{u \in \mathbb{H}_a}
	\mathbb{E} \left[ V^{\rho, T} (I_T (u) + W_T) + \frac{1}{2} \int_0^{\infty}
		\| u_t \|^2_{L^2} \mathd t \right] \label{eq:int-BD},
\end{equation}
where $I_t (u) \assign \int_0^t \dot{G}^{1 / 2}_s u_s \mathd s$ and
$\mathbb{H}_a$ is an appropriate space of predictable processes. As expected,
the representations~\eqref{eq:approx-fbsde} and~\eqref{eq:int-BD} are not
stable in the small-scale limit $T \rightarrow \infty$, and they require a
renormalisation of the potential $V^{\rho, T}$ involving diverging constants.
To overcome this problem, suppose that $F$ is a sufficiently nice scale
dependent function $F = (F_t)_{t \in [0, T]}$ such that $F_T = \mathD V^{\rho,
			T}$. By Ito's formula, solving the FBSDE \eqref{eq:approx-fbsde} is equivalent
to solving the FBSDE
\begin{equation}
	\begin{cases}
		Z_t = - \int_0^t \dot{G}_s (F_s (Z_s + W_s) + R_s) \mathd s, \\
		R_t =\mathbb{E}_t \int_t^T H_s (Z_s + W_s) \mathd s -\mathbb{E}_t \int_t^T
		\mathD F_s  \dot{G}_s R_s \mathd s,
	\end{cases}  \label{eq:int-FBSDE} \qquad t \in [0, T],
\end{equation}
where the functional $(H_t)_{t \in [0, T]}$ is given by
\[ H_t \assign \partial_t F_t + \frac{1}{2} \tmop{Tr} (\dot{G}_t \mathD^2 F_t)
	- \mathD F_t  \dot{G}_t F_t, \qquad t \in [0, T] . \]
The solution $X$ to~\eqref{eq:approx-fbsde} can then be obtained
from~\eqref{eq:int-FBSDE} with the identification $X_t = Z_t + W_t$. In this
representation, the limit $T \rightarrow \infty$ is associated to the
convergence of the integral over scales in the equation for the
\tmtextit{remainder} $R$. Constructing the measure~\eqref{eq:int-Gibbs-ms}
reduces to two tasks:
\begin{enumerate}
	\item Find an approximation $F$ for the effective force $\mathbb{E}_t
		      [\mathD V_{\tmop{SG}} (X_{\infty})]$ that makes the source term $H_s (W_s +
		      Z_s)$ of the backward equation in \eqref{eq:int-FBSDE} integrable as $s
		      \rightarrow \infty$, while preserving good continuity and growth properties.

	\item Control the associated FBSDE~\eqref{eq:int-FBSDE} uniformly in the
	      regularisations $T$ and $\rho$ and establish global existence for the
	      solutions to \eqref{eq:int-FBSDE}.
\end{enumerate}
The first task involves a good understanding of approximate solutions to the
well-known infinite dimensional and non-linear (backward) Polchinski renormalisation flow
equation (see
e.g.~{\cite{salmhoferRenormalization1999}} or the recent
review~{\cite{bauerschmidtStochasticDynamicsPolchinski}}),
\begin{equation}
	\begin{cases}
		\partial_t v_t + \frac{1}{2} \tmop{Tr} (\dot{G}_t \mathD^2 v_t) -
		\frac{1}{2} \mathD v_t  \dot{G}_t \mathD v_t = 0, \\
		v_{\infty} (\varphi) = V_{\tmop{SG}} (\varphi).
	\end{cases} \label{eq:int-flow-eq}
\end{equation}
Indeed, given a solution $v$ to \eqref{eq:int-flow-eq} and taking $F_t =
	\mathD v_t$ we would have $H_t = 0$ and therefore $R_t = 0$. The remainder $R$
allows for additional freedom in the choice for the scale interpolation of the
force $F_t$ and avoids a precise technical analysis of \eqref{eq:int-flow-eq}.

The second task requires good a priori estimates for the non-standard FBSDE
\eqref{eq:int-FBSDE}, which are uniform in the regularisation $T$. The
equation \eqref{eq:int-FBSDE} is in general nonlinear and thus solutions need
not be global so that this step is non-trivial and indeed the reason why the present
work is limited to the regime $\beta^2 < 6 \pi$. It would be very interesting
to better understand the solution theory for FBSDEs of the form
\eqref{eq:int-FBSDE} also in a more general setting for different models, that
is different choices of $F_t$.

Our main result is the following.

\begin{theorem}
	\label{thm:int-wp}Let $\beta^2 < 6 \pi$. For $\rho \in C^{\infty}_c
		(\mathbb{R}^2)$ or $\rho \equiv 1$ and $T \in [0, \infty]$, there is scale
	dependent function $F^{\rho, T} = (F^{\rho, T}_s)_{s \in [0, T]}$ such that
	up to first order (in $\bar{\lambda}$),
	$F^{\rho, T}_T$ corresponds perturbatively to the Wick-renormalised sine
	\[ F^{\rho, T}_t (W_{t\wedge T}) (x) = - \rho (x) \beta \lambda \llbracket \sin (\beta
		W_{t\wedge T} (x)) \rrbracket + O(\bar{\lambda}^2),\]
	and the associated FBSDE~\eqref{eq:int-FBSDE} has a solution~$(Z^{\rho, T},
		R^{\rho, T}) \in \mathbb{H}^{\infty} (L^{\infty}) \times \mathbb{H}^{\infty}
		(L^{\infty})$.

	If the volume is finite, that is $\rho \in C^{\infty}_c (\mathbb{R}^2)$, \
	or if the coupling constant $\bar{\lambda}$ is sufficiently small, $\tmop{Law}(X_\infty^\rho)=\nu_{\tmop{SG}}^\rho$ is unique. For $\rho = 1, T = \infty$ and any $\varepsilon > 0$,
	there is a version of the drift $Z = Z^{1, \infty}$ with terminal value
	$Z_{\infty} \in L^{\infty} (\mathd \mathbb{P} ; B_{p, p}^{2 - \beta^2 / 4 \pi -
				\varepsilon, - n})$, and the sine-Gordon measure is given as a random shift
	of the Gaussian free field $W_{\infty}$,
	\[ \nu_{\tmop{SG}} = \tmop{Law} (W_{\infty} + Z_{\infty}) . \]
\end{theorem}
It should be emphasised that while our analysis provides uniqueness only if
the coupling constant $\bar{\lambda}$ is small or the volume is finite, its
existence is guaranteed for any $\lambda \in \mathbb{R}$ also in the full
space: we obtain uniform bounds on the FBSDE for any $\lambda \in \mathbb{R}$
which imply tightness for the family $\nu_{\tmop{SG}}^{\rho, T} = \tmop{Law}
	(W_T + Z^{\rho, T}_T)$. Indeed, at the level of the approximate measures \eqref{eq:int-approx-measures}, Theorem \ref{thm:int-wp} implies following result.
\begin{corollary}
	For any $\lambda\in \mathbb R$, the family of measures $(\nu_{\tmop{SG}}^{\rho,T})_{T,\rho}$ defined in \eqref{eq:int-approx-measures} is tight on $H^{-\varepsilon}(\langle x\rangle^{-2})$. Moreover, there is a unique accumulation point if either of the conditions below is satisfied
	\begin{itemize}
		\item $\bar{\lambda}$ is sufficiently small. In this case $(\nu_{\tmop{SG}}^{\rho,T})_{T,\rho}$ converges $H^{-\varepsilon}(\langle x\rangle^{-2})$-weakly to a measure $\nu_{\tmop{SG}}$.
		\item $\rho\in C^\infty_c(\mathbb{R}^2)$ remains fixed. Then, as $T\to\infty$, the sequence $(\nu_{\tmop{SG}}^{\rho,T})_{T\geqslant 0}$ converges $H^{-\varepsilon}(\langle x\rangle^{-2})$-weakly to a limiting measure $\nu_{\tmop{SG}}^\rho$.
	\end{itemize}
\end{corollary}

To demonstrate the advantages of the representation, we transport some
properties of the free field $W_{\infty}$ to the sine-Gordon shift $W_{\infty}
	+ Z_{\infty}$. A neat application is the exponential decay of correlation via
a simple coupling argument as
in~{\cite{vecchiStochasticAnalysisSubcritical2025, gubinelliDecayCorrelationsStochastic2025}}.
In this setting, we can show that for the unique solution $Z_t$ to
\eqref{eq:int-FBSDE} at $T = \infty$, $\rho = 1$, the process $(X_t)_{t \in
			[0, \infty]} = (Z_t + W_t)_{t \in [0, \infty]}$ inherits the following decay
of correlations from $W$. Note that the theorem below includes $t = \infty$
and thus $\nu_{\tmop{SG}} = \tmop{Law} (X_{\infty})$.

\begin{theorem}
	\label{thm:decay-of-correlations}Let $\chi$ be a smooth function supported
	on $B_1 (0)$ and $x_1, x_2 \in \mathbb{R}^2$. If $\bar{\lambda}$ is sufficiently small,
	then there is a constant
	$\gamma \in (0, 1)$ depending only on the mass $m$ such that for any two
	bounded and Lipschitz observables $\mathcal{O}_1, \mathcal{O}_2 : H^{-
				\varepsilon, - n} \rightarrow \mathbb{R}$, it holds that
	\[ \left| \mathrm{Cov} [\mathcal{O}_1 (\chi \cdot X_t (\cdot + x_1)) ;
			\mathcal{O}_2 (\chi \cdot X_t (\cdot + x_2))]  \right| \lesssim \mathe^{- m \gamma | x_1 - x_2 |} . \]
	Here, the implicit constant depends only on the bounds and Lipschitz
	constants of the observables $\mathcal{O}_1$ and $\mathcal{O}_2$.
\end{theorem}

In the first region $\beta^2 < 4 \pi$, it is not difficult to see that the
finite volume sine-Gordon measure is absolutely continuous with respect to the
Gaussian free field. Using the FBSDE, we
can show that this is no longer the case beyond this threshold. To the best of
our knowledge, this is the first proof of this fact.

\begin{theorem}
	\label{thm:singularity}For $\beta^2 \in [4 \pi,6\pi)$, the finite volume
	sine-Gordon measure and the Gaussian free field are mutually singular.
\end{theorem}

As a result of this singularity, the control problem~\eqref{eq:int-BD} cannot
be transferred to the UV-limit verbatim, in contrast to the simpler setting
$\beta^2 < 4 \pi$ (treated in {\cite{barashkovStochasticControlApproach2022}}).
Building on the same ideas used for the change of variables in the FBSDE from
\eqref{eq:fbsde-intro} to \eqref{eq:int-FBSDE}, we reformulate the variational
problem \eqref{eq:int-BD} in terms of an (absolutely continuous) remainder.
This reformulation, combined with a localisation property of the limiting
measure, allows us to recover a variational problem for the Laplace transform
of $\nu_{\tmop{SG}}$ in the infinite volume (see \ref{thm:infinite-vol-var}) as long as $\bar{\lambda}$ is sufficiently small.

With this variational formulation at hand, we can show that the
limiting measure $\nu_{\tmop{SG}}$ defines a non-Gaussian EQFT and derive a
Laplace principle for the semi-classical limit $\hbar \rightarrow 0$. To make
this slightly more precise, let $(\mu^{\hbar})_{\hbar \in (0, 1)}$ be the
family of rescaled Gaussian free fields with covariance $\hbar (m^2 -
	\Delta)^{- 1}$. For now, formally define the measures
\[ \text{``} \nu_{\tmop{SG}}^{\hbar} (\mathd \varphi) \assign \Xi_{\hbar}^{-
		1} \exp (- \hbar^{- 1} V (\varphi)) \mu^{\hbar} (\mathd \varphi) \text{.''}
\]
We show the following theorem.

\begin{theorem}
	\label{thm:LDP}As $\hbar \rightarrow 0$, the family
	$\nu^{\hbar}_{\tmop{SG}}$ satisfies a Laplace principle with rate $\hbar^{-
			1}$ and rate function
	\begin{equation}
		I (\varphi) \assign \begin{cases}
			\lambda \int (\cos (\beta \varphi) - 1) + \frac{1}{2} \int \varphi (m^2
			- \Delta) \varphi, & \varphi \in H^1 (\mathbb{R}^2), \\
			\infty,            & \text{otherwise} .
		\end{cases} \label{eq:def-rf}
	\end{equation}
	More precisely, for any continuous and bounded $f : \mathcal{S} '
		(\mathbb{R}^2) \rightarrow \mathbb{R}$,
	\begin{equation}
		\lim_{\hbar \rightarrow 0} - \hbar \log \int_{\mathcal{S}' (\mathbb{R}^2)}
		\exp (- \hbar^{- 1} f (\varphi)) \nu_{\tmop{SG}}^{\hbar} (\mathd \varphi)
		= \inf_{\varphi \in H^1} \{ f (\varphi) + I (\varphi) \} . \label{eq:LDP}
	\end{equation}
\end{theorem}

Finally, we verify the
Osterwalder Schrader axioms for small couplings $\bar{\lambda}$.

\begin{remark}
	Our approach relies only on some general estimates for the heat kernel of
	the Laplacian (see Appendix~\ref{app:hk}) and can be easily extended with
	respect to the dimension of the underlying Euclidean space. In the general
	$d$-dimensional setting, the sine-Gordon theory can be considered with respect to the $d$-dimensional log-correlated Gaussian field and is subcritical for $\beta^2 / 2 \pi \in
		[0, 2 d)$ and the argument presented here allows to construct the
	sine-Gordon measure in $\beta^2 / 2 \pi \in [0, d + 1)$. In one dimension, this means we can cover the full subcritical
	regime, (see also
		{\cite{lacoinProbabilisticApproachUltraviolet2022}}, where a one-dimensional version of the model is considered). We can moreover generalise our results to the
	(compact) Riemannian manifold context, in analogy to the recent
	work~{\cite{bailleulPhi4_3MeasuresCompact2023}} on $\Phi^4_3$ on a compact
	Riemannian manifold. For the sake of clarity, we refrain from including
	these modifications. The required changes are minimal and we do not believe
	that the associated results would justify extending this contribution.
\end{remark}

\subsection{Related work}

The sine--Gordon model has been subject to many studies in the constructive
literature, covering finite or infinite volume interactions and allowing
various ranges for $\beta^2 \in [0, 8 \pi)$ and the coupling constant $\lambda
	\in \mathbb{R}$. However, the full mathematical understanding of this model is
still lacking and none of these works cover the theory on the full space
$\mathbb{R}^2$ for all $\beta^2 \in (0, 8 \pi)$ and all $\lambda \in
	\mathbb{R}$. A comprehensive review of the vast literature on the model can be found in the
paper~{\cite{bauerschmidtColemanCorrespondenceFree2023}} where in the reader
will also find a description of the correspondence with certain fermionic
Euclidean models.

We single out the pioneering work of Benfatto et
al.~{\cite{benfattoMassiveSineGordonEquation1982}} and Nicol{\'o} et
al.~{\cite{nicoloMassiveSineGordonEquation1986}} who establish existence of
the model for a finite volume interaction and small coupling constants in the
full subcritical range $\beta^2 < 8 \pi$ via a probabilistic method initiated
by the Roman school of Gallavotti and co-authors. A more modern account is the
martingale method of~{\cite{lacoinProbabilisticApproachUltraviolet2022}} which
covers the full subcritical regime in the case $d = 1$, in a bounded domain
but without restrictions on the coupling constant $\lambda \in \mathbb{R}$.

The more classical stochastic quantisation using the Langevin dynamics for the
sine-Gordon model has only been partially resolved after the preprint of
the present article appeared. In contrast to the $\Phi^4_3$ theory, there is no
obvious coercive term to ensure global in time existence and even though the
local solution theory was established already in
\cite{hairerDynamicalSineGordonModel2016,chandraDynamicalSineGordonModel2018},
the first global in time existence results for $4\pi+\varepsilon$ were obtained by Chandra, Feltes and
Weber \cite{chandraPrioriBounds2d2024} as a corollary of global existence for the generalised parabolic Anderson model in a finite volume
(see also the related paper \cite{shenGlobalWellposedness2D2025}).
Even more recently, Bringmann and Cao \cite{bringmannGlobalWellposednessDynamical2024}
managed to show global well-posedness for the Langevin dynamics on the torus for $\beta^2<6\pi$.

Due to the analytic treatability of the sine-Gordon interaction, there have
been several accounts based on renormalisation group ideas and a direct
analysis of the Polchinski flow equation \eqref{eq:int-flow-eq}. In this
regard, we want to mention the analysis of Brydges and
Kennedy~{\cite{brydgesMayerExpansionsHamiltonJacobi1987}}, where they lay the
foundations for this approach relying on a majorant method to establish
convergence of the Mayer expansion up to $\beta^2 < \frac{2}{3} 8 \pi$. More
recently, Bauerschmidt and
Bodineau~{\cite{bauerschmidtLogSobolevInequalityContinuum2021}} showed
convergence for the Mayer expansion up to $6 \pi$ which allows them to
establish a uniform log-Sobolev inequality for a lattice approximation of the
model. In a related work, Bauerschmidt and
Hofstetter~{\cite{bauerschmidtMaximumCouplingSineGordon2022}} use the solution
obtained from the Mayer expansion to construct a multiscale coupling between
the Gaussian free field and the sine-Gordon model \ and analyse the maximum of
the sine-Gordon measure. Similar ideas were applied by Barashkov, Gunaratnam
and Hofstetter~{\cite{barashkovMultiscaleCouplingMaximum2023}} to analyse the
maximum of the $P (\varphi)_2$ models in a bounded domain. These last two
papers are similar in spirit and complementary to ours, but rely on a direct
analysis of the Polchinski equation \eqref{eq:int-flow-eq} and focus on the
extremal analysis in a finite volume instead of a general analysis and
properties of the resulting EQFT.

Focusing now on the connection between the FBSDE and stochastic optimal
control, a direct precursor of the results presented here is the work of
Barashkov~{\cite{barashkovStochasticControlApproach2022}} (and the related PhD
thesis~{\cite{barashkovVariationalApproachGibbs2021}}), where the model is
studied in the first region $\beta^2 < 4 \pi$ on the full space $\mathbb{R}^2$
using a variational approach. This approach is based on the stochastic
control problem~\eqref{eq:int-BD} and was first applied to the $\Phi^4_3$
model in bounded volume in~{\cite{barashkovVariationalMethod$Phi^4_3$2020}}. The
more recent extension in~{\cite{barashkovVariationalMethodEuclidean2023}} to
the infinite volume limit for the polynomial and exponential interaction in
the $2$ dimensional setting relies on a weak formulation of the FBSDE we use
here. In the case of a Grassmannian field, the FBSDE approach has been
successfully applied in~{\cite{vecchiStochasticAnalysisSubcritical2025}} to
cover the full subcritical regime. This also includes the complete inductive
analysis of the corresponding approximate flow equation. After a first preprint
of this article got published, Duch constructed a marginal asymptotically free Fermionic theory
\cite{duchConstructionGrossNeveuModel2024} by solving the continuous renormalisation
group exactly.

Finally, we want to point out a general (tentative) axiomatic
framework~{\cite{bailleulWilsonIt^oDiffusions2023}} proposed by Bailleul,
Chevyrev and the first author. This framework provides a generalisation of
the coupling with the free field given by~\eqref{eq:fbsde-intro} to the
construction of random fields endowed with a Wilsonian scale-decomposition and a
stochastic dynamics associated to a Gaussian field. These so called
Wilson--Itô fields generate interesting questions ranging from the
characterisation of measures of the form~\eqref{eq:int-Gibbs-ms} via FBSDEs,
to locality properties, the structure of the pre-factorisation algebras
generated by the observables, or generalisations of the domain Markov
properties, some of which we hope to address in a future study.

\paragraph*{Acknowledgements}

M. G. was partly supported by UK Research and Innovation via the grant “StochFields” EP/Z534328/1. S. M. has been supported by the EPSRC Centre for Doctoral Training in
Mathematics of Random Systems: Analysis, Modelling and Simulation
(EP/S023925/1).
The authors thank Alexander Schill for pointing out a mistake in the proof of Lemma \ref{lem:verification} in an earlier version.
For the purpose of open access, the authors have applied a \href{https://creativecommons.org/licenses/by/4.0/}{CC BY} public copyright licence to
any author accepted manuscript arising from this submission.
This paper has been written with {\TeXmacs}
(\href{www.texmacs.org}{www.texmacs.org}).

\subsection{Notation and assumptions}\label{sec:notation}

Let us fix some general notation we will use throughout.
\begin{itemize}
	\item Let $\langle x \rangle \assign (1 + | x |^2)^{1 / 2}$, $x \in
		      \mathbb{R}^2$. We will often rely on the following inequality to commute
	      polynomial weights,
	      \begin{equation}
		      \langle x \rangle^k \langle y \rangle^{- k} \lesssim \langle x - y
		      \rangle^k, \quad k \in \mathbb{N}. \label{eq:jpb-triangle-ineq}
	      \end{equation}
	\item  For $\gamma \in (- 1, 1)$, we define the exponential weights
	      \[ w_{\gamma} (x) \assign \mathe^{\gamma m | x |} . \]
	\item For a non-negative function $w : \mathbb{R}^2 \rightarrow \mathbb{R}_+$, we define
	      the standard weighted Lebesgue, Sobolev and Besov spaces $L^p (w)$, $W^{s,
				      p} (w), H^s (w) = W^{s, 2} (w)$ and $B^s_{p, q} (w)$, $p, q \in [1, \infty],
		      s \in \mathbb{R}$ based on the measures $w (x) \mathd x$ on $\mathbb{R}^2$,
	      e.g. $L^p (w)$ is equipped with the norm
	      \[ \| f \|_{L^p(w)}^p = \| w \cdot f \|_{L^p}^p = \int_{\mathbb{R}^2} | fw |^p
		      = \int_{\mathbb{R}^2} | w (x) f (x) |^p \mathd x. \]
	      In the case of $w (x) = \langle x \rangle^k$ for some $k \in \mathbb{R}$, we
	      also write $L^{p, k} \assign L^p (\langle x \rangle^k)$ and analogously for
	      the Besov and Sobolev spaces. Throughout this paper, we will fix
	      a polynomial weight with $n$ sufficiently large so that $x \mapsto \langle x
		      \rangle^{- n} \in L^1 (\mathbb{R}^2)$.

	\item We denote by $\Delta_i = \varphi_i (\mathD)$ the Littlewood-Paley
	      blocks on $\mathbb{R}^d$ and by $K_i = \mathcal{F}^{- 1} (\varphi_i)$ their
	      associated $L^p$-kernels. We recall that then, for any $i \geqslant - 1$ and
	      $p \in [1, \infty)$,
	      \begin{equation}
		      \| K_i \|_{L^1} \lesssim 1, \quad \| K_i \|_{L^p} \lesssim 2^{2 i \frac{p
					      - 1}{p}} . \label{eq:LP-kernels}
	      \end{equation}
	      For any $\alpha \in \mathbb{R}$, $p, q \in [1, \infty]$ and $n$ we define
	      the usual Besov norms (with the usual modification for $q=\infty$)
	      \[ \| u \|^q_{B_{p, q}^{\alpha} (\langle x \rangle^{- n})} \assign \sum_{i
			      \geqslant - 1} 2^{\alpha iq} \| \Delta_i u \|^q_{L^{p, - n}}, \]
	      with the corresponding Besov spaces \[B^{\alpha}_{p, q} (\langle x \rangle^{-
			      n}) \assign \left\{ u \in \mathcal{S}' (\mathbb{R}^d) : \: \| u
		      \|_{B^{\alpha}_{p, q} (\langle x \rangle^{- n})} < \infty \right\}.\] For a
	      more detailed exposition, we refer to {\cite[Chapter
				      1]{bahouriFourierAnalysisNonlinear2011}}.

	\item For a collection of points $x_I = (x_i)_{i \in I}$ we denote its
	      \tmtextit{Steiner diameter}, that is the shortest tree connecting all points
	      in $x_I$, by $\tmop{St} (x_I)$. More precisely, we define
	      \begin{equation}
		      \tmop{St} (x_I) \assign \min_{x_J \supset x_I} \min_{\tau (x_J)} L (\tau),
		      \label{eq:def-St}
	      \end{equation}
	      where the second minimum runs over all trees $\tau (x_J)$ connecting the
	      points $x_J$ and $L (\tau)$ measures the length of the tree $\tau$ on
	      $\mathbb{R}^d$. We refer to {\cite{giulianiGentleIntroductionRigorous2021}}
	      for further details.

	\item We denote by $B = (B_t)_{t \geqslant 0}$ a cylindrical Brownian motion
	      on $L^2 (\mathbb{R}^2)$ and by $\mathbb{F}= (\mathcal{F}_t)_{t \geqslant 0}$
	      the augmentation of the filtration generated by $B$. All considerations are
	      with respect to this filtration and we will not explicitly mention it
	      elsewhere (i.e. adapted always means adapted to the filtration
	      $\mathbb{F}$). The conditional expectation with respect to $\mathcal{F}_t$
	      is denoted by $\mathbb{E}_t$. Given a measure $\nu$, we
	      write $\nu (f) \assign \int f \mathd \nu$ for the expectation under this
	      measure and if $\nu$ is a probability measure, we write
	      \[ \tmop{Cov}_{\nu} (f, g) \assign \nu (fg) - \nu (f) \nu (g) . \]
	\item For a Banach space $\mathcal{X}$, let $\mathbb{H}_a (\mathcal{X})$ be
	      the space of predictable processes taking values in $\mathcal{X}$ (no
	      integrability restrictions assumed). We also define the spaces, for any $p
		      \in [1, \infty]$
	      \begin{align*}
		      \mathbb{H}^p_T (\mathcal{X})                      & \assign \left\{ u \in \mathbb{H}_a
		      (\mathcal{X}) | \mathbb{E} \int_0^T \| u_s \|_{\mathcal{X}}^p \mathd s
		      < \infty \right\},                                                                     \\
		      \mathbb{H}^{\infty}_T (L^{\infty} (\mathbb{R}^2)) & \assign \{ u \in
		      \mathbb{H}_a (L^{\infty} (\mathbb{R}^2)) | \mathbbm{1}_{\{ t \leqslant
			      T \}} u_t \in L^{\infty} (\mathd t \otimes \mathd \mathbb{P} \otimes \mathd
		      x)_{\nosymbol} \}.
	      \end{align*}
	      If $T = \infty$, we may omit the subscript $T$ in the spaces above.

	\item We write $\rho \prec 1$ if $\rho$ is a smooth and compactly supported
	      function $\mathbb{R}^2 \rightarrow [0, 1]$ and analogously, we write $\rho
		      \preceq 1$ if $\rho \prec 1$ or $\rho \equiv 1$. For a family of spatial
	      cut-offs $(\rho_k)_k$ will write $\rho_k \rightarrow 1$ if $\rho\preceq 1$ and
	      $\lim_{k\to\infty}(1-\rho_k(x))=0$.

	\item We reserve $\delta \assign 1 - \beta^2 / 8 \pi > 0$ to denote the
	      distance to criticality of the sine-Gordon model in our normalisation. The
	      relevant thresholds for us, $\beta^2 < 4 \pi$, \ $\beta^2 < 6 \pi$ and
	      $\beta^2 < 8 \pi$, \ correspond to $\delta > \frac{1}{2}$, $\delta >
		      \frac{1}{4}$ and $\delta > 0$ respectively.
\end{itemize}
To study the Laplace transform of $\nu_{\tmop{SG}}$, we will have to consider
localised perturbations $g + V$ of the potential $V$ for functionals $g :
	\mathcal{S}' (\mathbb{R}^2) \rightarrow \mathbb{R}$. This localisation will be
quantified in terms of the semi-norms
\[ \begin{aligned}
		| g |_{1, p, k} \assign \sup_{\varphi \in L^{p, k}} \| \nabla g (\varphi)
		\|_{L^{p, k}}, \\
		| g |_{2, p, k} \assign \sup_{\phi_1, \phi_2 \in L^{p, k}}  \frac{\|
			\nabla g (\phi_1) - \nabla g (\phi_2) \|_{L^{p, k}}}{\| \phi_1 - \phi_2
			\|_{L^{p, k}}},
	\end{aligned}\]
where we drop the parameter $k$ if $k = 0$. We always assume that $g, \nabla g$ is
uniformly bounded, that is
\[ \sup_{\varphi \in \mathcal{S}' (\mathbb{R}^2)} \| \nabla g (\varphi)
	\|_{L^{\infty}} + |g(\varphi)| \leqslant L < \infty, \]
and that $g \in C^2_b (L^{2, - n}) \cap C^2_b (H^{- \varepsilon, - n})$, the
space of functions $L^{2, - n} \rightarrow \mathbb{R}$ with two continuous and
bounded derivatives with a continuous extension in $C^2_b (H^{- \varepsilon, -
			n})$.
The class of functions $g$ satisfying the assumptions above is large enough to
be rate function determining (for a proof see e.g. {\cite[Lemma
			9]{barashkovStochasticControlApproach2022}}).

\section{ Stochastic control set-up for Gibbs measures}\label{sec:CP}

In this section, we set up the general variational framework required to study
Gibbsian perturbations of the form~\eqref{eq:int-approx-measures} of a
Gaussian measure $\mu^T$ from a stochastic control perspective. More
precisely, for a functional $g : \mathcal{S}' (\mathbb{R}^2) \rightarrow
	\mathbb{R}$ satisfying the assumptions laid out in Section~\ref{sec:notation}
and suitable functions $U \in C^{\infty}_b (\mathbb{R})$ and a spatial cut-off
$\rho \prec 1$, we consider a generic perturbed potential
\begin{equation}
	V^g (\varphi) \assign (g + V) (\varphi) \assign \lambda g (\varphi) +
	\lambda \int_{\mathbb{R}^2} \rho (x) U (\varphi (x)) \mathd x,
	\label{eq:def-Vf}
\end{equation}
and study the generic Gibbs measures,
\begin{equation}
	\nu (\mathd \varphi) = \nu^V (\mathd \varphi) = \Xi_V^{- 1} \exp (- V
	(\varphi)) \mu^T (\mathd \varphi) . \label{eq:nu-general-pot}
\end{equation}
We agree to drop the superscript $g$ whenever $g = 0$. Note that the measures
$\nu_{\tmop{SG}}^{\rho, T}$ as defined in \eqref{eq:int-approx-measures} are
precisely of this form whenever $\rho\prec1, T<\infty$.

Before we can begin the analysis of the control problem, we have to construct
a suitable probability space. This requires a Brownian martingale $W$ with the
Gaussian free field as its terminal value.

\subsection{Scale decomposition}

Mainly for technical convenience and concreteness, we use a heat kernel
decomposition to interpolate the covariance of the free field as
\[ (m^2 - \Delta)^{- 1} = \int_0^{\infty} Q_t^2 \mathd t \quad \tmop{with}
	\quad Q_t \assign \left( \frac{1}{t^2} \mathe^{- (m^2 - \Delta) / t}
	\right)^{1 / 2} . \]
For a cylindrical Brownian motion $B$ on $L^2 (\mathbb{R}^2)$, we then define
the Brownian martingale $(W_t)_{t \geqslant 0}$ as the corresponding scale
interpolation of the Gaussian free field, that is
\[ W_t \assign \int_0^t Q_s \mathd B_s . \]
By construction, the measure $\mu^t \assign \tmop{Law} (W_t)$ has covariance,
\begin{equation}
	G_t (x, y) \assign G_t (x - y) \assign \int_0^t Q^2_s (x - y) \mathd s,
	\label{eq:def-G}
\end{equation}
where we abuse the notation to use the same symbol for the operator and its
associated kernel on $L^2 (\mathbb{R}^2)$. A standard computation shows that
the kernels are explicitly given by,
\begin{equation}
	\dot{G}_s^{1 / 2} (x) = Q_s (x) = \frac{1}{2 \pi} \mathe^{- m^2 / 2 s}
	\mathe^{- \frac{s}{2} | x |^2}, \quad \dot{G}_s (x) = \frac{1}{4 \pi s} \mathe^{-
		m^2 / s} \mathe^{- \frac{s}{4} | x |^2} \label{eq:G-kernel}, \qquad x \in
	\mathbb{R}^2 .
\end{equation}
In addition to the smoothing property of the heat kernel, we rely on the following properties of the scale interpolation
\begin{itemize}
	\item the kernel of $\dot{G}$ has a positive convolutional
	      square root $Q$, that is $\dot{G}_t = Q_t \ast Q_t$.
	\item the kernels decay exponentially fast on $\mathbb R^2$, which is required to show the decay of correlations in Section
	      \ref{sec:correlations},
	\item the kernels $\dot{G}_s$ are invariant under Euclidean transformations, which is used in our proof of the Euclidean invariance in Section \ref{sec:OS-Axioms}.
\end{itemize}
Apart from the properties above, the precise choice of the scale
interpolation is not important for us and we will only require elementary
bounds on the kernels, all of which we collect in Appendix~\ref{app:hk}.

A simple computation shows that the martingale $W$ serves as a smooth
approximation to the free field. Before we proceed, let us note this fact for
future reference. We postpone the proof to Appendix \ref{app:GFF-convergence}.
\begin{lemma}
	\label{lem:GFF-convergence}For any $\varepsilon > 0, p \in [1, \infty)$ and
	$n > 2$, the sequence $(W_t)_{t \geqslant 0}$ converges in $L^p (\mathd \mathbb{P} ;
		B_{p, p}^{- \varepsilon, - n})$ and almost surely to a random variable
	$W_{\infty} \sim \mu$, where $\mu$ is the Gaussian free field, that is the
	centred Gaussian measure on $\mathcal{S}' (\mathbb{R}^2)$ with covariance
	$(m^2 - \Delta)^{- 1}$. Moreover, for any $T < \infty$, the stopped process
	$(W_{t \wedge T})_{t \geqslant 0}$ is a Gaussian process taking values in
	the function space $L^{\infty, - n}$.
\end{lemma}

\subsection{The control problem}

With the scale interpolation $(W_t)_t$ of the free field, and thus the
probability space, constructed, we can return to the
measures~\eqref{eq:nu-general-pot}. The goal of this section is to establish
the connection between Gibbsian perturbations of a Gaussian and the stochastic
control problem which is the basis for the FBSDE formulation.
\begin{theorem}
	\label{thm:ctrl-FBSDE}{\tmdummy}
	\begin{enumeratealpha}
		\item  \label{prop:optimal-SDE}For any $T \in [0, \infty)$ and $\varphi
			\in \mathcal{S}' (\mathbb{R}^2)$, the FBSDE
		\begin{equation}
			Z^g_t (\varphi) = \varphi - \int_0^t \dot{G}_s \mathbb{E}_s [\nabla V^g
			(Z^g_T (\varphi) + W_T)] \mathd s, \label{eq:optimal-SDE}
		\end{equation}
		has a unique solution in $\mathbb{H}^{\infty}_T (L^{\infty})$.

		\item \label{thm:ctrl-form}The process $X^g_t = Z_t^g + W_t$, where
		$Z_t^g$ is the solution to~\eqref{eq:optimal-SDE}, satisfies $\tmop{Law}
			(X_T^0) = \nu$ and the pair
		\begin{equation}
			(\bar{u}^g_t, X_t^g) (\varphi) \assign (- Q_t \mathbb{E}_t [\nabla V^g
			(X_T^g (\varphi))], X_t^g (\varphi)), \label{eq:optimal-pair}
		\end{equation}
		is the unique optimiser for the stochastic control problem,
		\begin{equation}
			X_t (u ; \varphi) = Z_t (u ; \varphi) + W_t, \quad \text{where} \quad
			Z_t (u ; \varphi) \assign \varphi + \int_0^t Q_s u_s \mathd s
			\label{eq:ctrl-process}
		\end{equation}
		subject to the cost functional
		\begin{equation}
			\mathcal{V}^{V + g} (\varphi) \assign \inf_{u \in \mathbb{H}_a} J^{V^g}
			(u ; \varphi) \assign \inf_{u \in \mathbb{H}_a} \mathbb{E} \left[ V^g
				(Z_T (u ; \varphi) + W_T) + \frac{1}{2} \int_0^T \| u_s \|^2_{L^2}
				\mathd s \right] . \label{eq:def-value-func}
		\end{equation}
		In particular, the Laplace transform of $\nu$ satisfies the variational
		problem,
		\begin{equation}
			\mathcal{W}^V (g ; \varphi) \assign - \log \nu_{\varphi} (\mathe^{- g})
			= \inf_{u \in \mathbb{H}_a} J^{V + g} (u ; \varphi) - \inf_{u \in
				\mathbb{H}_a} J^V (u ; \varphi) . \label{eq:def-Wf}
		\end{equation}
	\end{enumeratealpha}
\end{theorem}

Let us agree to drop the dependence on the initial value $\varphi$ as long as
no ambiguities arise. This dependence on the initial value $\varphi$ will only
become relevant in Section \ref{sec:non-Gaussian} and  can safely be ignored for
the rest of the paper. \ We will arrive at Theorem \ref{thm:ctrl-FBSDE} in
several steps. We start with the variational description for exponential
functionals of Brownian motion by Bou{\'e} and Dupuis
(Lemma~\ref{lem:BD-form}). We then show that any optimally controlled process
has the correct law (Lemma~\ref{lem:dda-W}. \
Finally, we obtain necessary conditions on the optimal control
(Lemma~\ref{lem:EL-cond}) and use a verification theorem to show existence and
uniqueness of an optimal control (Lemma~\ref{lem:verification}) which will
imply that the optimal dynamics is indeed given by~\eqref{eq:optimal-SDE}.

We say a real valued random variable $Y$ is \tmtextit{tame} (with respect to
the probability measure $\mathbb{P}$) if there are H{\"o}lder conjugates $p, q
	> 1$ (that is $1 / p + 1 / q = 1$) such that
\[ \mathbb{E} [\exp (- qY)] +\mathbb{E} | Y |^p < \infty . \]
The assumptions on $g$ and the
boundedness of $V$ defined in \eqref{eq:def-Vf} imply that this condition is
always satisfied for $Y = V^g (W_t)$ and $t \in [0, \infty)$. Recall the the
variational formula from~{\cite{boueVariationalRepresentationCertain1998}} in
the more general version of {\cite{ustunelVariationalCalculationLaplace2014}}.

\begin{theorem}[Bou{\'e}--Dupuis]
	\label{lem:BD-form}Let $B$ be a cylindrical Brownian motion on a Hilbert
	space $H$ and let $W = \int_0^{\cdot} Q_t \mathd B_t$ be a Brownian motion
	on $\tilde{H}$ with covariance $G_t = \int_0^t Q_s^2 \mathd s : H
		\rightarrow \tilde{H}$ and define for $u \in \mathbb{H}_a$,
	\begin{equation}
		X_t (u) = Z_t (u) + W_t, \label{eq:BD-Xu} \quad \text{where} \quad Z_t
		(u) = \int_0^t \mathd s Q_s u_s .
	\end{equation}
	For any Borel-measurable functional $F : \tilde{H} \rightarrow \mathbb{R}$
	such that $F (W)$ is tame, it holds that
	\begin{equation}
		- \log \mathbb{E} [\mathe^{- F (W)}] = \inf_{u \in \mathbb{H}_a}
		\mathbb{E} \left[ F (X (u)) + \frac{1}{2} \int_0^{\infty} \| u_s
			\|^2_{L^2} \mathd s \right] \backassign \inf_{u \in \mathbb{H}_a} J^F (u)
		. \label{eq:BD-OC}
	\end{equation}
\end{theorem}

Our interest in this formula is justified by the following observation. If $g
	: \mathcal{S}' (\mathbb{R}^2) \rightarrow \mathbb{R}$ satisfies the
assumptions laid out in Section \ref{sec:notation}, then $V^g$ is tame and the
formula~\eqref{eq:BD-OC} provides a variational representation for the Laplace
transform of~\eqref{eq:nu-general-pot} via
\begin{equation}
	\mathcal{W}^V (g) = - \log \nu (\mathe^{- g}) = - \log \left(
	\frac{\mathbb{E} [\mathe^{- (g + V) (W_T)}]}{\mathbb{E} [\mathe^{- V
			(W_T)}]} \right) = \inf_{u \in \mathbb{H}_a} J^{V + g} (u) - \inf_{u \in
		\mathbb{H}_a} J^V (u) . \label{eq:def-WV}
\end{equation}
If the infimum is a minimum, it turns out that the control problem actually
provides a more direct description of the measure $\nu$ via the dynamics $X_t
	(u)$ given by~\eqref{eq:BD-Xu}. We recall Lemma~11
from~{\cite{barashkovVariationalMethodEuclidean2023}}, which is the key to
establish this relationship.

\begin{lemma}
	\label{lem:dda-W}Let $g : \mathcal{S}' (\mathbb{R}^2) \rightarrow
		\mathbb{R}$ be bounded and continuous. If for some $\alpha \in \mathbb{R}$
	the variational problem $\inf_{u \in \mathbb{H}_a} J_T^{\alpha g} (u)$ has a
	minimiser $\bar{u}^{\alpha g}$, then $\alpha \mapsto \mathcal{W}^V (\alpha
		g)$ satisfies
	\[ \frac{\mathd}{\mathd \alpha}  \mathcal{W}^V (\alpha g) =\mathbb{E} [g
			(X_T (\bar{u}^{\alpha g}))] . \]
	In particular, $\tmop{Law} (X_T (\bar{u})) = \nu$.
\end{lemma}

Next, we show a necessary condition for the optimal control, which will also
provide a candidate for the minimiser of~\eqref{eq:BD-OC} as feedback control.

\begin{lemma}
	\label{lem:EL-cond}If $\bar{u}^g \in \mathbb{H}_a$ is optimal for the
	control problem \eqref{eq:BD-OC}, then $\mathd t \otimes \mathd \mathbb{P}$-almost
	surely,
	\begin{equation}
		\bar{u}_t^g = - Q_t \mathbb{E}_t [\nabla V^g (X_T (\bar{u}^g))] .
		\label{eq:optimal-u}
	\end{equation}
\end{lemma}

\begin{proof}
	Standard stability results for SDEs imply that the solution $X (u)$ to
	\eqref{eq:ctrl-process} is differentiable in $u$. Similarly, the regularity
	assumed on $V$ and $g$ imply that also $J^{V + g} (u)$ is differentiable
	along all directions $\delta u \in \mathbb{H}^2_T (L^2)$. We compute
	\begin{align*}
		\nabla_{\varepsilon} X_t^{u, \delta u} & \assign \left.
		\frac{\mathd}{\mathd \varepsilon} \right|_{\varepsilon = 0} X_t (u +
		\varepsilon \delta u) = \int_0^t Q_s \delta u_s \mathd s, \\
		\nabla_{\varepsilon} J^{u, \delta u}   & \assign \left.
		\frac{\mathd}{\mathd \varepsilon} \right|_{\varepsilon = 0} J^{V + g}
		(u + \varepsilon \delta u) =\mathbb{E} \left[ \nabla V^g (X_T (u))
			\nabla_{\varepsilon} X^{u, \delta u}_T + \int_0^T u_s \delta u_s \mathd
			s \right].
	\end{align*}
	Since the control $u$ has to be adapted, we may insert a conditional
	expectation to find
	\begin{equation}
		\nabla_{\varepsilon} J^{u, \delta u}
		= \mathbb{E} \int_0^T (Q_s \mathbb{E}_s [\nabla V^g (X_T (u))] + u_s)
		\delta u_s \mathd s. \label{eq:EL-der}
	\end{equation}
	For an optimal control $u = \bar{u}^g$, it must hold for any direction
	$\delta u \in \mathbb{H}^2 (L^2)$ and $\varepsilon > 0$,
	\[ J^g (\bar{u}^g + \varepsilon \delta u_s) - J^g (\bar{u}^g) \geqslant 0.
	\]
	Moreover, since $\nabla Y_t^g (u) \assign \mathbb{E}_t [\nabla V^g (X_T
		(u))]$ does not depend on the direction $\delta u$, we arrive at the claimed
	first order condition for optimality
	\[ \bar{u}^g_t + Q_t \nabla Y^g_t (\bar{u}^g) = 0 \Leftrightarrow
		\bar{u}^g_t = - Q_t \nabla Y_t^g (\bar{u}^g) . \]
\end{proof}

Up until this point, we cannot guarantee existence of a minimiser. For the
potentials $V$ as defined in~\eqref{eq:def-Vf}, we can close this gap with a
verification theorem for feedback controls. Given a feedback control $u_t =
	\hat{u}_t (X_t (u))$, we say that the pair $(u, X (u))$ is
\tmtextit{admissible} if $X (u)$ is a strong solution to the
SDE~\eqref{eq:BD-Xu} controlled by $u$, that is $X (u)$ is a strong solution
to the SDE
\[ X_t = \int_0^t Q_s  \hat{u}_s (X_s) \mathd s + W_t, \quad t \in [0, T] . \]
\begin{lemma}
	\label{lem:verification}The feedback control~\eqref{eq:optimal-u} is optimal
	for the control problem~\eqref{eq:def-value-func}. Moreover, the Hamilton--Jacobi--Bellman equation
	\begin{equation}
		\begin{cases}
			\partial_t v_t + \frac{1}{2} \tmop{Tr} (\dot{G}_t \mathD^2 v_t) =
			\frac{1}{2} \mathD v_t \dot{G_t} \mathD v_t, & t \in [0, T], \\
			v_T = V + g,
		\end{cases}  \label{eq:HJB}
	\end{equation}
	has a unique bounded solution $v_t$ and $\bar{u}^g$ defined
	in~\eqref{eq:optimal-u} satisfies
	\begin{equation}
		\bar{u}^g_t = - Q_t \nabla v^g_t (X_t (\bar{u}^g)) . \label{eq:u-via-HJB}
	\end{equation}
\end{lemma}

\begin{proof}
	Let us first show the statement for a bounded function $g$ and write $v =
		v^g$, $V = V^g$. The Hamilton--Jacobi--Bellman equation (HJB-equation; for
	short) associated to the control problem~\eqref{eq:def-value-func} is given
	by
	\begin{equation}
		\begin{cases}
			\partial_t v_t + \inf_{a \in L^2} \left\{ \frac{1}{2} \tmop{Tr}
			(\dot{G}_t \mathD^2 v_t) + \langle \mathD v_t, Q_t a \rangle_{L^2} +
			\frac{1}{2} \| a \|_{L^2}^2 \right\} = 0, \label{eq:gen-HJB} \\
			v_T = V.
		\end{cases}
	\end{equation}
	see e.g.~{\cite[Section~2.5.1]{fabbriStochasticOptimalControl2017}}. Solving
	the quadratic optimisation problem in~\eqref{eq:gen-HJB} we find that the
	optimum is attained at $a = - Q_t \mathD v_t$ so that the
	PDE~\eqref{eq:gen-HJB} reduces to~\eqref{eq:HJB}. Define the function
	\begin{equation}
		v_t (\varphi) \assign - \log \mathbb{E} [\exp (- V (\varphi + W_T - W_t))]
		. \label{eq:def-v-HJB}
	\end{equation}
	Since $V$ is bounded and smooth by assumption, the representation
	in~\eqref{eq:def-v-HJB} implies that also $v$ is smooth and bounded, say $v
		\in C^1_b ([0, T], C^2_b (\mathbb{R}^2))$. We readily verify by a direct
	computation that $v$ is a solution to~\eqref{eq:HJB}.

	Having found a solution to the HJB equation \eqref{eq:gen-HJB}, we have
	access to the verification theorem (see
	e.g.~{\cite[Theorem~2.36]{fabbriStochasticOptimalControl2017}}): if the
	feedback control $\bar{u}$ as defined in~\eqref{eq:u-via-HJB} is admissible
	and satisfies for almost every $s \in [0, T]$ $\mathbb{P}$-almost surely,
	\begin{equation}
		\bar{u}_s \in \tmop{argmin}_{a \in L^2} \left\{ \frac{1}{2} \tmop{Tr}
		(\dot{G}_s \mathD^2 v_s (X_s (\bar{u}))) + \langle \mathD v_s (X_s
		(\bar{u})), Q_s a \rangle + \frac{1}{2} \| a \|^2_{L^2} \right\}, \label{eq:verification}
	\end{equation}
	it follows that $\bar{u}$ is optimal for the control problem. By the same
	reasoning as before for the HJB-equation, the unique $L^2$-optimiser of
	\eqref{eq:verification} is given by $\bar{u}_s = - Q_s \nabla v_s (X_s
		(\bar{u}))$. Since $V$ is bounded, we see from~\eqref{eq:def-v-HJB} that the
	solution $\exp (- v_t (\varphi))$ is bounded away from $0$ and the gradient
	is given by
	\begin{equation}
		\nabla v_t (\varphi) = \frac{-\mathbb{E} [\nabla V (\varphi + W_T - W_t)
				\exp (- V (\varphi + W_T - W_t))]}{\exp (- v_t (\varphi))} .
		\label{eq:gradv-HJB}
	\end{equation}
	Hence, the gradient $\nabla v_t$ inherits the Lipschitz continuity from $V$
	and $\nabla V$. As a result, the standard fixed point argument for SDEs with
	bounded Lipschitz coefficients shows that the pair $(\bar{u}, X (\bar{u}))$
	is admissible for the control problem. Finally, expanding the function $f_s
		\assign \nabla v_s$ along the flow of the optimally controlled process $X =
		X (\bar{u})$ using Ito's formula and the fact that $v_s$
	solves~\eqref{eq:HJB}, yields
	\begin{equation}
		\begin{aligned}
			f_t (X_t)
			 & =\mathbb{E}_t \left[ \nabla V (X_T) - \int_t^T \left( \partial_s
				f_s + \frac{1}{2} \tmop{Tr} \dot{G}_s \mathD^2 f_s - \frac{1}{2} \mathD
			(f_s  \dot{G}_s f_s) \right) (X_s) \mathd s \right]                 \\
			 & =\mathbb{E}_t [\nabla
			V (X_T)] .
		\end{aligned} \label{eq:flow-exact-sol}
	\end{equation}
	which is the missing equality
	\begin{equation}
		\bar{u}_t = - Q_t \nabla v_t (X_t (\bar{u})) = - Q_t \mathbb{E}_t [\nabla
		V (X_T (\bar{u}))] . \label{eq:v-V}
	\end{equation}
\end{proof}

\begin{proof}[Proof of Theorem \ref{thm:ctrl-FBSDE}.]
	To see that~\eqref{eq:optimal-SDE} has a unique solution, note that by
	\eqref{eq:v-V}, the SDE \eqref{eq:optimal-SDE} is equivalent to
	\eqref{eq:BD-Xu} with the feedback control $\bar{u}_t = - Q_t \nabla v_t
		(X_t (\bar{u}))$. By Lemma \ref{lem:verification} this control is
	admissible, i.e. there is a unique strong solution. By~\eqref{eq:def-WV},
	the variational problem for the Laplace transform is a direct consequence of
	Lemma~\ref{lem:BD-form}. Lemma~\ref{lem:EL-cond} and~\ref{lem:verification}
	imply combined that the pair defined in~\eqref{eq:optimal-pair} is optimal
	for the control problem. Moreover, the condition~\eqref{eq:optimal-u} is
	necessary and since the solution to the SDE~\eqref{eq:optimal-SDE} is
	unique, the pair~$(\bar{u}^g, X (\bar{u}^g))$ defined
	by~\eqref{eq:optimal-pair} is the unique optimiser
	for~\eqref{eq:def-value-func}. Finally, Lemma~\ref{lem:dda-W} shows that the solution $X$ to~\eqref{eq:optimal-SDE}
	for $g = 0$ has the desired law,
	\[ \tmop{Law} (X_T) = \nu . \]
\end{proof}

\begin{remark}
	\label{rem:ctrl-prob}{\tmdummy}
	\begin{enumeratealpha}
		\item Compared to the more general setting considered
		in~{\cite{barashkovVariationalMethodEuclidean2023}}, the fact that the
		potential is Lipschitz and bounded allows us to directly use the solution
		to the HJB-equation~\eqref{eq:HJB} and enables the verification theorem.
		This means that we do not need to relax the variational problem to ensure
		existence of a minimiser. The difference is only a technical one and not
		crucial to our analysis: the subsequent analysis could be carried out
		verbatim for a relaxed version of the control problem, by possibly
		enlarging the underlying filtration.

		\item \label{rem:HJB-vs-condexp}We should emphasise the difference between
		the two formulas
		\begin{equation}
			\bar{u}^g_t = - Q_t \nabla v^g_t (X_t (\bar{u}^g)),
			\label{eq:u-via-HJB-2}
		\end{equation}
		via the solution $v^g$ to~\eqref{eq:HJB} and
		\begin{equation}
			\bar{u}^g_t = - Q_t \mathbb{E}_t [\nabla V^g (X_T (\bar{u}^g))],
			\label{eq:u-via-condexp}
		\end{equation}
		via the stochastic maximum principle. The PDE~\eqref{eq:HJB} is not only
		non-linear but also infinite dimensional. The only reason we were able to
		easily show well-posedness here are the explicit
		formulas~\eqref{eq:def-v-HJB} and~\eqref{eq:gradv-HJB} for $v$ and its
		gradient. Both rely on the boundedness and Lipschitz continuity of $V^g$
		and its gradient. In our main application of interest, where $V = V^{\rho,
					T}$, both of these properties disappear as the regularisations $\rho$ and
		$T$ are removed. As a result, this strategy does not readily transfer to
		the unregularised setting.

		In contrast, the formula~\eqref{eq:u-via-condexp} yields the entirely
		self-contained forward-backward dynamics~\eqref{eq:optimal-SDE}. This
		FBSDE is an appealing candidate for a stochastic quantisation equation for
		the measures $\nu^{\rho, T}$ that we can also transfer to the limit $\rho
			\rightarrow 1, T \rightarrow \infty$. Controlling~\eqref{eq:optimal-SDE}
		uniformly in both regularisations is the objective of the next section.
	\end{enumeratealpha}
\end{remark}

\subsection{The effective FBSDE}

Motivated by the issues highlighted in Remark
\ref{rem:ctrl-prob}-\ref{rem:HJB-vs-condexp}, we move to a reformulation of
the~FBSDE~\eqref{eq:optimal-SDE}, which is stable in the $\rho \rightarrow 1$,
$T \rightarrow \infty$ limit and which can be studied without relying on a
direct analysis of the PDE~\eqref{eq:HJB}. This means we do not have access to
the exact solution of \eqref{eq:HJB}. In place of the exact solution, we look
for a scale dependent function $(F_t)_t$ such that the error, or remainder,
$R$ defined by
\begin{equation}
	R_t \assign \mathbb{E}_t [\mathD V (X_T)] - F_t (X_t), \label{eq:Ansatz-R}
\end{equation}
is small in a suitable sense. For $V = V^{\rho, T}$, we would like the
bounds to also be uniform in $\rho \prec 1$ and $T < \infty$.
While we should keep this goal in mind, the ideas are more general and we
therefore first develop them for a generic function $\mathD V$. Similarly to the
computation in~\eqref{eq:flow-exact-sol}, we develop the function $F$ along
the flow of the SDE \eqref{eq:optimal-SDE} and obtain a BSDE for the remainder
$R$,
\begin{align*}
	R_t
	 & =\mathbb{E}_t [F_T (X_T) - F_t (X_t)] \\
	 & =\mathbb{E}_t  \int_t^T H_s
	(X_s) \mathd s -\mathbb{E}_t  \int_t^T \mathD F_s (X_s)  \dot{G}_s R_s
	\mathd s +\mathbb{E}_t \int_t^T \mathD F_s (X_s) \mathd W_s,
\end{align*}
where
\begin{equation}
	H_t (\varphi) = \left( \partial_t F_t + \frac{1}{2} \tmop{Tr} (\dot{G}_t
		\mathD^2 F_t) - \frac{1}{2} \mathD (F_t  \dot{G}_t F_t) \right) (\varphi) .
	\label{eq:def-H}
\end{equation}
Since the stochastic integral is a martingale, it vanishes under the
conditional expectation. Allowing again a small perturbation $g$ in the
potential, the optimal dynamics in~\eqref{eq:optimal-SDE} can equivalently be
described by the FBSDE
\begin{equation}
	\begin{cases}
		X_t = \varphi + W_t - \int_0^t \dot{G}_s (F_s (X_s) + R_s) \mathd s, \\
		R_t =\mathbb{E}_t \left[ \nabla g (X_T) + \int_t^T H_s (X_s) \mathd s -
			\int_t^T \mathD F_s (X_s)  \dot{G}_s R_s \mathd s \right] .
	\end{cases} \label{eq:FBSDE-generic}
\end{equation}
If there is no additional perturbation added to the potential $V$, choosing $F = \nabla v$ for  $v$ the exact solution to \eqref{eq:HJB}, we
recover $R = 0$. Even in this unperturbed case, the remainder proves to be useful as it allows us the freedom to
choose the function $F$, and let the remainder $R$ compute the error resulting
from this approximation. We, therefore, set out to find a systematic way to
construct functions $F$ for which the error term $H$ is small in the next
section.

\begin{remark}
	Observe that we really treat the function $g$ in \eqref{eq:FBSDE-generic} as
	a perturbation: we only develop the unperturbed gradient $\nabla V$ along
	the flow. The error due to $g$ is collected entirely in the terminal
	condition for the remainder $R$. This means that we only have to analyse the
	flow equation for the unperturbed periodic potential $\nabla V$.
	This distinction is of technical importance to us -- in the subsequent analysis, we will use the periodicity of $\nabla V$ and $V$ to our advantage
	and including a generic perturbation $g$ would break this symmetry.
\end{remark}

\section{Analysis of the flow equation}\label{sec:flow-eq}

In this section, we inductively derive the bounds on the coefficients of the
FBSDE~\eqref{eq:FBSDE-generic} \ using a truncated version of the
renormalisation flow equation
\begin{equation}
	\partial_t F_t + \frac{1}{2} \tmop{Tr} (\dot{G}_t \mathD^2 F_t) -
	\frac{1}{2} \mathD (F_t  \dot{G}_t F_t) = 0 \label{eq:F-flow}, \quad
	\text{subject to\quad} F_T = \mathD V^T .
\end{equation}
\subsection{Truncating the
	flow}\label{rem:relevant-terms-b-leq-4}\label{sec:truncated-flow}

Heuristically, we expect that successive Picard iterations of the flow
equation~\eqref{eq:F-flow} improve the the approximation. Accordingly, we
define an iterative scheme starting from $F^{[0]} \assign 0$ and define
$F^{[\ell]}$ for $\ell > 0$ as the solution to the equation
\begin{equation}
	\partial_t F^{[\ell]}_t + \frac{1}{2} \tmop{Tr} (\dot{G}_t \mathD^2
	F^{[\ell]}_t) = \sum_{\ell' + \ell'' = \ell} \frac{1}{2} \mathD
	(F_t^{[\ell']}  \dot{G}_t F^{[\ell'']}_t), \label{eq:fe-picard}
\end{equation}
subject to the terminal conditions
\begin{equation}
	F^{[\ell]}_T (\varphi) = \begin{cases}
		\nabla V^T (\varphi), & \text{for } \ell = 1, \\
		0,                    & \text{otherwise,}
	\end{cases} \label{eq:fe-terminal-conditions}
\end{equation}
for a suitable potential $V^T$ to be determined later. The initial condition
$F^{[0]} \equiv 0$ ensures that~\eqref{eq:fe-picard} is triangular in $\ell$
and we can solve~\eqref{eq:fe-picard} as a linear PDE with a source term.
Proceeding in this way, we define the $\ell^{\ast}$-th order approximation
$F^{[\leqslant \ell^{\ast}]}_s \assign \sum_{\ell \leqslant \ell^{\ast}}
	F^{[\ell]}_s$. With this choice for $F$ in the FBSDE~\eqref{eq:FBSDE-generic},
the generator of the backward equation as defined in~\eqref{eq:def-H} reduces
to
\begin{align*}
	H^{[\leqslant \ell^{\ast}]}_s
	 & \assign \partial_s F^{[\leqslant
					\ell^{\ast}]}_s + \frac{1}{2} \tmop{Tr} (\dot{G}_s \mathD^2 F^{[\leqslant
		\ell^{\ast}]}_s) - \frac{1}{2} \mathD (F^{[\leqslant \ell^{\ast}]}_s
	\dot{G}_s F^{[\leqslant \ell^{\ast}]}_s) \\
	 & = - \frac{1}{2}
	\sum_{\tmscript{\begin{array}{c}
				                \ell' + \ell'' > \ell^{\ast} \\
				                \ell', \ell'' \leqslant \ell^{\ast}
			                \end{array}}} \mathD (F_s^{[\ell']}  \dot{G}_s F^{[\ell'']}_s) .
	\label{eq:appr-H}
\end{align*}
The estimates on the flow equation will rely on the following simple Lemma.
\begin{lemma}
	\label{lem:lambda-integrability}Let $\lambda_t = \lambda
		\mathe^{\frac{\beta^{2}}{2} G_t (0)}$, $\bar{\lambda}_t:=|\lambda_t|$ and $\delta = 1 - \frac{\beta^2}{8 \pi} >
		0$. Then, for any $n \in \mathbb{N}$ and $\alpha > 1 - n \delta$,
	\begin{equation}
		\int_t^{\infty} \bar{\lambda}_s^n \langle s \rangle^{- n} \langle s \rangle^{-
			\alpha} \mathd s \lesssim_n \bar{\lambda}_t^n \langle t \rangle^{- (n - 1) -
			\alpha} . \label{eq:int-alpha}
	\end{equation}
	In particular, for $n \delta > 1$ we can choose $\alpha = 0$ and
	\begin{equation}
		\int_t^{\infty} \bar{\lambda}_s^n \langle s \rangle^{- n} \mathd s \lesssim_n
		\bar{\lambda}_t^n \langle t \rangle^{- (n - 1)} . \label{eq:int-rem}
	\end{equation}
\end{lemma}

\begin{proof}
	With the heat kernel estimate~\eqref{eq:cov-0-bound} from
	Lemma~\ref{lem:hk-basic-bounds}, we see that $\bar{\lambda}_t \asymp  C \bar{\lambda} (t \vee
		1)^{1 - \delta}$ for some $C > 0$. Now the claim follows from $\delta > 0
		\Leftrightarrow \beta^2 < 8 \pi$.
\end{proof}

Let us take a moment to heuristically explain how successive iterations
of~\eqref{eq:fe-picard} should improve in~$\ell$. Starting from the first
order approximation $\ell = 1$, the bilinear term does not give any
contributions, and the linear equation~\eqref{eq:fe-picard} computes the usual
Wick-ordering. In the specific case of the cosine interaction, this means more
concretely that
\begin{equation}
	F^{[1]}_t (\varphi) = - \lambda_t \beta \sin (\beta \varphi)
	\label{eq:F1-explicit}, \qquad \text{where\quad} \bar{\lambda}_t \assign \bar{\lambda}
	\mathe^{\frac{\beta^2}{2} G_t (0)} \lesssim \bar{\lambda} C \langle t
	\rangle^{\beta^2 / 8 \pi} = \bar{\lambda} C \langle t \rangle^{1 - \delta} .
\end{equation}
Here, we absorbed the coupling constant $\lambda = \lambda_0$ into the
renormalisation constant $\lambda_t$. The estimates on $\lambda_t$ are a
direct consequence of basic heat kernel estimates
(Lemma~\ref{lem:hk-basic-bounds}). We directly read off the bounds,
\begin{equation}
	\| \mathD F_t^{[1]} (\varphi) \|_{L^{\infty}} + \| F_t^{[1]} (\varphi)
	\|_{L^{\infty}} \lesssim \bar{\lambda}_t \lesssim \bar{\lambda} \langle t \rangle^{1 -
		\delta} . \label{eq:F1-scaling}
\end{equation}
Due to the form of the non-linearity of the flow equation \eqref{eq:fe-picard}
and Lemma~\ref{lem:lambda-integrability}, we can expect the bound
\begin{equation}
	\| \mathD F^{[\ell]}_t (\varphi) \|_{L^{\infty}} + \| F^{[\ell]}_t (\varphi)
	\|_{L^{\infty}} \lesssim \bar{\lambda}_t^{\ell} \langle t \rangle^{- (\ell - 1)}
	\lesssim \bar{\lambda}^{\ell} \langle t \rangle^{1 - \ell \delta},
	\label{eq:F-scaling-guess}
\end{equation}
to propagate inductively. Indeed, assuming that the
bound~\eqref{eq:F-scaling-guess} holds for all $\ell', \ell'' < \ell$, we
obtain from Young's inequality and the estimate $\| \dot{G}_s \|_{L^1}
	\lesssim \langle s \rangle^{- 2}$ that,
\begin{equation}
	\| \mathD (F_t^{[\ell']}  \dot{G}_t F^{[\ell'']}_t) \|_{L^{\infty}}
	\leqslant \| \mathD F_t^{[\ell']} \|_{L^{\infty}} \| \dot{G_t} \|_{L^1} \|
	F_t^{[\ell'']} \|_{L^{\infty}} \lesssim \bar{\lambda}_t^{\ell' + \ell''} \langle t
	\rangle^{- (\ell' + \ell'')} \label{eq:DFGF-young} .
\end{equation}
Since $\dot{G}$ is positive, formally integrating out the linear part in
\eqref{eq:fe-picard} and passing to the mild formulation (see the next section
for details), this suggests
\[ \| F^{[\ell]}_t (\varphi) \|_{L^{\infty}} \lesssim \int_t^T
	\bar{\lambda}_s^{\ell} \langle s \rangle^{- (\ell - 2)} \langle s \rangle^{- 2}
	\mathd s \lesssim \int_t^T \bar{\lambda}_s^{\ell} \langle s \rangle^{-\ell} \mathd
	s . \]
Hence, Lemma~\ref{lem:lambda-integrability} propagates the
bound~\eqref{eq:F-scaling-guess} only if $\ell \delta > 1$. Otherwise, we will
have to improve our analysis and introduce additional regularisations to
propagate the bounds from one level to the next. We therefore refer to the
terms with $\ell > 1 / \delta$ as \tmtextit{irrelevant} and $\ell \leqslant 1
	/ \delta$ as \tmtextit{relevant}. To obtain uniform bounds on the remainder
$R$ in~\eqref{eq:FBSDE-generic}, the source term in \eqref{eq:FBSDE-generic}
$H$ should contain only irrelevant terms. The estimates~\eqref{eq:DFGF-young}
suggest that
\begin{equation}
	\| H_t^{[\leqslant \ell^{\ast}]} (\varphi) \|_{L^{\infty}} \lesssim
	\sum_{\tmscript{\begin{array}{c}
				\ell' + \ell'' > \ell^{\ast} \\
				\ell', \ell'' \leqslant \ell^{\ast}
			\end{array}}} \bar{\lambda}_t^{\ell' + \ell''} \langle t \rangle^{- (\ell' +
		\ell'')} \lesssim \bar{\lambda}_t^{\ell^{\ast}} \langle t \rangle^{-\ell^{\ast}},
	\label{eq:H-scaling}
\end{equation}
which is integrable in $t$ from $\infty$ for $\ell^{\ast} > 1 / \delta$ by
Lemma~\ref{lem:lambda-integrability}. The number of relevant terms depends on
the parameter $\beta^2$. If
\begin{equation}
	\beta^2 < \beta_{\ell^{\ast}}^2 \assign \left(
	\frac{\ell^{\ast}}{\ell^{\ast} + 1} \right) 8 \pi, \label{eq:thresholds}
\end{equation}
then only terms at the levels $\ell \leqslant \ell^{\ast}$ are relevant. At
$\beta^2 = 8 \pi$, the number of relevant terms is infinite and the model
reaches criticality. In the subcritical regime, $\beta^2 < 8 \pi$, we see that
the number of relevant terms is finite, but grows arbitrary large as we
approach the critical value $\beta^2 = 8 \pi$.

Indeed, for the first region, $\beta^2 < \beta_1^2 = 4 \pi$, only the first
level $\ell = 1$ is relevant and we can gather all higher order terms in the
remainder. Outside the first region, we have to deal with two related issues:
\begin{enumeratealpha}
	\item due to \eqref{eq:H-scaling}, the terms $\ell < \ell^{\ast}$ cannot be
	included in the equation for $R$, so that we have to
	iterate~\eqref{eq:fe-picard} at least up to $\ell^{\ast}$;

	\item the heuristic considerations suggest that the
	bound~\eqref{eq:F-scaling-guess} cannot naively propagate through the flow
	equation on its own and these terms require renormalisation.
\end{enumeratealpha}
The goal of our subsequent analysis is to deal with both difficulties and
recover estimates to replace~\eqref{eq:F-scaling-guess}
and~\eqref{eq:H-scaling} beyond this first threshold $\beta^2 < 4 \pi$.

Since our analysis of the FBSDE is limited to the regime $\beta^2 < \beta_3^2
	= 6 \pi$, we develop the ideas for the flow equation only up to this
threshold, where $\ell^{\ast} = 3$ is sufficient. We still emphasise that the
inductive reasoning produces (possibly field dependent) bounds on the
truncated flow in the entire subcritical regime $\beta^2 < 8 \pi$.

\subsection{The Fourier representation}

To proceed with the iteration defined in \eqref{eq:fe-picard} and finally
obtain estimates on $F_s^{[\ell]}$, we restrict our attention to a suitable
parametrised space of functions $\mathcal{S}' (\mathbb{R}^2) \rightarrow
	\mathcal{S}' (\mathbb{R}^d)$. Here, we use the periodicity of the potential to
our advantage and pass to a Fourier representation
following~{\cite{brydgesMayerExpansionsHamiltonJacobi1987}}. For a $\frac{2
		\pi}{\beta}$-periodic functional $V : \mathbb{R}_+ \times \mathbb{R}^2
	\rightarrow \mathbb{R}$ we introduce the formal power series
\begin{equation}
	V_t (\varphi) = \sum_{\ell = 0}^{\infty} V_t^{[\ell]} (\varphi),
	\label{eq:formal-fourier}
\end{equation}
where with $\xi = (\sigma, x) \in \{ - 1, 1 \} \times \mathbb{R}^2$ and
$\xi_{1 : \ell} = (\xi_1, \ldots, \xi_{\ell})$, we define
\begin{equation}
	V_t^{[\ell]} (\varphi) \assign \sum_{\sigma_i \in \{ - 1, 1 \}^{\ell}}
	\int_{(\mathbb{R}^2)^{\ell}} \mathd x_{1 : \ell} f_t^{[\ell]} (\xi_{1 :
		\ell}) \mathe^{i \beta \sigma_1 \varphi (x_1)} \ldots \mathe^{i \beta
		\sigma_{\ell} \varphi (x_{\ell})} . \label{eq:fourier-rep-V}
\end{equation}
Since the level $\ell$ is determined uniquely by the number of arguments
$\xi_{1 : \ell}$, we may drop the superscript $\ell$ in $f^{[\ell]}$ without
introducing ambiguities. For brevity of the subsequent notation, we write
$[\ell]:=\{1,\dots,\ell\}$ and introduce
the following shorthand for the integrals and the exponential fields,
\[
	\int \mathd \xi f (\xi) \assign \sum_{\sigma = \pm 1} \int_{\mathbb{R}^2}
	\mathd xf (\sigma, x),  \quad \psi_x^{\sigma} \assign \mathe^{i \beta
		\sigma \varphi (x)}, \quad  \psi (\xi_{1 : \ell}) \assign \prod_{i =
		1}^{\ell} \psi_{x_i}^{\sigma_i},
\]
and for a set $I\subset [\ell]$ we write $\xi_I=(\xi_i)_{i\in I}$.
Finally, define the covariance matrix
\begin{equation}
	\Gamma_{t, s} (\xi_{1 : \ell}) \assign - \frac{\beta^2}{2}  \sum_{i, j} \sigma_i
	\sigma_j (G_s - G_t) (x_i - x_j), \quad t \leqslant s. \label{eq:def-W}
\end{equation}
With this notation and basic set-up, we can rewrite the flow
equation~\eqref{eq:fe-picard} in terms of the coefficients $f$. Since any
additive shift of the potential $V_T$ by a constant does not affect the force,
the terminal condition~\eqref{eq:fe-terminal-conditions} translates to
\begin{equation}
	f_T^{[\ell], T} (\xi_{1 : \ell}) = \begin{cases}
		\frac{\lambda_T}{2}, & \ell = 1, \\
		0,                   & \ell > 1.
	\end{cases} \label{eq:ic-fe}
\end{equation}
The functional $V^{[\ell]}$ satisfies
the truncated flow equation~\eqref{eq:fe-V} below at level $\ell$ if and only
if, modulo positive combinatorial coefficients which we gather in $C_{ | I_1 |,
			| I_2 |}$,
\begin{equation}
	f_t^{[\ell], T} (\xi_{1 : \ell}) = - \beta^2 \sum_{I_1 \dot{\cup} I_2 = [\ell]} C_{ | I_1 |,
			| I_2 |} \int_t^T \mathd s \mathe^{\Gamma_{t, s} (\xi_{1 : \ell})} f_s
	(\xi_{I_1}) \Big[ \sum_{\substack{{i \in I_1}\\{j \in I_2}}} \sigma_i \sigma_j
	\dot{G}_s (x_i - x_j) \Big] f_s  (\xi_{I_2}) . \label{eq:fe-v-coeffients}
\end{equation}
We include a proof of the relevant implication in Lemma
\ref{lem:equivalence-fe} at the end of this section.
Instead of controlling the functions $F$ and $V$ directly, we now want to
inductively derive estimates on these kernels $f^{[\ell]}$. Of course,
eventually we will be able to transfer these estimates back to $F$ and $V$ in
a straightforward manner (see Section~\ref{sec:definitions}).

Before we proceed and derive bounds on the kernels $f$, some remarks about the
setup seem appropriate.

\begin{remark}
	\label{rem:flow-eq}{\tmdummy}

	\begin{enumeratealpha}
		\item We are primarily interested in the flow equation for the force.
		However, for the variational description in Section \ref{sec:Var+LDP}, we
		will have to work at the level of the potential as well. Since the
		equations for the force $F$ are readily obtained by differentiating the
		equations for $V$, we prefer to use it as a starting point. Up to an
		additive constant, both descriptions are equivalent on the finite volume
		and $F^{[\ell]} = \mathD V^{[\ell]}$ satisfies~\eqref{eq:fe-picard} if and
		only if $V^{[\ell]}$ satisfies the Picard scheme for~\eqref{eq:HJB}, that
		is
		\begin{equation}
			\partial_s V^{[\ell]}_s + \frac{1}{2} \tmop{Tr} (\dot{G}_s \mathD^2
			V^{[\ell]}_s) = \frac{1}{2}\sum_{\ell' + \ell'' = \ell} (\mathD
			V_s^{[\ell']}  \dot{G}_s \mathD V^{[\ell'']}_s) . \label{eq:fe-V}
		\end{equation}
		We nonetheless emphasise that we never rely on the fact that $F$ is the
		gradient of a potential in our analysis.

		\item The coefficients $f^{[\ell]}$ are symmetric in their arguments
		$\xi_{1 : \ell}$, i.e. for any permutation $\pi$ of $[\ell]$,
		\begin{equation}
			f^{[\ell]}_t (\xi_1, \ldots, \xi_{\ell}) = f^{[\ell]}_t (\xi_{\pi (1)},
			\ldots, \xi_{\pi (\ell)}) . \label{eq:f-symm}
		\end{equation}
		\item \label{rem:translation-invariance}If $f^{[1]}$ is translation
		(respectively rotation) invariant, we inductively see
		from~\eqref{eq:fe-v-coeffients} and the Euclidean invariance of the heat
		kernel $\dot{G}$ that also the kernels $f^{[\ell]}$ at the higher levels
		$\ell > 1$ are translation (respectively rotation) invariant.
		Correspondingly, if $f^{[1]}$ is invariant under complex conjugation (that
		is with $\bar{\xi} = (- \sigma, x)$ we have $f^{[1]} (\xi) = f^{[1]}
			(\bar{\xi})$) then also $f^{[\ell]} (\xi_{1 : \ell}) = f^{[\ell]}
			(\bar{\xi}_{1 : \ell})$ is true for any $\ell > 1$.

		\item We always consider truncations
		\[ V_t^{[\leqslant \ell^{\ast}]} = \sum_{\ell = 0}^{\ell^{\ast}}
			V_t^{[\ell]}, \quad \text{and\quad} F_t^{[\leqslant \ell^{\ast}]} =
			\sum_{\ell = 0}^{\ell^{\ast}} F_t^{[\ell]}, \]
		of \eqref{eq:formal-fourier} for some $\ell^{\ast} < \infty$. Therefore,
		we are not concerned with questions of convergence as $\ell^{\ast}
			\rightarrow \infty$. We will refer to the
		truncated series $F^{[\leqslant \ell^{\ast}]}_s$ as the
			{$\ell^{\ast}$-th order approximation}, even though we do not
		provide quantitative estimates on the convergence of the series
		$\sum_{\ell} F^{[\ell]}_s (\varphi)$. This can at least be motivated by
		the observation that~\eqref{eq:formal-fourier} is a formal power series in
		the coupling constant $\lambda$, which formally solves the PDE \eqref{eq:fe-V}. The fact that the representation is not
		unique \ (both with respect to the summands in~\eqref{eq:formal-fourier}
		and the coefficients in~\eqref{eq:fourier-rep-V}) does not cause any
		inconvenience for us.

		\item The representation \eqref{eq:formal-fourier} is also known as Mayer
		expansion in the literature and its convergence was already studied
		in~{\cite{brydgesMayerExpansionsHamiltonJacobi1987}} and more recently in
		a series of
		papers~{\cite{bauerschmidtLogSobolevInequalityContinuum2021,bauerschmidtMaximumCouplingSineGordon2022,bauerschmidtColemanCorrespondenceFree2023,kroschinskyMayerSeriesTwoDimensional2019}}
		for the sine-Gordon model. In contrast to our analysis, these results
		construct the exact solution to the flow equation~\eqref{eq:fe-V} in different parameter regimes $\beta,\lambda$ by showing that the formal
		series~\eqref{eq:formal-fourier} converges.
	\end{enumeratealpha}
\end{remark}
\begin{lemma}\label{lem:equivalence-fe}
	If $(f^{[\ell]})_{\tmscript{\begin{array}{c}
						t \in [0, T]
					\end{array}}}$ for $\ell \leqslant \ell^{\ast}$ satisfy
	\eqref{eq:fe-v-coeffients}, then $V^{[\ell]}_t$ as defined in
	\eqref{eq:fourier-rep-V} satisfies \eqref{eq:fe-V} for $\ell\leqslant\ell^\ast$.
\end{lemma}
\begin{proof}
	Suppose that $f_t^{[\ell]}$ satisfies \eqref{eq:fe-v-coeffients} and define
	$V^{[\ell]}_t$ according to \eqref{eq:fourier-rep-V}. We compute,

	\[ \frac{1}{2} \tmop{Tr} (\dot{G}_s \mathD^2 V^{[\ell]}_s (\varphi)) = -
		\frac{\beta^2}{2}  \sum_{i, j} \int \mathd x_i \int \mathd x_j \sigma_i
		\sigma_j \dot{G}_s (x_i - x_j) f_t^{[\ell]} (\xi_{1 : \ell}) \psi (\xi_{1
			: \ell}), \]
	and with $|I_1| = \ell'$, $|I_1| = \ell''$,
	\[ \begin{aligned}
			\frac{1}{2} & \mathD V_s^{[\ell']} (\varphi) \dot{G}_s \mathD
			V_s^{[\ell'']} (\varphi)                                      \\
			            & = \frac{1}{2} \sum_{i \in I_1} \int \mathd
			x_{I_1} f_t^{[\ell']} (\xi_{I'}) \sum_{j \in I_2} \int \mathd x_{I_2}
			\sigma_i \sigma_j\dot{G}_s (x_i - x_j) f_t^{[\ell'']} (\xi_{I_2}) \psi (\xi_{I_1}) \psi
			(\xi_{I_2}) .
		\end{aligned} \]
	Differentiating the equation for $f^{[\ell]}$ \eqref{eq:fe-v-coeffients} we see that
	\[ \begin{aligned}
			\partial_t & f_t^{[\ell]} (\xi_{1 : \ell}) - \frac{\beta^2}{2}  \sum_{i,
				j \in [\ell]} \sigma_i \sigma_j \dot{G}_s (x_i - x_j) f_t^{[\ell]}
			(\xi_{1 : \ell})                                                                            \\
			           & = \sum_{\tmscript{\begin{array}{c}
						                               I_1 \dot{\cup} I_2 = [\ell] \\
					                               \end{array}}} C (| I_1 |, | I_2 |) f_t^{[\ell']} (\xi_{I_1})
			\left[ \sum_{i \in I_1} \sum_{j \in I_2}
				\sigma_i \sigma_j  \dot{G}_s (x_i - x_j) \right] f_t^{[\ell'']} (\xi_{I_2}).
		\end{aligned}
	\]
	Inserting this into
	\[ \partial_t V_t^{[\ell]} (\varphi) = \sum_{\sigma_i \in \{ - 1, 1
			\}^{\ell}} \int_{(\mathbb{R}^2)^{\ell}} \mathd x_{1 : \ell} \partial_t
		f_t^{[\ell]} (\xi_{1 : \ell}) \mathe^{i \beta \sigma_1 \varphi (x_1)}
		\ldots \mathe^{i \beta \sigma_{\ell} \varphi (x_{\ell})}, \]
	yields the claim.
\end{proof}

\subsection{Estimates on the Fourier
	coefficients}\label{sec:flow-coefficients}

In this section, we derive our main estimates on the kernels $f$ defined
in~\eqref{eq:fe-v-coeffients} to control the flow under the conditional
expectation in~\eqref{eq:optimal-SDE}. For $\varsigma \in (0, 1)$ and some
kernel $\kappa$ to be chosen later (see \eqref{eq:def-k} below), we will be using
the norms
\begin{equation}
	\interleave f \interleave_t \assign \sup_{\xi_1}  \int \mathd \xi_{2 : \ell}
	| f (\xi_{1 : \ell}) \kappa_t (\xi_{1 : \ell}) \omega_{\varsigma} (x_{1 : \ell})
	|  \quad \text{where } \omega_{\varsigma} (x_{1 : \ell}) \assign
	\mathe^{\varsigma m (\tmop{St} (x_{1 : \ell}))}, \label{eq:def-exp-St}
\end{equation}
for the Fourier kernels (see \eqref{eq:def-St} for the definition of the
Steiner diameter $\tmop{St} (x_I)$). If $\kappa_t \equiv 1$ does not depend on $t$,
we may drop the subscript $t$. Since the coefficients $f$ are symmetric in
their arguments (see also~\eqref{eq:f-symm}), the point $\xi_1$ is not special in any way
and the supremum could have been taken over any other $\xi_k$ instead. The
exponential tree weights $\omega_{\varsigma}$ allow us to quantify the decay
of the coefficients at large separation between the points $x_1, \ldots,
	x_{\ell}$, which we require to show decay of correlations in Section
\ref{sec:correlations}. As
\[ \omega_{\varsigma} (x_{I_1 \cup I_2}) \leqslant \omega_{\varsigma}
	(x_{I_1}) \omega_{\varsigma} (x_{I_2}) \mathe^{\varsigma m d (x_{I_1},
		x_{I_2})}, \quad \text{where\quad} d (x_{I_1}, x_{I_2}) \assign \min_{x_i
		\in x_{I_i}} | x_1 - x_2 |, \]
these norms work nicely with the flow equation for the
coefficients~\eqref{eq:fe-v-coeffients} provided we choose $\varsigma \in (0,
	1)$. Indeed, since the convolution $\dot{G}_s$ in~\eqref{eq:fe-v-coeffients}
always contracts along $(x_i - x_j)$ for $i \in I_1$ and $j \in I_2$, Young's
convolution inequality implies that for $\kappa_t \equiv 1$,
\begin{align}
	\begin{split}
		         & \sup_{\xi_1}  \int \mathd \xi_{2 : \ell} \omega_{\varsigma} (x_{I_1 \cup
			I_2}) \left| f^{[\ell']}_t (\xi_{I_1}) \left[ \sum_{i \in I_1} \sum_{j \in
				I_2} \sigma_i \sigma_j  \dot{G}_t (x_i - x_j) \right] f^{[\ell'']}_t
		(\xi_{I_2}) \right|                                                                 \\
		\lesssim & \interleave f_t^{[\ell]} \interleave
		\interleave f_t^{[\ell'']} \interleave  \| \dot{G}_t \|_{L^1
				(w_{\varsigma})} . \label{eq:fe-young-ineq}
	\end{split}
\end{align}
We now want to make the heuristic bounds \eqref{eq:F-scaling-guess} precise and for an appropriate choice
for $\kappa_t$, we will show that for any $\ell\leqslant \ell^\ast$,
\begin{equation}
	\interleave f_t^{[\ell]} \interleave_t
	\lesssim \bar{\lambda}_t^{\ell}\langle t\rangle^{-(\ell-1)}\label{eq:f-ell-all}.
\end{equation}

To motivate our set-up going forward, consider
again~\eqref{eq:fe-v-coeffients}. Because $\dot{G}_s$ is a positive definite
kernel, it follows immediately from the definition~\eqref{eq:def-W} of $\Gamma$,
\begin{equation}
	\Gamma_{t, s} (\xi_{1 : \ell}) \leqslant 0 \label{eq:W-leq-0}, \quad \text{for }
	t \leqslant s,
\end{equation}
and consequently $\mathe^{\Gamma_{t, s} (\xi_{1 : \ell})} \leqslant 1$. Applying
this estimate in \eqref{eq:fe-v-coeffients} for $\ell = 2$ yields, with the
convolution inequality~\eqref{eq:fe-young-ineq}, the estimates on the first
order term in \eqref{eq:F1-explicit} and the heat kernel estimates from Lemma
\ref{lem:hk-exp} using the assumption $\varsigma < 1$,
\[ \interleave f^{[2]}_t \interleave \lesssim \int_t^T \mathd s \interleave
	f_s^{[1]} \interleave  \interleave f_s^{[1]} \interleave  \| \dot{G}_s
	\|_{L^1 (\omega_{\varsigma})} \lesssim \int_t^T \mathd s \bar{\lambda}_s^2
	\langle s \rangle^{- 2}, \]
which is not integrable from $\infty$ unless $2 \delta > 1$ ($\Leftrightarrow
	\beta^2 < 4 \pi$). Therefore, we need additional help to propagate uniform
bounds along the flow. This help will partially come from the structure of the
covariance matrix $\Gamma_{t, s}$, and partially from the choice of $\kappa_t$ in the
definition of the norm \eqref{eq:def-exp-St}. To this end, define
\[ q (\xi_{1 : \ell}) \assign \sum^{\ell}_{k = 1} \sigma_k, \]
the \tmtextit{charge of $\xi_{1 : \ell}$}. We will call a contribution $\xi_{1
		: \ell}$ \tmtextit{neutral} if $q (\xi_{1 : \ell}) = 0$ and \tmtextit{charged}
otherwise. The relevance of the charge is best illustrated by the improved
estimates on the covariance matrix $\Gamma_{t, s}$. If $\xi_{1 : \ell}$ is
charged, the exponential factor in~\eqref{eq:fe-v-coeffients} can help bring
down the scale. As a pleasant side effect, these estimates will also imply
that including an additional odd level, that is going from $\ell = 2 k$ to
$\ell + 1$, introduces no new difficulties to the analysis. To not
interrupt the flow of ideas, we postpone the mostly technical proof to
Appendix~\ref{app:charged-cov-bound}.

\begin{lemma}
	\label{lem:charged-cov-bound} Suppose that $\xi_{1 : \ell}$ is charged. Then
	there is a constant $C > 0$ such that for all $s \geqslant t$,
	\begin{equation}
		\Gamma_{t, s} (\xi_{1 : \ell}) \leqslant \frac{\beta^2}{2} (G_t (0) - G_s
		(0)) + C \label{eq:charged-cov-bound},
	\end{equation}
	and in particular
	\[ \mathe^{\Gamma_{t, s} (\xi_{1 : \ell})} \lesssim \bar{\lambda}_t \bar{\lambda}_s^{- 1} .
	\]
\end{lemma}

\begin{remark}
	\label{rem:charge}For neutral contributions, $q (\xi_{1 : \ell}) = 0$, the
	point-wise bound $\mathe^{\Gamma_{t, s} (\xi_{1 : \ell})} \leqslant 1$ is sharp:
	If $x_i = 0$ for all $i = 1, \ldots, \ell$, then we have $\Gamma_{t, s} (\xi_{1 :
			\ell}) = 0$. As a result, point-wise estimates on the linear propagator
	$\mathe^{\Gamma_{t, s} (\xi_{1 : \ell})}$ cannot help to transport estimates for
	the kernels $f$ along the flow of \eqref{eq:fe-v-coeffients}. Conversely, if
	$| q (\xi_{1 : \ell}) | > 1$, then it follows from the proof of Lemma
	\ref{lem:charged-cov-bound} (see \eqref{eq:W-charged-split}) that we could
	iterate the same procedure until only the neutral part remains and extract
	more terms from the diagonal. In other words, the tighter bound
	\[ \Gamma_{t, s} (\xi_{1 : \ell}) \leqslant \frac{\beta^2}{8 \pi} | q (\xi_{1 :
			\ell}) | (\log (t \vee 1) - \log (s \vee 1)) + C, \]
	is also true. For our purposes, the bound~\eqref{eq:charged-cov-bound} will
	always be sufficient.
\end{remark}

With Lemma~\ref{lem:charged-cov-bound}, the integrability estimates from
Lemma~\ref{lem:lambda-integrability} for $\alpha = 1$ show that the charged
contributions no longer pose a problem for us, allowing to set $\kappa_t \equiv 1$
in this case. However, for the neutral contributions, this norm is too strong
and we will have to rely on the kernel $\kappa_t$.

Recall that $\dot{G}$ is exponentially concentrated on $| x | \lesssim t^{- 1
			/ 2}$, so that (see Lemma~\ref{lem:hk-exp}),
\begin{equation}
	\int_{\mathbb{R}^2} \mathd x | x |^{2 \alpha} \dot{G_t} (x) w_{\varsigma}
	(x) \lesssim \langle t \rangle^{- 2 - \alpha} . \label{eq:G-scaling}
\end{equation}
Combined with Lemma~\ref{lem:lambda-integrability}, we expect that introducing
an additional zero of order $2 \alpha$ in $x_i - x_j$ whenever $\sigma_i = -
	\sigma_j$ should help to propagate a bound on a regularised version of the
kernel $f^{[\ell]}$. Of course, this regularisation comes at a price we have
to pay later. For now, let us ignore this issue and discuss how we can define
a regularised version of the kernels which allow to propagate the bounds for
the neutral contributions. With $\delta_{i j} x = x_1 - x_2$, $c \in \left( 0,
	\frac{1}{4} \right)$ and $\alpha \in [0, 1)$ to be chosen later, we introduce
the (rotation and translation invariant) kernels
\begin{equation}
	\kappa_t (\xi_{1 : \ell}) \assign \begin{cases}
		t^{\alpha} | \delta_{12} x |^{2 \alpha} \mathe^{ct | \delta_{12} x |^2}, &
		\ell = 2 \text{ and } q (\xi_1, \xi_2) = 0,                                                  \\
		1,                                                                       & \text{otherwise.}
	\end{cases} \label{eq:def-k}
\end{equation}
The increment $| \delta_{12} x |^{2 \alpha}$ ensures the integrability from
$\infty$ thanks to \eqref{eq:G-scaling}, the additional exponential weight in
the kernel is included for technical reasons that will become clear later and
the factor $t^{\alpha}$ is included for convenience. Given a charge $q \in
	\mathbb{Z}$, we will also use the notation
\begin{equation}
	f^{[\ell] (q)}_t (\xi_{1 : \ell}) \assign \mathbbm{1}_{\{ q (\xi_{1 : \ell})
			= q \}} f^{[\ell]}_t (\xi_{1 : \ell}), \label{eq:def-f0}
\end{equation}
with analogous notations for the potential
\begin{equation}
	V^{[\ell] (q)} (\varphi) \assign \sum_{\sigma_i \in \{ - 1, 1 \}^{\ell}}
	\int_{(\mathbb{R}^2)^{\ell}} \mathd x_{1 : \ell} f_t^{[\ell] (q)} (\xi_{1 :
		\ell}) \mathe^{i \beta \sigma_1 \varphi (x_1)} \ldots \mathe^{i \beta
		\sigma_{\ell} \varphi (x_{\ell})}, \label{eq:def-f0-2}
\end{equation}
and the force $F^{[\ell], (q)} = \mathD V^{[\ell], (q)}$ to consider the
bounds for charged and neutral contributions separately.

We can now proceed with the estimates on the regularised kernels for the $2$-point contributions, whose analysis already contains all additional
difficulties resulting from neutral contributions.

\begin{lemma}
	\label{lem:f2-kernel-scaling}For any $\delta > 0$ and $\alpha > (1 - 2
		\delta) \vee 0$,
	\begin{equation}
		\interleave f_t^{[2]} \interleave_t \lesssim \bar{\lambda}_t^2 \langle t
		\rangle^{- 1} . \label{eq:f2-bounds}
	\end{equation}
	Moreover, the kernels $f^{[2]}_t$ inherit the concentration to $| x_1 - x_2
		| \leqslant \langle t \rangle^{- 1 / 2}$ from $\dot{G}$. More precisely,
	letting
	\[ \tilde{f}_t (\xi_1, \xi_2) = f_t (\xi_1, \xi_2) | \delta_{12} x |^{2
				\kappa}, \]
	for some $\kappa \geqslant 0$, it holds that
	\begin{equation}
		\interleave \tilde{f}_t \interleave_t \lesssim \langle t \rangle^{-
			\kappa} \interleave f_t \interleave_t . \label{eq:coeff-concentration}
	\end{equation}
\end{lemma}

\begin{proof}
	By definition~\eqref{eq:fe-v-coeffients},
	\[ f^{[2]}_t (\xi_1, \xi_2) = C \int_t^T \mathd s \mathe^{\Gamma_{t, s} (\xi_1,
			\xi_2)} f^{[1]}_s (\xi_1) \sigma_1 \sigma_2  \dot{G}_s (x_1 - x_2)
		f^{[1]}_s (\xi_2), \]
	where for $\ell = 1$,
	\begin{equation}
		| f^{[1]}_s (\xi) | \lesssim \bar{\lambda} \mathe^{\frac{1}{2} \beta^2 G_s (0)}
		= \bar{\lambda}_s = \bar{\lambda}_s^{\ell} \langle s \rangle^{- (\ell - 1)} .
		\label{eq:v1}
	\end{equation}
	We only show the bound \eqref{eq:coeff-concentration}, as
	\eqref{eq:f2-bounds} follows directly by letting $\kappa = 0$. We deal with
	the two cases, charged and neutral, separately. If $(\xi_1, \xi_2)$ is
	charged, we use Young's inequality, Lemma \ref{lem:charged-cov-bound} and
	the basic estimate \eqref{eq:G-scaling} for $\alpha = 0$, to conclude for
	any $\delta > 0$,
	\begin{align*}
		\interleave \tilde{f}_t^{[2] (\pm 2)} \interleave_t
		 & \lesssim \sup_{\xi_1}
		\bar{\lambda}_t  \int_t^T \mathd s \bar{\lambda}_s^{- 1} \interleave f_s^{[1]}
		\interleave_t^2  \| | x |^{\kappa}  \dot{G}_s \|_{L^1 (w_{\varsigma})}
		\lesssim \bar{\lambda}_t  \int_t^T \mathd s \bar{\lambda}_s \langle s \rangle^{- 2 -
		\kappa}                                                              \\
		 & \lesssim \bar{\lambda}_t^{\ell} \langle t \rangle^{- (\ell - 1) -
			\kappa} .
	\end{align*}
	If $(\xi_1, \xi_2)$ is neutral, we have to be more careful. By the
	definition~\eqref{eq:def-W} of $\Gamma_{t, s}$, we can absorb the renormalisation
	constants $\bar{\lambda}_s = \bar{\lambda} \mathe^{\frac{\beta^2}{2} G_s (0)}$ coming
	from $f^{[1]}_s$ through,
	\[ \Gamma_{t, s} (\xi_1, \xi_2) + \beta^2 G_s (0) = \beta^2 G_t (0) - \beta^2 G_t
		(x_1 - x_2) + \beta^2 G_s (x_1 - x_2) . \]
	Instead of the worst-case scaling $\beta^2 G_s (0)$, for which only
	point-wise estimates are possible, this means we only have to deal with
	$\beta^2 G_s (x_1 - x_2)$. Here, combining the averaging in space with the
	regularisation from the kernels $\kappa$ defined in \eqref{eq:def-k} allows
	us to estimate the integral uniformly. Indeed, using the above we obtain
		{\footnotesize\begin{align*}
				 & \interleave \tilde{f}_t^{[2] (0)} \interleave_t                    \\
				 & =  C \sup_{\xi_1}
				\left| \int \mathd \xi_2 \kappa_t (\xi_1, \xi_2) | x_1 - x_2 |^{\kappa}
				\omega_{\varsigma} (x_1, x_2)  \int_t^T \mathd s \mathe^{\Gamma_{t, s}
					(\xi_1, \xi_2)} f^{[1]}_s (\xi_1) f^{[1]}_s (\xi_2)  \dot{G}_s (x_1 -
				x_2) \right|                                                          \\
				 & \lesssim  \sup_{x_1} \left| \int_t^T \mathd s \int_{\mathbb{R}^2}
				\mathd x_2 \omega_{\varsigma} (x_1, x_2)  \dot{G}_s (x_1 - x_2) | x_1 -
				x_2 |^{2 \alpha + \kappa} t^{\alpha} \mathe^{ct | x_1 - x_2 |^2}
				\mathe^{\Gamma_{t, s} (\xi_1, \xi_2) + \beta^2 G_s (0)} \right|       \\
				 & \lesssim  \mathe^{\beta^2 G_t (0)} t^{\alpha}  \int_{\mathbb{R}^2}
				\mathd x | x |^{2 \alpha + \kappa} \mathe^{ct | x |^2 + \varsigma m | x
					|} \int_t^T \mathd s \dot{G}_s (x) \mathe^{\beta^2 G_s  (x) - \beta^2
				G_t (x)}                                                              \\
				 & \lesssim  \mathe^{\beta^2 G_t (0)} t^{\alpha}  \int_{\mathbb{R}^2}
				\mathd x | x |^{2 \alpha + \kappa} \mathe^{ct | x |^2 + \varsigma m | x
					|} (\mathe^{\beta^2 (G_{\infty} - G_t) (x)} - 1),
			\end{align*}}where we used that $\dot{G_s}$ has a positive kernel in the last
	inequality to replace $G_T$ by $G_{\infty}$. Choosing $\alpha > (1 - 2
		\delta) \vee 0$ we have access to \eqref{eq:hk-scaling} to compute the
	integral over $x$ above and obtain
	\[ \interleave \tilde{f}^{[2] (0)}_t \interleave_t \lesssim \bar{\lambda}_t^2
		t^{\alpha}  \int_t^T \mathd s \langle s \rangle^{- 2 - \alpha - \kappa}
		\lesssim \bar{\lambda}_t^2 \langle t \rangle^{- 1 - \kappa} = \bar{\lambda}_t^{\ell}
		\langle t \rangle^{- (\ell - 1) - \kappa} . \]
\end{proof}

With the dipole under control, we can propagate bounds on the (charged) subsequent contributions essentially for free using Lemma \ref{lem:charged-cov-bound}. With some additional work, we will show improved estimates that do not rely on the kernel $\kappa$ in the following lemma.

\begin{lemma}
	\label{lem:f3-kernel-scaling}For $\alpha \leqslant 1 / 2$ and $\delta > 1 /
		4$, it holds that
	\[ \interleave f^{[3]}_t \interleave \lesssim \bar{\lambda}_t^3 \langle t
		\rangle^{- 2} . \]
\end{lemma}

\begin{proof}
	By the definition \eqref{eq:fe-v-coeffients} of the coefficients the kernel
	$f^{[3]}$ is given by a linear combination of functions of the form
	\[ \bar{f}_t (\xi_1, \xi_2, \xi_3) = C \int_t^T \mathd s \mathe^{\Gamma_{t, s}
			(\xi_{1 : 3})} f^{[1]}_s (\xi_1) f^{[2]}_s (\xi_2, \xi_3) \sigma_1
		[\sigma_2 \dot{G_s} (x_1 - x_2) + \sigma_3 \dot{G_s} (x_1 - x_3)],\]
	obtained by considering all the permutations of the arguments $(\xi_i)_{i =
				1, 2, 3}$. If the $2$-point contribution $(\xi_2, \xi_3)$ is charged,
	applying \eqref{eq:fe-young-ineq} immediately implies the bound on
	$f^{[3]}$,
	\[ \interleave \bar{f}_t \interleave \lesssim \bar{\lambda}_t  \int_t^T \mathd s
		\bar{\lambda}_s^{- 1} \langle s \rangle^{- 2} \interleave f_s^{[1]} \interleave
		\interleave f_s^{[2] (\pm)} \interleave \lesssim \bar{\lambda}_t^3 \langle t
		\rangle^{- 2} = \bar{\lambda}_t^{\ell} \langle t \rangle^{- (\ell - 1)} . \]
	Otherwise, if $(\xi_2, \xi_3)$ is neutral, we only have uniform bounds on
	$\kappa_t f_t^{[2]}$ but not on $f^{[2]}_t$. Therefore, we insert $1 = \kappa_t \kappa_t^{-
			1}$ and absorb $\kappa_t^{- 1}$ with the convolution $\dot{G}_t$ to obtain the
	bounds on $\bar{f}_t$. Here, we compute with
	Lemma~\ref{lem:hk-weighted-Lipschitz}, using the assumption $\alpha =
		\frac{1}{2}$, and
	\[ \omega_{\varsigma} (x_{1 : 3}) \leqslant \omega_{\varsigma} (x_2, x_3)
		\omega_{\varsigma} (x_1, x_2) . \]
	Thus, writing $\delta_{i j} x \assign x_i - x_j$,
	\begin{align}
		\begin{split}
			         & \sup_{\xi_1}  \int \mathd \xi_{2 : 3} \omega_{\varsigma} (x_{1 : 3})
			\frac{| \dot{G_s} (x_1 - x_2) - \dot{G_s} (x_1 - x_3) |}{s^{\alpha} |
				\delta_{23} x |^{2 \alpha}} \mathe^{- ct | \delta_{23} x |^2} f_s^{[1]}
			(\xi_1) (\kappa_s f_s^{[2]}) (\xi_{2,} \xi_3)                                   \\
			\lesssim & \sup_{\xi_1}  \int \mathd \xi_{2 : 3} (| \delta_{12} x | + |
			\delta_{23} x |) s^{- 1 / 2} \mathe^{- ct | \delta_{12} x |^2 +
				\varsigma m | \delta_{12} x | - m^2 / s} f_s^{[1]} (\xi_1)
			(\omega_{\varsigma} \kappa_s f_s^{[2]}) (\xi_{2,} \xi_3)                        \\
			\lesssim & \interleave f_s^{[1]} \interleave  \int \mathd \xi_{2 : 3}
			\mathe^{- cs | \delta_{12} x |^2 + \varsigma m | \delta_{12} x | - m^2 /
				s} [| \delta_{12} x | + | \delta_{23} x |] s^{- 1 / 2} (\omega_{\varsigma}\kappa_s f^{[2]}_s)
			(\xi_2, \xi_3).\label{eq:dg-dx}
		\end{split}
	\end{align}
	By Young's convolution inequality, the last integral can be estimated as
	\begin{align*}
		\begin{split}
			         & \sup_{\xi_1} \left| \int \mathd (\xi_2, \xi_3) (\omega_{\varsigma} \kappa_s f_s^{[2]}) (\xi_2, \xi_3) \mathe^{- cs | \delta_{12} x
				|^2 + \varsigma m | \delta_{12} x | - m^2 / s}  [| \delta_{12} x | + |
			\delta_{23} x |] \right|                                                                                                                      \\
			\lesssim & \sup_{x_1 \in \mathbb{R}^2}  \int \mathd x_2 \mathe^{- cs |
				\delta_{12} x |^2 + \varsigma m | \delta_{12} x | - m^2 / s} |
			\delta_{12} x |  \interleave f_s^{[2]} \interleave_s                                                                                          \\
			         & + \sup_{x_1 \in
				\mathbb{R}^2}  \int \mathd x_2 \mathe^{- cs | \delta_{12} x |^2 +
				\varsigma m | \delta_{12} x | - m^2 / s} \interleave f^{[2]}_s |
			\delta_{23} x |\interleave_s .
		\end{split}
	\end{align*}
	Using the scaling properties of $\kappa_s f_s^{[2]}$ from
	Lemma~\ref{lem:f2-kernel-scaling} and evaluating the Gaussian integral with
	Lemma \ref{lem:hk-exp} using $\varsigma < 1$ and choosing $c \in \left( 0,
		\frac{1}{4} \right)$ in \eqref{eq:def-k} sufficiently close to $1 / 4$, we
	arrive at the required claim
	\[ \interleave \bar{f}_t \interleave \lesssim \bar{\lambda}_t  \int_t^T \mathd s
		\bar{\lambda}_s^2 \langle s \rangle^{- 3} \lesssim \bar{\lambda}_t^3 \langle t
		\rangle^{- 2} . \]
\end{proof}

\begin{remark}
	\label{rem:h-estimates}The proof above more generally shows that with
	$\alpha \leqslant 1 / 2$, $\delta > 1 / 4$ and $| I_1 | + | I_2 | = \ell >
		2$, the following bounds hold,
	\begin{equation}
		\sup_{\xi_1}  \int \mathd \xi_{2 : \ell} \omega_{\varsigma} (x_{1 : \ell})
		| f_t (\xi_{I_1}) | \left| \sum_{i \in I_1} \sum_{j \in I_2} \sigma_i
		\sigma_j  \dot{G}_t (x_i - x_j) \right| | f_t  (\xi_{I_2}) | \lesssim
		(\bar{\lambda}_t \langle t \rangle^{- 1})^{\ell} \label{eq:h-estimates}.
	\end{equation}
	We will rely on this later to obtain estimates on $H$ in section \ref{sec:F-estimates}.
\end{remark}

\subsection{The renormalised problem}\label{sec:definitions}

Given a spatial cut-off $\rho \prec 1$ and a UV cut-off $T < \infty$, let
\begin{equation}
	V_t^{[\leqslant \ell^{\ast}], \rho, T} (\varphi) = \sum_{\ell \leqslant
		\ell^{\ast}} [V^{[\ell], \rho, T}_t (\varphi) - c^{[\ell], \rho, T}],
	\label{eq:def-V}
\end{equation}
for some suitable renormalisation constants $c^{[\ell], \rho, T}$ to be chosen
later. Here, we denote by $V^{[\ell], \rho, T}$ the $\ell$-th order
contribution as defined via its Fourier expansion as in
\eqref{eq:fourier-rep-V} subject to the condition
\[ V_T^{[1], \rho, T} (\varphi) = \int_{\mathbb{R}^2} \mathd x \rho (x)
	\lambda_T \cos (\beta \varphi) . \]
We use the analogous definition and notation for the force $F^{[\leqslant
					\ell^{\ast}], \rho, T}_t = \tmop{D} V_t^{[\leqslant \ell^{\ast}], \rho, T}$ and the
remainder $H^{[\leqslant \ell^{\ast}], \rho, T}$ defined in \eqref{eq:def-H}.
For $\beta^2 < 6 \pi$ and $\ell^{\ast} = 3$, we transfer the bounds we
obtained for the Fourier coefficients $f^{[\ell]}$ in the previous section to
the truncated potential $V^{[\leqslant \ell^{\ast}]}$ and the truncated force
$F^{[\leqslant \ell^{\ast}]}$. In this step, we have to pay the price for the
regularisation with the kernels $\kappa$ defined in \eqref{eq:def-k}. For $\beta^2
	< 6 \pi$, these kernels only appear at level $\ell = 2$, and by definition,
\[ \begin{aligned}
		V^{[2], \rho, T}_t (\varphi) & =
		\int \mathd \xi_{1 : 2} f_t^{[2], \rho, T} (\xi_1,
		\xi_2) \psi (\xi_1) \psi (\xi_2) .
	\end{aligned} \]
If $(\xi_1, \xi_2)$ is charged, then $\kappa_t (\xi_1, \xi_2) = 1$ and it follows
from $| \psi (\xi) | = 1$ combined with \eqref{eq:f2-bounds} from Lemma
\ref{lem:f2-kernel-scaling},
\[ | V^{[\ell], \rho, T}_t (\varphi) | \lesssim \interleave f^{[2]}_t
	\interleave_t \lesssim \bar{\lambda}_t^2 \langle t \rangle^{- 1} . \]
If $(\xi_1, \xi_2)$ is neutral, we have by a Taylor expansion
\begin{equation}
	\psi (\xi_1) \psi (\xi_2) = 1 + \psi (\xi_1) (x_2 - x_1)  \int_0^1 \mathd
	\vartheta \nabla_x \psi (x_1 + \vartheta (x_2 - x_1)) .
	\label{eq:exp-taylor}
\end{equation}
Therefore, choosing
\begin{equation}\label{eq:def-counterterm}
	c^{[2], \rho, T} \assign  \int \mathd \xi_{1 : 2}
	\int_0^T \mathd s \partial_s{f}_s^{[2] (0), \rho, T} (\xi_1, \xi_2),
\end{equation}
it follows that
\begin{align*}
	V^{[2] (0), \rho, T}_T (\varphi) = &
	\int_{(\mathbb{R}^2)^2} \mathd
	x_{1 : 2} (\kappa_T f_T^{[2] (0), \rho, T}) (\xi_1, \xi_2) \kappa_T (\xi_1,
	\xi_2)^{- 1} (x_2 - x_1) \psi (\xi_1)\times                                    \\
	                                   & \times \int_0^1 \mathd \vartheta \nabla_x
	\psi (x_1 + \vartheta (x_2 - x_1)) + c^{[2], \rho, T},
\end{align*}
and thus,
\begin{equation}\label{eq:V2-bound}
	\begin{aligned}
		| V_T^{[2] (0), \rho, T} (\varphi) - c^{[2], \rho .T} | & \lesssim
		\sup_{\xi_1, \xi_2} | \kappa_T^{-1} (\xi_1, \xi_2) (x_1 - x_2) |  \interleave
		f_T^{[2], \rho, T} \interleave_T  \| \nabla \varphi \|_{L^{\infty}}                                                          \\
		                                                        & \lesssim_{\rho}  \bar{\lambda}_T^2 \langle T \rangle^{- 1}  \left(
		\left\langle T \right\rangle^{- 1 / 2} \| \nabla \varphi
		\|_{L^{\infty}} \right) .
	\end{aligned}
\end{equation}
We summarise the bounds in the following Lemma for later reference.
\begin{lemma}
	\label{lem:F2-W1-bounds}For $\ell^{\ast} = 3$, there is a choice for
	$c^{[\ell], \rho, T}$ such that for any $\rho \preceq  1$, and $T <
		\infty$,
	\begin{equation}
		\begin{aligned}
			| V_T^{\rho, T} (\varphi) | \lesssim                & \| \rho \|_{L^1}
			\sum_{\tmscript{\begin{array}{c}
						                \ell \leqslant \ell^{\ast} \\
						                n \leqslant \lfloor{\ell / 2}\rfloor
					                \end{array}}} \bar{\lambda}_t^{\ell} \langle T \rangle^{- (\ell - 1)}  \left(
			1 + \left\langle T \right\rangle^{- 1 / 2} \| \nabla
			\varphi \|_{L^{\infty}} \right)^n,                                     \\
			\| F_t^{\rho, T} (\varphi) \|_{L^{\infty}} \lesssim &
			\sum_{\tmscript{\begin{array}{c}
						                \ell \leqslant \ell^{\ast} \\
						                n \leqslant \lfloor{\ell / 2}\rfloor
					                \end{array}}} \bar{\lambda}_t^{\ell} \langle t \rangle^{- (\ell - 1)}  \left(
			1 + \left\langle t \right\rangle^{- 1 / 2} \| \nabla
			\varphi \|_{L^{\infty}} \right)^n .
		\end{aligned} \label{eq:F2-W1-bounds}
	\end{equation}
\end{lemma}
\begin{proof}
	The estimate on $V$ follows directly by inserting the bound \eqref{eq:V2-bound} and the estimates \eqref{eq:f-ell-all} on the coefficients in \eqref{eq:def-V}.
	In the same way, we can start from
	\[\|F^{[\ell]}_t(\varphi)\|\lesssim \sup_x| \int \mathd \xi_{1:\ell} f^{[\ell]}_t(\xi_{1:\ell})\big(\sum_{j=1}^{\ell}\sigma_j\delta_x(x_j)\big) \psi(\xi_{1:\ell})|,\]
	which with \eqref{eq:f-ell-all} yields the desired estimate for all terms except for $\ell=2,q=0$. For $F^{[2](0)}$, we follow the same steps as for $V^{[2](0)}$, with the important distinction that the constant term
	absorbed by \eqref{eq:def-counterterm} vanishes since $f(\xi_1,\xi_2)=f(\xi_2,\xi_1)$ which implies
	\[\int \mathd \xi_{1:\ell} \sigma_1 f_t^{[2](0)}(\xi_{1:2})\big(\sum_{j=1}^{\ell}\sigma_j\delta_x(x_j)\big)=0.\]
\end{proof}
\begin{remark}
	\label{rem:b<gtr>6}If the order of the approximation $\ell^{\ast}$ and the
	smoothing $\alpha \in (0, 1)$ is chosen appropriately ($\alpha > 1 - 2
		\delta$, according to \eqref{eq:thresholds} and Lemma
	\ref{lem:lambda-integrability}), then by modifying \eqref{eq:exp-taylor},
	one can show that these estimates generalise in the full subcritical regime
	$\beta^2 < 8 \pi$  and $0 < t < T \leqslant \infty$ to bounds of the form
	\begin{equation}
		\begin{aligned}
			| V_T^{\rho, T} (\varphi) |                   & \lesssim_{\rho}
			\sum_{\tmscript{\begin{array}{c}
						                \ell \leqslant \ell^{\ast} \\
						                n \leqslant \lfloor{\ell / 2}\rfloor
					                \end{array}}} \bar{\lambda}_t^{\ell} \langle T \rangle^{- (\ell - 1)}  \left(
			1 + \left\langle T \right\rangle^{- \alpha} \| \varphi
			\|_{B_{\infty, \infty}^{2 \alpha}} \right)^n, & \quad \rho\prec1,     \\
			\| F_t^{\rho, T} (\varphi) \|_{L^{\infty}}    & \lesssim
			\sum_{\tmscript{\begin{array}{c}
						                \ell \leqslant \ell^{\ast} \\
						                n \leqslant \lfloor{\ell / 2}\rfloor
					                \end{array}}} \bar{\lambda}_t^{\ell} \langle t \rangle^{- (\ell - 1)}  \left(
			1 + \left\langle t \right\rangle^{- \alpha} \| \varphi
			\|_{B_{\infty, \infty}^{2 \alpha}} \right)^n, & \quad \rho\leqslant1.
		\end{aligned} \label{eq:F-field-dep}
	\end{equation}
	However, the field dependency in the estimates \eqref{eq:F2-W1-bounds} and
	\eqref{eq:F-field-dep} means that we are currently not able to control the
	FBSDE uniformly in the UV-cut-off which restricts our analysis to the regime
	$\beta^2 < 6 \pi$. What saves our analysis in this case is the observation
	that for the FBSDEs \eqref{eq:optimal-SDE} and \eqref{eq:FBSDE-generic} the
	force $F_s$ only appears in combination with the heat kernel. Indeed, it
	turns out that the smoothing
	properties of the heat kernel are enough to recover field independent bounds for $Q_s
		F_s$ and $\mathD F_s Q_s$.
\end{remark}

\begin{remark}
	The truncated solutions still satisfy for any function $\varphi$ and any $T
		< \infty$,
	\begin{equation}
		V^{\rho, T}_T (\varphi) = \int_{\mathbb{R}^2} (\lambda_T \cos (\beta
		\varphi) - c^{\rho, T}) \rho (x) \mathd x, \quad F^{\rho, T}_t (\varphi)
		(x) = - \beta \lambda_T \rho (x) \sin (\beta \varphi (x)) . \label{eq:FTT}
	\end{equation}
	Here, the renormalisation constant $\lambda_t = \lambda
		\mathe^{\frac{\beta^2}{2} G_t (0)}$ is the usual Wick-ordering and $c^{\rho,
				T} \assign \sum_{\ell} c^{[\ell], \rho, T}$ is the additive renormalisation
	resulting from higher order corrections.
\end{remark}

\subsection{Estimates on the force}\label{sec:F-estimates}

From now on we will always assume that $\beta^2 < 6 \pi$ and that in the
definition~\eqref{eq:def-k} of the kernel $\kappa_t$ we fix $c \in (0, 1 / 4)$
sufficiently close to $1 / 4$ and $\alpha = 1 / 2$. Since we only deal with
the case $\ell^{\ast} = 3$, let us also agree to suppress the dependence on
$\ell^{\ast}$ for $V, F$ and $H$ writing e.g. \ $F \assign F^{[\leqslant
					\ell^{\ast}]} = F^{[\leqslant 3]}$. Our goal in this section is to recover
field independent bounds on all coefficients of \eqref{eq:FBSDE-leq3}, that is
on $Q_s F_s$, $\mathD F_s Q_s$ and $H_s$.

\begin{lemma}
	\label{lem:F2-uniform-bounds}For any $\varphi \in \mathcal{S}'
		(\mathbb{R}^2)$,
	\[ \| Q_t F_t^{[2] (0)} (\varphi) \|_{L^{\infty}} \lesssim (\bar{\lambda}_t
		\langle t \rangle^{- 1})^2 . \]
\end{lemma}

\begin{proof}
	We follow exactly the same strategy as in the proof of
	Lemma~\ref{lem:f3-kernel-scaling}, where we now require bounds on
	\begin{equation}
		\sup_{x \in \mathbb{R}^2}  \int \mathd \xi_1 \int \mathd \xi_2 | \kappa_t
		f_t^{[2] (0)} (\xi_1, \xi_2) |  \frac{| Q_t (x_1 - x) - Q_t (x_2 - x) |}{|
			x_1 - x_2 |} \mathe^{- ct | x_1 - x_2 |^2} . \label{eq:QF2}
	\end{equation}
	Thanks to the translation invariance, we can apply
	Lemma~\ref{lem:hk-weighted-Lipschitz} for $Q_t$ and absorb the increment $|
		x_1 - x_2 |^{- 1}$,
	\begin{equation}
		\frac{| Q_t (x_1 - x) - Q_t (x_2 - x) |}{| x_1 - x_2 |} \mathe^{- ct | x_1
			- x_2 |^2} \lesssim t (| x_1 | + | x_1 - x_2 |) \mathe^{- \frac{c}{2} t |
			x_1 |^2} \mathe^{- m^2 / 2 s} . \label{eq:d12Q}
	\end{equation}
	Using this in~\eqref{eq:QF2} we get from Young's convolution inequality and
	the scaling properties of the kernels ${f}^{[2]}$ (see
	Lemma~\ref{lem:f2-kernel-scaling}),
	\begin{align*}
		\| Q_t F_t^{[2] (0)} (\varphi) \|_{L^{\infty}} \lesssim & t \int
		\mathd \xi_1 \int \mathd \xi_2 (| x_1 | + | x_1 - x_2 |) t^{- 1 / 2}
		\mathe^{- \frac{c}{2} t | x_1 |^2 - m^2 / 2 t} | \kappa_t f_t^{[2] (0)}
		(\xi_1, \xi_2) |                                                                                                             \\
		\lesssim                                                & t^{1 / 2} \int \mathd x_1 \mathe^{- \frac{c}{2} t | x_1
			|^2 - m^2 / 2 t} \sup_{\xi_1}  \int \mathd \xi_2 | x_1 - x_2 | | \kappa_t
		f_t^{[2] (0)} (\xi_1, \xi_2) |                                                                                               \\
		                                                        & +  t^{1 / 2} \int \mathd x_1 | x_1 | \mathe^{- \frac{c}{2} t | x_1
			|^2 - m^2 / 2 t} \sup_{\xi_1}  \int \mathd \xi_2 | \kappa_t f_t^{[2] (0)}
		(\xi_1, \xi_2) |                                                                                                             \\
		\lesssim                                                & \bar{\lambda}_t^2 \langle t \rangle^{- 2} .
	\end{align*}

\end{proof}

For the remaining levels, the estimates on the coefficients transfer directly
to the force. To remove the cut-offs later, we will also have to control the
dependence of the approximate solution $F$ on these parameters. Therefore, let
us again keep track of this dependence by writing $F^T$ for the solution to
the flow equation on $[0, T]$ with terminal
conditions~\eqref{eq:fe-terminal-conditions} at $T$ and in the same way
$F^{\rho}$ for $\rho \preceq  1$. In the estimates it is assumed that the
suppressed parameters coincide.

\begin{proposition}
	\label{prop:force-estimates}$E = L^{\infty}$ or $E = L^{2, k}$ for any $k \in \mathbb{Z}$. For any $\varphi \in \mathcal{S}'
		(\mathbb{R}^2)$, $R \in L^{\infty} (\mathbb{R}^2) $, $\rho, \rho_1, \rho_2
		\leqslant 1$ and $T, T_1, T_2 < \infty$, the following estimates apply.
	\begin{enumeratealpha}
		\item \label{prop:GF-bounds}\tmtextbf{(Uniform boundedness)} For any $p\in[1,\infty]$,
		\begin{equation}
			\begin{aligned}
				\| Q_t F_t (\varphi) \|_{L^\infty}          & \lesssim\bar{\lambda}_t
				\langle t \rangle^{- 1},                                                         \\
				\| \mathD F_t (\varphi) Q_t R \|_{L^\infty} & \lesssim
				\bar{\lambda}_t \langle t \rangle^{- 1} \| R \|_{L^\infty},                      \\
				\| H_t (\varphi) \|_{L^\infty}              & \lesssim(\bar{\lambda}_t \langle t
				\rangle^{- 1})^4 .
			\end{aligned} \label{eq:GF-bdd}
		\end{equation}
		\item \label{prop:GF-Lipschitz}\tmtextbf{(Uniform Lipschitz condition)}
		\begin{equation}
			\begin{aligned}
				\| Q_t F_t  (\varphi) - Q_t F_t (\tilde{\varphi})
				\|_E                                          & \lesssim  \bar{\lambda}_t \langle t \rangle^{- 1} (\| \varphi -
				\tilde{\varphi} \|_E),                                                                                          \\
				\| (\mathD F_t (\varphi) Q_t - \mathD F_t
				(\tilde{\varphi}) Q_t) R \|_E                 & \lesssim  \bar{\lambda}_t \langle t
				\rangle^{- 1} (\| \varphi - \tilde{\varphi} \|_E  \| R
				\|_{L^{\infty}}),                                                                                               \\
				\| H_t (\varphi) - H_t (\tilde{\varphi}) \|_E & \lesssim  (\bar{\lambda}_t
				\langle t \rangle^{- 1})^4 \| \varphi - \tilde{\varphi} \|_E                                                    \\
			\end{aligned} \label{eq:GF-Lipschitz}
		\end{equation}
		\item \tmtextbf{(Dependence on $T$)}\label{prop:fe-T-dependence} There is
		an $\varepsilon > 0$ depending only on $\beta^2$ such that,
		\begin{equation}
			\begin{aligned}
				\sup_{t\geq0}\bar{\lambda}_t^{-1}\langle t\rangle^{1-2\varepsilon}\| Q_t (F^{T_1}_t - F_t^{T_2}) (\varphi) \|_{L^\infty} & \lesssim
				\bar{\lambda} \langle T_1 \wedge T_2 \rangle^{- \varepsilon},                                                                                                                       \\
				\sup_{t\geq0}\bar{\lambda}_t^{-1}\langle t\rangle^{1-2\varepsilon}\| (\mathD F^{T_1}_t - \mathD F_t^{T_2}) (\varphi) Q_t R
				\|_{L^\infty}                                                                                                            & \lesssim \bar{\lambda} \langle T_1 \wedge T_2 \rangle^{-
				\varepsilon}  \| R \|_{L^\infty},                                                                                                                                                   \\
				\sup_{t\geq0}\bar{\lambda}_t^{-1}\langle t\rangle^{1-2\varepsilon}\| H^{T_1}_t - H_t^{T_2} (\varphi) \|_{L^\infty}       & \lesssim
				\bar{\lambda}^4 \langle T_1 \wedge T_2 \rangle^{- \varepsilon} .
			\end{aligned} \label{eq:fe-T-dependence}
		\end{equation}
		\item \tmtextbf{(Dependence on $\rho$)} \label{prop:fe-rho-dependence}For
		any $n > 2$, it holds that
		\begin{equation}
			\begin{aligned}
				\| Q_t (F^{\rho_1}_t - F_t^{\rho_2}) (\varphi) \|_{L^{2, - n}} &
				\lesssim \bar{\lambda}_t \langle t \rangle^{- 1} \| \rho_1 - \rho_2
				\|_{L^{2, - n}},                                                                                                     \\[2ex]
				\| (\mathD F^{\rho_1}_t - \mathD F_t^{\rho_2}) (\varphi) Q_t R
				\|_{L^{2, - n}}                                                & \lesssim \bar{\lambda}_t \langle t \rangle^{- 1} \|
				\rho_1 - \rho_2 \|_{L^{2, - n}}  \| R \|_{L^{\infty}},                                                               \\
				\| (H^{\rho_1}_t - H_t^{\rho_2}) (\varphi) \|_{L^{2, - n}}     & \lesssim
				(\bar{\lambda}_t \langle t \rangle^{- 1})^4 \| \rho_1 - \rho_2 \|_{L^{2, -
								n}} .
			\end{aligned} \label{eq:fe-rho-dependence}
		\end{equation}
	\end{enumeratealpha}
	In particular, for $E=L^\infty$ or $E=L^{2,-n}$ we obtain estimates that are uniform in the volume cut-off $\rho$.
\end{proposition}

\begin{proof}

	\begin{enumeratealpha}
		\item For all contributions other than for $F^{[2] (0)}$ this follows
		directly from the bounds on the kernels $f$ and $| \psi (\xi_{1 : \ell}) |
			\leqslant 1$. In fact, in these cases, we obtain the better bound,
		\[ \| F_t^{[\ell]} (\varphi) \|_{L^{\infty}} \lesssim \bar{\lambda}_t^{\ell}
			\langle t \rangle^{- (\ell - 1)}, \]
		so that
		\[ \| Q_t F_t^{[\ell]} (\varphi) R \|_{L^{\infty}} \leqslant \| Q_t
			\|_{L^1} \| F^{[\ell]}_t (\varphi) \|_{L^{\infty}} \| R \|_{L^{\infty}}
			\lesssim \bar{\lambda}_t^{\ell} \langle t \rangle^{- (\ell - 1)} \| R
			\|_{L^{\infty}} \lesssim \bar{\lambda}_t . \]

		For $F^{[2] (0)}_t$, this bound was shown in
		Lemma~\ref{lem:F2-uniform-bounds}. For the derivative, note that for any
		test function $R \in L^{\infty} (\mathbb{R}^2)$
		\[ \begin{aligned}
				\| \mathD F^{[\ell]}_t (\varphi) Q_t R \|_{L^{\infty}} & \lesssim
				\sup_{\xi_1} \left| \int \mathd \xi_{2 : \ell} \int \mathd
				yf^{[\ell]}_t (\xi_{1 : \ell}) \sum_{k \leqslant \ell} i \beta
				\sigma_k Q_t (x_k - y) R (y) \psi (\xi_{1 : \ell}) \right|                                                                \\
				                                                       & \lesssim  \| R \|_{L^{\infty}} \sup_{\xi_1} \int \mathd \xi_{2 :
					\ell} \int \mathd y \left| f^{[\ell]}_t (\xi_{1 : \ell}) \sum_{k
					\leqslant \ell} i \beta \sigma_k Q_t (x_k - y) \right|,
			\end{aligned} \]
		and now the same reasoning as for $F_t$ applies for the integral on the
		right-hand side.

		Finally, the estimates on $H$ follow from the estimates above and
		\[ \begin{aligned}
				\| H_t (\varphi) \|_{L^{\infty}} & \leqslant  \frac{1}{2}
				\sum_{\ell' + \ell'' > 3} \| \mathD (F^{[\ell]}_t  \dot{G}_t
				F^{[\ell'']}_t) (\varphi) \|_{L^{\infty}}                                                          \\
				                                 & \lesssim  \sum_{\ell' + \ell'' > 3} \| \mathD F_t^{[\ell']} Q_t
				(Q_t F^{[\ell'']}_t) \|_{L^{\infty}} \lesssim \sum_{\ell' + \ell'' >
					3} (\bar{\lambda}_t \langle t \rangle^{- 1})^{\ell' + \ell''} .
			\end{aligned} \]
		\item The Lipschitz bounds follow as above in part \ref{prop:GF-bounds},
		combined with the observation that thanks to the boundedness of the
		complex exponential fields it holds that, writing $\tilde{\psi} (\xi) =
			\tilde{\psi} (\sigma, x) = \mathe^{i \sigma \beta \tilde{\varphi} (x)}$,
		\[ | \psi (\xi_{1 : \ell}) - \tilde{\psi} (\xi_{1 : \ell}) | \leqslant
			\sum_k | \psi (\xi_k) - \tilde{\psi} (\xi_k) | \leqslant \sum_k |
			\varphi (\xi_k) - \tilde{\varphi} (\xi_k) |. \]
		\item This follows from Lemma~\ref{lem:coeff-T-dependence} in the same way
		as Proposition \ref{prop:stability}-\ref{prop:GF-bounds} followed from
		Lemma~\ref{lem:f2-kernel-scaling} and~\ref{lem:f3-kernel-scaling}.

		\item We only show the estimates on $Q F$ as the others are a direct
		consequence as illustrated in the proof of part \ref{prop:GF-bounds}.
		Again, except for $F^{[2] (0)}_t$, these estimates follow immediately from
		the convolution inequalities (see Lemma \ref{lem:hk-scaling}),
		\[ \begin{aligned}
				\| Q_t (F_t^{[\ell], \rho_1} - F^{[\ell], \rho_2}_t) (\varphi)
				\|_{L^{2, - n}} & \leqslant  \| Q_t \|_{L^1} \| (F_t^{[\ell],
				\rho_1} - F^{[\ell], \rho_2}_t) (\varphi) \|_{L^{2, - n}}                      \\
				                & \lesssim  \langle t \rangle^{- 1} \| (F_t^{[\ell], \rho_1} -
				F^{[\ell], \rho_2}_t) (\varphi) \|_{L^{2, - n}},
			\end{aligned} \]
		and
		\begin{align*}
			\| (F_t^{[\ell], \rho_1} - F^{[\ell], \rho_2}_t) (\varphi) \|_{L^{2,
			- n}}^2 & =  \left\| \int \mathd \xi_{2 : \ell} (\rho_1 - \rho_2)
			(\xi_{1 : \ell}) f^{[\ell]}_t (\xi_{1 : \ell}) \right\|_{L^2_{x_1}
			(\langle x \rangle^{- n})}^2                                        \\
			        & \lesssim \| \rho_1 - \rho_2 \|_{L^{2, - n}}^2 \interleave
			f_t^{[\ell]} \interleave^2 .
		\end{align*}
		The remaining estimate on $F^{[2], (0)}_t$ follows in the same way as
		before, using $Q_t$ to absorb the increment via
		Lemma~\ref{lem:hk-weighted-Lipschitz} as in~\eqref{eq:QF2}
		and~\eqref{eq:d12Q} and then following the same steps as above.
	\end{enumeratealpha}
\end{proof}

\begin{remark}
	\label{rem:force-estimates}
	Proposition \ref{prop:force-estimates}-\ref{prop:fe-T-dependence}
	implies that we can define $Q_s F_s^{\infty} (\varphi)$, $\mathD
		F_s^{\infty} (\varphi) Q_s$ and $H_s^{\infty} (\varphi)$ as the
	$L^{\infty}$-limit of $Q_s F_s^T (\varphi)$, $\mathD F^T_s (\varphi) Q_s$
	and $H^T_s (\varphi)$, respectively as $T \rightarrow \infty$.
\end{remark}

\subsection{Estimates on the potential}

The same arguments we used previously for the gradient show the following
estimates on the remainder
\begin{equation}
	\mathcal{H}^{\rho, T}_t (\varphi) \assign \left( \partial_t V^{\rho, T}_t +
	\frac{1}{2} \tmop{Tr} \dot{G}_t \mathD^2 V^{\rho, T}_t - \frac{1}{2} \mathD
	V^{\rho, T}_t \dot{G_t} \mathD V^{\rho, T}_t \right)(\varphi) \label{eq:def1-H},
\end{equation}
at the level of the potential. We should emphasise that in contrast to the
estimates on $F$, they of course only apply in the finite volume, that is for
$\rho \prec 1$. The results of this section will be used only later on to recover
the variational description for the unregularised measures in
Section~\ref{sec:Var+LDP} and we invite the reader skip them at first reading.

\begin{proposition}
	\label{prop:H-estimates}Given a set $A \subset \mathbb{R}^2$, denote its
	Lebesgue measure by $| A |$. For any spatial cut-off $\rho \prec 1$, UV cut-off
	$T, T_1, T_2 \leqslant \infty$, and $\varphi, \tilde{\varphi} \in
		\mathcal{S}' (\mathbb{R}^2)$, it holds that
	\begin{enumeratealpha}
		\item \tmtextbf{\label{prop:H-Lipschitz}(Lipschitz estimates)}
		\[ | \mathcal{H}^{\rho}_t (\varphi) -\mathcal{H}^{\rho}_t
			(\tilde{\varphi}) | \lesssim | \tmop{supp} (\rho) |  (\bar{\lambda}_t
			\langle t \rangle^{- 1})^4  \| \varphi - \tilde{\varphi}
			\|_{L^{\infty}}, \quad t \in [0, T] . \]
		\item \tmtextbf{(Dependence on the regularisation)
		}\label{prop:H-r,T-dependence}There is an $\varepsilon > 0$ such that
		\[ | (\mathcal{H}_t^{\rho, T_1} -\mathcal{H}^{\rho, T_2}_t) (\varphi) |
			\lesssim | \tmop{supp} (\rho) | \langle T_1 \wedge T_2 \rangle^{-
				\varepsilon}, \quad t \in [0, T], \]
		and
		\begin{equation}
			| \mathcal{H}_t^{\rho, T} (\varphi) -\mathcal{H}^{\rho, T}_t
			(\tilde{\varphi}) | \lesssim \| \rho \|_{L^{2, - n}} \| \varphi -
			\tilde{\varphi} \|_{L^{2, n}} \label{eq:H-supp-rho} . \quad t \in [0, T]
			.
		\end{equation}
	\end{enumeratealpha}
\end{proposition}

\begin{proof}
	By the definition \eqref{eq:def1-H} of $\mathcal{H}^{\rho, T}$above, we see
	that $\mathcal{H}^{\rho, T}$ has the Fourier series representation
	\begin{equation}
		\mathcal{H}_t^{\rho, T} (\varphi) \assign \sum_{\ell = 4}^6 \int \mathd
		\xi_{1 : \ell} \rho (\xi_{1 : \ell}) h^T_t (\xi_{1 : \ell}) \psi (\xi_{1 :
			\ell}), \label{eq:def-pot-remainder}
	\end{equation}
	where we used the notation $\rho (\xi_{1 : \ell}) = \prod_{k \leqslant \ell}
		\rho (x_k)$ and for $\ell \in \{ 4, 5, 6 \}$, we defined
	\begin{equation}
		h^T_t (\xi_{1 : \ell}) \assign \frac{1}{2}
		\sum_{\tmscript{\begin{array}{c}
					I_1 \cup I_2 = [\ell]
				\end{array}}} C (| I_1 |, | I_2 |) f_t^T (\xi_{I_1}) \left[ \sum_{i \in
				I_1} \sum_{j \in I_2} \sigma_i \sigma_j  \dot{G}_t (x_i - x_j) \right]
		f_t^T (\xi_{I_2}), \label{eq:def-h-coefficients}
	\end{equation}
	for a positive combinatorial constant $C (| I_1 |, | I_2 |)$. Moreover, by
	Remark \ref{rem:h-estimates}, we have the estimate
	\begin{equation}
		\interleave h^{[\ell], T}_t \interleave \lesssim (\bar{\lambda}_t \langle t
		\rangle^{- 1})^{\ell} . \label{eq:h-coefficients-scaling}
	\end{equation}
	\begin{enumeratealpha}
		\item This follows directly from the definition of $\mathcal{H}^{\rho}$,
		the bounds on the kernels~\eqref{eq:h-coefficients-scaling} and simple
		rearrangements, using again the boundedness of $| \psi (\xi) | = 1$,
		\[ \begin{aligned}
				          & \left| \sum_{\ell = 4}^6 \int \mathd \xi_{1 : \ell} \rho (\xi_{1 :
					\ell}) h^T_t (\xi_{1 : \ell}) (\psi (\xi_{1 : \ell}) - \tilde{\psi}
				(\xi_{1 : \ell})) \right|                                                      \\
				\leqslant & \sum_{\ell = 4}^6 \int \mathd x_1 \rho (x_1)  \int \mathd
				\xi_{2 : \ell} h^T_t (\xi_{1 : \ell}) | \psi (\xi_{1 : \ell}) -
				\tilde{\psi} (\xi_{1 : \ell}) |                                                \\
				\leqslant & | \tmop{supp} (\rho) |  \sum_{\ell = 4}^6 \sup_{\xi_1}
				\int \mathd \xi_{2 : \ell} | h^T_t (\xi_{1 : \ell}) |  \| \varphi -
				\tilde{\varphi} \|_{L^{\infty}},                                               \\
				\lesssim  & | \tmop{supp} (\rho) |  \| \varphi - \tilde{\varphi}
				\|_{L^{\infty}}  \sum_{\ell = 4}^6 \bar{\lambda}_t^{\ell} \langle t
				\rangle^{- \ell} .
			\end{aligned} \]
		\item This proof is in complete analogy to Proposition
		\ref{prop:force-estimates}-\ref{prop:fe-T-dependence}
		and~\ref{prop:fe-rho-dependence} using the same ideas as in Proposition
		\ref{prop:H-estimates}-\ref{prop:H-Lipschitz} above.
	\end{enumeratealpha}
\end{proof}

\section{ Analysis of the FBSDE}\label{sec:FBSDE}

With good approximate solutions to the flow under the conditional expectation
and the renormalisation sorted out, we can return to the FBSDE
\begin{equation}
	\begin{cases}
		X_t^{\rho, T, g} = \varphi + W_t - \int_0^t \dot{G}_s (F^{\rho, T}_s
		(X_s^{\rho, T, g}) + R_s^{\rho, T, g}) \mathd s, \\
		R^{\rho, T, g}_t =\mathbb{E}_t \left[ \nabla g (X_T^g) + \int_t^T H^{\rho,
					T}_s (X_s^{\rho, T, g}) \mathd s - \int_t^T \mathD F^{\rho, T}_s
			(X_s^{\rho, T, g})  \dot{G}_s R_s^{\rho, T} \mathd s \right],
	\end{cases} \label{eq:FBSDE-leq3}
\end{equation}
with $F$ and $H$ as defined in Section~\ref{sec:definitions}.

We will often work with $Z^{\rho, T, g} \assign X_t^{\rho, T, g} - (W_t +
	\varphi)$ directly to obtain deterministic bounds on the drift $Z$. To lighten
the notation, we leave the dependence of the solution on $T$ and $\rho$
implicit whenever possible and fix the perturbation $g$. Unless
explicitly stated otherwise, all estimates are uniform in the parameters
$\rho$ and $T$.

We will furthermore always implicitly assume that the solution
to~\eqref{eq:FBSDE-leq3} is extended to the positive half line $[0, \infty)$
in the standard way, that is
\[ (X_t^T, R_t^T) \assign (X_{T \wedge t}^T, R_{T \wedge t}^T), \quad t \in
	[0, \infty) . \]
\subsection{Well-posedness for the FBSDE}

As a first step, we show well-posedness for the FBSDE~\eqref{eq:FBSDE-leq3}
with the regularisations in place. We follow a standard Picard-iteration for
the solution map $\Gamma (z) = Z^z$ defined by
\begin{equation}
	\begin{cases}
		Z^z_t = -\int_0^t \mathd s \dot{G}_s (F_s (z_s + W_s) + R^z_s), \\
		R^z_t =\mathbb{E}_t \left[ \nabla g (z_s + W_T) + \int_t^T \mathd s H_s
			(z_s + W_s) - \int_t^T \mathd s \mathD F_s (z_s + W_s)  \dot{G}_s R^z_s
			\right] .
	\end{cases} \label{eq:FBSDE-sol-map}
\end{equation}
Standard well-posedness for decoupled Lipschitz FBSDEs ensures the existence
of a unique solution $(Z^z, R^z) \in \mathbb{H}^{\infty}_T (L^{\infty}) \times
	\mathbb{H}^{\infty}_T (L^{\infty})$ to \eqref{eq:FBSDE-sol-map} for any $z \in
	\mathbb{H}^{\infty}_T (L^{\infty})$. The only term in~\eqref{eq:FBSDE-sol-map}
that cannot be estimated in a linear fashion immediately from Proposition
\ref{prop:force-estimates} is the term $\mathD F_s (z_s + W_s)  \dot{G}_s R^z_s$
in the backward equation. The next Lemma ensures that also this term stays
bounded and does not cause any issues.

\begin{lemma}
	\label{lem:B-to-B}For all in $z \in \mathbb{H}^{\infty}_T (L^{\infty})$ and
	$\lambda \in \mathbb{R}$ (not necessarily small), there are constants \(C_{\bar{\lambda}}, C_{g, \bar{\lambda}}\) non-decreasing in \(\bar{\lambda}\)  
	\[ \sup_t \| R^z_t \|_{L^{\infty}} \lesssim C_{\bar{\lambda}}( | g |_{1, \infty} +
		\bar{\lambda}_t^4 \langle t \rangle^{- 3} )< \infty, \quad \text{and} \quad
		\sup_t  \| Z^z_t \|_{L^{\infty}} \leqslant C_{g, \bar{\lambda}} \]
\end{lemma}

\begin{proof}
	From the definition of $R^z$, the regularity of $W_t \in L^{2, - n}$ for any
	$t < \infty$ and the bounds on the flow from Proposition
	\ref{prop:force-estimates}-\ref{prop:GF-bounds},
	\[ \begin{aligned}
			\| R_t^z \|_{L^{\infty}} & \leqslant  \bar{\lambda} | g |_{1, \infty} +
			\int_t^T \mathd s \| H_s (z_s + W_s) \|_{L^{\infty}} + \bar{\lambda} \int_t^T
			\mathd s \| \mathD F_s (z_s)  \dot{G}_s R^z_s \|_{L^{\infty}},                                             \\
			                         & \lesssim  \bar{\lambda} | g |_{1, \infty} + \int_t^T \mathd s \bar{\lambda}_s^4
			\langle s \rangle^{- 4} + \int_t^T \mathd s \bar{\lambda}_s \langle s
			\rangle^{- 2} \| R^z_s \|_{L^{\infty}} .
		\end{aligned} \]
	By a backward version of Gronwall's inequality, this implies with $4 \delta >
		1$ (see Lemma~\ref{lem:lambda-integrability}),
	\begin{equation}
		\| R_t^z \|_{L^{\infty}} \lesssim \left( \bar{\lambda} | g |_{1, \infty} +
		\int_t^T \mathd s \bar{\lambda}_s^4 \langle s \rangle^{- 4} \right)
		\mathe^{\bar{\lambda} \int_t^T \mathd s \langle s \rangle^{- 1 - \delta}}
		\lesssim \bar{\lambda} | g |_{1, \infty} + \bar{\lambda}_t^4 \langle t \rangle^{- 3} .
		\label{eq:exp-R}
	\end{equation}
	Thus, using the bound just derived for $R^z$ in the equation for $Z^z$,
	\[ \| Z^z_t \| \leqslant \int_0^t \mathd s \| \dot{G}_s (F_s (z_s + W_s) +
		R^z_s) \|_{L^{\infty}} \lesssim \int_0^{\infty} \mathd s \langle s
		\rangle^{- 2} (\bar{\lambda}_s + \| R_s^z \|_{L^{\infty}}) \lesssim \bar{\lambda} C_g
		. \]
\end{proof}

With this issue resolved, we are in a position to show
that~\eqref{eq:FBSDE-sol-map} defines a contraction on $H^{\infty}_T
	(L^{\infty})$.

\begin{proposition}
	\label{prop:contraction}For $\bar{\lambda}$ sufficiently small, the map $\Gamma :
		\mathbb{H}^{\infty}_T (L^{\infty}) \rightarrow \mathbb{H}^{\infty}_T
		(L^{\infty}) ; z \mapsto Z^z$ is a contraction.
\end{proposition}

Since $Z$ uniquely determines the solution $(X, R)$ to~\eqref{eq:FBSDE-leq3}
via
\[ Z \mapsto (\varphi + Z_t + W_t, R^Z_t)_{t \geqslant 0}, \]
this immediately implies the existence of a unique solution to
\eqref{eq:FBSDE-leq3} when combined with the regularity of the stopped
Brownian motion $(W_t)_{t \in [0, T]} \in \mathbb{H}^2_T (L^{2, - n})$.

\begin{corollary}
	\label{cor:well-posedness}For any $\rho \preceq  1, T < \infty$, $\bar{\lambda}$ sufficiently small and $p \in
		[1, \infty)$ the FBSDE~\eqref{eq:FBSDE-leq3} has a unique solution
	\[ (X_t, R_t)_{t \geqslant 0} = (\varphi + Z_t + W_t, R_t)_{t \geqslant 0}
		\in \mathbb{H}^p_T (L^{p, - n}) \times \mathbb{H}^{\infty}_T (L^{\infty})
		. \]
\end{corollary}

\begin{proof}[Proof of Proposition~\ref{prop:contraction}]
	Let $z_1, z_2 \in \mathbb{H}^{\infty}_T (L^{\infty})$ and consider the FBSDE
	for the difference $(\delta Z, \delta R) = (Z^{z_1}, R^{z_1}) - (Z^{z_2},
		R^{z_2})$ given by
	\begin{equation}
		\begin{cases}
			\delta Z_t = - \int_0^t \mathd s \dot{G}_s (F_s (X^{z_1}_s) - F_s
			(X^{z_2}_s) + \delta R_s) \\
			\delta R_t =\mathbb{E}_t \left[ \delta_z \nabla g + \int_t^T \mathd s
				\delta_z H_s - \int_t^T \mathd s (\mathD F_s (X^{z_1}_s) \dot{G_s}
				R^{z_1}_s - \mathD F_s (X^{z_2}_s) \dot{G_s} R^{z_2}_s) \right],
		\end{cases} \label{eq:FBSDE-spread}
	\end{equation}
	where we use the shorthand $X_s^z = \varphi + z_s + W_s$ so that $\delta X_s
		= \delta z_s$ and
	\[ \delta_z \nabla g \assign \nabla g (X^{z_1}_s) - \nabla g (X^{z_2}_s),
		\quad \delta_z H_s \assign H_s (X^{z_1}_s) - H_s (X^{z_2}_s) . \]
	To deal with the bilinear term in the backward equation, we combine the
	estimates from Proposition \ref{prop:force-estimates}-\ref{prop:GF-bounds}
	with the boundedness of $R^z$ provided by Lemma~\ref{lem:B-to-B} to conclude
	\begin{align*}
		\| (\mathD F_s (X^{z_1}_s) - \mathD F_s (X^{z_2}_s)) \dot{G_s} R^{z_2}_s
		\|_{L^{\infty}}
		 & \lesssim \| R^{z_2}_s \|_{L^{\infty}} \| [\mathD F_s
		(X^{z_1}_s) - \mathD F_s (X^{z_2}_s)]  \dot{G}_s \|_{L^{\infty}} \\
		 & \lesssim
		\bar{\lambda}_s \langle s \rangle^{- 2} \| \delta z_s \|_{L^{\infty}} .
	\end{align*}
	The remaining terms in the backward equation can all be estimated directly
	using Proposition \ref{prop:force-estimates}-\ref{prop:GF-bounds} and the
	Lipschitz continuity of $\nabla g$,
	\begin{align*}
		\| \delta R_t \|_{L^{\infty}}\lesssim \int_t^T \mathd s \bar{\lambda}_s^4 \langle s \rangle^{- 4} \|
		\delta z_s \|_{L^{\infty}} + \int_t^T \mathd s \bar{\lambda}_s \langle s
		\rangle^{- 2} \| \delta R \|_{L^{\infty}} + \int_t^T \mathd s \bar{\lambda}_s
		\langle s \rangle^{- 2} \| \delta z_s \|_{L^{\infty}},
	\end{align*}
	which implies by Gronwall's inequality,
	\[ \sup_t \| \delta R_t \|_{L^{\infty}} \leqslant C \bar{\lambda} \sup_t \| \delta
		z_t \|_{L^{\infty}} . \]
	Using this estimate on $R$ in the equation for the forward component with
	the Lipschitz estimates from Proposition
	\ref{prop:force-estimates}-\ref{prop:GF-bounds}, we obtain
	\[ \begin{aligned}
			\sup_t \| \delta Z_t \|_{L^{\infty}} & \leqslant  C \int_0^T \mathd s
			\| \dot{G}_s (F_s (X^{z_1}_s) - F_s (X^{z_2}_s) + \delta R_s)
			\|_{L^{\infty}}                                                                                             \\
			                                     & \leqslant  C \bar{\lambda} \int_0^T \mathd s \langle s \rangle^{- 2}
			\bar{\lambda}_s \| \delta z_s \|_{L^{\infty}} + \int_0^T \mathd s \langle s
			\rangle^{- 2} \| \delta R \|_{L^{\infty}}                                                                   \\
			                                     & \leqslant  C \bar{\lambda} \sup_s \| \delta z_s \|_{L^{\infty}} + C
			\sup_s \| \delta R_s \|_{L^{\infty}}                                                                        \\
			                                     & \leqslant  C \bar{\lambda} \sup_s \| \delta z_s \|_{L^{\infty}},
		\end{aligned} \]
	which yields the required contraction for $\bar{\lambda}$ small enough.
\end{proof}

\begin{remark}
	\label{rem:FBSDE-6p}We crucially rely on the uniform, field independent
	estimates on the approximate force obtained in
	Proposition~\ref{prop:force-estimates} which hold only up to $6 \pi$.
	Indeed, assuming only a weaker estimate like~\eqref{eq:F-field-dep}, it is
	unclear how to obtain suitable replacements for the a priori estimates of
	Lemma~\ref{lem:B-to-B} and rule out explosion in finite time. Even linear
	growth in $\mathD F_t (\varphi)$ would require an additional argument as any
	trivial estimate for the backward equation results in an exponential
	dependence on $\| Z \|_{L^{\infty}}$ in the equation for the remainder
	through \eqref{eq:exp-R}.
\end{remark}

\subsection{Stability properties}

In this section, we show that the FBSDE~\eqref{eq:FBSDE-leq3} is stable in both regularisations $\rho$ and $T$,
provided that the coupling constant $\bar{\lambda}$ is chosen sufficiently small. We
summarise these properties below.

\begin{proposition}
	\label{prop:stability}For $\bar{\lambda}$ sufficiently small and $n > 2$ (so
	that $x \mapsto \langle x \rangle^{- n} \in L^1 (\mathbb{R}^2)$), the
	following stability estimates hold.
	\begin{enumeratealpha}
		\item \tmtextbf{Dependence on the spatial cut-off:}
		\label{prop:rho-convergence}Let $\rho_1, \rho_2 \leqslant 1$ and denote
		the associated solutions to \eqref{eq:FBSDE-leq3} by $(Z^{\rho_1},
			R^{\rho_1})$ and $(Z^{\rho_2}, R^{\rho_2})$respectively. Then, the
		difference between the solution $(\delta_{\rho} Z, \delta_{\rho} R)
			\assign (Z^{\rho_1}, R^{\rho_1}) - (Z^{\rho_2}, R^{\rho_2})$ satisfies
		\[ \sup_t \| \delta_{\rho} Z_t \|_{L^{2, - n}} + \sup_t \| \delta_{\rho}
			R_t \|_{L^{2, - n}} \lesssim \bar{\lambda} \| \rho_1 - \rho_2 \|_{L^{2, - n}}
			. \]

		\item \tmtextbf{Dependence on the UV cut-off:}\label{prop:T-convergence}
		Let $T_1, T_2 < \infty$ and denote the associated solutions
		to~\eqref{eq:FBSDE-leq3} by $(Z^{T_1}, R^{T_1})$ and $(Z^{T_2}, R^{T_2})$
		respectively. Then, the difference between the solution $(\delta_T Z,
			\delta_T R) \assign (Z^{T_1}, R^{T_1}) - (Z^{T_2}, R^{T_2})$ satisfies for
		some $\varepsilon > 0$,
		\[ \sup_t \| \delta_T Z_t \|_{L^{\infty}} + \sup_t \| \delta_T R_t
			\|_{L^{\infty}} \lesssim \langle T \rangle^{- \varepsilon} . \]
		\item \tmtextbf{Dependence on local
			perturbations:}\label{prop:local-perturbations} Let $(Z^{\rho, T, g},
			R^{\rho, T, g})$ be the unique solution to~\eqref{eq:FBSDE-leq3}. It holds
		that
		\[ \sup_t \| Z^{\rho, T, g}_t - Z_t^{\rho, T, 0} \|_{L^{2, n}} + \sup_t \|
			R_t^{\rho, T, g} - R_t^{\rho, T, 0} \|_{L^{2, n}} \lesssim \bar{\lambda} |
			g |_{1, 2, n} . \]
	\end{enumeratealpha}
\end{proposition}

\begin{proof}

	\begin{enumeratealpha}
		\item We follow essentially the same argument as before for the proof of
		Proposition \ref{prop:contraction}, writing the FBSDE for the difference
		as
		\begin{equation}
			\begin{cases}
				\delta_{\rho} Z_t = - \int_0^t \mathd s \dot{G}_s (F_s^{\rho_1}
				(X^{\rho_1}_s) - F_s^{\rho_2} (X_s^{\rho_2}) + \delta_{\rho} R_s) \\
				\delta_{\rho} R_t =\mathbb{E}_t \left[ \delta_{\rho} \nabla g +
					\int_t^T \mathd s \left(\delta_{\rho} H_s + \mathD
					F_s^{\rho_1} \left( X_s^{\rho_1} \right)  \dot{G}_s R_s^{\rho_1} -
					\mathD F_s^{\rho_2} (X_s^{\rho_2})  \dot{G}_s R_s^{\rho_2} \right)
					\right]
			\end{cases} \label{eq:FBSDE-rho-spread}
		\end{equation}
		where $\delta_{\rho} \nabla g = \nabla g (X^{\rho_1}_T) - \nabla g
			(X^{\rho_1}_T)$ and $\delta_{\rho} H_s = H_s^{\rho_1} (X^{\rho_1}_s) -
			H_s^{\rho_2} (X^{\rho_2}_s)$. Using the estimates from Proposition
		\ref{prop:force-estimates}-\ref{prop:fe-rho-dependence} in the FBSDE for
		the difference~\eqref{eq:FBSDE-rho-spread}, we obtain
		\[ \begin{aligned}
				\sup_t \| \delta_{\rho} Z_t \|_{L^{2, - n}} & \lesssim  \int_0^T
				\mathd s \bar{\lambda}_s \langle s \rangle^{- 2} (\| \delta_{\rho} Z_s
				\|_{L^{2, - n}} + \| \rho_1 - \rho_2 \|_{L^{2, - n}} + \|
				\delta_{\rho} R_s \|_{L^{2, - n}})                                                                                   \\
				                                            & \lesssim  \bar{\lambda} (\sup_s \| \delta_{\rho} Z_s \|_{L^{2, - n}} +
				\| \rho_1 - \rho_2 \|_{L^{2, - n}}) + \sup_s \| \delta_{\rho} R_s
				\|_{L^{2, - n}},
			\end{aligned} \]
		and
		\[ \begin{aligned}
				\sup_t \| \delta_{\rho} R_t \|_{L^{2, - n}} \lesssim & \int_0^T
				\mathd s \bar{\lambda}_s \langle s \rangle^{- 2} (\| \delta_{\rho} Z_s
				\|_{L^{2, - n}} + \| \rho_1 - \rho_2 \|_{L^{2, - n}})                                                                  \\
				                                                     & + \int_0^T \mathd s \bar{\lambda}_s \langle s \rangle^{- 2} [\|
					\rho_1 - \rho_2 \|_{L^{2, - n}} + \| \delta_{\rho} Z_s \|_{L^{2, -
				n}} + \| \delta_{\rho} R_s \|_{L^{2, - n}}]                                                                            \\
				\lesssim                                             & \bar{\lambda} (\sup_s \| \delta_{\rho} Z_s \|_{L^{2, - n}} +
				\sup_s \| \delta_{\rho} R_s \|_{L^{2, - n}} + \| \rho_1 - \rho_2
				\|_{L^{2, - n}}),
			\end{aligned} \]
		which yields the claim after choosing $\bar{\lambda}$ sufficiently small and
		rearranging.

		\item For concreteness, let $T_2 < T_1$. The difference between the two
		solutions solves the FBSDE,
		\[ \begin{cases}
				\delta_T Z_t = - \int_0^t \mathd s \dot{G}_s (F_s^{T_1} (X^{T_1}_s) -
				F_s^{T_2} (X^{T_2}_s) + \delta_T R_s)            \\
				\delta_T R_t =\mathbb{E}_t \left[ \delta_T \nabla g +
				\mathbb{E}_t\int_t^T \delta_T H_s\mathd s\right] \\
				\phantom{\delta_T R_t =}+\int_t^T\left( -( \mathD F_s^{T_1} \left(
				X_s^{T_1} \right)  \dot{G}_s R_s^{T_1} - \mathD F_s^{T_2} (X_s^{T_2})
				\dot{G}_s R_s^{T_2}) \right) \mathd s,
			\end{cases} \]
		where $\delta_T \nabla g = \nabla g (X^{T_1}_{T_1}) - \nabla g (X^{T_2}_{T_2})$
		and $\delta_T H_s = H_s^{T_1} (X^{T_1}_s) - H_s^{T_2} (X^{T_2}_s)$. The
		only difference to the estimates before is the additional tail in the
		backward equation. Other than that we proceed as before. For the forward
		equation, splitting up the differences as in the proof of part
		\ref{prop:rho-convergence}, we have from Proposition
		\ref{prop:force-estimates}-\ref{prop:GF-bounds}
		and \ref{prop:force-estimates}-\ref{prop:fe-T-dependence} for $\varepsilon$ sufficiently small,
		\[ \begin{aligned}
				\| \delta_T Z_t \|_{L^{\infty}} & \lesssim  \bar{\lambda} \sup_s \|
				\delta_T Z_s \|_{L^{\infty}} + \| \delta_T R_s \|_{L^{\infty}} +
				\langle T_2 \rangle^{- \varepsilon}.
			\end{aligned} \]
		For the backward equation, we split the terms as
		\[ \begin{aligned}
				\delta_T R_t = & \nabla g (X^{T_1}_{T_1}) - \nabla g
				(X^{T_2}_{T_2})                                                                \\
				               & +\mathbb{E}_t  \int_t^{T_2} \mathd s (H_s^{T_1} (X^{T_1}_s) -
				H_s^{T_2} (X^{T_2}_s))                                                         \\
				               & -\mathbb{E}_t  \int_t^{T_2} \mathd s (\mathD
				F_s^{T_1} (X_s^{T_1})  \dot{G}_s R_s^{T_1} - \mathD F_s^{T_2}
				(X_s^{T_2})  \dot{G}_s R_s^{T_2})                                              \\
				               & +\mathbb{E}_t  \int_{T_2}^{T_1} \mathd s H_s^{T_1}
				(X^{T_1}_s) -\mathbb{E}_t  \int_{T_2}^{T_1} \mathd s \mathD
				F_s^{T_1} (X_s^{T_1}  \dot{G}_s R_s^{T_1}),
			\end{aligned} \]
		we proceed similarly on $[0, T_2]$ to obtain from Proposition
		\ref{prop:force-estimates}-\ref{prop:fe-T-dependence}
		and \ref{prop:force-estimates}-\ref{prop:GF-bounds},
		\[ \begin{aligned}
				\int_t^{T_2} \mathd s \| H_s^{T_1} (X^{T_1}_s) - H_s^{T_2}
				(X^{T_2}_s) \|_{L^{\infty}} & \lesssim  \bar{\lambda} \sup_s \| \delta_T
				Z_s \|_{L^{\infty}} + \bar{\lambda} \langle T_2 \rangle^{- \varepsilon},
			\end{aligned} \]
		and
		\begin{align*}
			\int_t^{T_2} & \mathd s \| \mathD F_s^{T_1} (X_s^{T_1})  \dot{G}_s
			R_s^{T_1} - \mathD F_s^{T_2} (X_s^{T_2})  \dot{G}_s R_s^{T_2}
			\|_{L^{\infty}}                                                    \\&\lesssim \bar{\lambda} (\sup_s \| \delta_T Z_s
			\|_{L^{\infty}} + \sup_s \| \delta_T R_s \|_{L^{\infty}} + \langle T_2
			\rangle^{- \varepsilon}).
		\end{align*}
		The integrals on $[T_2, T_1]$ can be estimated using just the boundedness
		of the coefficients provided by Proposition
		\ref{prop:force-estimates}-\ref{prop:GF-bounds},
		\begin{align*}
			\int_{T_2}^{T_1} \mathd s \| H_s^{T_1} (X^{T_1}_s) \|_{L^{\infty}} +
			\int_{T_2}^{T_1} \mathd s \| \mathD F_s^{T_1} (X_s^{T_1})  \dot{G}_s
			R_s^{T_1} \|_{L^{\infty}}
			 & \lesssim \bar{\lambda} \int_{T_2}^{T_1} \mathd s
			\langle s \rangle^{- 4 \delta} + \langle s \rangle^{- 1 - \delta} \\
			 & \lesssim \bar{\lambda} \langle T_2 \rangle^{- \varepsilon}.
		\end{align*}
		Finally, by the regularity assumed on $g$,
		\[ \| \nabla g (X^{T_1}_{T_1}) - \nabla g (X^{T_2}_{T_2}) \|_{L^{\infty}}
			\lesssim \bar{\lambda} | g |_{2, \infty}  \| X^{T_1}_{T_2} - X^{T_2}_{T_2}
			\|_{L^{\infty}} \lesssim \bar{\lambda} \sup_s \| \delta X_s \|_{L^{\infty}} .
		\]
		Combining all of the above yields the claim for $\bar{\lambda}$ small enough
		after rearranging.

		\item This proof is straightforward and does not require any new arguments
		compared to e.g. the proof of part \ref{prop:rho-convergence}. Indeed, now
		the FBSDE for the difference is
		\[ \begin{cases}
				\delta_g Z_t = - \int_0^t \mathd s \dot{G}_s (F_s (X^g_s) - F_s
				(X^0_s) + \delta_g R_s) \\
				\delta_g R_t = \nabla g (X^g_T) +\mathbb{E}_t  \int_t^T \mathd s
				\delta_g H_s - \mathbb{E}_t  \int_t^T \mathd s \left( \mathD F_s
				(X^g_s)  \dot{G}_s R^g_s - \mathD F_s  (X_s^0)  \dot{G}_s R_s^0
				\right),
			\end{cases} \]
		where again $\delta_g H_s = H_s (X^g_s) - H_s (X^0_s)$. The same steps as
		in part~\ref{prop:rho-convergence} and~\ref{prop:T-convergence} imply the
		claim with the Lipschitz estimates from Proposition
		\ref{prop:force-estimates}-\ref{prop:GF-Lipschitz}.
	\end{enumeratealpha}
\end{proof}

\subsection{Recovering the EQFT}

Throughout this section, we assume that $\bar{\lambda}$ is chosen small enough for
Corollary~\ref{cor:well-posedness} and Proposition \ref{prop:stability} to
apply. Then, for any $\rho \prec 1$ and $T < \infty$, we denote by $(X^{\rho, T},
	R^{\rho, T})$ the unique solution to~\eqref{eq:FBSDE-leq3}. We show the
following refined version of Theorem~\ref{thm:int-wp}.

\begin{theorem}
	\label{thm:T,rho-convergence}As the cut-offs are removed, the family $\{
		(Z^{\rho, T}, R^{\rho, T}) \}_{\rho \preceq  1, T < \infty}$ converges in
	$\mathbb{H}^2 (L^{2, - n}) \times \mathbb{H}^2 (L^{2, - n})$ to a unique
	limit $(Z, R) \in \mathbb{H}^{\infty} (L^{\infty}) \times
		\mathbb{H}^{\infty} (L^{\infty})$, that is
	\[ \lim_{\tmscript{\begin{array}{c}
					\rho \rightarrow 1 \\
					T \rightarrow \infty
				\end{array}}} \sup_t \{ \| Z^{\rho, T}_t - Z_t \|_{L^{2, - n}} + \|
		R^{\rho, T}_t - R_t \|_{L^{2, - n}} \} = 0. \]
	The limiting process is the unique solution to \eqref{eq:FBSDE-leq3} with $T=\infty, \rho=1$.
	Moreover, for any $\varepsilon > 0$, $p \in [0, \infty)$ and $\varphi \in
		B^{0 - \varepsilon, - n}_{p, p}$, there is a version of the drift process
	$Z$ with terminal value $Z_{\infty} \in L^{\infty} \left( \mathd \mathbb{P} ; \;
		B_{p, p}^{2 - \beta^2 / 4 \pi - \varepsilon, - n} \right)$, so that
	\begin{equation}
		X_{\infty} = Z_{\infty} + (\varphi + W_{\infty}) \in L^{\infty} \left(
		\mathd \mathbb{P} ; \; B_{p, p}^{2 - \beta^2 / 4 \pi - \varepsilon, - n} \right) +
		L^p (\mathd \mathbb{P} ; B_{p, p}^{0 - \varepsilon, - n}) .
		\label{eq:phi-infty-regularity}
	\end{equation}
	In particular, for any $\varepsilon > 0$, the family
	$(\nu_{\tmop{SG}}^{\rho, T})_{\rho, T}$ has a unique weak limit in $H^{-
				\varepsilon, - n}$ as $\rho \rightarrow 1$ and $T \rightarrow \infty$ which
	we denote by $\nu_{\tmop{SG}}$. It is given as a random shift of the
	Gaussian free field,
	\[ \tmop{Law} (Z_T^{\rho, T} + W_T) = \nu^{\rho, T}_{\tmop{SG}} \longrightarrow
		\nu_{\tmop{SG}} = \tmop{Law} (Z_{\infty} + W_{\infty}) . \]
\end{theorem}

For $\beta^2 < 4 \pi$, we obtain $Z_{\infty} \in H^{1 +, - n}$ and in the
finite volume, that is for $\rho \prec 1$, the same argument in the unweighted
spaces implies $Z^{\rho}_{\infty} \in H^1$. In this case, the Wick ordered cosine $\llbracket \cos (\beta W_t) \rrbracket$
converges in $H^{\alpha}$ for any $\alpha < - \beta^2 / 4 \pi\in (-1,0)$, and we can define
all products on the right-hand side of
\begin{equation}
	\llbracket \cos (\beta (Z_{\infty} + W_{\infty})) \rrbracket \assign \cos
	(\beta Z_{\infty}) \llbracket \cos (\beta W_{\infty}) \rrbracket - \sin
	(\beta Z_{\infty}) \llbracket \sin (\beta W_{\infty}) \rrbracket,
	\label{eq:trig-id}
\end{equation}
so that the partition function $\Xi^{\rho} =\mathbb{E} [\exp (- \bar{\lambda}
		V^{\rho, \infty} (Z_{\infty} + W_{\infty}))]$ stays bounded. Consequently, we
recover that the law of the shift $\nu_{\tmop{SG}}^{\rho} = \tmop{Law}
	(Z_{\infty}^{\rho} + W_{\infty})$ is absolutely continuous with respect to the
Gaussian free field $\mu = \tmop{Law} (W_{\infty})$. For $\beta^2 \geqslant 4
	\pi$, this is no longer the case (see Theorem \ref{thm:singularity} and
Theorem \ref{thm:Ueps-div}) and indeed Theorem~\ref{thm:T,rho-convergence}
only ensures the regularity $Z^{\rho}_{\infty} \in H^{2 - \beta^2 / 4 \pi -}
	(\mathbb{R}^2) = H^{1 / 2 +} (\mathbb{R}^2)$, which we conjecture to be
optimal for $\beta^2<6\pi$. This regularity no longer allows to define the products on the right-hand side of~\eqref{eq:trig-id}.
\begin{proof}[Proof of Theorem \ref{thm:T,rho-convergence}.]
	The fact that the limit exists follows from Proposition
	\ref{prop:stability}-\ref{prop:rho-convergence}
	and~\ref{prop:T-convergence}. Note that all constants are uniform in $\rho
		\leqslant 1$ and $T \leqslant \infty$, so the order in which we take the
	limits is irrelevant. Denote the limiting processes of $(Z^{\rho, T},
		R^{\rho, T})_{\rho, T}$ by $(Z, R)$ and let $X \assign \varphi + Z + W$.
	Then, the aforementioned convergence results transfer the bound from
	Lemma~\ref{lem:B-to-B} to the limit so that,
	\[ \| Z_{\infty} \|_{L^{\infty}} + \| R_{\infty} \|_{L^{\infty}} \leqslant
		\sup_t (\| Z_t \|_{L^{\infty}} + \| R_t \|_{L^{\infty}}) \lesssim 1. \]
	The convergence of $W_t$ to the Gaussian free field in $H^{0 -, - n}$ is the
	content of Lemma~\ref{lem:GFF-convergence} and
	\[ \tmop{Law} (Z_T^{\rho, T} + W_T) = \nu_{\tmop{SG}}^{\rho, T}, \]
	was already shown in Theorem \ref{thm:ctrl-FBSDE}-\ref{thm:ctrl-form}. By
	Gaussian hypercontractivity and the Besov embeddings,
	for any $p \in [0, \infty]$ there is a version
	of the free field such that $W_{\infty} \in B_{p, p}^{0 -, - n}$ holds and it
	remains to show only that the drift $Z_t$ has a terminal value $Z_{\infty}$
	with the required regularity. Thanks to Lemma \ref{lem:I(u)-regularity}, for
	any $\alpha \in (0, 1)$ and $\varepsilon > 0$,
	\begin{equation}
		\| Z_{\infty} \|_{B_{p, p}^{\alpha}} = \left\| \int_0^{\infty} \mathd s
		\dot{G}_s (F_s (X_s) + R_s) \right\|_{B_{p, p}^{\alpha, - n}} \lesssim
		\sup_s \langle s \rangle^{\alpha / 2 + \varepsilon} \| Q_s (F_s(X_s) + R_s)
		\|_{L^{\infty}} . \label{eq:Z-Besov-reg}
	\end{equation}
	Using Proposition \ref{prop:force-estimates} and Lemma \ref{lem:B-to-B} we
	know
	\[ \| Q_s (F_s + R_s) \|_{L^{\infty}} \lesssim \langle s \rangle^{- 1}
		\bar{\lambda}_s + \langle s \rangle^{- 1} \lesssim s^{- \delta} = \langle s\rangle^{\beta^2 /
			8 \pi - 1} . \]
	Therefore, we can choose $\varepsilon > 0$ small enough for
	\eqref{eq:Z-Besov-reg} to be finite provided
	\[ \alpha < 2 - \beta^2 / 4 \pi = 2 \delta . \]
\end{proof}

As a direct consequence of Theorem~\ref{thm:T,rho-convergence}, we also get
exponential moments for the limiting measure.

\begin{corollary}
	\label{cor:exp-integrability} For any $\varepsilon > 0$, there is a constant
	$\gamma > 0$ such that
	\[ \int_{\mathcal{S}' (\mathbb{R}^2)} \mathe^{\gamma \| \varphi \|_{H^{-
							\varepsilon, - n}}^2} \nu_{\tmop{SG}} (\mathd \varphi) < \infty . \]
\end{corollary}

\begin{proof}
	By Fernique's theorem (see e.g.~{\cite[Theorem
		2.8.5.]{bogachevGaussianMeasures1998}}) the Gaussian free field $W_{\infty}$
	has squared exponential moments in $H^{- \varepsilon, - n}$ for some $\gamma
		> 0$. Combined with the bounds on $Z_{\infty}$ from Theorem
	\ref{thm:T,rho-convergence},
	\[ \int_{\mathcal{S}' (\mathbb{R}^2)} \mathe^{\gamma \| \varphi \|_{H^{-
							\varepsilon, - n}}^2} \nu_{\tmop{SG}} (\mathd \varphi) \lesssim
		\mathbb{E} [\exp (\gamma \| W_{\infty} \|_{H^{- \varepsilon, - n}}^2 +
			\gamma \| Z_{\infty} \|_{L^{2, - n}}^2)] < \infty . \]
\end{proof}

For future reference, let us also note the following regularity property of
the solution.

\begin{lemma}
	\label{lem:W-Halpha}For any $\alpha \in (0, 1)$, $\varepsilon > 0$ and $p
		\in [1, \infty]$, the solution to the FBSDE \eqref{eq:FBSDE-leq3} satisfies
	\[ \mathbb{E} \left[\sup_t\, \langle t \rangle^{- \alpha / 2} \| W_t \|_{B^{\alpha
							- \varepsilon, - n}_{p, p}}\right] \lesssim 1, \quad \sup_t\,  \langle t
		\rangle^{- \alpha / 2 + \delta} \| Z_t \|_{B^{\alpha - \varepsilon, -
						n}_{p, p}} \lesssim 1. \]
	In particular, $\sup_t  \langle t \rangle^{- \alpha / 2} \| X_t
		\|_{B^{\alpha - \varepsilon, - n}_{p, p}} < \infty$ almost surely.
\end{lemma}

\begin{proof}
	For the estimate on the field $(W_t)_{t \geqslant 0}$, we refer to
	Appendix~\ref{app:pf-lem-W-Halpha}. The estimate on $Z_t$ follows in the
	same way as \eqref{eq:Z-Besov-reg}. Combining the estimates for $Z$ and $W$
	yields the estimate for $(X_t)_{t \geqslant 0}$.
\end{proof}

\subsection{Uniqueness for the finite volume measure}\label{sec:finite-vol}

The convergence to a \tmtextit{unique} measure in Theorem \ref{thm:T,rho-convergence} requires the
coupling constant $\bar{\lambda}$ to be sufficiently small to close the argument for
the coupled forward backward system. In this section, we show that the limiting measure is still unique as long as the volume remains finite. We do
not expect this to be true for large values of the coupling constant in the
infinite volume. Let us fix $g = 0$ and a volume cut-off $\rho \prec 1$ left
implicit throughout this section.

\begin{proposition}
	Let $\bar{\lambda} \in \mathbb{R}, \hspace{0.17em} T \in [0, \infty], \rho
		\prec 1$ and let $(Z,R)=(Z^{\rho}, R^{\rho})$ be any accumulation point of $(Z^{\rho,T},R^{\rho,T})_{T\in [0,\infty)}$. Then, $\tmop{Law(Z_\infty+W_\infty)}$ is unique.
\end{proposition}

\begin{proof}
	Let  O  be bounded and continuous $H^{- \varepsilon, - n} (\mathbb{R}^2)
		\rightarrow \mathbb{R}$ for some $\varepsilon > 0$. The claim follows once we show that
	$X_T^{\rho, T} = Z_T^{\rho, T} + W_T$ satisfies
	\begin{equation}
		\lim_{T \rightarrow \infty} \mathbb{E} [\mathcal{O} (X^T_T)] =
		\frac{\mathbb{E} \left[ \mathcal{O} (\bar{X}_{\infty}) \mathe^{- \int_{0
					}^{\infty} \mathcal{H}_s (\bar{X}_s) \mathd s} \right]}{\mathbb{E} \left[
				\mathe^{- \int_0^{\infty} \mathcal{H}_s (\bar{X}_s) \mathd s} \right]},
		\label{eq:finite-vol-law}
	\end{equation}
	where $\bar{X}$ is the unique strong solution to
	\begin{equation}
		\bar{X}_t = - \int_0^t \dot{G}_s F_s (\bar{X}_s) \mathd s + W_t .
		\label{eq:XF-SDE}
	\end{equation}
	First, let us note that the estimates on $\dot{G}F$ (bounded and Lipschitz) from Proposition \ref{prop:force-estimates} guarantee existence of a unique strong solution to \eqref{eq:XF-SDE}. The same is true for $\bar{X}^T$ the solution to
	\begin{equation}\label{eq:XFT-SDE}
		\bar{X}^T_t=-\int_0^t\dot{G}_sF_s^T(\bar{X}^T_s)\mathd s +W_t\qquad t\in[0,T].
	\end{equation}
	Moreover, the estimates on $F^T,F$ in Proposition \ref{prop:force-estimates} and the arguments used in the proof of Theorem \ref{thm:T,rho-convergence} show that $\bar{X}^T$ converges to $\bar{X}$ and that $\bar{X}$ has a terminal value $\bar{X}_\infty$. More precisely, with $\bar{Z}=\bar{X}-W$,
	\begin{equation}\label{eq:Xbar-conv}
		\lim_{T\to\infty}\sup_{s\in[0,T]}\|\bar{X}_s^T-\bar{X}_s\|_{L^\infty(\mathbb{P}; L^\infty)}=0.
	\end{equation}
	By definition of $X$, we know that
	\[ \mathbb{E} [\mathcal{O} (X_T^T)] = \frac{\mathbb{E} [\mathcal{O} (W_T)
				\mathe^{- V_T^T (W_T)}]}{\mathbb{E} [\mathe^{- V_T^T (W_T)}]} =
		\frac{\mathbb{E} \left[ \mathcal{O} (W_T) \mathe^{- V_0^T (W_0) - \int_0^T
					\mathcal{H}_s^T (W_s) \mathd s} \mathcal{E}_{0, T}
				\right]}{\mathbb{E} \left[ \mathe^{- V_0^T (W_0) - \int_0^T \mathcal{H}_s^T
					(W_s) \mathd s} \mathcal{E}_{0, T} \right]}, \]
	where
	\[ \mathcal{E}_{t, T} = \exp \left\{ - \int_{t \wedge T}^T F_s^T (W_s) \mathd
		W_s - \frac{1}{2} \int_{t \wedge T}^T F_s^T (W_s) \dot{G}_s F_s^T (W_s)
		\mathd s \right\}.
	\]
	The estimates on $F$ from Proposition \ref{prop:force-estimates} show that $(\mathcal{E}_{t, T})_{t\geq 0}$ is a
	martingale. By Girsanov's theorem,
	\[ W_t^F \assign W_t + \int_0^t \dot{G}_s F_s^T (W_s) \mathd s, \]
	is a Brownian motion under $\mathd \mathbb{P}^{F^T} = \mathcal{E}_{0, T}
		\mathd \mathbb{P}$. Therefore, using that $V_0^T (W_0) = V_0^T (0)$ is
	deterministic,
	\[ \mathbb{E} [\mathcal{O} (X_T^T)] = \frac{\mathbb{E} \left[ \mathcal{O}
				(\bar{X}_T^T) \mathe^{- \int_0^T \mathcal{H}_s^T (\bar{X}_s^T) \mathd s}
				\right]}{\mathbb{E} \left[ \mathe^{- \int_0^T \mathcal{H}_s^T (\bar{X}^T_s)
					\mathd s} \right]},\]
	where $\bar{X}^T$ is the unique strong solution to
	\eqref{eq:XFT-SDE}.
	Combining the convergence of $\bar{X}^T_T$ from \eqref{eq:Xbar-conv} with the convergence of $\mathcal{H}_s^T$ to $\mathcal{H}_s$ from Proposition \ref{prop:H-estimates}, we conclude \eqref{eq:finite-vol-law} and thus the claim.
\end{proof}
\begin{remark}
	Let again $\bar{X}$ be the unique strong solution to \eqref{eq:XF-SDE} and denote the set of all $\bar{X}$ adapted processes by $\mathcal{A}(\bar{X})$. Consider the $\mathbb{R}$-valued BSDE
	\begin{equation}
		\mathcal{R}_t = \int_t^{\infty} \mathcal{H}_s (\bar{X}_s) \mathd s -
		\frac{1}{2} \int_t^\infty \| \dot{G}_s^{1 / 2} \bar{R}_s \|^2_{L^2} \mathd s -
		\int_t^\infty \bar{R}_s \mathd W_s . \label{eq:qBSDE}
	\end{equation}
	The estimates on $\mathcal{H}$ and $\bar{X}$, imply that \eqref{eq:qBSDE}
	has a unique bounded solution $(\mathcal{R}, \bar{R})$ which is also adapted to $\bar{X}$. Moreover, standard
	comparison arguments for $\mathbb{R}$-valued BSDEs imply that the control
	$\bar{R}$ is the unique optimiser for the following relaxed version of the finite volume control problem discussed in section \ref{sec:Var+LDP},
	\[ \mathcal{V}^{w} = \inf_{R \in \mathcal{A} (\bar{X})}
		\mathbb{E}^{\mathbb{P}^{R}} \left[ V_0 (\bar{X}_0) +
			\int_0^{\infty} \mathcal{H}_s (\bar{X}_s) \mathd s + \frac{1}{2}
			\int_0^{\infty} \| \dot{G}_s^{1 / 2} R_s \|^2_{L^2} \mathd s \right], \]
	where
	\[ \mathd \mathbb{P}^{R} = \exp \left\{ \int_0^{\infty} R_s \mathd
		W_s - \frac{1}{2} \int_0^{\infty} \| \dot{G}_s^{1 / 2} R_s \|^2_{L^2}
		\mathd s \right\} \mathd \mathbb{P},\]
	and $\tmop{Law}^{\mathbb {P}^{\bar{R}}}(\bar{X}_\infty)=\nu_{\tmop{SG}}$.
\end{remark}

\section{Decay of correlations}\label{sec:correlations}
Using the scale-by-scale coupling via~\eqref{eq:FBSDE-leq3}, a coupling method
allows us to transfer the decay of correlations from the massive free field to
the sine-Gordon measure and establish Theorem~\ref{thm:decay-of-correlations}.
We follow mostly~{\cite{gubinelliDecayCorrelationsStochastic2025}} but similar
arguments can be found e.g.
in~{\cite{vecchiStochasticAnalysisSubcritical2025}} and the idea appears to
originate in~{\cite{funakiReversibleMeasuresMultidimensional1991}}.

For simplicity and to not distract from the main ideas, let $\mathcal{O}_1,
	\mathcal{O}_2 : H^{- \varepsilon, - n} \rightarrow \mathbb{R}$ be two
Lipschitz and bounded observables. Given a smooth bump function $\chi$
supported on $B_1 (0)$ we want to show that for some $c>0$, with $X=X_\infty$
the solution to \eqref{eq:FBSDE-leq3}
\[ \tmop{Cov} (\mathcal{O}_1 (\chi \cdot X (\cdot + x_1)),
	\mathcal{O}_2 (\chi \cdot X(\cdot + x_2))) \lesssim \mathe^{- c | x_1 - x_2 |} .
\]
Let us agree on some notation to use throughout this proof. We denote by
$\mathfrak{l} \assign | x_1 - x_2 |$ the distance between the two points of
interest. For $i = 1, 2$, let $D_i (r)$ be the open ball of radius $r > 0$
centred at $x_i$, where we drop the argument in case $r =\mathfrak{l}/ 2$.
Given a smooth bump function $\vartheta^{(i)}$ supported on $D_i (\mathfrak{l}/ 4)$
such that $\vartheta (x) \equiv 1$ on $D_i (\mathfrak{l}/ 8)$, we define the
exponential weights
\[ q^{(i)} (x) = \mathe^{- \gamma m | x - x_i |}  \quad \text{and} \quad
	\bar{q}^{(i)} (x) = \vartheta^{(i)} (x) q^{(i)} (x), \]
In order for the heat kernels $Q$ and $G$ to work nicely with the weights
$q^{(i)}$, we always assume that $\gamma \in (0, 1)$. To set up the coupling
argument, let $W^{(0)} \assign W$ and $D_0 \assign \mathbb{R}^2$ and define
the identically distributed Brownian motions $W^{(1)}, W^{(2)}$ with
covariance
\[ \mathbb{E} [W_t^{(i)} (x) W_t^{(j)} (y)] = \int_0^t \mathd s \int \mathd z
	Q_s (z - x) \mathbbm{1}_{D_i \cap D_j} (z) Q_s (z - y), \]
so that $W^{(1)}$ and $W^{(2)}$ are independent and $W^{(0)} \approx W^{(i)}$
near $x_i$. Denote by $X^{(i)}$ the solution to the FBSDE
\eqref{eq:FBSDE-leq3} with $T = \infty, \rho = 1$ and $g = 0$ driven by
$W^{(i)}$. Then, the solutions $X^{(1)}$ and $X^{(2)}$ have the same law as $X$
and inherit the
independence from their driving noise. Inserting the $X^{(i)}$ for $X$, we
find that
\[ \begin{aligned}
		  & \tmop{Cov} (\mathcal{O}_1 (\chi \cdot X (\cdot + x_1)), \mathcal{O}_2
		(\chi \cdot X (\cdot + x_2)))                                                 \\
		= & \mathbb{E} [\mathcal{O}_1 (\chi \cdot X (\cdot + x_1))  \mathcal{O}_2
		(\chi \cdot X (\cdot + x_2))]                                                 \\
		  & -\mathbb{E} [\mathcal{O}_1 (\chi \cdot X^{(1)}
			(\cdot + x_1))] \mathbb{E} [\mathcal{O}_2 (\chi \cdot X^{(2)} (\cdot +
		x_2))]                                                                        \\
		= & \phantom{+}\mathbb{E} [\mathcal{O}_2 (\chi \cdot X^{(2)}
			(\cdot + x_2))\;(\mathcal{O}_1 (\chi \cdot X (\cdot + x_1)) -
		\mathcal{O}_1 (\chi \cdot X^{(1)} (\cdot + x_1)))]                            \\
		  & +\mathbb{E} [\mathcal{O}_1 (\chi \cdot X (\cdot + x_1)) \; (\mathcal{O}_2
			(\chi \cdot X (\cdot + x_2)) - \mathcal{O}_2 (\chi \cdot X^{(2)} (\cdot +
			x_2)))],
	\end{aligned} \]
where we denote $X^{(i)} = X^{(i)}_{\infty}$. Thus, for any $\alpha < 0$ and
$p \in [1, \infty]$,
\[ \begin{aligned}
		|                                       & \tmop{Cov}(\mathcal{O}_1 (\chi \cdot X (\cdot + x_1)), \mathcal{O}_2
		(\chi \cdot X (\cdot + x_2))) |                                                                                \\
		\lesssim_{\mathcal{O}_1, \mathcal{O}_2} &
		\mathbb{E} \left[\| \chi \cdot (X (\cdot + x_1) - X^{(1)} (\cdot + x_1))
		\|_{B^{\alpha, - n}_{p, p}}\right]                                                                             \\
		                                        & \times
		\mathbb{E} \left[\| \chi \cdot (X (\cdot + x_1) - X^{(2)} (\cdot + x_1))
			\|_{B^{\alpha, - n}_{p, p}}\right].
	\end{aligned}\]
It remains to estimate $\| \chi (X (\cdot + x_i) - X^{(i)} (\cdot + x_i))
	\|_{B_{p, p}^{\alpha, - n}}$, for $i=1,2$. If $x_1, x_2$ are close, say $\mathfrak{l}
	\leqslant 8$, then we use the boundedness of the observables to conclude
\[ | \tmop{Cov} (\mathcal{O}_1 (\chi \cdot X (\cdot + x_1)), \mathcal{O}_2
	(\chi \cdot X (\cdot + x_2))) | \lesssim 1. \]
If on the other hand $\mathfrak{l}> 8$, then $D_i (1) \subset D_i
	(\mathfrak{l}/ 8)$ and thus $\vartheta^{(i)} \equiv 1$ on $\tmop{supp} (\chi (\cdot
	- x_i)) \subset D_i (1)$ so that $(\bar{q}^{(i)})^{- 1} \leqslant \mathe^{\gamma m}$
on $D_i$. Consequently,
\[ \| \chi \cdot X (\cdot + x_i) - \chi \cdot X^{(i)} (\cdot + x_i) \|_{B_{p,
					p, \ell}^{\alpha}} \leqslant \mathe^{m \gamma}  \| \bar{q}^{(i)} (X - X^i)
	\|_{L^p (D_i (1))} \lesssim \mathe^{m \gamma (1 -\mathfrak{l}/ 8)}, \]
where the last inequality follows from Lemma \ref{lem:FBSDE-decay-est} below.
The remainder of this section will be devoted to its proof.

\begin{lemma}
	\label{lem:FBSDE-decay-est}Let $\mathfrak{l}> 8$ and $|\bar{\lambda}|$ sufficiently small.
	The solutions $X^{(i)}$ to
	the FBSDE \eqref{eq:FBSDE-leq3} driven by $W^{(i)}$ satisfy for some $\gamma
		< 1$,
	\[ \mathbb{E} \left[\sup_t \| \bar{q}^{(i)} (X_t - X_t^{(i)}) \|_{L^{\infty}}\right]
		\lesssim \mathe^{- \gamma m\mathfrak{l}/ 8} . \]
\end{lemma}

\begin{proof}
	The FBSDE for the difference is given by
		{\footnotesize\[  \begin{aligned}
					X_t - X_t^{(i)} = & -\int_{0^{\nosymbol}}^t \mathd s \dot{G}_s (F_s
					(X_s^{\nosymbol})-F_s (X_s^{(i)})) - \int_0^t \mathd s \dot{G}_s (R_s-R_s^{(i)}) + W_t - W_t^{(i)}, \\
					R_t - R_t^{(i)} = & \int_t^{\infty} \mathd s [H_s (X_s) - H_s
							(X_s^{(i)})]- \left(\int_t^{\infty} \mathd s \mathD F_s (X^{\nosymbol}_s)
					\dot{G}_s R_s - \mathD F_s (X_s^{(i)}) \dot{G_s} R_s^{(i)}\right) .
				\end{aligned}  \]}
	The estimate on the Gaussian part $W_t - W_t^{(i)}$ is proven in Lemma
	\ref{lem:corr-W-Wi} below. For the drift, $Z_t - Z_t^{(i)} \assign (X_t -
		X_t^{(i)}) - (W_t - W_t^{(i)})$, we apply Lemma \ref{lem:rho-G} and \ref{lem:Lip-corr} to find
	\begin{equation}
		\begin{aligned}
			\sup_t\, & \| \bar{q}^{(i)} (Z_t - Z_t^{(i)}) \|_{L^{\infty}}            \\
			\lesssim &
			\int_0^{\infty} \mathd s \bar{\lambda}_s \langle s \rangle^{- 2} (\mathe^{-
				\gamma m\mathfrak{l}/ 8} \| X_s - X_s^{(i)} \|_{L^{\infty, - n} (D_i^c
					(\mathfrak{l}/ 8))} + \| \bar{q}^{(i)} (X_s - X_s^{(i)})
			\|_{L^{\infty}})                                                         \\
			         & + \int_0^{\infty} \mathd s \langle s \rangle^{- 2} (\mathe^{-
				\gamma m\mathfrak{l}/ 8} \| R_s - R_s^{(i)} \|_{L^{\infty, - n} (D_i^c
					(\mathfrak{l}/ 8))} + \| \bar{q}^{(i)} (R_s - R_s^{(i)})
			\|_{L^{\infty}}) .
		\end{aligned} \label{eq:Z-Zi}
	\end{equation}
	Similarly for the remainder,
	\begin{equation}
		\begin{aligned}
			         & \| \bar{q}^{(i)} (R_t - R_t^{(i)}) \|_{L^{\infty}}                  \\
			\lesssim & \int_t^{\infty} \mathd s (\langle s \rangle^{- 4} \bar{\lambda}_s^4
			+ \langle s \rangle^{- 2} \bar{\lambda}_s) (\mathe^{- \gamma m\mathfrak{l}/ 8}
			\| X_s - X_s^{(i)} \|_{L^{\infty, - n} (D_i^c (\mathfrak{l}/ 8))} + \|
			\bar{q}^{(i)} (X_s - X_s^{(i)}) \|_{L^{\infty}})                               \\
			         & + \int_t^{\infty} \mathd s \bar{\lambda}_s \langle s \rangle^{- 2}
			(\mathe^{- \gamma m\mathfrak{l}/ 8} \| R_s - R_s^{(i)} \|_{L^{\infty, -
							n} (D_i^c (\mathfrak{l}/ 8))} + \| \bar{q}^{(i)} (R_s - R_s^{(i)})
			\|_{L^{\infty}}) .
		\end{aligned} \label{eq:R-Ri}
	\end{equation}
	In the region $D_i^c (\mathfrak{l}/ 8)$, we cannot expect the difference $W
		- W^{(i)}$ to behave any better than the Brownian motion $W_t$ itself. We
	therefore control the solution in this region using the uniform bounds on
	the drift from Lemma \ref{lem:B-to-B} combined with the bounds on the
	Brownian motion $(W_t)_t$ from Lemma \ref{lem:W-Halpha}. This implies for
	any $\varepsilon > 0$,
	\[ \begin{aligned}
			\| R_s - R_s^{(i)} \|_{L^{\infty, - n} (D_i^c (\mathfrak{l}/ 8))} &
			\leqslant  2\| R_s \|_{L^{\infty}} \lesssim 1,                                \\
			\mathbb{E} [\sup_s  \langle s \rangle^{- \varepsilon} \| X_s -
				X_s^{(i)} \|_{L^{\infty, - n} (D_i^c (\mathfrak{l}/ 8))}]
			                                                                  & \leqslant
			2\mathbb{E} [\sup_s \| Z_s \|_{L^{\infty}}] + 2\mathbb{E} [\sup_s
				\langle s \rangle^{- \varepsilon} \| W_s \|_{L^{\infty, - n}}] \lesssim
			1,
		\end{aligned} \]
	where the estimate on $(W_t)_t$ follows from Lemma \ref{lem:W-Halpha} and
	\[ \mathbb{E} \left[\sup_{t \geqslant 1} \| t^{- \varepsilon / 2} W_t
			\|^{\nosymbol}_{L^{\infty}}\right] \lesssim \mathbb{E} \left[\sup_{t \geqslant 1}
			\| t^{- \varepsilon / 2} W_t \|^{\nosymbol}_{B_{\infty, \infty}^{\alpha,
							- n}}\right], \quad \text{} \alpha \in (0, \varepsilon) . \]
	Moreover, with Lemma \ref{lem:corr-W-Wi} below,
	\[ \mathbb{E} \left[\sup_{s \geqslant 0} \| \bar{q}^{(i)} (X_s - X_s^{(i)})
			\|_{L^{\infty}}\right] \lesssim \mathbb{E} \left[\sup_{s \geqslant 0} \|
			\bar{q}^{(i)} (Z_s - Z_s^{(i)}) \|_{L^{\infty}}\right] + \mathe^{- \gamma
			m\mathfrak{l}/ 8} . \]
	With these estimates, the equation \eqref{eq:R-Ri} reduces to
	\[ \begin{aligned}
			\mathbb{E}\| \bar{q}^{(i)} (R_t - R_t^{(i)}) \|_{L^{\infty}}
			\lesssim & \mathe^{- \gamma m\mathfrak{l}/ 8} +\mathbb{E} \left[\sup_{s
			\geqslant 0} \| \bar{q}^{(i)} (Z_s - Z_s^{(i)}) \|_{L^{\infty}}\right]  \\
			         & +
			\int_t^{\infty} \mathd s \bar{\lambda}_s \langle s \rangle^{- 2} \mathbb{E}
			\| \bar{q}^{(i)} (R_s - R_s^{(i)}) \|_{L^{\infty}},
		\end{aligned} \]
	which implies by Gronwall's lemma
	\[ \mathbb{E} \left[\| \bar{q}^{(i)} (R_t - R_t^{(i)}) \|_{L^{\infty}}\right] \lesssim
		\mathe^{- \gamma m\mathfrak{l}/ 8} +\mathbb{E} \left[\sup_{s \geqslant 0} \|
			\bar{q}^{(i)} (Z_s - Z_s^{(i)}) \|_{L^{\infty}}\right] . \]
	Taking expectation \eqref{eq:Z-Zi} these bounds
	imply
	\[ \mathbb{E} \left[\sup_t \| \bar{q}^{(i)} (Z_t - Z_t^{(i)}) \|_{L^{\infty}}\right]
		\lesssim \int_0^{\infty} \mathd s \bar{\lambda}_s \langle s \rangle^{- 2 +
			\varepsilon} \mathe^{- \gamma m\mathfrak{l}/ 8} + \bar{\lambda} \mathbb{E}
		\left[\sup_t \| \bar{q}^{(i)} (Z_t - Z_t^{(i)}) \|_{L^{\infty}}\right], \]
	so that for $\bar{\lambda}$ sufficiently small after rearranging
	\[ \mathbb{E} \left[\sup_t \| \bar{q}^{(i)} (Z_t - Z_t^{(i)}) \|_{L^{\infty}}\right]
		\lesssim \mathe^{- \gamma m\mathfrak{l}/ 8} . \]
\end{proof}

\begin{lemma}
	\label{lem:rho-G}For any function $v$, it holds that
	\begin{equation}
		\| q^{(i)}  \dot{G}_s v \|_{L^{\infty}} \lesssim \langle s \rangle^{- 2}
		(\mathe^{- \gamma m\mathfrak{l}/ 8} \| v \|_{L^{\infty, - n} (D_i^c
				(\mathfrak{l}/ 8))} + \| \bar{q}^{(i)} v \|_{L^{\infty}}) \label{eq:rho-G}
		.
	\end{equation}
\end{lemma}

\begin{proof}
	First observe that since $\vartheta^{(i)}$ is compactly supported, the weight
	$\langle x \rangle^{- n}$ is uniformly bounded away from zero and thus
	$\vartheta^{(i)} (x) \lesssim \vartheta^{(i)} (x) \langle x \rangle^{- n}$. Combined
	with the triangle inequality and the estimate \eqref{eq:jpb-triangle-ineq}
	on the polynomial weights, we have
	\begin{equation}
		\bar{q}^{(i)} (x) = \vartheta^{(i)} (x) \mathe^{- \gamma m | x - x_i |} \lesssim
		\vartheta^{(i)} (x) \langle x - y \rangle^n \mathe^{\gamma m | x - y |}
		\mathe^{- \gamma m | y - x_i |} \langle y \rangle^{- n} .
		\label{eq:rho-est}
	\end{equation}
	Thus, for any function $v$,
	\[ \begin{aligned}
			\| q^{(i)}  \dot{G}_s v \|_{L^{\infty}} & = \sup_x \left| \int \mathd
			y \bar{q}^{(i)} (x)  \dot{G}_s (x - y) v (y) \right|                                                   \\
			                                        & \lesssim  \sup_x  \int \mathd y \langle x - y \rangle^n
			\mathe^{\gamma m | x - y |}  \dot{G}_s (x - y) | \mathe^{- \gamma m | y
			- x_i |} \langle y \rangle^{- n} v (y) |                                                               \\
			                                        & \lesssim  \sup_x | \mathe^{- \gamma m | x - x_i |} \langle x
			\rangle^{- n} v (x) | \sup_x  \int \mathd y \langle y - x \rangle^n
			\mathe^{\gamma m | y - x |}  \dot{G}_s (y - x) .
		\end{aligned} \]
	If $x \in D_i^c (\mathfrak{l}/ 8)$, then $| x - x_i | >\mathfrak{l}/ 8$ so
	that
	\[ \mathbbm{1}_{D_i^c (\mathfrak{l}/ 8)} (x) | \mathe^{- \gamma m | x - x_i
			|} \langle x \rangle^{- n} v (x) | \leqslant \mathe^{- \gamma
			m\mathfrak{l}/ 8} \| v (x) \|_{L^{\infty, - n} (D_i^c (\mathfrak{l}/ 8))}
		. \]
	On the other hand, since $\mathbbm{1}_{D_i (\mathfrak{l}/ 8)}
		\leqslant \vartheta^{(i)}$, for $x \in D_i (\mathfrak{l}/ 8)$ we can insert
	$\vartheta^{(i)}$ to find
	\[ \mathbbm{1}_{D_i (\mathfrak{l}/ 8)} (x) | \mathe^{- \gamma m | x - x_i |}
		\langle x \rangle^{- n} v (x) | \lesssim | \mathe^{- \gamma m | x - x_i
			|} \vartheta^{(i)} (x) v (x) | = \| \bar{q}^{(i)} v \|_{L^{\infty}} . \]
	Thus, the estimates on $\dot{G}$ from Lemma~\ref{lem:hk-exp} conclude with
	the assumption that $\gamma < 1$.
\end{proof}

\begin{lemma}
	\label{lem:Lip-corr}In the same notation as before, the following Lipschitz
	estimates apply
		{\footnotesize
			\begin{align}
				\| \bar{q}^{(i)}  \dot{G}_s (F_s (\varphi) - F_s (\varphi^{(i)}))
				\|_{L^{\infty}}                                                        & \lesssim  \bar{\lambda}_s \langle s \rangle^{- 2} (\mathe^{-
					\gamma m\mathfrak{l}/ 8}  \| \varphi - \varphi^{(i)} \|_{L^{\infty, - n}
						(D_i^c (\mathfrak{l}/ 8))} + \| \bar{q} (\varphi - \varphi^{(i)})
				\|_{L^{\infty}}) \label{eq:fwd-est}                                                                                                   \\
				\| \bar{q}^{(i)} (H_s (\varphi) - H_s (\varphi^{(i)})) \|_{L^{\infty}} &
				\lesssim \bar{\lambda}^4_s \langle s \rangle^{- 4} (\mathe^{- \gamma
					m\mathfrak{l}/ 8}  \| \varphi - \varphi^{(i)} \|_{L^{\infty, - n} (D_i^c
						(\mathfrak{l}/ 8))} + \| \bar{q} (\varphi - \varphi^{(i)})
				\|_{L^{\infty}}) \label{eq:H-est}                                                                                                     \\
				\| \bar{q}^{(i)} \mathD F_s (\varphi)  \dot{G}_s R \|_{\nosymbol
				L^{\infty}}                                                            & \lesssim  \bar{\lambda}_s \langle s \rangle^{- 2} (\mathe^{-
					\gamma m\mathfrak{l}/ 8} \| R \|_{L^{\infty, - n} (D_i^c (\mathfrak{l}/
						8))} + \| \bar{q}^{(i)} R \|_{L^{\infty}}) \label{eq:bwd-est} .
			\end{align}}
\end{lemma}

\begin{proof}
	Regarding~\eqref{eq:fwd-est} and~\eqref{eq:H-est}, we only show how to
	commute the weight through a generic term of the Ansatz for the
	force~\eqref{eq:fourier-rep-V}. The optimal, field independent bounds on
	$F_s$ are then obtained in the same way as in the proof of
	Proposition~\ref{prop:force-estimates}. In this case, we need to estimate
	expressions of the form
	\[ \sup_{x_1} \left| \int \mathd x_{2 : \ell} f (x_{1 : \ell}) [\mathe^{i
					\beta \varphi (x_{2 : \ell})} - \mathe^{i \beta \varphi^{(i)} (x_{1 :
						\ell})}]  \bar{q}^{(i)} (x_1) \right|, \]
	where $f$ is one of the (potentially regularised) force coefficients and
	$x_{1 : \ell} \in (\mathbb{R}^2)^{\ell}$ for some $\ell$. To this end,
	thanks to the boundedness of the complex exponential function, it is
	sufficient to estimate the terms
	\[ \sup_{x_1}  \left| \int \mathd x_{2 : \ell} f (x_{1 : \ell}) | \varphi
		(x_k) - \varphi^{(i)} (x_k) |  \bar{q}^{(i)} (x) \right|, \quad k = 1,
		\ldots, \ell . \]
	Using \eqref{eq:rho-est} we obtain,
	\begin{equation}
		\begin{aligned}
			         & f (x_{1 : \ell}) [\mathe^{i \beta \varphi (x_{2 : \ell})} - \mathe^{i
			\beta \varphi^{(i)} (x_{1 : \ell})}]  \bar{q}^{(i)} (x_1)                        \\
			\lesssim & f (x_{1 : \ell}) \vartheta^{(i)} (x) \langle x - x_k \rangle^n
			\mathe^{\gamma m | x - x_k |} (\varphi_s - \varphi_s^i) (x_k) [\mathe^{-
						\gamma m\mathfrak{l}/ 8} \langle x_k^{- m} \rangle \mathbbm{1}_{D_i^c
			<      (\mathfrak{l}/ 8)} (x_k)                                                  \\
			         & + \bar{q} (x_k) \mathbbm{1}_{D_i (\mathfrak{l}/
			8)} (x_k)] .                                                                     \\
			\lesssim & [\mathe^{- \gamma m\mathfrak{l}/ 8}  \| \varphi_s -
					\varphi_s^i \|_{L^{\infty, - n} (D_i^c (\mathfrak{l}/ 8))} + \|
					(\varphi_s - \varphi_s^i)  \bar{q} \|_{L^{\infty}}] f (x_{1 : \ell})
			\mathe^{\gamma m | x - x_k |} \langle x - x_k \rangle^n .
		\end{aligned} \label{eq:exp-sep}
	\end{equation}
	From here, thanks to the exponential decay of the force in the separation of
	the points, the estimates on $f^{[\ell]}_s$ obtained in Section
	\ref{sec:flow-coefficients} conclude since by definition of the Steiner
	weights $\omega_{\varsigma}$ we have for $\varsigma \in (\gamma, 1)$,
	\[ \sup_{x_1} \left| \int \mathd x_{2 : \ell} f (x_{1 : \ell})
		\mathe^{\gamma m | x - x_k |} \langle x - x_k \rangle^n \right| \lesssim
		\interleave f \interleave . \]
	Applying exactly the same reasoning as in the proof of Lemma~\ref{lem:rho-G}
	to~\eqref{eq:bwd-est}, we obtain
	\begin{align*}
		| \bar{q} (x) & (\mathD F_s (X)  \dot{G}_s R) (x) | \\
		              & \lesssim \int \mathd y
		\mathD F_s (X) (x, y) \langle x - z \rangle^n \mathe^{\gamma m | x - y |}
		\int \mathd z \dot{G_s} (y - z) q^{(i)} (z) \langle z \rangle^{- n} R
		(z),
	\end{align*}
	which implies the claim after splitting up the integral in between $D_i^c
		(\mathfrak{l}/ 8)$ and $D_i (\mathfrak{l}/ 8)$.
\end{proof}

\begin{lemma}
	\label{lem:corr-W-Wi}For i=1,2 and $\mathfrak{l}> 8$, it holds that
	\[ \mathbb{E} [\sup_t \| \bar{q}^{(i)} (W_t - W_t^{(i)}) \|_{L^{\infty}}]
		\lesssim \mathe^{- \gamma m\mathfrak{l}/ 8} . \]
\end{lemma}

\begin{proof}
	First note that since $D_i$ is compact, the restricted weight $q|_{D_i}$ is
	of order $1$ in $D_i$, that is $1 \lesssim q|_{D_i} \lesssim 1$. Therefore,
	by Besov embeddings,
	\begin{equation*}
		\begin{aligned}
			\| \bar{q} (W_t - W_t^{(i)}) \|_{L^{\infty}}
			\,\lesssim\, & \| \vartheta^{(i)} (W_t -
			W_t^{(i)}) \|_{L^{\infty}}                          \\
			\,\lesssim\, & \| \vartheta^{(i)} (W_t - W_t^{(i)})
			\|_{B^{\alpha - \delta}_{\infty, \infty}}
			\,\lesssim\,  \| \vartheta^{(i)} (W_t -
			W_t^{(i)}) \|_{B_{p, p}^{\alpha + \delta}},
		\end{aligned}
	\end{equation*}
	provided $0 < \delta < \alpha$ and $p > d / \delta$. Following the same
	logic as in the proof of Lemma \ref{lem:W-Halpha}, we have for any $\alpha
		\in \mathbb{R}$, $p \in [1, \infty)$ thanks to Gaussian hypercontractivity,
	\begin{equation*}
		\begin{aligned}
			\mathbb{E} [\sup_t \| \vartheta^{(i)} (W_t - W_t^{(i)}) \|_{B_{p, p}^{\alpha}}]
			\lesssim & \mathbb{E} [\sup_t \| \vartheta^{(i)} (W_t - W_t^{(i)}) \|_{B_{p,
			p}^{\alpha}}^p]^{1 / p}                                                      \\
			\lesssim & \mathbb{E} [\sup_t \| \vartheta^{(i)} (W_t -
				W_t^{(i)}) \|_{H^{\alpha}}^2]^{1 / 2} .
		\end{aligned}
	\end{equation*}
	Interpolating between $L^2$ and $H^1$, it is therefore sufficient to show
	that
	\[ \mathbb{E} \left[\sup_t \| \vartheta^{(i)} (W_t - W_t^{(i)}) \|^2_{H^1}\right] \lesssim
		\mathe^{- m \gamma \mathfrak{l}/ 8} . \]
	Here, we compute similarly to the argument in Lemma~\ref{lem:hk-exp} using
	now the separation $d (D_i^c, D_i (\mathfrak{l}/ 4)) >\mathfrak{l}/ 4$,
	\begin{align*}
		          & \mathbb{E} \left[\| \vartheta^{(i)} (W_t - W_t^{(i)})
		\|_{{H^1}}^2\right]                                                                                \\
		\leqslant & \int_0^t \mathd s \int_{D_i (\mathfrak{l}/ 4)} \mathd x
		\int_{D_i^c} \mathd z (| \nabla Q_s (x - z) |^2 + | Q_s (x - z) |^2)                               \\
		\leqslant & \sup_{\tmscript{\begin{array}{l}
					                            x \in D_i (\mathfrak{l}/ 4) \\
					                            z \in D_i^c
				                            \end{array}}} \mathe^{- \frac{m}{2} \gamma | x - z |}  \int_0^t \mathd
		s \int_{D_i (\mathfrak{l}/ 4)} \mathd x \int_{D_i^c} \mathd z
		\mathe^{\frac{m}{2} \gamma | x - z |} (| \nabla Q_s (x - z) |^2 + | Q_s
		(x - z) |^2)                                                                                       \\
		\lesssim  & \mathe^{- m \gamma \mathfrak{l}/ 8}  \int_0^t \mathd s
		\int_{D_i (\mathfrak{l}/ 4)} \mathd x \int_{D_i^c} \mathd z (s | x - z
		| + 1) \mathe^{- 2 s | x - z |^2 + \frac{m}{2} \gamma | x - z | - m^2 /
		s}                                                                                                 \\
		\lesssim  & \mathe^{- m \gamma \mathfrak{l}/ 8} .
	\end{align*}
	Finally, the maximal martingale inequalities allow to take the supremum
	inside the expectation and we arrive at the claim.
\end{proof}

\section{Singularity for $\beta^2 \geqslant 4 \pi$}\label{sec:singularity}
We use the FBSDE~\eqref{eq:FBSDE-leq3} to show Theorem \ref{thm:singularity},
that is that the finite volume sine-Gordon measure and the Gaussian free field
are mutually singular for $\beta^2 \geqslant 4 \pi$. Our proof relies on the
asymptotics for the regularised cosine potential. It is similar in spirit to
the method used in~{\cite{barashkovF34MeasureGirsanovs2021}}, but does not rely
on a change of measure. We also refer
to~{\cite{ohStochasticQuantization$Phi^3_3$model2021}}, where the authors show
singularity of the $\Phi^3_3$ measure using a variational problem.

Before we go into more detail, let us discuss the heuristics behind our argument. To show that absolute continuity is violated it is enough to find an event $S$ such that
$\mu(S)=0$ but $\nu_{\tmop{SG}}^{\rho}(S)>0$. Good candidates for events $S$ are limits of
approximations of $W$ and $X$, chosen such that the almost sure limits are different.
If such an event exists, then we also get the reverse singularity result for free: since $\mu(S^c)=1$ and $\nu_{\tmop{SG}}^{\rho}(S^c)=0$,
both measures are singular with respect to each other.

A natural choice is an appropriate rescaling of the cosine itself.
For $W$, Gaussian hypercontractivity suggests that as $t\to\infty$ the martingale behaves like
\[
	\llbracket \cos(\beta W_t(0))\rrbracket-1 =
	\int_{0}^{t} -\beta\llbracket \sin(\beta W_s(0))\rrbracket\mathd W_s(0) \asymp t^{\frac{1}{2}-\delta}\vee1.
\]
For the interacting field $X_t=W_t+Z_t$, the Wick ordered cosine is no longer a martingale as we pick up
an additional drift term,
\[\begin{aligned}
		\llbracket \cos(\beta X_t(0))\rrbracket-1
		= \int_{0}^{t} -\beta\llbracket \sin(\beta X_s(0))\rrbracket\dot{Z}_s(0)\mathd s
		+ \int_{0}^{t} -\beta\llbracket \sin(\beta X_s(0))\rrbracket\mathd W_s(0).
	\end{aligned}
\]
For the martingale part, we expect the same asymptotics as in the Gaussian case.
However, the drift should behave like
\[\int_{0}^{t} -\beta\llbracket \sin(\beta X_s(0))\rrbracket\dot{Z}_s(0)\mathd s
	\asymp \int_{0}^{t} \langle s\rangle^{2(1-\delta)}\langle s \rangle^{-2}\mathd s
	\asymp t^{1-2\delta}\vee 1.\]
Note that there is no blow up for $\delta>\frac{1}{2}\iff\beta^2<4\pi$. However,
for $\delta\leqslant\frac{1}{2}$, the drift and martingale blow up in
the ultraviolet limit at  different rates and so we are able to find a rescaling of the Wick ordered
cosine which extracts only the divergence coming from the drift. In other words, we
want to find $c(t)$ such that as $t\to\infty$,
\[c(t)\llbracket \cos(\beta W_t(0))\rrbracket-1\to 0,\qquad
	c(t)\llbracket \cos(\beta X_t(0))\rrbracket-1\to \infty.\]

To make the above heuristics precise, we have to make some adjustments.
First, instead of the pointwise cosine, we will consider an average in space.
Moreover, we will have to formulate all of the above only in terms of the terminal
values $X_\infty$ and $W_\infty$ correspond to the physically meaningful EQFT measures.
The bulk of this section then consists in showing that the divergence heuristics
above are precise and that the drift and the martingale can be separated by tuning
$c(t)$. The result is the following theorem.

\begin{theorem}
	\label{thm:Ueps-div}For
	\[\gamma \in \begin{cases}
			(1 / 2, 1),                                         & \delta = 1 / 2                      \\
			2 (1 / 2 - \delta \vee 1 - 3 \delta, 1 - 2 \delta), & \delta \in(\frac{1}{4},\frac{1}{2})
		\end{cases},\]
	define
	\[ r (\varepsilon) = \begin{cases}
			\log (\varepsilon^{- 2} \vee 1)^{- \gamma}, & \text{$\delta = 1 / 2$
			}                                                                                          \\
			{\varepsilon^{\gamma}} , \,                 & \text{$\delta \in(\frac{1}{4},\frac{1}{2})$}
		\end{cases},\]
	and let $\chi^{\varepsilon} = \varepsilon^{- 2} \chi (\cdot / \varepsilon)$
	be a standard mollifier with radially symmetric and compactly supported
	Fourier transform $\hat{\chi}^{\varepsilon} = \chi (\varepsilon \cdot)$.
	Define the observable
	\begin{equation}
		U^{\varepsilon} (\varphi) = \int r (\varepsilon) \left(
		\mathe^{\frac{\beta^2}{2} G^{\varepsilon} (0)} \cos (\beta
				(\chi^{\varepsilon} \ast \varphi) (x)) - 1 \right) \rho (x) \mathd x,
		\label{eq:def-Ueps}
	\end{equation}
	where $G^{\varepsilon} = \tmop{Cov} (\chi^{\varepsilon} \ast W_{\infty})$ is
	the covariance of the mollified Gaussian free field. Then, there is a
	subsequence $\varepsilon_n \rightarrow 0$ such that
	\begin{equation}
		| U^{\varepsilon_n} (Z_{\infty} + W_{\infty}) | \overset{n \rightarrow
			\infty}{\longrightarrow} \infty \quad \text{while } \quad
		U^{\varepsilon_n} (W_{\infty}) \overset{n \rightarrow
			\infty}{\longrightarrow} 0, \label{eq:Ueps-div}
	\end{equation}
	where $Z_{\infty} = Z_{\infty}^{\rho}$ solves~\eqref{eq:FBSDE-leq3}.
\end{theorem}
Before we prove this statement, let us note the following consequence.

\begin{corollary}
	For $\delta \leqslant 1 / 2$, that is $\beta^2 \geqslant 4 \pi$, the finite
	volume sine-Gordon measure and the Gaussian free field are mutually
	singular.
\end{corollary}

\begin{proof}
	For some $\alpha > 0$ arbitrarily small, define the event
	\[ S = \left\{ \varphi \in H^{- \alpha} (\mathbb{R}^2) : \; \lim_{n
			\rightarrow \infty} U^{\varepsilon_n} (\varphi) = 0 \right\}, \]
	where $(\varepsilon_n)_{n \in \mathbb{N}}$ is a suitable but fixed subsequence and \
	$U^{\varepsilon}$ is defined as in \eqref{eq:def-Ueps}. It follows from
	Theorem \ref{thm:Ueps-div} that $\nu_{\tmop{SG}}^{\rho} (S) = 0 = \mu(S^c)$ while
	$\mu (S) = 1 = \nu_{\tmop{SG}}^{\rho} (S^c)$, which implies the claim.
\end{proof}

\begin{proof}[Proof of Theorem~\ref{thm:Ueps-div}]
	Let $(U_t^{\varepsilon})_{t \geqslant 0}$ be the scale interpolation of the functional
	$U^{\varepsilon}$ such that $(U_t^{\varepsilon} (W_t))_t$ is a martingale,
	that is
	\begin{equation}
		U_t^{\varepsilon} (\varphi) = \int r (\varepsilon) \left(
		\mathe^{\frac{\beta^2}{2} G_t^{\varepsilon} (0)} \cos (\beta
				(\chi^{\varepsilon} \ast \varphi) (x)) - 1 \right) \rho (x) \mathd x,
		\label{eq:def-Ueps-t}
	\end{equation}
	where $G_t^{\varepsilon} = \tmop{Cov} (\chi^{\varepsilon} \ast W_t)$. For
	convenience, we always assume $\varepsilon < 1$ and we also write
	$\lambda^{\varepsilon}_t \assign \mathe^{\frac{\beta^2}{2} G^{\varepsilon}_t
			(0)}$, $\bar{\lambda}^{\varepsilon}_t\assign |\lambda^{\varepsilon}_t|$, $\varphi^{\varepsilon} = \chi^{\varepsilon} \ast \varphi$, and
	$W^{\varepsilon} = \chi^{\varepsilon} \ast W$. It follows from Ito's formula
	that
	\[ \begin{aligned}
			U^{\varepsilon} (Z_{\infty} + W_{\infty}) & = \int_0^{\infty} \mathd
			s \left( \partial_s U^{\varepsilon}_s + \frac{1}{2} \tmop{Tr}
			(\dot{G}^{\varepsilon}_s \mathD^2 U^{\varepsilon}_s) \right) (Z_s +
			W_s)                                      & (\Iota^{\varepsilon})                                           \\
			                                          & + \int_0^{\infty} \mathd s \mathD U^{\varepsilon}_s (Z_s + W_s)
			\dot{Z}_s                                 & (\Iota \Iota^{\varepsilon})                                     \\
			                                          & + \int_0^{\infty} \mathD U^{\varepsilon}_s (Z_s + W_s)
			\mathd W_s .                              & (\Iota \Iota \Iota^{\varepsilon})
		\end{aligned} \]
	Thanks to the choice of the interpolation~\eqref{eq:def-Ueps-t}, the term
	$(\Iota^{\varepsilon})$ vanishes for all $\varepsilon$.

	\tmtextbf{$(\Iota \Iota \Iota^{\varepsilon})$} \ Let for $\varphi\in\{X,W\}$,
	\[ M^{\varepsilon}_{\infty} = \int_0^{\infty} \mathD
		U^{\varepsilon}_s (\varphi_s) \mathd W_s = - \beta r (\varepsilon)
		\int_0^{\infty} \bar{\lambda}^{\varepsilon}_s \int \mathd x \rho (x) \sin
		(\beta \varphi_s^{\varepsilon} (x)) \mathd W_s^{\varepsilon} (x), \]
	so that
		{\small\begin{align*}
				 & \mathbb{E} | M_{\infty}^{\varepsilon} |^2 \\
				 & = \beta^2 r (\varepsilon)^2
				\mathbb{E} \int_0^{\infty}  (\bar{\lambda}_s^{\varepsilon})^2  \int \mathd x_1
				\rho (x_1)  \int \mathd x_2 \rho (x_2) \sin (\beta \varphi^{\varepsilon}
					(x_1)) \sin (\beta \varphi^{\varepsilon} (x_2)) \mathd \langle
				W^{\varepsilon} (x_1) W^{\varepsilon} (x_2) \rangle_s,
			\end{align*}}
	where we compute
	\[ \mathd \langle W^{\varepsilon} (x_1) W^{\varepsilon} (x_2) \rangle_s =
		\int \mathd y_1 \int \mathd y_2 \chi_{\varepsilon} (x_1 - y_1) \dot{G}_s
		(y_1 - y_2) \chi_{\varepsilon} (x_2 - y_2) \mathd s. \]
	Using $\chi^{\varepsilon} = \varepsilon^{- 2} \chi (\cdot / \varepsilon)$
	and $\| \chi^{\varepsilon} \|_{L^p} \lesssim \varepsilon^{- 2 (1 - 1 / p)}$
	we obtain with Young's convolution inequalities,
	\[ \begin{aligned}
			\mathbb{E} | M_{\infty}^{\varepsilon} |^2 & \leqslant  \beta^2 r
			(\varepsilon)^2 \int_0^{\infty} \mathd s (\bar{\lambda}_s^{\varepsilon})^2
			\int \mathd x_1 \rho (x_1)  \int \mathd x_2 \rho (x_2) \times                                              \\
			                                          & \times\int \mathd y_1
			\int \mathd y_2 \chi_{\varepsilon} (x_1 - y_1) \dot{G}_s (y_1 - y_2)
			\chi_{\varepsilon} (x_2 - y_2)                                                                             \\
			                                          & \leqslant  \beta^2 r (\varepsilon)^2  \int_0^{\infty} \mathd s
			(\bar{\lambda}_s^{\varepsilon})^2  \| \chi^{\varepsilon} \|_{L^1} \|
			\chi^{\varepsilon} \|_{L^1} \| \dot{G}_s \|_{L^1}                                                          \\
			                                          & \lesssim  r (\varepsilon)^2  \int_0^{\infty} \mathd s
			(\bar{\lambda}_s^{\varepsilon})^2 s^{- 2} \mathe^{- m^2 / s}.
		\end{aligned} \]
	Using Lemma~\ref{lem:Geps} below,
	\[ \begin{aligned}
			\mathbb{E} | M_{\infty}^{\varepsilon} |^2 & \lesssim  r
			(\varepsilon)^2  \int_0^{\varepsilon^{- 2}} \mathd ss^{- 2} \mathe^{-
				m^2 / s} s^{2 (1 - \delta)} + r (\varepsilon)^2 \varepsilon^{- 4 (1 -
			\delta)} \int_{\varepsilon^{- 2}}^{\infty} s^{- 2} \mathd s                                                                    \\
			                                          & \lesssim  r (\varepsilon)^2 \begin{cases}
				                                                                        \log (\varepsilon^{- 2} \vee 1) + 1, & \delta = 1 / 2, \\
				                                                                        \varepsilon^{- 2 + 4 \delta} + 1     & \delta < 1 / 2.
			                                                                        \end{cases}
		\end{aligned} \]
	Combined, choosing $\gamma > 1 / 2$ implies for $\delta = 1 / 2$,
	\[ \mathbb{E} | M_{\infty}^{\varepsilon} |^2 \lesssim r (\varepsilon)^2
		(\log (\varepsilon^{- 2}) + 1) = \log (\varepsilon^{- 2})^{- 2 \gamma}
		(\log (\varepsilon^{- 2}) + 1) \overset{\varepsilon \rightarrow
			0}{\longrightarrow} 0. \]
	Otherwise, if $\delta < 1 / 2$,
	\[ \mathbb{E} | M_{\infty}^{\varepsilon} |^2 \lesssim \varepsilon^{2 \gamma
			- 2 + 4 \delta} \overset{\varepsilon \rightarrow 0}{\longrightarrow} 0,
	\]
	provided $\gamma > 2 (1 / 2 - \delta) $. Passing to a subsequence, this
	implies $M^{\varepsilon_n}_{\infty} \rightarrow 0$ almost surely and
	consequently also $\sup_n | M^{\varepsilon_n} | < \infty$ almost surely.
	For the Gaussian, we can set $Z\equiv 0$ to find
	$U^{\varepsilon_n} (W_{\infty}) = M_{\infty}^{\varepsilon_n}$, which
	gives the second claim in \eqref{eq:Ueps-div}.

	\tmtextbf{$(\Iota \Iota^{\varepsilon})$} To get started, we split this term
	into the two parts,
	\[ \begin{aligned}
			(\Iota \Iota^{\varepsilon}) = & - \beta r (\varepsilon)
			\int_0^{\infty} \mathd s \int \mathd x \rho (x) \sin (\beta
			\varphi^{\varepsilon} (x)) \chi_{\varepsilon} \ast \dot{G}_s F_s^{[1]}
			(\varphi_s)                                                                      \\
			                              & - \beta r (\varepsilon) \int_0^{\infty} \mathd s
			\lambda_s^{\varepsilon} \int \mathd x \rho (x) \sin (\beta
			\varphi^{\varepsilon} (x)) \chi_{\varepsilon} \ast \dot{G}_s [(F_s  -
			F_s^{[1]}) (\varphi_s) + R_s]                                                    \\
			\backassign                   & (\Iota \Iota^{\varepsilon}_1) + (\Iota
			\Iota_{\geqslant 2}^{\varepsilon}) .
		\end{aligned} \]
	\tmtextbf{$(\Iota \Iota^{\varepsilon}_{\geqslant 2})$} We claim that under
	the assumptions of Theorem~\ref{thm:Ueps-div}, this term is uniformly
	bounded in $\varepsilon > 0$ and $\Omega$. Indeed, using again the estimate
	on $\lambda^{\varepsilon}_s$ from Lemma~\ref{lem:Geps}, we compute with
	Proposition~\ref{prop:force-estimates} and the a priori estimates on the
	remainder (Proposition~\ref{lem:B-to-B}),
	\[ \begin{aligned}
			(\Iota \Iota_{\geqslant 2}^{\varepsilon}) & \leqslant  r (\varepsilon)
			\int_0^{\infty} \mathd s \bar{\lambda}_s^{\varepsilon} \left| \beta \int
			\mathd x \rho (x) \sin (\beta \varphi^{\varepsilon} (x))
			(\chi_{\varepsilon} \ast \dot{G}_s) [(F_s  - F_s^{[1]}) (\varphi_s) +
			R_s] \right|                                                                                                    \\
			                                          & \lesssim  r (\varepsilon) \| \chi_{\varepsilon} \|_{L^1}
			\int_0^{\infty} \mathd s \bar{\lambda}_s^{\varepsilon} \| \dot{G}_s [(F_s  -
			F_s^{[1]}) (\varphi_s) + R_s] \|_{L^{\infty}}                                                                   \\
			                                          & \lesssim  r (\varepsilon) \left[ 1 + \int_1^{\varepsilon^{- 2} \vee
					1} \mathd s \langle s \rangle^{(1 - \delta)} \langle s \rangle^{- 2}
				\langle s \rangle^{1 - 2 \delta} \right] + r (\varepsilon)
			\varepsilon^{- 2 (1 - \delta)} \int_{\varepsilon^{- 2}}^{\infty} \mathd
			s \langle s \rangle^{- 2} \langle s \rangle^{1 - 2 \delta} .                                                    \\
			                                          & \lesssim  r (\varepsilon) ((\langle s \rangle^{1 - 3 \delta}) |_{s
					= 1}^{s \equiv \varepsilon^{- 2}} + 1) + r (\varepsilon) \varepsilon^{-
			2 + 6 \delta}                                                                                                   \\
			                                          & \lesssim  r (\varepsilon) (\varepsilon^{- 2 + 6 \delta} + 1) + r
			(\varepsilon) \varepsilon^{- 2 + 6 \delta}.
		\end{aligned} \]
	For $\delta = 1 / 2$, we see that $- 2 + 6 \delta = 1 > 0$ so that
	$\sup_{\varepsilon > 0} (\Iota \Iota^{\varepsilon}_{\geqslant 2}) < \infty$.
	For $\delta < 1 / 2$, the assumption $\gamma > 2 (1 - 3 \delta)$ implies the
	analogous bound.

	\tmtextbf{$(\Iota \Iota^{\varepsilon}_1)$} We show that for $\gamma$ small
	enough according to the assumptions, this term can be split into a divergent
	term, and uniformly bounded almost surely finite term. To get started note
	that
	\[ U^{\varepsilon}_t (\varphi) = \frac{\lambda_t^{\varepsilon}}{2} \int
		\rho (x) \left( \sum_{\sigma = \pm 1} (\mathe^{i \sigma \beta
				\varphi^{\varepsilon} (x)} - 1) \right) \mathd x, \]
	so that in the same way as in Section~\ref{sec:flow-eq}, we find
	\begin{align*}
		  & \mathD U_t (\varphi_t) \dot{G}_t F^{[1]}_t (\varphi_t) \\
		= & -
		\sum_{\sigma_1, \sigma_2 \in \{ \pm 1 \}} \int \mathd x_1 \rho (x_1) \int
		\mathd x_2 \lambda_t^{\varepsilon} \lambda_t \mathe^{i \beta (\sigma_1
			\varphi^{\varepsilon}_t (x_1) + \sigma_2 \varphi_t (x_2))} \sigma_1
		\sigma_2 \beta^2 (\chi^{\varepsilon} \ast \dot{G}_t ) (x_1 - x_2) .
	\end{align*}
	Motivated by the renormalisation constant produced by the neutral
	contribution (c.f. Section~\ref{sec:definitions}), we treat the summands for
	the charged case,
	{\small\[ \mathcal{C}^{\varepsilon}_t = - \sum_{\sigma_1 = \sigma_2 \in \{ \pm 1
					\}} \int \mathd x_1 \rho (x_1) \int \mathd x_2 \lambda_t^{\varepsilon}
				\lambda_t \mathe^{i \beta (\sigma_1 \varphi^{\varepsilon}_t (x_1) +
					\sigma_2 \varphi_t (x_2))} \sigma_1 \sigma_2 \beta^2 (\chi^{\varepsilon}
				\ast \dot{G}_t ) (x_1 - x_2), \]}
	and the neutral case
		{\small\[ \mathcal{N}^{\varepsilon}_t = - \sum_{\sigma_1 = - \sigma_2 \in \{ \pm 1
					\}} \int \mathd x_1 \rho (x_1) \int \mathd x_2 \lambda_t^{\varepsilon}
				\lambda_t \mathe^{i \beta (\sigma_1 \varphi^{\varepsilon}_t (x_1) +
					\sigma_2 \varphi_t (x_2))} \sigma_1 \sigma_2 \beta^2 (\chi^{\varepsilon}
				\ast \dot{G}_t ) (x_1 - x_2), \]}
	separately.

	\tmtextbf{($\mathcal{C}^{\varepsilon}_t$).} We start by rewriting this sum
	again as a trigonometric function,
	\[ \mathcal{C}^{\varepsilon}_t = r (\varepsilon) \int \mathd z \beta^2
		(\chi^{\varepsilon} \ast \dot{G}_t ) (z) \lambda_t
		\lambda_t^{\varepsilon} \int \mathd x \rho (x) \cos (\beta (\varphi_t (z
				- x) + \varphi_t^{\varepsilon} (x))), \]
	where we can add and subtract $\varphi_t (x)$ to obtain using the
	trigonometric identities,
	\[ \begin{aligned}
			\cos (\beta (\varphi_t (z - x) + \varphi_t^{\varepsilon} (x))) = &
			\cos (\beta (\varphi_t (z - x) + \varphi_t (x))) \cos (\beta
			(\varphi_t^{\varepsilon} (x) - \varphi_t (x)))                                                                                    \\
			                                                                 & - \sin (\beta (\varphi_t (z - x) + \varphi_t (x))) \sin (\beta
				(\varphi_t^{\varepsilon} (x) - \varphi_t (x))) .
		\end{aligned} \]
	Since both of these terms are estimated in the exact same way, let us only
	consider the contribution coming from the cosine. Here, we apply the
	trigonometric identities again, now for $\varphi_t = Z_t + W_t$, to rewrite
	\[ \begin{aligned}
			\cos (\beta (\varphi_t^{\varepsilon} (x) - \varphi_t (x))) = & \cos
			(\beta (Z_t^{\varepsilon} (x) - Z_t (x))) \cos (\beta
			(W_t^{\varepsilon} (x) - W_t (x)))                                                                                          \\
			                                                             & - \sin (\beta (Z_t^{\varepsilon} (x) - Z_t (x))) \sin (\beta
			(W_t^{\varepsilon} (x) - W_t (x))) .
		\end{aligned} \]
	Use the trivial estimate $| \cos (\beta (W_t^{\varepsilon} (x) - W_t (x))) |
		\leqslant 1$ for the contribution from the GFF while we use the additional
	regularity $\sup_t \| Z_t \|_{B_{\infty, \infty}^{2 \delta -}} < \infty$
	(see Theorem \ref{thm:T,rho-convergence}) in the drift $Z_t$ to get the
	improved bound
	\[ \sup_x  | \cos (\beta (Z_t^{\varepsilon} (x) - Z_t (x))) | \lesssim
		\varepsilon^{\gamma_1} \| Z_t \|_{B^{\gamma_2}_{\infty, \infty}}, \quad
		\text{provided } \gamma_1 < 2 \delta . \]
	It remains to deal with $\int \mathd x \rho (x) \cos (\beta (\varphi_t (z -
				x) + \varphi_t (x)))$, for which we follow the same procedure,
	\[ \begin{aligned}
			\cos (\beta (\varphi_t (z - x) + \varphi_t (x))) = & \cos (\beta (Z_t
			(z - x) + Z_t (x))) \cos (\beta (W_t (z - x) + W_t (x)))                                                               \\
			                                                   & - \sin (\beta (Z_t (z - x) + Z_t (x))) \sin (\beta (W_t (z - x) +
			W_t (x))),
		\end{aligned} \]
	so that for any $s > 0$,
	\[ \begin{aligned}
			         & \int \mathd x \rho (x) \llbracket \cos (\beta (\varphi_t (z - x) +
			\varphi_t (x))) \rrbracket                                                    \\
			\lesssim & \| \cos (\beta (Z_t (z - \cdot) + Z_t (\cdot))) \|_{B^s_{p,
							p}}  \| \llbracket \cos (\beta (W_t (z - \cdot) + W_t (\cdot)))
			\rrbracket \rho (\cdot) \|_{B^{- s}_{q, q}}                                   \\
			         & + \| \sin (\beta (Z_t (z - \cdot) + Z_t (\cdot))) \|_{B^s_{p,
							p}}  \| \llbracket \sin (\beta (W_t (z - \cdot) + W_t (\cdot)))
			\rrbracket \rho (\cdot) \|_{B^{- s}_{q, q}}
		\end{aligned} \]
	Here, we defined the Wick ordered cosine with respect to the Gaussian $W_t
		(z - x) \pm W_t (x)$ in the usual way,
	\begin{equation}
		\llbracket \cos (\beta (\varphi_t (z - x) \pm \varphi_t (x))) \rrbracket
		\assign \mathe^{\frac{\beta^2}{2} \mathbb{E} [| W_t (z - x) \pm W_t (x)
				|^2]} \cos (\beta (\varphi_t (z - x) \pm \varphi_t (x))),
		\label{eq:def-Wick-sum}
	\end{equation}
	and similarly for the sine. In what follows we will focus on the cosine term.
	The sine can be dealt with in the same way.
	It follows from Lemma~\ref{lem:sing-wick-estimates} below that for any
	$\gamma_1 > 0$, $\gamma_3 > 2 - 3 \delta$ and $\alpha < 2 \delta$ sufficiently
	close to $2 \delta$,
	\[ \sup_{z, \varepsilon} \mathe^{- \beta^2 G_t (z)} t^{- \gamma_3}  | z
		|^{\gamma_1 / 2} \| \llbracket \cos (\beta (W_t (z - \cdot) + W_t
		(\cdot))) \rrbracket \rho (\cdot) \|_{B^{-\alpha}_{q, q}} < \infty \quad a.s.
	\]
	Moreover, from Theorem~\ref{thm:T,rho-convergence}, $\sup_t \| \cos (\beta
		(Z_t (z - \cdot) + Z_t (\cdot))) \|_{B^\alpha_{p, p}} \lesssim 1$ for any $\alpha < 2
		\delta$ so that by a Kolmogorov argument,
	\[ \mathbb{M}^c \assign \sup_{z, \varepsilon} | z |^{\gamma_1} \mathe^{-
			\beta^2 G_t (z)} t^{- \gamma_3}  \int \mathd x \rho (x)  \llbracket \cos
		(\beta (\varphi_t (z - x) + \varphi_t (x))) \rrbracket < \infty \quad
		a.s. \]
	Combined, this implies
	\[ \begin{aligned}
			| \mathcal{C}^{\varepsilon}_t | & \leqslant  \left| r (\varepsilon)
			\int \mathd z \beta^2 (\chi^{\varepsilon} \ast \dot{G}_t ) (z)
			\bar{\lambda}_t \bar{\lambda}_t^{\varepsilon} \int \mathd x \rho (x) \cos (\beta
			(\varphi_t (z - x) + \varphi_t^{\varepsilon} (x))) \right|                                            \\
			                                & \lesssim  r (\varepsilon) \varepsilon^{\gamma_2} \| \dot{G}_t (z) /
			| z |^{\gamma_1} \|_{L^1 (\mathd z)} \| \chi^{\varepsilon} \|_{L^1}
			\bar{\lambda}_t \bar{\lambda}_t^{\varepsilon} \bar{\lambda}_t^{- 2} t^{\gamma_3}
			\mathbb{M}^c                                                                                          \\
			                                & \lesssim  r (\varepsilon) \varepsilon^{\gamma_2} \langle t
			\rangle^{- 2 + \gamma_1 / 2 + \gamma_3} \bar{\lambda}_t^{\varepsilon}
			\bar{\lambda}_t^{- 1} \mathbb{M}^c .
		\end{aligned} \]
	Integrating over the scales and using the usual estimate for
	$\lambda^{\varepsilon}_t$ for $c \in (0, 1)$ we find
	\[ \begin{aligned}
			\int_0^{\infty} | \mathcal{C}^{\varepsilon}_t | \mathd t &
			\lesssim_{\mathbb{M}^c}                                  & r (\varepsilon) \varepsilon^{\gamma_2}
			\int_0^{\infty} \mathd t \langle t \rangle^{- 3 + \gamma_1 / 2 +
				\gamma_3 + \delta} \bar{\lambda}_s^{\varepsilon}
		\end{aligned} . \]
	For $\delta = 1 / 2$, we find for $\gamma_3 > 2 - 3 \delta$ and $\gamma_1 >
		0$ sufficiently small for some $\bar{\gamma} \in (0, 1 / 2)$,
	\[ \int_0^{\infty} \mathd t \langle t \rangle^{- 3 + \gamma_1 / 2 +
			\gamma_3 + \delta} \langle t \rangle^{1 - \delta} \leqslant
		\int_0^{\infty} \mathd t \langle t \rangle^{- 3 / 2 + \bar{\gamma}} <
		\infty, \]
	so that for any $\gamma > 0$,
	\[ \sup_{\varepsilon \in (0, 1)} \int_0^{\infty} |
		\mathcal{C}^{\varepsilon}_t | \mathd t \lesssim C (\mathbb{M}^c)
		\sup_{\varepsilon \in (0, 1)} \log (\varepsilon^{- 2})^{- \gamma}
		\varepsilon^{2 \delta} < \infty, \quad a.s. \]
	For $\delta < 1 / 2$, choosing $\gamma_1 > 0$, $\gamma_3 > 2 - 3 \delta$
	sufficiently small, we find for some $\bar{\gamma}$
	\[ \begin{aligned}
			 & \int_0^{\infty} | \mathcal{C}^{\varepsilon}_t | \mathd t         \\
			 & \lesssim  r (\varepsilon) \varepsilon^{\gamma_2} \int_0^{\infty}
			\mathd t \langle t \rangle^{- 3 + \gamma_1 / 2 + \gamma_3 + \delta}
			\bar{\lambda}_s^{\varepsilon}                                       \\
			 & = r (\varepsilon) \varepsilon^{\gamma_2} \int_0^{\varepsilon^{-
					2}} \langle t \rangle^{- 2 + \gamma_1 / 2 + \gamma_3} + r (\varepsilon)
			\varepsilon^{\gamma_2} \varepsilon^{- 2 (1 - \delta)}
			\int_{\varepsilon^{- 2}}^{\infty} \langle t \rangle^{- 3 + \gamma_1 / 2
			+ \gamma_3 + \delta}                                                \\
			 & \lesssim  r (\varepsilon) \varepsilon^{2 \delta} (1 +
			\varepsilon^{- 2 + 6 \delta - \bar{\gamma}}),
		\end{aligned} \]
	so that the right-hand side is uniformly bounded provided $\gamma > 2 (1 - 4 \delta)$.

	\tmtextbf{($\mathcal{N}^{\varepsilon}_t$).} In the same way as for
	$\mathcal{C}^{\varepsilon}_t$, we rewrite $\mathcal{N}^{\varepsilon}_t$ in
	terms of the trigonometric functions
	\[ \mathcal{N}^{\varepsilon}_t = r (\varepsilon) \int \mathd z \beta^2
		(\chi^{\varepsilon} \ast \dot{G}_t ) (z) \lambda_t
		\lambda_t^{\varepsilon} \int \mathd x \rho (x) \cos (\beta (\varphi_t (z
				- x) - \varphi_t^{\varepsilon} (x))), \]
	In contrast to the charged contribution, the Wick ordering
	\eqref{eq:def-Wick-sum} now introduces a divergent contribution instead.
	Therefore, we split $\mathcal{N}^{\varepsilon}_t$ once more as
	\[ \begin{aligned}
			\mathcal{N}^{\varepsilon}_t = & r (\varepsilon) \int \mathd z \beta^2
			(\chi^{\varepsilon} \ast \dot{G}_t ) (z) \lambda_t
			\lambda_t^{\varepsilon}  \int \mathd x \rho (x)                                                  \\
			                              & + r (\varepsilon) \int \mathd z \beta^2 (\chi^{\varepsilon} \ast
			\dot{G}_t ) (z) \lambda_t \lambda_t^{\varepsilon} \int \mathd x \rho
			(x) (\cos (\beta (\varphi_t (z - x) - \varphi_t^{\varepsilon} (x))) -
			1)                                                                                               \\
			\backassign                   & c^{\varepsilon}_t + r (\varepsilon) \int \mathd z
			\beta^2 (\chi^{\varepsilon} \ast \dot{G}_t ) (z) \lambda_t
			\lambda_t^{\varepsilon} \int \mathd x \rho (x) (\cos (\beta (\varphi_t
						(z - x) - \varphi_t^{\varepsilon} (x))) - 1) .
		\end{aligned} \]
	We claim that under the assumptions on $r (\varepsilon)$, the constant
	$c^{\varepsilon}_t$ diverges, while the difference $\sup_{\varepsilon \in (0, 1)} |
		\mathcal{N}^{\varepsilon}_t - c_t^{\varepsilon} |$ is almost surely finite.
	Indeed, from the asymptotics of $G^{\varepsilon}$ in Lemma \ref{lem:Geps}
	below, it follows that for any $c > 0$, and a constant $C$ allowed to change
	from line to line,
	\[ \begin{aligned}
			c_t^{\varepsilon} & = r (\varepsilon) \int \mathd z \beta^2
			(\chi^{\varepsilon} \ast \dot{G}_t ) (z) \lambda_t
			\lambda_t^{\varepsilon} = Cr (\varepsilon) \| \chi^{\varepsilon}
			\|_{L^1} \| \dot{G}_t \|_{L^1}  (\lambda_t \lambda_t^{\varepsilon} + O
			(1))                                                                       \\
			                  & = Cr (\varepsilon) \langle t \rangle^{- 2}  (\lambda_t
			\lambda_t^{\varepsilon} + O (1)),
		\end{aligned} \]
	where we used that $\| \chi^{\varepsilon} \|_{L^1} = \| \chi \|_{L^1}$. We
	again split the integral over the scales at $\varepsilon^{- 2}$ to extract
	the divergent contribution, using the bounds $\lambda_t^{\varepsilon}$ from
	Lemma~\ref{lem:Geps},
	\[ \begin{aligned}
			\int_0^{\varepsilon^{- 2}} c_t^{\varepsilon} \mathd t & = \beta^2 r
			(\varepsilon)  \int_0^{\varepsilon^{- 2}} \mathd t \langle t \rangle^{-
			1 - \delta} \langle t \rangle^{1 - \delta} + O (1),                                                                              \\
			                                                      & = \beta^2 r (\varepsilon)  \int_0^{\varepsilon^{- 2}} \mathd t
			\langle t \rangle^{- 2 \delta} + O (1)                                                                                           \\
			                                                      & = \begin{cases}
				                                                          C \log (\varepsilon^{- 2})^{- \gamma} (\log (\varepsilon^{- 2}) + O
				                                                          (1)),                                                & \delta = 1 / 2, \\
				                                                          Cr (\varepsilon) (1 + \varepsilon^{- 2 + 4 \delta}), & \delta < 1 /
				                                                          2,
			                                                          \end{cases}
		\end{aligned} \]
	while
	\[ \int_{\varepsilon^{- 2}}^{\infty} c_t^{\varepsilon} \mathd t = Cr
		(\varepsilon) \varepsilon^{- 2 (1 - \delta)}
		\int^{\infty}_{\varepsilon^{- 2}} \langle t \rangle^{- 1 - \delta} \mathd
		t = Cr (\varepsilon) \varepsilon^{- 2 + 2 \delta} = \begin{cases}
			Cr (\varepsilon),                      & \delta = 1 / 2, \\
			C \varepsilon^{\gamma - 2 + 2 \delta}, & \delta < 1 / 2.
		\end{cases} \]
	Combined this implies with the assumptions $\gamma < 1$, in case $\delta = 1
		/ 2$ and $\gamma < 2 (1 - 2 \delta)$, that
	\[ \int_0^{\infty} c_t^{\varepsilon} \mathd t = \left\{\begin{array}{lcl}
			C \log (\varepsilon^{- 2})^{- \gamma} (\log (\varepsilon^{- 2}) + O
			(1)),                                                & \delta = 1 / 2 \\
			Cr (\varepsilon) (1 + \varepsilon^{- 2 + 4 \delta}), & \delta < 1 / 2
		\end{array} \right\} \overset{\varepsilon \rightarrow 0}{\longrightarrow}
		\infty . \]
	It remains to show that $\sup_{\varepsilon \in (0, 1)} \int_0^{\infty} |
		\mathcal{N}^{\varepsilon}_t - c_t^{\varepsilon} | \mathd t < \infty$ almost
	surely. To this end, we Taylor expand the cosine as
		{\small\[ (\cos (\beta (\varphi_t (z - x) - \varphi_t^{\varepsilon} (x))) - 1) = |
				\varphi_t (z - x) - \varphi^{\varepsilon}_t (x) |^2 \int_0^1 \mathd
				\vartheta \cos (\vartheta \beta ((\varphi_t (z - x) -
				\varphi_t^{\varepsilon} (x)))) .\]}
	Recall that for any $\alpha \in (0, 1)$, $\sup_t \left\| \langle t
		\rangle^{- \alpha / 2} {\varphi_t}  \right\|_{B^{\alpha -}_{\infty, \infty}}
		< \infty$ almost surely by Lemma \ref{lem:W-Halpha}, so that for any
	$\gamma_1$
	\[ | \varphi_t (z - x) - \varphi^{\varepsilon}_t (x) | \lesssim \left( | z
		|^2 \| \varphi_t \|_{B^2_{\infty, \infty}} + \varepsilon^{\gamma_1} \|
		\varphi_t \|_{B^{\gamma_1}_{\infty, \infty}} \right) \lesssim \mathbb{X}
		(| z |^2 \langle t \rangle^{1 +} + \varepsilon^{\gamma_1} \langle t
		\rangle^{\gamma_1 / 2 +}), \]
	where $\mathbb{X} \assign \sup_t \| \langle t \rangle^{- \gamma_1 / 2 -}
		\varphi_t \|_{B^{\gamma_1}_{\infty, \infty}} \vee \sup_t \| \langle t
		\rangle^{- 1 -} \varphi_t \|_{B^2_{\infty, \infty}}$ is almost surely
	finite.

	Regarding the cosine term, we proceed similarly to the charged case,
	repeatedly applying the trigonometric identities
	\[ \begin{aligned}
			  & \cos (\vartheta \beta ((\varphi_t (z - x) - \varphi_t^{\varepsilon}
			(x))))                                                                  \\
			= & \cos (\vartheta \beta ((\varphi_t (z - x) - \varphi_t
			(x)))) \cos (\vartheta \beta ((\varphi_t (x) - \varphi_t^{\varepsilon}
			(x))))                                                                  \\
			  & - \sin (\vartheta \beta ((\varphi_t (z - x) - \varphi_t (x))))
			\sin (\vartheta \beta ((\varphi_t (x) - \varphi_t^{\varepsilon} (x))))
			.
		\end{aligned} \]
	As before, we restrict our attention to the cosine term, with the analysis for
	the sines being analogous. The difference due to the mollification can be
	repeated verbatim to obtain
	\[ \cos (\vartheta \beta ((\varphi_t (x) - \varphi_t^{\varepsilon} (x))))
		\lesssim \varepsilon^{\gamma_2} \| Z_t \|_{B^{\gamma_2}_{\infty,
						\infty}}, \]
	knowing that $\sup_t \| Z_t \|_{B^{\gamma_2}_{\infty, \infty}} < \infty$
	almost surely provided $\gamma_2 < 2 \delta $. For the remaining term, we
	again insert the Wick ordering and apply the trigonometric identities for
	$\varphi = Z + W$ to obtain
	\begin{equation*}
		\begin{aligned}
			          & \int \mathd x \rho (x) \llbracket \cos (\vartheta \beta ((\varphi_t (z -
			x) - \varphi_t (x)))) \rrbracket                                                     \\
			\leqslant & \| \llbracket \cos (\vartheta \beta  (W_t (z - \cdot) - W_t
			(\cdot))) \rrbracket \rho (x) \|_{L^1}  \| \cos (\vartheta \beta (Z_t (z
			- \cdot) - Z_t (\cdot)) ) \|_{L^{\infty}} .
		\end{aligned}
	\end{equation*}
	It follows from Lemma~\ref{lem:hk-basic-bounds} and Lemma~\ref{lem:Z-Linfty}
	below that
	\[ \| \cos (\vartheta \beta (Z_t (z - \cdot) - Z_t (\cdot)) )
		\|_{L^{\infty}} \lesssim \| Z_t (z - \cdot) - Z_t (\cdot) \|_{L^{\infty}}
		\lesssim t^{1 / 2 - \delta} | z | . \]
	Combined with Lemma \ref{lem:sing-wick-estimates}, for $\delta' = 1 - (\beta
		\vartheta)^2 / 8 \pi$, $\gamma_3 > - 1 + 4 (1 - \delta')$ and $\gamma_4 > 1
		/ 2$
	\[ \int \mathd x \rho (x) \llbracket \cos (\vartheta \beta ((\varphi_t (z -
		x) - \varphi_t (x)))) \rrbracket \leqslant t^{\gamma_4 + 1 / 2 - \delta}
		| z |^{- \gamma_3 + 1} \mathe^{- \vartheta \beta^2 G_t (z)} \mathbb{M}^n,
	\]
	where
	\[ \mathbb{M}^n \assign \sup_{t, z} t^{- \gamma_4} | z |^{\gamma_3}
		\mathe^{\vartheta \beta^2 G_t (z)} \| \rho (\cdot) \llbracket \cos
		(\vartheta \beta (W_t (z - \cdot) - W_t (\cdot))) \rrbracket \|_{L^1}  <
		\infty, \quad a.s. \]
	Combined, for some implicit random but almost surely finite constant
	depending on $\mathbb{X}, \mathbb{M}^n$ and $\| Z_t
		\|_{B^{\gamma_2}_{\infty, \infty}}$,
	\begin{equation}
		\begin{aligned}
			         & | \mathcal{N}^{\varepsilon}_t - c_t^{\varepsilon} |  \\
			\lesssim & r (\varepsilon) \varepsilon^{\gamma_2} \int \mathd z
			(\chi^{\varepsilon} \ast \dot{G}_t ) (z) \bar{\lambda}_t
			\bar{\lambda}_t^{\varepsilon} (| z |^2 \langle t \rangle^{1 +} +
			\varepsilon^{\gamma_1} \langle t \rangle^{\gamma_1 / 2 +}) t^{- \gamma_4
					+ 1 / 2 - \delta} | z |^{- \gamma_3 + 1} \mathe^{- (\beta \vartheta)^2
				G_t (0)} .
		\end{aligned}  \label{eq:N-c}
	\end{equation}
	Since $2 - 4 (1 - \delta) > - d$, we can choose $\tilde{\gamma}_3 = -
		\gamma_3 + 1 < 2 - 4 (1 - \delta')$ sufficiently large so that
	$\tilde{\gamma}_3 > - d$ and
	\[ \varepsilon^2 \int (\chi^{\varepsilon} \ast \dot{G}_t ) (z) | z
		|^{\tilde{\gamma}_3} \lesssim \varepsilon^{2 + \tilde{\gamma}_3} \langle
		t \rangle^{- 2}, \]
	and in the same way
	\[ \int (\chi^{\varepsilon} \ast \dot{G}_t ) (z) | z |^{\tilde{\gamma}_3 +
				2} \lesssim \langle t \rangle^{- 2 - \tilde{\gamma}_3 / 2 - 1} . \]
	Inserting these bounds in \eqref{eq:N-c}, we obtain
	\[ \begin{aligned}
			| \mathcal{N}^{\varepsilon}_t - c_t^{\varepsilon} | \lesssim & r
			(\varepsilon) \varepsilon^{\gamma_2} \int_0^1 \mathd \vartheta \int
			\mathd z (\chi^{\varepsilon} \ast \dot{G}_t ) (z) | z |^{-
					\tilde{\gamma}_3 + 2} \bar{\lambda}_t \bar{\lambda}_t^{\varepsilon} t^{\gamma_4 + 1
			/ 2 - \delta + 1} \mathe^{- (\beta \vartheta)^2 G_t (0)}                                                                       \\
			                                                             & + r (\varepsilon) \varepsilon^{\gamma_2} \varepsilon^{\gamma_1}
			\int_0^1 \mathd \vartheta \int \mathd z (\chi^{\varepsilon} \ast
			\dot{G}_t ) (z) | z |^{- \tilde{\gamma}_3} \bar{\lambda}_t
			\bar{\lambda}_t^{\varepsilon} \langle t \rangle^{\gamma_1 / 2} t^{\gamma_4 +
			1 / 2 - \delta} \mathe^{- (\beta \vartheta)^2 G_t (0)}                                                                         \\
			\backassign                                                  & (I^a_t) + (I^b_t) .
		\end{aligned} \]
	Integrating over the scales with the estimate on $\lambda^{\varepsilon}_t$
	from Lemma \ref{lem:Geps}, we obtain for the first term after using the
	conditions on $\gamma_i$, $i = 1, \ldots, 4$, we have for some $\bar{\gamma}
		> 0$ arbitrarily small,
	\[ \begin{aligned}
			 & \int_0^{\varepsilon^{- 2}} (I^a_t) \mathd t                        \\ & \lesssim
			\int_0^{\varepsilon^{- 2}} \mathd t r (\varepsilon)
			\varepsilon^{\gamma_2} \int_0^1 \mathd \vartheta \int \mathd z
			(\chi^{\varepsilon} \ast \dot{G}_t ) (z) | z |^{- \tilde{\gamma}_3 + 2}
			t^{2 (1 - \delta)} t^{\gamma_4 + 1 / 2 - \delta + 1} \mathe^{- (\beta
			\vartheta)^2 G_t (0)}                                                 \\
			 & \lesssim  r (\varepsilon) \varepsilon^{\gamma_2}
			\int_0^{\varepsilon^{- 2}} t^{- 3 \delta} t^{\bar{\gamma} / 2} \mathd
			t                                                                     \\
			 & \lesssim  r (\varepsilon) \varepsilon^{\gamma_2} (\varepsilon^{- 2
				+ 6 \delta - \bar{\gamma}} + 1) ,
		\end{aligned} \]
	and
	\[ \begin{aligned}
			 & \int_{\varepsilon^{- 2}}^{^{\infty}} (I^a_t) \mathd t
			\\
			 & \lesssim  r
			(\varepsilon) \varepsilon^{\gamma_2} \varepsilon^{- 2 (1 - \delta)}
			\int^{\infty}_{\varepsilon^{- 2}} \mathd t \int_0^1 \mathd \vartheta
			\int \mathd z (\chi^{\varepsilon} \ast \dot{G}_t ) (z) | z |^{-
					\tilde{\gamma}_3 + 2} t^{1 - \delta} t^{\gamma_4 + 1 / 2 - \delta + 1}
			\mathe^{- (\beta \vartheta)^2 G_t (0)}                                 \\
			 & \lesssim  r (\varepsilon) \varepsilon^{\gamma_2} \varepsilon^{- 2 +
				6 \delta - \bar{\gamma}} .
		\end{aligned} \]
	Choosing $\gamma_i$, $i = 1, \ldots, 4$ such that $\bar{\gamma} < \gamma_2 <
		2 \delta$, we see that this term is uniformly bounded in $\varepsilon$ in
	the case $\delta = 1 / 2$. Similarly, choosing $\bar{\gamma}$ sufficiently
	small and $\gamma_2$ sufficiently large, the condition $\gamma > 2 (1 - 3
		\delta) > 2 (1 - 4 \delta)$ implies the boundedness for $\delta < 1 / 2$.

	For the second term, we argue similarly, again using the conditions on
	$\gamma_i$ we have for some arbitrarily small $\bar{\gamma} > 0$,
	\[ \sup_{\varepsilon \in (0, 1)} \int_0^{\infty} (I_t^b) \mathd t \lesssim
		\sup_{\varepsilon \in (0, 1)} r (\varepsilon) (\varepsilon^{\gamma_2}
		\varepsilon^{- 2 + 6 \delta - \bar{\gamma}} + 1) < \infty, \]
	due to the assumptions on $\gamma$. \
\end{proof}

\begin{lemma}
	\label{lem:Geps}Using the notation introduced in \eqref{eq:def-Ueps-t}, it
	holds that
	\begin{equation}
		\begin{aligned}
			G_t^{\varepsilon} (0) = \frac{1}{4 \pi} \log ((\varepsilon^{- 2} \wedge
			t) \vee 1) + O (1)
		\end{aligned} . \label{eq:Geps-log}
	\end{equation}
\end{lemma}

\begin{proof}
	By definition of $G^{\varepsilon}$, it holds that
	\[ G_t^{\varepsilon} (0) =\mathbb{E} [| W_t^{\varepsilon} (0) |^2]
		=\mathbb{E} | \langle W_t, \chi_{\varepsilon} \rangle |^2 = \langle
		\chi_{\varepsilon}, G_t \chi_{\varepsilon} \rangle . \]
	For $t > \varepsilon^{- 2}$, passing to Fourier space this implies with \[G_t
		= \int_0^t \frac{\mathd s}{s^{- 2}} \mathe^{- (m^2 - \Delta) / s} =
		\mathe^{- (m^2 - \Delta) / t} (m^2 - \Delta)^{- 1},\]
	\[ \langle \chi^{\varepsilon}, G_t \chi^{\varepsilon} \rangle = \int \mathd
		\xi \frac{| \hat{\chi}^{\varepsilon} (\xi) |^2}{m^2 + | \xi |^2} - \int
		\mathd \xi \frac{| \hat{\chi}^{\varepsilon} (\xi) |^2}{m^2 + | \xi |^2}
		(1 - \mathe^{- (m^2 + | \xi |^2) / t}) . \]
	The second term is uniformly bounded using $t < \varepsilon^{- 2}$, and
	moreover as $(1 - \mathe^{- \varepsilon^2 (m^2 + | \xi |^2)}) \rightarrow 0$
	as $\varepsilon \rightarrow 0$, vanishes in the limit by dominated
	convergence. For the first term, using $\tmop{supp} \hat{\chi} \subset B_1
		(0)$, we find
	\[ \int \mathd \xi \frac{| \hat{\chi}^{\varepsilon} (\xi) |^2}{m^2 + | \xi
			|^2} = \frac{1}{2 \pi} \int_0^{\varepsilon^{- 1}} \frac{r}{m^2 + r^2}
		\mathd r = O (1) + \frac{1}{2 \pi} \int_1^{\varepsilon^{- 1}}
		\frac{\mathd r}{r} = \frac{1}{4 \pi} \log (\varepsilon^{- 2} \vee 1) + O
		(1) . \]
	In the case $t < \varepsilon^{- 2}$, we use
	\[ \int \mathd x (\chi^{\varepsilon} (x) - \delta (x)) G_t (x) \lesssim
		\varepsilon^{\alpha} \| G_t \|_{B^{\alpha}_{\infty, \infty}} \lesssim
		t^{- \alpha / 2} t^{\alpha / 2} = O (1), \]
	to obtain,
	\[ \begin{aligned}
			G_t^{\varepsilon} (0) & = \int \mathd x \chi^{\varepsilon} (x) \int
			\mathd y \chi^{\varepsilon} (y) G_t (x - y)                                             \\
			                      & = G_t (0) + \int \mathd x (\chi^{\varepsilon} (x) - \delta (x))
			G_t (x) + \int \mathd x \chi^{\varepsilon} (x) \int \mathd y
			(\chi^{\varepsilon} (y) - \delta (y)) G_t (x - y)                                       \\
			                      & = \frac{1}{4 \pi} \log (t \vee 1) + O (1) .
		\end{aligned} \]

\end{proof}

\begin{lemma}
	\label{lem:Z-Linfty}For any $z \in \mathbb{R}^2$, $t \in [0, \infty)$,
	\[ \| Z_t (z - \cdot) - Z_t (\cdot) \|_{L^{\infty}} \lesssim t^{1 / 2 -
				\delta} | z | . \]
\end{lemma}

\begin{proof}
	This follows directly from~\eqref{eq:hk-g0-gx} in
	Lemma~\ref{lem:hk-basic-bounds}, and the FBSDE~\eqref{eq:FBSDE-leq3} for
	$Z = X - W$.
\end{proof}

\begin{lemma}
	\label{lem:sing-wick-estimates}\tmtextbf{(N)} For any $\gamma_1 > 1 / 2$,
	$\gamma_2 > - 1 + 4 (1 - \delta)$ it holds that
	\[ \sup_{t, z} t^{- \gamma_1} | z |^{\gamma_2} \mathe^{\beta^2 G_t (z)} \|
		\llbracket \cos (\beta (W_t (z - \cdot) - W_t (\cdot))) \rrbracket \rho
		(\cdot) \|_{L^1}  < \infty, \quad a.s. \]
	\tmtextbf{(C)} For any $\gamma_1 > 0$, $\gamma_2 > 2 - 3 \delta$ and $s < 2
		\delta$ sufficiently large, it holds that
	\[ \sup_{t, z} \| \mathe^{- \beta^2 G_t (z)} | z |^{\gamma_1 / 2} t^{-
				\gamma_2} \llbracket \cos (\beta (W_t (\cdot - z) + W_t (\cdot)))
		\rrbracket \rho (\cdot) \|_{B^{- s}_{p, p} (\mathd x)} < \infty, \quad
		a.s. \]
\end{lemma}

The proof of this lemma is given on page~\pageref{proof-lemma-NC} at the end
of Appendix~\ref{app:wick-cos}.

\section{Variational description and large deviations}\label{sec:Var+LDP}
\subsection{Finite volume}

If $\beta^2 < 4 \pi$, the variational description in the finite volume is
essentially a direct consequence of the convergence of the Wick-ordered cosine
and the refinement of the Bou{\'e}--Dupuis formula (Lemma~\ref{lem:BD-form})
from~{\cite{ustunelVariationalCalculationLaplace2014}}. Beyond the first
threshold, the apparent singularity of the sine-Gordon measure means that both
the renormalised potential, and the quadratic part in the cost functional
$J^{\rho, g} \assign J^{V^{\rho} + g}$ as defined in~\eqref{eq:def-value-func}
cannot be expected to stay bounded as $T \rightarrow \infty$. To overcome this
difficulty, we follow the same strategy as for the FBSDE and introduce a
change of variables that isolates the singular part of the control from a more
regular remainder. In these new variables, we can again recover uniform
estimates and pass to the limit for any coupling constant $\bar{\lambda}$.
Throughout this section, we assume that $\rho \prec 1$ is fixed and suppress the
dependency on $\rho$ and $g$ whenever no ambiguities arise.

Translating the same ideas as before now to the level of the variational
problem, we begin by developing the potential along the flow. This yields by
Ito's formula
\begin{equation}
	V_T (X_T^u) = V_t (X_t^u) + \int_t^T \left[ \left( \partial_s V_s +
		\frac{1}{2} \tmop{Tr} \dot{G}_s \mathD^2 V_s \right) (X_s^u) + \mathD V_s
		(X^u_s) Q_s u_s \right] \mathd s + \tmop{mart.}, \label{eq:Vu-ito-1}
\end{equation}
where we use the shorthand
\[ X_t^u \assign X_t (u) = \varphi + \int_0^t Q_s
	u_s \mathd s + W_t . \]
Again, we want to use the fact that $V$ approximately solves the flow
equation~\eqref{eq:fe-V}. Adding the missing terms we can insert the remainder
$\mathcal{H}$ as defined in~\eqref{eq:def1-H} and rewrite~\eqref{eq:Vu-ito-1}
as
\begin{equation}
	V_T (X_T^u) = V_t (X_t^u) + \int_t^T \mathd s \left\{ \mathcal{H}_s (X_s^u)
	+ \mathD V_s (X^u_s) Q_s u_s + \frac{1}{2} (\mathD V_s  \dot{G}_s \mathD
	V_s) (X_s^u) \right\} + \text{mart.} \label{eq:Vu-ito-2} .
\end{equation}
Since $\mathcal{H}$ \ is integrable in the scale parameter $t$ from $\infty$,
it remains to deal with the quadratic terms. Using the notation
\[ z^u_s \assign - Q_s F_s (X^u_s), \]
for the singular part of the control, the variational
problem~\eqref{eq:def-value-func} becomes upon inserting~\eqref{eq:Vu-ito-2}
for $V_T$ and completing the square,
\begin{equation}
	\begin{aligned}
		 & \inf_{u \in \mathcal{A}} J_T^g (u)                                    \\ &= \inf_{u \in \mathbb{H}_a}
		\mathbb{E} \left[ g (X_T^u) + V_0 (X_0^u) + \int_0^T \mathd s \left\{
			\mathcal{H}^T_s \left( X_s^u \right) - \langle z_s^u, u_s \rangle +
			\frac{1}{2} \| z^u_s \|_{L^2}^2 + \frac{1}{2} \| u_s \|^2_{L^2} \right\}
		\right],                                                                 \\
		 & = \inf_{u \in \mathbb{H}_a} \mathbb{E} \left[ g (X_T^u) + V_0 (X_0^u)
			+ \int_0^T \mathd s \left\{ \mathcal{H}^T_s (X_s^u) + \frac{1}{2} \| u_s -
			z^u_s \|^2_{L^2} \right\} \right] .
	\end{aligned} \label{eq:def-J-infty}
\end{equation}
Importantly, this reformulation no longer imposes square integrability on the
control $u$ but only on $u - z^u$, which heuristically corresponds to the
(more regular) remainder. We take this as an invitation to introduce the
change of variables
\begin{equation}
	r_t^u \assign u_t - z^u_t . \label{eq:def-ru}
\end{equation}
The following Lemma ensures that this change of variables does not affect the
variational problem~\eqref{eq:def-J-infty}.

\begin{lemma}
	\label{lem:u(r)}For any $r \in \mathbb{H}^2 (L^2)$, $T \leqslant \infty$ and
	$\rho \prec 1$, there is a unique solution $\hat{Z}^r \in \mathbb{H}^{\infty}
		(L^{\infty})$ to the SDE
	\begin{equation}
		\hat{Z}^r_t = \int_0^t Q_s (r_s - Q_s F^{\rho, T}_s (\hat{Z}^r_s + W_s))
		\mathd s. \label{eq:def-Xr}
	\end{equation}
	In particular, with $\hat{X} = \hat{Z} + W$, defining $u^r_t \assign - Q_t
		F^{\rho, T}_t (\hat{X}^r_t) + r_t$, the control $u^r$ is admissible for the
	finite-horizon control problem \eqref{eq:def-J-infty} and we have $\hat{X}^r
		\equiv X^{u^r}$ and $r_t = u^r_t - z_t^{u^r}$ almost surely.
\end{lemma}

\begin{proof}
	The estimates on the approximate solution to the flow equation imply that
	$Q_t F_t^{\rho, T}$ is globally Lipschitz and bounded, uniformly in $t$. The
	result now follows from standard well-posedness for SDEs with Lipschitz
	coefficients.
\end{proof}

This allows yet another reformulation of~\eqref{eq:def-J-infty} in terms of
the remainder. To avoid confusion with the infinite volume control problem
later, let us make the dependence on $\rho$ explicit again and define the cost
functional
\begin{equation}
	\hat{J}_T^{\rho, g} (r) \assign \mathbb{E} \left[ g (\hat{X}_T^r) +
		V^{\rho}_0 (\hat{X}_0^r) + \int_0^T \mathd s \left\{ \mathcal{H}^{\rho, T}_s
		(\hat{X}_s^r) + \frac{1}{2} \| r_s \|^2_{L^2} \right\} \right]
	\label{eq:def-J-r},
\end{equation}
with $\hat{X}^r$ defined as the unique solution to \eqref{eq:def-Xr}. Observe
that in contrast to $J^{\rho, g}_T$, the functional $\hat{J}^{\rho, g}_T$
satisfies for some $C = C_g > 0$,
\begin{equation}
	\hat{J}^{\rho, g}_T (r) \geqslant - C + \frac{1}{2}  \int_0^T \mathd s \|
	r_s \|^2_{L^2}, \label{eq:lower-bound-Jhat}
\end{equation}
which we immediately verify from~\eqref{eq:def-J-r} and the estimates on
$\mathcal{H}$ (see Proposition
\ref{prop:force-estimates}-\ref{prop:H-r,T-dependence}). In particular, the
cost functional $\hat{J}_T^{\rho, g}$ makes sense also at $T = \infty$.
From~\eqref{eq:lower-bound-Jhat} we see that $\hat{J}^{\rho, g}_T (r) =
	\infty$, whenever $r \nin \mathbb{H}^2_T (L^2)$. We can thus enlarge the set
over which we take the infimum and use Lemma~\ref{lem:u(r)} to see
for any $T < \infty$,
\[ \inf_{r \in \mathbb{H}_a}  \hat{J}_T^{\rho, g} (r) = \inf_{r \in
		\mathbb{H}^2 (L^2)}  \hat{J}_T^{\rho, g} (r) = \inf_{u \in \mathbb{H}_a}
	J^{\rho, g}_T (u) . \]
Therefore, the relation between the FBSDE~\eqref{eq:FBSDE-leq3} and the variational
problem~\eqref{eq:def-value-func} obtained in Theorem
\ref{thm:ctrl-FBSDE}-\ref{thm:ctrl-form} is also valid for the UV-limit,
provided we renormalise the form of the cost functional.

\begin{theorem}
	\label{thm:finite-vol-var}Denote by $(Z^{\rho, g}, R^{\rho, g})$ the unique
	solution to~\eqref{eq:FBSDE-leq3} with $\rho \prec 1$ and $T = \infty$. Then,
	\begin{equation}
		r^{\rho, g}_t \assign - Q_t R^{\rho, g}_t \label{eq:optimal-remainder},
	\end{equation}
	is admissible and optimal for the control problem~\eqref{eq:def-J-r} at $T
		= \infty$.
\end{theorem}

\begin{proof}
	Let us fix a cut-off $\rho \prec 1$ and leave the dependence implicit for this
	proof. Thanks to Lemma~\ref{lem:B-to-B}, the candidate for the optimal
	control for the control problem defined in~\eqref{eq:optimal-remainder}
	satisfies
	\[ \mathbb{E} \int_0^{\infty} \| r^g_s \|^2_{L^2} \mathd s < \infty, \]
	and thus $r^g_s \in \mathbb{H}^2 (L^2 (\mathbb{R}^2))$ so that
	\[ \inf_{r \in \mathbb{H}^2 (L^2)}  \hat{J}_{\infty}^g (r) \leqslant
		\hat{J}_{\infty}^g (r^g) . \]
	It remains to show the reverse inequality. To this end, let $\bar{u}^{T, g}
		= - Q_t (R^{T, g}_t + F_t (X^{T, g}_t))$ be the optimal control for the
	control problem ~\eqref{eq:def-value-func}. Then, for any finite $T$,
	(writing $\bar{z}^{T, g} \equiv z^{\bar{u}^{T, g}}$ and $X^{T, g} = X^{u^{T,
						g}} = Z^{T, g} + W$), it holds that
	\[ \bar{r}_t^{T, g} \assign \bar{u}^{T, g}_t - \bar{z}_t^{T, g} = - Q_t (R^{T, g}_t + F_t
		(X^{T, g}_t)) + Q_t F_t (X_t^{T, g}) = - Q_t R_t^{T, g}, \]
	is optimal for the control problem~\eqref{eq:def-J-infty} as shown in
	Theorem \ref{thm:ctrl-FBSDE}-\ref{thm:ctrl-form}. On the finite volume
	Proposition \ref{prop:force-estimates}-\ref{prop:H-Lipschitz}, implies for
	$\varphi \in \mathcal{S}' (\mathbb{R}^2)$ and $\varepsilon,
		\tilde{\varepsilon} > 0$ sufficiently small,
	\[ \begin{aligned}
			\int_0^{\infty} | (\mathcal{H}^{\infty}_t -\mathbbm{1}_{\{ t \leqslant
			T \}} \mathcal{H}^T_t) (\varphi) | \mathd t & \lesssim
			\int_T^{\infty} | \mathcal{H}^{\infty}_t (\varphi) | \mathd t +
			\int_0^T | (\mathcal{H}^{\infty}_t -\mathcal{H}^T_t) (\varphi) | \mathd
			t                                                                                                           \\
			                                            & \lesssim_{\rho}  \langle T \rangle^{1 - 4 \delta} + \langle T
			\rangle^{- \varepsilon} \underbrace{\int_0^T \langle t \rangle^{- 4
					\delta + \tilde{\varepsilon}} \mathd t}_{< \infty} .
		\end{aligned} \]
	Combined with the convergence of $X_T^g \rightarrow X_{\infty}^g$ from
	Theorem \ref{thm:T,rho-convergence} and the continuity of $g$, this implies
	for any fixed $r \in \mathbb{H}^2 (L^2)$,
	\[ \hat{J}^g_{\infty} (r) = \lim_{T \rightarrow \infty}  \hat{J}_T^g (r) .
	\]
	By the optimality of $r^{T, g}$ for $T < \infty$ this implies for any $r \in
		\mathbb{H}^2 (L^2)$,
	\begin{equation}
		\liminf_{T \rightarrow \infty}  \hat{J}_T^g (r^{T, g}) = \liminf_{T
			\rightarrow \infty} \inf_{r \in \mathbb{H}^2 (L^2)}  \hat{J}^g_T (r)
		\leqslant \lim_{T \rightarrow \infty}  \hat{J}^g_T (r) =
		\hat{J}_{\infty}^g (r) . \label{eq:liminf-leq-Jinfty}
	\end{equation}
	From the continuity of $\nabla g$ and the convergence of the solution
	$(Z^{T, g}, R^{T, g}) \rightarrow (Z^g, R^g)$ derived in Theorem
	\ref{thm:T,rho-convergence}, we immediately get in the $T\to\infty$ limit
	\[ \mathbb{E} \int_0^{\infty} \| r^g_t - r_t^{T, g} \|_{L^2}^2 \mathd t
		=\mathbb{E} \int_0^{\infty} \| Q_t (R_t^{T, g} - R_t^g) \|^2_{L^2} \mathd
		t \lesssim \mathbb{E} \int_0^{\infty} \mathd t \langle t \rangle^{- 2}
		\langle T \rangle^{- \varepsilon} \rightarrow 0,\]
	Therefore, by Fatou's Lemma and the continuity of $g$ and $V_0$,
	\begin{equation}
		\hat{J}_{\infty}^g (r^g) =\mathbb{E} \left[ g (X_{\infty}^g) + V_0 (X_0^g)
			+ \int_0^{\infty} \mathcal{H}^{\infty}_s (X^g_s) \mathd s +
			\int_0^{\infty} \| r^g_t \|_{L^2}^2 \mathd t \right] \leqslant \liminf_{T
			\rightarrow \infty}  \hat{J}_T^g (r^{T, g}) . \label{eq:Jr-leq-liminf}
	\end{equation}
	Combining~\eqref{eq:liminf-leq-Jinfty} and~\eqref{eq:Jr-leq-liminf} we
	obtain the missing inequality,
	\[ \hat{J}_{\infty}^g (r^g) \leqslant \liminf_{T \rightarrow \infty}
		\hat{J}_T^g (r^{T, g}) \leqslant \inf_{r \in \mathbb{H}^2 (L^2)}
		\hat{J}_{\infty}^g (r) . \]
\end{proof}

\begin{remark}
	The boundedness of the cosine interaction is the reason we have good bounds
	over the optimisers to the control problem~\eqref{eq:def-value-func}
	uniformly in $T$. This allows us to bypass the technically more involved
	$\Gamma$-convergence for the cost functionals $\hat{J}^T \rightarrow J$ to
	remove the small-scale regularisation $T$ as was instead necessary
	in~{\cite{barashkovVariationalMethod$Phi^4_3$2020}} in the case of the $\Phi^4_3$
	model on a bounded domain.
\end{remark}

The variational description for the Laplace transform is now an immediate
consequence of the description in Theorem~\ref{thm:finite-vol-var}.

\begin{corollary}
	The variational problem for the Laplace transform~\eqref{eq:def-Wf} also
	holds for $T = \infty$, that is,
	\[ - \log \nu_{\tmop{SG}}^{\rho} (\mathe^{- g}) = \mathcal{W}^{\rho} (g)
		\assign \inf_{r \in \mathbb{H}^2 (L^2)}  \hat{J}_{\infty}^{\rho, g} (r) -
		\inf_{r \in \mathbb{H}^2 (L^2)}  \hat{J}_{\infty}^{\rho, 0} (r) . \]
\end{corollary}

\begin{proof}
	From the weak convergence of $\nu_{\tmop{SG}}^{\rho, T} \rightarrow
		\nu_{\tmop{SG}}^{\rho}$ in $H^{- \varepsilon}$,
	{\small{\[ - \log \nu_{\tmop{SG}}^{\rho} (\mathe^{- g}) = \lim_{T \rightarrow
						\infty} - \log \nu_{\tmop{SG}}^{\rho, T} (\mathe^{- g}) = \lim_{T
						\rightarrow \infty}  \mathcal{W}_T^{\rho} (g) = \lim_{T \rightarrow
						\infty}  \mathcal{V}_T^{V^{\rho} + g} - \lim_{T \rightarrow \infty}
					\mathcal{V}_T^{V^{\rho}} = \mathcal{W}^{\rho} (g) , \]}}
	where we used~\eqref{eq:def-W} \ and Theorem~\ref{thm:finite-vol-var} to
	justify the last two equalities.
\end{proof}

\begin{remark}
	It should be emphasised that the change of variables~\eqref{eq:def-ru} makes
	the extension to $T = \infty$ possible: passing to the remainder term $r^u =
		u - z^u$ allows us to incorporate the singular part $z^u$ of the control
	into the flow equation remainder $\mathcal{H}^{\rho, T}_s$ while optimising
	only over absolutely continuous shifts $r^u$. Indeed, while we have $z^u \in
		\mathbb{H}_T^2 (L^2)$ for any $u \in \mathbb{H}_T^2 (L^2)$, our estimates on
	$z^u$ only allow
	\[ \| z^u_t \|_{L^2}^2 \lesssim \langle t \rangle^{- 2 \delta}, \]
	which is not sufficient to conclude $z^u \in \mathbb{H} (L^2)$ unless
	$\delta > 1 / 2$ ($\Leftrightarrow \beta^2 < 4 \pi$). In contrast, the
	estimates on the optimal FBSDE (see e.g.~Lemma~\ref{lem:B-to-B}) suggests
	that the remainder $r=z^u-u$ remains square-integrable for the whole
	subcritical regime $\delta > 0$ (provided of course that an appropriate
	approximate solution $V$ to the flow equation is used).
\end{remark}

\begin{remark}
	\label{rem:gradJ=R}Differentiating~\eqref{eq:def-J-infty} with respect to
	the initial value $X_0 = \varphi$, we obtain a formula for the gradient of
	the value function in terms of the solution to the optimal
	FBSDE~\eqref{eq:FBSDE-leq3}
	\begin{equation}
		\nabla \mathcal{V}^{V + g} (\varphi) = \nabla J^g (\bar{u}^g ; \varphi) =
		(\nabla g + F_0) (\varphi) + R^g_0 (\varphi) . \label{eq:gradV=F+R}
	\end{equation}
\end{remark}

\subsection{Infinite volume}

We finally want to remove the restriction to the finite volume. Of course, the
potential will not be meaningful without a spatial cut-off. What saves the
variational problem for the Laplace transform in the infinite volume are the
localisation properties we derived earlier: since the effect of a local
perturbation only has a localised effect on the optimal control by Proposition
\ref{prop:stability}-\ref{prop:local-perturbations} we are able to show that
the functional
\[ \hat{\mathcal{J}}^{g, \rho} (v) \assign \hat{J}^{g, \rho} (v +
	\bar{r}^{\rho}) - \hat{J}^{0, \rho} (\bar{r}^{\rho}), \]
stays meaningful in the infinite volume limit, at least if the functional $g$
is sufficiently localised and the coupling constant $\bar{\lambda}$ is small enough.
This change of variables follows the same idea we used for the finite volume
variational problem: it again allows us to absorb the singular part in a
normalisation while we only optimise along the absolutely continuous directions,
which in this case corresponds to controls in $\mathbb{D}=\mathbb{H}^2 (L^{2,
			n})$.

The aim is to show the following.

\begin{theorem}
	\label{thm:infinite-vol-var}Let $R^0$ be the backward component of the
	solution to the FBSDE~\eqref{eq:FBSDE-leq3} for $g = 0$, $\rho = 1, T =
		\infty$ and define $\bar{r} \assign -QR^0$. Then, with
	\[ \hat{\mathcal{J}}^g (v) =\mathbb{E} \left[ g (\hat{X}_{\infty}^{\bar{r} +
					v}) + \int_0^{\infty} \mathcal{H}_s^1 (\bar{r}_s, v_s) \mathd s +
			\frac{1}{2} \int_0^{\infty} \| v_s \|^2_{L^2} \mathd s + \int_0^{\infty}
			\langle \bar{r}_s, v_s \rangle_{L^2} \mathd s \right], \]
	the Laplace transform of the infinite volume sine-Gordon measure
	$\nu_{\tmop{SG}}$ satisfies the variational problem
	\[ \mathcal{W} (g) \assign - \log \nu_{\tmop{SG}} (\mathe^{- g}) = \inf_{v
			\in \mathbb{D}}  \hat{\mathcal{J}}^g (v) . \]
	Here, the functional $\mathcal{H}_s^{\rho} (\bar{r}_s, v_s)$ is defined for
	any $\rho \preceq  1$ in terms of~\eqref{eq:def-h-coefficients}, in
	complete analogy to~\eqref{eq:def-pot-remainder},
	\[ \mathcal{H}_t^1 (\bar{r}_t, v_t) \assign \sum_{\ell = 4}^6 \int \mathd
		\xi_{1 : \ell} h_t (\xi_{1 : \ell}) [\psi^{\bar{r} + v}_t -
				\psi^{\bar{r}}_t] (\xi_{1 : \ell}), \]
	with $\psi^{\bar{r} + v}_t (\xi_{1 : \ell}) \assign \exp \left( i \beta
		\sum_{k = 1}^{\ell} \sigma_k  \hat{X}_t^{\bar{r} + v} (x_k) \right)$.
\end{theorem}

\begin{proof}
	\tmtextit{Restriction to $\mathbb{D}_g$:} Motivated by Proposition
	\ref{prop:stability}-\ref{prop:local-perturbations}, we expect that the
	regular part of the control is captured nicely by the domain,
	\begin{equation}
		\mathbb{D}_g =\mathbb{D}_g (C) \assign \left\{ v \in \mathbb{H}_a : \;
		\mathbb{E} \int_0^{\infty} \| v_s \|^2_{L^{2, n}} \mathd s \leqslant C | g
		|_{1, 2, n} \right\}, \label{eq:Df-r}
	\end{equation}
	provided $C > 0$ is chosen sufficiently large. We first show convergence of
	the restricted variational problem
	\[ \widehat{\mathcal{W}}_g^{\rho} (g) \assign \inf_{v \in \mathbb{D}_g (C)}
		\hat{\mathcal{J}}^{g, \rho} (v) \overset{\rho \rightarrow
			1}{\longrightarrow} \mathcal{W} (g) \assign - \log \nu_{\tmop{SG}}
		(\mathe^{- g}) = \inf_{v \in \mathbb{D}_g}  \hat{\mathcal{J}}^g (v) . \]
	We claim that this restriction does not change the finite volume variational
	problem, that is for any $\rho \prec 1$,
	\begin{equation}
		\inf_{v \in \mathbb{D}_g}  \hat{\mathcal{J}}^{g, \rho} (v) =
		\widehat{\mathcal{W}}_g^{\rho} (g) = \mathcal{W}^{\rho} (g) = \inf_{v \in
			\mathbb{H}^2 (L^2)}  \hat{\mathcal{J}}^{g, \rho} (v) . \label{eq:Wf=W}
	\end{equation}
	We know from Theorem~\ref{thm:finite-vol-var} that $\bar{r}^{g, \rho}_t =
		- Q_t R_t^{g, \rho}$ is optimal for the variational
	problem~\eqref{eq:def-J-r}, and from Proposition
	\ref{prop:force-estimates}-\ref{prop:local-perturbations} that $\| R_t^{g,
				\rho} - R_t^{0, \rho} \|_{L^{2, n}} \lesssim | g |_{1, 2, n}$. Thus, for
	some constant $C_g > 0$,
	\[ \mathbb{E} \int_0^{\infty} \| \overline{r_s}^{g, \rho} - \bar{r}_s^{\rho}
		\|_{L^{2, n}}^2 \mathd s \lesssim \mathbb{E} \int_0^{\infty} \langle s
		\rangle^{- 2} \| R_t^{g, \rho} - R_t^{0, \rho} \|^2_{L^{2, n}} \leqslant
		C_g | g |_{1, 2, n}^2 . \]
	But then $\bar{v}^{g, \rho} = \bar{r}^{g, \rho} - \bar{r}^{\rho} \in
		\mathbb{D}_g$ for $C$ sufficiently large this implies \eqref{eq:Wf=W} for
	any $C \geqslant C_g$.

	\tmtextit{Convergence:} We show that uniformly on $\mathbb{D}_g$,
	\[ \begin{aligned}
			\hat{\mathcal{J}}^{g, \rho} (v) & = \mathbb{E} \left[ g
				(\hat{X}_{\infty}^{\bar{r}^{\rho} + v}) + \int_0^{\infty}
				\mathcal{H}_s^{\rho} (\bar{r}^{\rho}_s, v_s) \mathd s + \frac{1}{2}
				\int_0^{\infty} \| \bar{r}^{\rho}_s + v_s \|^2_{L^2} \mathd s -
			\frac{1}{2} \int_0^{\infty} \| \bar{r}^{\rho}_s \|^2_{L^2} \right]                                   \\
			                                & \overset{\rho \rightarrow 1}{\longrightarrow}  \mathbb{E} \left[ g
				(\hat{X}_{\infty}^{\bar{r} + v}) + \int_0^{\infty} \mathcal{H}_s^1
				(\bar{r}_s, v_s) \mathd s + \frac{1}{2} \int_0^{\infty} \| v_s
				\|^2_{L^2} \mathd s + \int_0^{\infty} \langle \bar{r}_s, v_s
				\rangle_{L^2} \mathd s \right] .
		\end{aligned} \]
	We proceed term by term. Going left to right, we start by estimating
	\[ | g (\phi_1) - g (\phi_2) | = \int_0^1 (\nabla g (\phi_1 + \vartheta
		(\phi_2 - \phi_1))) (\phi_2 - \phi_1) \mathd \vartheta \lesssim | g |_{1,
				2, n} \| \phi_2 - \phi_1 \|_{L^{2, n}} . \]
	The convergence $g (\hat{X}_{\infty}^{\rho, \bar{r}^{\rho} + v}) \rightarrow
		g (\hat{X}_{\infty}^{\bar{r} + v})$ then follows from
	Lemma~\ref{lem:X-rho-convergence} below. For the remainder term, we write,
	\[ | \mathcal{H}_s^1 (\bar{r}_s, v_s) -\mathcal{H}_s^{\rho}
		(\bar{r}^{\rho}_s, v_s) | \leqslant | \mathcal{H}_s^1 (\bar{r}_s, v_s)
		-\mathcal{H}_s^1 (\bar{r}^{\rho}_s, v_s) | + | \mathcal{H}_s^1
		(\bar{r}^{\rho}_s, v_s) -\mathcal{H}_s^{\rho} (\bar{r}^{\rho}_s, v_s) | .
	\]
	For the first term, the definition of $\mathcal{H}^1$ and the estimates on
	the coefficients $h$ in~\eqref{eq:h-coefficients-scaling} imply,
	\[ \begin{aligned}
			| \mathcal{H}_s^1 (\bar{r}_t, v_t) -\mathcal{H}_t^1 (\bar{r}^{\rho}_t,
			v_t) | & \leqslant  \sum_{\ell = 4}^6 \int \mathd \xi_{1 : \ell} h_t
			(\xi_{1 : \ell}) [\psi^{\bar{r} + v}_t - \psi^{\bar{r}}_t -
					(\psi^{\bar{r}^{\rho} + v}_t - \psi^{\bar{r}^{\rho}}_t)] (\xi_{1 :
			\ell})                                                                   \\
			       & \lesssim  \interleave h_t \interleave (\| \delta_v
			\hat{X}_t^{\bar{r} + v} - \delta_v  \hat{X}_t^{\rho, \bar{r}^{\rho} +
			v} \|_{L^{2, n}})                                                        \\
			       & \lesssim  \bar{\lambda}_t^4 \langle t \rangle^{- 4} \| \delta_v
			\hat{X}_t^{\bar{r} + v} - \delta_v  \hat{X}_t^{\rho, \bar{r}^{\rho} +
					v} \|_{L^{2, n}} .
		\end{aligned} \]
	Since $\bar{\lambda}_t^4 \langle t \rangle^{- 4} \in L^1 (\mathbb{R}_+)$, the
	desired convergence will follow from Lemma \ref{lem:X-rho-convergence}
	below. For the second term,
	\[ \begin{aligned}
			| \mathcal{H}_s^1 (\bar{r}^{\rho}_t, v_t) -\mathcal{H}_s^{\rho}
			(\bar{r}^{\rho}_t, v_t) | & \leqslant  \sum_{\ell = 4}^6 \int \mathd
			\xi_{1 : \ell} | 1 - \rho (\xi_{1 : \ell}) |  | h_t (\xi_{1 : \ell})
			[\psi^{\bar{r}^{\rho} + v}_t - \psi^{\bar{r}^{\rho}}_t] (\xi_{1 :
			\ell}) |                                                                                       \\
			                          & \leqslant  \| 1 - \rho \|_{L^{2, - n}} \interleave h_t \interleave
			\| \hat{X}_t^{\bar{r}^{\rho} + v} - \hat{X}_t^{\bar{r}^{\rho}}
			\|_{L^{2, n}}                                                                                  \\
			                          & \lesssim  \| 1 - \rho \|_{L^{2, - n}} \bar{\lambda}_t^4 \langle t
			\rangle^{- 4}  \| v_t \|_{L^{2, n}} .
		\end{aligned} \]
	Hence,
	\[ \lim_{\rho \rightarrow 1} \sup_{v \in \mathbb{D}_g} \mathbb{E}
		\int_0^{\infty} | \mathcal{H}_s^1 (\bar{r}_s, v_s) -\mathcal{H}_s^1
		(\bar{r}^{\rho}_s, v_s) | \mathd s = 0. \]
	Finally, for the quadratic terms, we expand the square to find,
	\[ \frac{1}{2} \| \bar{r}^{\rho}_s + v_s \|^2_{L^2} - \frac{1}{2} \|
		\bar{r}^{\rho}_s \|^2_{L^2} = \frac{1}{2} \| v_s \|^2_{L^2} + \langle
		\bar{r}^{\rho}_s, v_s \rangle_{L^2} . \]
	Consequently, for any $v \in \mathbb{D}_g$,
	\[ \mathbb{E} \int_0^{\infty} \mathd s \langle \bar{r}^{\rho}_s -
		\bar{r}_s, v_s \rangle_{L^2} \lesssim \left( \mathbb{E} \int_0^{\infty}
		\| \bar{r}^{\rho}_s - \bar{r}_s \|^2_{L^{2, - n}} \mathd s \right)^{1 /
			2} \left( \mathbb{E} \int_0^{\infty} \| v_s \|^2_{L^{2, n}} \mathd s
		\right)^{1 / 2}, \]
	which with $\bar{r}_t - \bar{r}^{\rho}_t = - Q_t (R_t - R_t^{\rho})$ and the
	estimates on $Q$ in Lemma~\ref{lem:hk-scaling} and Proposition
	\ref{prop:stability}-\ref{prop:local-perturbations} converges to $0$
	uniformly on $\mathbb{D}_g$.

	\tmtextit{Recovering the full domain}: Finally, we show that for any $C
		\geqslant C_g$,
	\[ \inf_{v \in \mathbb{D}_g (C)}  \hat{\mathcal{J}}^g (v) = \inf_{v \in
			\mathbb{D}}  \hat{\mathcal{J}}^g (v) . \]
	Since $\mathbb{D}_g (C) \subset \mathbb{D}$, clearly $\inf_{v \in
			\mathbb{D}_g (C)}  \hat{\mathcal{J}}^g (v) \leqslant \inf_{v \in \mathbb{D}}
		\hat{\mathcal{J}}^g (v)$. For the reverse inequality, let $\bar{v} \in
		\mathbb{D}$ and let $\bar{C} > \| \bar{v} \|^2_{\mathbb{D}} \assign \|
		\bar{v} \|^2_{\mathbb{H}^2 (L^{2, n})}$ so that $\bar{v} \in
		\hat{\mathbb{D}}_g (\bar{C})$ and thus by the argument used to show
	convergence on $\mathbb{D}_f (C)$,
	\[ \mathcal{J}^g (\bar{v}) \geqslant \inf_{v \in \mathbb{D}_g (\bar{C})}
		\mathcal{J}^g (v) =\mathcal{J}^g (v^{\ast}) = \inf_{v \in \mathbb{D}_g
			(C)} \mathcal{J}^g (v) . \]
	Taking the infinitum over $\bar{v} \in \mathbb{D}$ in the inequality above
	then yields the claim.
\end{proof}

We still have to supplement the following two convergence results to finish up
the proof of Theorem~\ref{thm:infinite-vol-var}.

\begin{lemma}
	\label{lem:X-rho-convergence}Using the notation defined
	in~\eqref{eq:def-Xr}, it holds uniformly in $v \in \mathbb{D}_g$,
	\[ \lim_{\rho \rightarrow 1} \sup_{s \geqslant 0} \| \hat{X}_s^{\bar{r} + v}
		- \hat{X}_s^{\bar{r}^{\rho} + v} \|_{L^{2, - n}} = 0. \]
\end{lemma}

\begin{proof}
	This follows immediately from Proposition \ref{prop:force-estimates} following the arguments in Proposition
	\ref{prop:stability}-\ref{prop:rho-convergence} and Proposition
	\ref{prop:stability}-\ref{prop:fe-rho-dependence}.
\end{proof}

\subsection{Large deviations}\label{sec:LDP}

We apply the variational problem for the Laplace transform just derived in
Theorem~\ref{thm:infinite-vol-var} to show the Laplace principle from
Theorem~\ref{thm:LDP} for the limiting measure $\nu_{\tmop{SG}}$. More
precisely, we want to study the family of rescaled measures are formally given
by
\begin{equation}
	\text{``}\nu_{\tmop{SG}}^{\hbar} (\mathd \varphi) = \Xi_{\hbar}^{- 1} \exp
	(\hbar^{- 1} V (\varphi)) \mu^{\hbar} (\mathd \varphi)\text{,''}
	\label{eq:def-SGh}
\end{equation}
in the limit $\hbar \rightarrow 0$. Here $V (\varphi) = \lambda \int \mathd x
	\cos (\beta \varphi (x))$ denotes as before the cosine interaction and
$\mu^{\hbar}$ is the Gaussian measure with covariance $\hbar (m^2 - \Delta)^{-
		1}$. Taking the Wick-ordering with respect to $\mu^{\hbar}$ and the obvious
modification to the interpolation $G_s^{\hbar} \assign \hbar G_s$, the same
derivation as before yields a description for $\nu_{\tmop{SG}}^{\hbar}$ via
the rescaled FBSDE,
\begin{equation}
	\begin{cases}
		Z^{\hbar}_t = - \int_0^t \mathd s \dot{G}_s^{\hbar} (F^{\hbar}_s
		(Z^{\hbar}_s + \hbar^{1 / 2} W_s)) - \int_0^t \mathd s \dot{G}_s
		R^{\hbar}_s, \\
		R^{\hbar}_t =\mathbb{E}_t \int_t^{\infty} \mathd s \hbar H^{\hbar}_s
		(Z^{\hbar}_s + \hbar^{1 / 2} W_s) -\mathbb{E}_t \int_t^{\infty} \mathd s
		\hbar \mathD F^{\hbar}_s (Z^{\hbar}_s + \hbar^{1 / 2} W_s)
		\dot{G}^{\hbar}_s R^{\hbar}_s,
	\end{cases} \label{eq:FBSDE-h}
\end{equation}
where $F^{\hbar} = F^{\hbar, \infty}$ and $F^{\hbar, T}$ is the approximate
solution to the flow equation~\eqref{eq:fe-picard} with covariance $G^{\hbar}$
and the rescaled initial data,
\[ F_T^{[1], T, \hbar} (\varphi) \assign - \beta \hbar^{- 1} \lambda_0
	\mathe^{\frac{\hbar \beta^2}{2} G_T (0)} \sin (\beta \varphi) = - \hbar^{-
		1} \lambda_t^{\hbar} \beta \sin (\beta \varphi) . \]
At least when $\hbar \in [0, 1]$, we have $\bar{\lambda}^{\hbar}_t \leqslant
	\bar{\lambda}_t$ and the well-posedness of~\eqref{eq:FBSDE-h} follows in the exact
same way as the well-posedness of~\eqref{eq:FBSDE-leq3} in
Proposition~\ref{prop:contraction}. Moreover, rescaling the analysis of
Theorem~\ref{thm:T,rho-convergence}, we see that also the drift $Z^{\hbar}$
has a terminal value with regularity
\begin{equation}
	Z^{\hbar}_{\infty} \in L^{\infty} (\mathd \mathbb{P} ; H^{2 - \hbar \beta^2 / 4 \pi,
			- n}) \subset L^{\infty} (\mathd \mathbb{P} ; H^{2 - \beta^2 / 4 \pi, - n}) .
	\label{eq:Zh-infty}
\end{equation}
Thus, the same reasoning as in Theorem \ref{thm:T,rho-convergence} applies and
can use the solution to the FBSDE~\eqref{eq:FBSDE-h} to make the formal
definition~\eqref{eq:def-SGh} precise. Define the measures
$\nu_{\tmop{SG}}^{\hbar}$ as a random shift of the rescaled Gaussian free
field,
\[ \nu_{\tmop{SG}}^{\hbar} \assign \tmop{Law} (Z_{\infty}^{\hbar} \noplus
	\noplus + \hbar^{1 / 2} W_{\infty}) . \]
Since we are only interested in the limit $\hbar \rightarrow 0$, we can limit
our considerations to a small neighbourhood of $0$. This has the advantage
that the measure $\nu_{\tmop{SG}}^{\hbar}$ will essentially behave like the
sine-Gordon measure with parameter $\beta \hbar^{1 / 2}$ (see
also~{\cite[Section C]{colemanQuantumSineGordonEquation1975}}): if $\hbar$ is
sufficiently small (say $\hbar < \hbar_0$ where $\beta^2 \hbar_0 < 4 \pi$),
then, by~\eqref{eq:Zh-infty}, the measure $\nu_{\tmop{SG}}^{\hbar}$ is a
Girsanov shift of the free field. More concretely this means we can carry out
the analysis of~\eqref{eq:FBSDE-h} by relying on the convergence of the
Wick-ordered cosine illustrated in \eqref{eq:trig-id}, as in the absolutely
continuous first region $\beta^2 < 4 \pi$ already covered
in~{\cite{barashkovStochasticControlApproach2022}}. It only remains to check
that this approach is compatible with our definition of the measures
via~\eqref{eq:FBSDE-h}. This is in part resolved by the following Lemma.

\begin{lemma}
	\label{lem:Zhbar-wick}Let $\hbar < \hbar_0$. The solution $Z^{\hbar}$
	to~\eqref{eq:FBSDE-h} satisfies
	\begin{equation}
		Z^{\hbar}_t = \beta \lambda_0  \int_0^t \mathd s \dot{G}_s \mathbb{E}_s
		[\llbracket \sin (\beta (Z_{\infty}^{\hbar} + \hbar^{1 / 2} W_{\infty}))
		\rrbracket] \label{eq:Zhbar-wick} = \beta \lambda_0 I_t
		(\bar{u}^{\hbar}_t),
	\end{equation}
	where $\llbracket \sin (\varphi + \hbar^{1 / 2} W_{\infty}) \rrbracket$ is
	defined for any $\varphi \in H^{1, - n}$ via~\eqref{eq:def-Vinfty} below.
\end{lemma}

\begin{proof}
	Let us first introduce again the approximate FBSDEs with the cut-off $T <
		\infty$. Then, we know from the definition of $F^{T, \hbar}$ and $R^{T,
				\hbar}$ that
	\begin{equation}
		F^{T, \hbar}_s (X^{T, \hbar}_s) + R^{T, \hbar}_s =\mathbb{E}_s [\nabla
		V^{T, \hbar}_T (\beta (Z_T^{T, \hbar} + \hbar^{1 / 2} W_t))] .
		\label{eq:Fh+Rh=cond}
	\end{equation}
	For $\hbar < \hbar_0$, it follows from Lemma~\ref{lem:I(u)-regularity} and
	the convergence of $Z^T_T \rightarrow Z_{\infty}$ in $\mathbb{H}^{\infty}
		(L^{\infty})$,
	\[ \begin{aligned}
			\lim_{T \rightarrow \infty} \| Z_T^{T, \hbar} - Z^{\hbar}_{\infty}
			\|_{H^{1, - n}}^2 = 0.
		\end{aligned} \]
	Moreover, we can use the trigonometric identities to rewrite
	\begin{equation}
		\llbracket \sin (\beta (Z^{\hbar}_T + \hbar^{1 / 2} W_T)) \rrbracket
		\assign \cos (\beta Z^{\hbar}_T) \llbracket \sin (\beta \hbar^{1 / 2} W_T)
		\rrbracket + \sin (\beta Z^{\hbar}_T) \llbracket \cos (\beta \hbar^{1 / 2}
		W_T) \rrbracket, \label{eq:sin-as}
	\end{equation}
	and similarly for the cosine. By Lemma \ref{lem:wick-conv}, the Wick-ordered
	sine (cosine) $\llbracket \sin (\beta \hbar^{1 / 2} W_T) \rrbracket$
	converge in $H^{- 1 + \varepsilon, - n}$. Thus, the products on the
	right-hand side of~\eqref{eq:sin-as} stay well-defined in the limit as $T
		\rightarrow \infty$ and consequently \ $\llbracket \sin (\beta (Z^{T,
				\hbar}_T + \hbar^{1 / 2} W_T)) \rrbracket$ and $\llbracket \cos (\beta
			(Z^{T, \hbar}_T + \hbar^{1 / 2} W_T)) \rrbracket$ converge in $L^2 (\mathd \mathbb{P}
		; H^{- 1 + \varepsilon, - n})$ and almost surely to a well-defined limit
	which we denote by
	\begin{equation}
		\llbracket \sin (\beta (Z^{\hbar}_{\infty} + \hbar^{1 / 2} W_{\infty}))
		\rrbracket \assign \cos (\beta Z^{\hbar}_{\infty}) \llbracket \sin (\beta
		\hbar^{1 / 2} W_{\infty}) \rrbracket + \sin (\beta Z^{\hbar}_{\infty})
		\llbracket \cos (\beta \hbar^{1 / 2} W_{\infty}) \rrbracket,
		\label{eq:def-Vinfty}
	\end{equation}
	and respectively
	\[ \llbracket \cos (\beta (Z^{\hbar}_{\infty} + \hbar^{1 / 2} W_{\infty}))
		\rrbracket \assign \cos (\beta Z^{\hbar}_{\infty}) \llbracket \cos (\beta
		\hbar^{1 / 2} W_{\infty}) \rrbracket + \sin (\beta Z^{\hbar}_{\infty})
		\llbracket \sin (\beta \hbar^{1 / 2} W_{\infty}) \rrbracket . \]
	By uniqueness of the limit, we can pass to the limit $T \rightarrow \infty$
	in~\eqref{eq:Fh+Rh=cond} to conclude
	\[ F_t^{\hbar} (Z_t^{\hbar} + \hbar^{1 / 2} W_t) + R_t^{\hbar} = - \beta
		\lambda \mathbb{E}_t \llbracket \sin (\beta (Z^{\hbar}_{\infty} +
			\hbar^{1 / 2} W_{\infty})) \rrbracket, \]
	which immediately implies~\eqref{eq:Zhbar-wick}.
\end{proof}

Applying the same argument as in Lemma~\ref{lem:Zhbar-wick} to the cost
functional $\hat{\mathcal{J}}^{\hbar}$ we can undo the change of variables to
the remainder in~\eqref{eq:def-J-infty} and~\eqref{eq:def-ru} provided $\hbar
	< \hbar_0$. In this case, we obtain the cost functionals
\begin{equation}
	\mathcal{J}^{g, \hbar} (w) \assign \mathbb{E} [g (I_{\infty}
		(\bar{u}^{\hbar} + w) + W^{\hbar}_s) + \hbar V_{\infty}^{\hbar} (u^{\hbar},
		w) \mathd s \nobracket + \nobracket \mathcal{E} (w, u^{\hbar})]
	\label{eq:def-Jhbar},
\end{equation}
where $W^{\hbar} = \hbar^{1 / 2} W$ is the rescaled Brownian motion,
$\bar{u}^{\hbar}_t \assign Q_t \hbar F^{\hbar}_t (Z^{\hbar}_t + \hbar^{1 / 2}
	W_t) + Q_t R_t^{\hbar}$ is the candidate for optimal control,
\[ V_{\infty}^{\hbar} (u, w) \assign \lambda \int_{\mathbb{R}^2} \mathd x
	(\llbracket \cos (\beta (I_{\infty} (u + w) + W^{\hbar}_{\infty}))
	\rrbracket - \llbracket \cos (\beta (I_{\infty} (u) + W^{\hbar}_{\infty}))
	\rrbracket) (x), \]
and
\[ \mathcal{E} (w, u) \assign \frac{1}{2}  \int_0^{\infty} \| w_s \|^2_{L^2}
	\mathd s + \int_0^{\infty} \langle w_s, u_s \rangle_{L^2} \mathd s. \]
Since the functional $\mathcal{J}^{g,\hbar}$ depends on $\hbar$ also through the
optimal control $\bar{u}^{\hbar}_t = - Q_t \hbar F^{\hbar}_t (Z^{\hbar}_t +
	\hbar^{1 / 2} W_t) - Q_t R_t^{\hbar}$, we have to first identify the limit of
$\bar{u}^{\hbar}$ before we can find the limiting candidate for
$\mathcal{J}^{g,0}$.

\begin{lemma}
	\label{lem:uhbar-to-0}With $\bar{u}^{\hbar}_t = - Q_t \hbar F^{\hbar}_t
		(Z^{\hbar}_t + \hbar^{1 / 2} W_t) - Q_t R_t^{\hbar}$, it holds that,
	\begin{equation}
		\lim_{\hbar \rightarrow 0} \mathbb{E} \int_0^{\infty} \| \bar{u}^{\hbar}_t
		\|_{L^{2, - n}}^2 \mathd t = 0. \label{eq:uhbar-to-0}
	\end{equation}
\end{lemma}

\begin{proof}
	We use the linear flow approximation $\tilde{F} \assign F^{[1]}$ for the
	SDE~\eqref{eq:Zhbar-wick} to obtain again a FBSDE. With $X^{\hbar}_t =
		Z_t^{\hbar} + \hbar^{1 / 2} W_t$, this results in the FBSDE,
	\[ \begin{cases}
			Z^{\hbar}_t = \int_0^t \mathd s \dot{G}_s (\lambda_s^{\hbar} \beta \sin
			(\beta X^{\hbar}_s) - \tilde{R}_s^{\hbar}), \\
			\tilde{R}_t^{\hbar} =\mathbb{E}_t  \int_t^{\infty} \mathd s
			(\lambda_s^{\hbar})^2 \beta \cos (\beta X^{\hbar}_s)  \dot{G}_s \sin
			\left( \beta (X^{\hbar}_s) \right) -\mathbb{E}_t  \int_t^{\infty}
			\lambda_s^{\hbar} \cos (\beta X^{\hbar}_s) \dot{G_s}
			\tilde{R}^{\hbar}_s \mathd s.
		\end{cases} \]
	Thus, the same arguments as before show using $\hbar \beta^2 < 4 \pi$,
	\[ \| R^{\hbar}_t \|_{L^{2, - n}} \lesssim (\bar{\lambda}^{\hbar}_t)^2 \langle t
		\rangle^{- 1} \sup_s \| Z^{\hbar}_s \|_{L^{2, - n}} \lesssim \bar{\lambda}^2
		\sup_s \| Z^{\hbar}_s \|_{L^{2, - n}} . \]
	Using this estimate in the equation for $Z^{\hbar}$,
	\begin{align*}
		\| Z^{\hbar}_t \|_{L^{2, - n}}
		 & \lesssim \int_0^t \mathd s
		\bar{\lambda}_s^{\hbar} \langle s \rangle^{- 2} \| Z^{\hbar}_s \|_{L^{2, - n}}
		+ \hbar^{1 / 2} \int_0^t \mathd s \langle s \rangle^{- 2}  \| W_s
		\|_{L^{2, - n}}               \\ &+ \bar{\lambda}^2  \int_0^t \mathd s \langle s \rangle^{- 2}
		\sup_r \| Z^{\hbar}_r \|_{L^{2, - n}} .
	\end{align*}
	Keeping in mind that
	\[ \mathbb{E} \| W_s \|^2_{L^{2, - n}} = G_s (0) \lesssim \log (s \vee 1) +
		1, \]
	we rearrange and take expectation to find for $\bar{\lambda}$ sufficiently small,
	\[ \mathbb{E} \| Z^{\hbar}_t \|_{L^{2, - n}} \lesssim \hbar^{1 / 2}
		\int_0^t \mathd s \langle s \rangle^{- 2} \mathbb{E} [\| W_s \|^2_{L^{2,
							- n}}]^{1 / 2} \lesssim \hbar^{1 / 2} \int_0^t \langle s \rangle^{- 2}
		(\log (s \vee 1) + 1) \lesssim \hbar . \]
	Putting everything together,
	\[ \begin{aligned}
			\mathbb{E} \int_0^{\infty} \| \bar{u}^{\hbar}_t \|^2_{L^{2, - n}}
			\mathd t & = \mathbb{E} \int_0^{\infty} \lambda_0 \beta \| Q_t \sin
			((\beta Z_t^{\hbar} + \hbar^{1 / 2} W_t)) \|_{L^{2, - n}}^2 \mathd t                \\
			         & \lesssim  \bar{\lambda} \int_0^{\infty} \langle t \rangle^{- 2}
			(\mathbb{E} \| Z^{\hbar}_t \|^2_{L^{2, - n}} + \hbar \mathbb{E} \| W_t
			\|^2_{L^{2, - n}}) \mathd t                                                         \\
			         & \lesssim  \bar{\lambda} \hbar + \hbar \int_0^{\infty} \mathd t \langle t
			\rangle^{- 2} \log (t \vee 1)                                                       \\
			         & \lesssim  \bar{\lambda} \hbar .
		\end{aligned} \]

\end{proof}

With the limiting optimal control sorted out, we can now show convergence as
$\hbar \rightarrow 0$ in the exact same way as for the case $\beta^2 < 4 \pi$
and we refer the reader to~{\cite[Section
	5]{barashkovStochasticControlApproach2022}} for details.

\section{Osterwalder--Schrader axioms}\label{sec:OS-Axioms}
The Osterwalder--Schrader axioms (OS axioms; for short), as introduced
in~{\cite{osterwalderAxiomsEuclideanGreens1973,osterwalderAxiomsEuclideanGreens1975}},
provide sufficient conditions under which the (Euclidean) Schwinger functions
define a relativistic QFT satisfying the Wightman axioms. We only briefly
introduce the aspects that are immediately relevant to our discussion, for a
more detailed exposition we refer to Chapter~6
in~{\cite{glimmQuantumPhysicsFunctional2012}} or Section~5
in~{\cite{gubinelliPDEConstructionEuclidean2021}}. For a Radon measure $\nu$
on \ $\mathcal{S}' (\mathbb{R}^2)$ for $n \in \mathbb{N}$ and $f_1, \ldots,
	f_n \in \mathcal{S} (\mathbb{R}^2)$, we define the associated Schwinger
functions $S^{\nu}_n \in (\mathcal{S} (\mathbb{R}^2))^{\otimes n}$, by
\begin{equation}
	S^{\nu}_n (f_1 \otimes \ldots \otimes f_n) \assign \int_{\mathcal{S}'
		(\mathbb{R}^2)} \langle \varphi, f_1 \rangle \ldots \langle \varphi, f_n
	\rangle \nu (\mathd \varphi) . \label{eq:def-schwinger-func}
\end{equation}
We say that $\nu$ satisfies the OS axioms, if its associated Schwinger
functions~\eqref{eq:def-schwinger-func} satisfy the OS-Axioms. We already
reformulate the axioms as conditions on the measures $\nu$ instead of the
Schwinger functions above. \ It is easy to verify that the conditions on $\nu$
below imply the OS axioms for the Schwinger
functions~\eqref{eq:def-schwinger-func}.
\begin{enumerate}
	\item (\tmtextbf{OS1-Regularity}) There is a Schwartz-norm $\| \cdot \|_S$
	      and a $\gamma > 0$ such that
	      \[ \int_{\mathcal{S}' (\mathbb{R}^2)} \mathe^{\gamma \| \varphi \|^2_S} \nu
		      (\mathd \varphi) < \infty . \]
	\item (\tmtextbf{OS2-Euclidean invariance}) The measure $\nu$ is invariant
	      under the action of the Euclidean group. More precisely, for any
	      $\mathcal{G} = (\mathcal{R}, a) \in O (2) \times \mathbb{R}^2$, it holds
	      that $\nu = \mathcal{G}_{\#} \nu$, where $\mathcal{G}_{\#} \nu (\cdot)
		      \assign \nu (\mathcal{G}^{- 1} (\cdot))$ denotes the push forward measure of
	      $\nu$ under $\mathcal{G}$.

	\item (\tmtextbf{OS3-Reflection positivity}) Define the reflection $\Theta : \mathbb{R}^2  \rightarrow \mathbb{R}^2, (x_0, x_1) \mapsto (- x_0, x_1)$
	      along the first coordinate axis. Then, for any exponential observable of the form
	      $\mathcal{O} (\varphi) = \prod_{i = 1}^n c_i \exp \{ \langle \varphi, f^i
		      \rangle \}$ for $f_i \in \mathcal{S} (\mathbb{R}^2)$ with support on
	      $\left\{ (x_0, x_1) \in \mathbb{R}^2 ; \; x^0 \geqslant 0 \right\}$,
	      \[ \int_{\mathcal{S}' (\mathbb{R}^2)} \overline{ (\Theta \mathcal{O})
			      (\varphi)}  \mathcal{O} (\varphi) \nu (\mathd \varphi) \geqslant 0. \]
	      Here, for $z \in \mathbb{C}$, we denote the complex conjugate by $\bar{z}$
	      and extended the reflection map $\Theta$ to functions $f \in \mathcal{S}
		      (\mathbb{R}^2)$ and the observables $\mathcal{O}$ via
	      \[ \Theta f (x_0, x_1) \assign f (- x_0, x_1), \quad \Theta \mathcal{O}
		      (\varphi) \assign \prod_{i = 1}^n \exp \{ \langle \varphi, \Theta f_i
		      \rangle \} . \]
\end{enumerate}
If the measures $\nu$ satisfies the conditions above, then the reconstruction
theorem~{\cite{osterwalderAxiomsEuclideanGreens1975}} (see also~{\cite[Theorem
	6.1.3]{glimmQuantumPhysicsFunctional2012}}) ensures the existence of a
Wightman theory corresponding to the measure $\nu$.
\begin{theorem}
	\label{thm:OS-axioms}
	Every accumulation point of $(\tmop{Law}(X^{\rho,T}_T))_{\rho,T}$ satisfies \textbf{OS1} and \textbf{OS3}. If moreover, $\bar{\lambda}$ is sufficiently small, then also \textbf{OS2} holds and the measure is not Gaussian.
\end{theorem}
The regularity property
for $\nu_{\tmop{SG}}$ was already shown in
Corollary~\ref{cor:exp-integrability}. The remaining claims are proven in the next three sections.

\subsection{Euclidean invariance}

The Euclidean invariance in this setting is a straightforward consequence of
the uniqueness obtained in Theorem~\ref{thm:T,rho-convergence}.

\begin{proposition}
	The joint law of $(Z_{\infty}, W_{\infty})$ is invariant under the action of
	the Euclidean group defined by
	\[ \mathcal{G} f (x) = f (\mathcal{R} (x - \mathcal{R}^{- 1} a))  \quad
		\text{for} \quad \mathcal{G} = (\mathcal{R}, a) \in O (2) \times
		\mathbb{R}^2 . \]
\end{proposition}

\begin{proof}
	Since the kernels $f^{[\ell]}_s$ are translation and rotation invariant (see
	Remark \ref{rem:flow-eq}-\ref{rem:translation-invariance}),
	\[ \mathcal{G} F_s (X_s) = F_s (\mathcal{G} X_s) . \]
	Moreover, immediately from the definition of $\dot{G}_s$, \ we have
	$\mathcal{G} (\dot{G_s} f) = \dot{G_s} (\mathcal{G} f)$. Therefore, for any
	$\rho \preceq  1$ and $T \leqslant \infty$, the transformed solution
	$\mathcal{G} X^{\rho, T}$ satisfies the equation,
	\begin{align*}
		\mathcal{G} X_t^{\rho, T} & = - \int_0^t \mathd s \mathcal{G}  \dot{G}_s
		(F^{\rho, T}_s (X_s^{\rho, T}) + R_s^{\rho, T}) + \mathcal{G} W_t        \\
		                          & = -
		\int_0^t \mathd s \dot{G_s} (F^{\mathcal{G} \rho, T}_s (\mathcal{G}
		X_s^{\rho, T}) + \mathcal{G} R_s^{\rho, T}) + \mathcal{G} W_t .
	\end{align*}
	With the same reasoning,
	\[ \mathcal{G} R_t^{\rho, T} =\mathbb{E}_t  \int_t^T \mathd s H^{\mathcal{G}
				\rho, T}_s (\mathcal{G} X_s^{\rho, T}) -\mathbb{E}_t  \int_t^T \mathd s
		\mathD F^{\mathcal{G} \rho, T}_s (\mathcal{G} X_s^{\rho, T})  \dot{G}_s
		\mathcal{G} R_s^{\rho, T} . \]
	In other words, $(\tilde{X}^{\rho, T}, \tilde{R}^{\rho, T}) \assign
		\mathcal{G} (X^{\rho, T}, R^{\rho, T})$ is a solution to
	\[ \begin{cases}
			\tilde{X}^{\rho, T}_t = \tilde{W}_t - \int_0^t \dot{G}_s
			(F^{\mathcal{G} \rho, T}_s (\tilde{X}^{\rho, T}_s) + \tilde{R}_{s,
			T}^{\rho}) \mathd s, \\
			\tilde{R}_t^{\rho, T} =\mathbb{E}_t  \int_t^T H^{\mathcal{G} \rho, T}_s
			(\tilde{X}_s^{\rho, T}) \mathd s -\mathbb{E}_t  \int_t^T \mathD
			F^{\mathcal{G} \rho, T}_s (\tilde{X}_s^{\rho, T})  \dot{G}_s
			\tilde{R}_s^{\rho, T} \mathd s,
		\end{cases} \]
	where $\tilde{W} \assign \mathcal{G} W = \int_0^{\cdot} Q_s \mathd
		(\mathcal{G} B_s)$ is again a Brownian motion with the same covariance as
	$W$. By the uniqueness of the solution to \eqref{eq:FBSDE-leq3} (see
	Corollary \ref{cor:well-posedness}) we then have
	\[ \tmop{Law} (\tilde{X}^{\mathcal{G}^{- 1} \rho, T}, \tilde{W}) =
		\tmop{Law} (X^{\rho, T}, W) . \]
	In other words, the joint law is invariant under the action of the Euclidean
	group $\mathcal{G}$ provided that $\rho = \mathcal{G} \rho$, which holds
	only when the weight $\rho$ is flat, that is $\rho \propto 1$. In this case,
	we have for any $T \leqslant \infty$,
	\[ \tmop{Law} (\tilde{X}^T_T, \tilde{W}_T) = \tmop{Law} (\mathcal{G} (X^T_T,
		W_T)) = \tmop{Law} (X^T_T, W_T) . \]
\end{proof}

\subsection{Reflection positivity}\label{sec:RP}

To show that $\nu_{\tmop{SG}}$ is reflection positive, we show that it is the
weak limit of reflection positive measures. We cannot use the approximating
sequence $\nu^{\rho, T}_{\tmop{SG}}$because the small scale regularisation for
$T < \infty$ mollifies the measure in all directions and consequently breaks
reflection positivity. Instead, we will construct a new sequence of reflection
positive measures $\nu_{\tmop{SG}}^{\varepsilon, \rho}$ such that
$\nu_{\tmop{SG}}^{\varepsilon, \rho} \rightarrow \nu_{\tmop{SG}}^{\rho}$ for
any spatial cut-off $\rho \prec 1$. Since weak limits of reflection positive
measures are reflection positive and since $\nu_{\tmop{SG}}$ is the weak limit
of the finite volume measures $\nu_{\tmop{SG}}^{\rho}$, this will prove the
claim. Throughout this section, we fix a symmetric cut-off $\rho$ and suppress
the dependency whenever it does not lead to ambiguities.

To preserve reflection positivity, we cannot mollify in the direction of physical time, and we instead mollify along only one of the coordinate axes. For $\eta \in C^{\infty}_c (\mathbb{R})$
supported on $| x | < 1$, define the family of mollifiers with
$\tilde{\eta}_{\varepsilon} = \varepsilon^{- 1} \eta (\cdot \varepsilon^{-
		1})$ on $\mathbb{R}^1$ and introduce the corresponding mollifiers
$\eta_{\varepsilon} = \delta_0 \otimes \widetilde{\eta_{\varepsilon}}$ on
$\mathbb{R}^2$. Then, using a variant of Theorem
\ref{thm:ctrl-FBSDE}-\ref{thm:ctrl-form}, we can define the measures,
\[ \nu_{\tmop{SG}}^{\varepsilon, \rho, T} = \tmop{Law} (X^{\varepsilon, \rho,
		T}_{\infty}), \]
where
\begin{equation}
	X^{\varepsilon, \rho, T}_t = - \int_0^t \mathd s \dot{G}^{\varepsilon}_s
	(F^{\varepsilon, \rho, T}_s (X^{\varepsilon, \rho, T}_s) + R^{\varepsilon, \rho, T}_s) +
	W^{\varepsilon}_t = - \int_0^t \mathd s \dot{G}_s^{\varepsilon} \mathbb{E}_s
	[\nabla V_T^{\varepsilon} (X_s^{\varepsilon, \rho, T})] + W_t^{\varepsilon} .
	\label{eq:X-eps-T}
\end{equation}
Here, we defined $G^{\varepsilon}_t \assign Q_t^{\varepsilon} \ast
	Q^{\varepsilon}_t$ with $Q_t^{\varepsilon} = \eta^{\varepsilon} \ast Q_t$, and
obtain $F^{\varepsilon}$ and $W^{\varepsilon}$ \ as before by replacing $G$ by
its mollification $G^{\varepsilon}$. Then, denoting $\mu^{\varepsilon, \rho, T}
	\assign \tmop{Law} (W_T^{\varepsilon})$, the same argument as in Theorem
\ref{thm:ctrl-FBSDE}-\ref{thm:ctrl-form} shows that
\[ \tmop{Law} (X^{\varepsilon, \rho, T}_T) \propto \exp \left( -
	\lambda_T^{\varepsilon} \int_{\mathbb{R}^2} \cos (\beta \varphi) \right)
	\mu^{\varepsilon, \rho, T} (\mathd \varphi), \]
where $\lambda^{\varepsilon}_t = \lambda \mathe^{- \frac{\beta^2}{2}
		G_t^{\varepsilon} (0)}$. The point is now that the additional convolution with
$\eta_{\varepsilon}$ ensures that the measures $\mu^{\varepsilon} =
	\mu^{\varepsilon, \infty}$ are supported on a function space. Indeed, we
compute for any $\varepsilon > 0$,
\[ G^{\varepsilon}_T (x) = \int_0^T \eta_{\varepsilon} \ast \dot{G_s} (x)
	\mathd s \leqslant \int_0^{\infty} \eta_{\varepsilon} \ast \dot{G}_s (x)
	\mathd s = G^{\varepsilon}_{\infty} (x) = \frac{1}{4 \pi} \log \left(
	\frac{1}{| x |^2 \vee \varepsilon} \right) + r_{\varepsilon} (x), \]
where $r_{\varepsilon}$ is bounded uniformly in $x \in \mathbb{R}^2$ and
$\varepsilon > 0$. In particular, the Wick-ordering with respect to the
Gaussian measure $\mu^{\varepsilon, \rho, T}$ is given by
\[ \llbracket \sin (\beta W_T) \rrbracket = \lambda_T^{\varepsilon} \sin
	(\beta W_T), \quad \text{where} \quad \lambda^{\varepsilon}_T = \lambda
	\mathe^{\frac{\beta^2}{2} G_T^{\varepsilon} (0)} \lesssim \lambda
	\varepsilon^{- 1}, \]
which is not only bounded uniformly in $T$ but also converges to a limit at $T
	= \infty$ with $\lambda^{\varepsilon} \assign \lambda_{\infty}^{\varepsilon} =
	\mathe^{\frac{\beta^2}{2} G_{\infty} (0)}$. Therefore, the same argument as
used in Theorem~\ref{thm:ctrl-FBSDE}-\ref{thm:ctrl-form} implies that for any
$\varepsilon > 0$, the SDE~\eqref{eq:X-eps-T} is meaningful also for $T =
	\infty$, with
\begin{equation}
	X^{\varepsilon, \rho}_t = - \int_0^t \mathd s \dot{G}_s^{\varepsilon} \mathbb{E}_s
	[\nabla V_{\infty}^{\varepsilon} (X^{\varepsilon, \rho}_{\infty})] +
	W_{\infty}^{\varepsilon}, \label{eq:SDE-V-RP}
\end{equation}
has a unique solution for $\bar{\lambda}$ sufficiently small. Theorem
\ref{thm:ctrl-FBSDE}-\ref{thm:ctrl-form} also implies that the law of
$X^{\varepsilon, \rho}_{\infty}$ is absolutely continuous with respect to
$\mu^{\varepsilon}$ and we define
\begin{equation}
	\nu_{\tmop{SG}}^{\rho, \varepsilon} \assign \tmop{Law}
	(X^{\varepsilon, \rho}_{\infty}) \propto \exp \left( - \lambda^{\varepsilon}
	\int_{\mathbb{R}^2} \mathd x \rho (x) \cos (\beta \varphi (x)) \right)
	\mu^{\varepsilon} (\mathd \varphi) . \label{eq:lem-rp-seq}
\end{equation}
For these measures, reflection positivity will follow directly from the
reflection positivity of $\mu^{\varepsilon}$.

\begin{lemma}
	For any $\varepsilon > 0$ and $\rho \prec 1$, the measures
	$\nu_{\tmop{SG}}^{\rho, \varepsilon}$ defined in~\eqref{eq:lem-rp-seq} are
	reflection positive.
\end{lemma}

\begin{proof}
	Denote the projection on the positive half-plane $\mathbb{R}_+ \times
		\mathbb{R}$ by $\pi_+$ and let again
	\[ \Theta f (x_0, x_1) \assign f (- x_0, x_1), \]
	be the reflection around the first coordinate axis. We first show that the
	Gaussian measure $\mu^{\varepsilon} = \tmop{Law} (W_{\infty}^{\varepsilon})$
	is reflection positive. Since a Gaussian measure is reflection positive if
	and only if its covariance is reflection positive (see e.g.~{\cite[Theorem
		6.2.2.]{glimmQuantumPhysicsFunctional2012}}), it is sufficient to check that
	for any function $f \in L^2 (\mathbb{R}^2)$,
	\begin{equation}
		G^{\varepsilon} (\pi_+ f, \Theta \pi_+ f) = \langle \eta_{\varepsilon}
		\ast \pi_+ f, (m^2 - \Delta)^{- 1} \Theta \eta_{\varepsilon} \ast \pi_+ f
		\rangle_{L^2} \geqslant 0, \label{eq:GFF-RP}
	\end{equation}
	where $G^{\varepsilon} = G^{\varepsilon}_{\infty}$ is the covariance
	$\mu^{\varepsilon}$. Because $\eta_{\varepsilon}$ leaves the first
	coordinate invariant, the convolution with $\eta_{\varepsilon}$ commutes
	with the projection $\pi_+$. The reflection positivity of $(m^2 - \Delta)^{-
				1}$ now implies \eqref{eq:GFF-RP}.

	To see that the measures $(\nu_{\tmop{SG}}^{\varepsilon,
			\rho})_{\varepsilon > 0}$ defined by~\eqref{eq:lem-rp-seq} are also
	reflection positive, we split the potential between the two half-planes $\{
		x_0 \geqslant 0 \}$ and $\{ x_0 < 0 \}$, as
	\[ V_{\varepsilon}^{\rho, \pm} (\varphi) \assign \lambda^{\varepsilon}
		\int_{\mathbb{R}_{\pm} \times \mathbb{R}} \rho (x) \cos (\beta \varphi
			(x)) \mathd x, \]
	so that
	\[ \nu_{\tmop{SG}}^{\rho, \varepsilon} (\mathd \varphi) = \exp (-
		(V_{\varepsilon}^{\rho, +} (\varphi) + V_{\varepsilon}^{\rho, -}
		(\varphi))) \mu^{\varepsilon} (\mathd \varphi) . \]
	For the symmetric cut-off $\rho$, the reflection $\Theta$ acts on this
	decomposition as $\Theta V_{\varepsilon}^{\rho, \pm} =
		V_{\varepsilon}^{\rho, \mp}$. Consequently, we have for any exponential
	observable $\mathcal{O}$ supported on the positive half plane as defined in
	\tmtextbf{(OS3)},
	\[ \begin{aligned}
			\int_{\mathcal{S}' (\mathbb{R}^2)} \mathcal{O} (\varphi) \Theta
			\overline{\mathcal{O} (\varphi)} \nu_{\tmop{SG}}^{\varepsilon, \rho}
			(\mathd \varphi) & = \int_{\mathcal{S}' (\mathbb{R}^2)} \mathcal{O}
			(\varphi) \Theta \mathcal{O} (\varphi) \mathe^{- V_{\varepsilon}^{\rho,
						+} (\varphi)} \mathe^{- V_{\varepsilon}^{\rho, -} (\varphi)}
			\mu^{\varepsilon} (\mathd \varphi)                                            \\
			                 & = \int_{\mathcal{S}' (\mathbb{R}^2)} \mathcal{O} (\varphi)
			\mathe^{- V_{\varepsilon}^{\rho, +} (\varphi)} \Theta (\mathcal{O}
				(\varphi) \mathe^{- V_{\varepsilon}^{\rho, +} (\varphi)})
			\mu^{\varepsilon} (\mathd \varphi) .
		\end{aligned} \]
	Since $V^{\rho, +}$ is supported on the positive half plane $\mathbb{R}_+
		\times \mathbb{R}$, the last integral is non-negative as a result of the
	reflection positivity of $\mu^{\varepsilon}$. In other words, for any
	symmetric cut-off $\rho \prec 1$, also $\nu_{\tmop{SG}}^{\varepsilon, \rho}$ is
	reflection positive.
\end{proof}

Having established reflection positivity for $\nu_{\tmop{SG}}^{\varepsilon,
		\rho}$ for any $\varepsilon > 0$, we want to extract a subsequence which
converges to the desired limiting measure $\nu_{\tmop{SG}}^{\rho}$ to conclude
this proof. That is, it remains to show that for any $\alpha > 0$, there is a
sequence $\varepsilon_N \downarrow 0$ such that
\begin{equation}
	\sup_t \mathbb{E} \| X_t^{\varepsilon_N} - X_t \|^2_{H^{- \alpha, - n}}
	\rightarrow 0. \label{eq:Xeps-conv}
\end{equation}
Adapting the definitions~\eqref{eq:FTT} to the current situation, we see that
with the usual remainder $R^{\varepsilon}$, the FBSDE for the difference is
given by
\begin{equation}
	\begin{cases}
		X_t^{\varepsilon} - X_t = - \int_0^t \mathd s [\dot{G_s^{\varepsilon}}
		(F_s^{\varepsilon} (X_s^{\varepsilon}) + R_s^{\varepsilon}) - \dot{G}_s
		(F_s (X_s) + R_s)] + W^{\varepsilon}_t - W_t, \\
		R_t^{\varepsilon} - R_t = \mathbb{E}_t\int_t^{\infty}
		(H^{\varepsilon}_s (X^{\varepsilon, \rho}_s) - H_s (X_s)-(\mathD F_s^{\varepsilon, \rho}
		(X^{\varepsilon, \rho}_s)  \dot{G}^{\varepsilon}_s R^{\varepsilon}_s - \mathD
		F_s (X_s)  \dot{G}_s R_s))\mathd s.
	\end{cases} \label{eq:FBSDE-eps}
\end{equation}
By definition of $G^{\varepsilon}$, it holds for any $\alpha > 0$,
\[ \lim_{\varepsilon \rightarrow 0} \sup_t \mathbb{E} \| W^{\varepsilon}_t -
	W_t \|_{H^{- \alpha, - n}}^2 = 0. \]
For some subsequence $\varepsilon_N \downarrow 0$ and any $\varphi \in
	\mathcal{S}' (\mathbb{R}^2)$, Lemma~\ref{lem:f-eps-convergence} combined with
Proposition~\ref{prop:force-estimates}-\ref{prop:GF-bounds} implies,
\[ \| Q_s (F_s^{\varepsilon_N} - F_s) (\varphi) \|_{L^{\infty}}^2 + \| Q_s
	(\mathD F_s^{\varepsilon_N} - \mathD F_s) (\varphi) \|_{L^{\infty}}^2
	\lesssim N^{- 1} \bar{\lambda}_s \langle s \rangle^{- 1}, \]
and
\[ \| (H^{\varepsilon}_s - H_s) (\varphi) \|_{L^{\infty}}^2 \lesssim N^{- 1}
	(\bar{\lambda}_s \langle s \rangle^{- 1})^4 . \]
Following the (by now standard) procedure for the FBSDE~\eqref{eq:FBSDE-eps}
yields~\eqref{eq:Xeps-conv} and thus concludes the proof.

\begin{remark}
	A slight modification of the argument allows to show reflection positivity
	for any accumulation point of $(\nu_{\tmop{SG}}^{\rho, T})_{T \geqslant 0}$.
	Therefore, reflection positivity holds also without the smallness assumption
	on the coupling constant $\bar{\lambda}$.
\end{remark}

\subsection{Non-Gaussianity}\label{sec:non-Gaussian}

We want to show now that for $\bar{\lambda}$ small the measure $\nu_{\tmop{SG}}$ is
non-Gaussian. For any Gaussian measure $\nu$ with support on a Hilbert space
$\mathcal{H}$ with dual pairing $\langle\cdot,\cdot\rangle$, mean $b \in \mathcal{H}^\ast$, covariance $C$, Cameron-Martin space
$H_{\tmop{CM}}(\nu)$ it holds for $\psi\in \mathcal{H}$,
\begin{equation}
	\frac{1}{2}\|C \psi\|_{H_{\tmop{CM}}(\nu)}^2=\frac{1}{2} \|C^{1/2} \psi\|^2_{\mathcal{H}} = \log\int \exp(-\langle \varphi,\psi\rangle)
	\nu(\mathd \varphi) + \langle b,\psi \rangle.\label{eq:MGF-Gaussian}
\end{equation}
Showing that the generating function of $\nu_{\tmop{SG}}$ does not satisfy
\eqref{eq:MGF-Gaussian} for will therefore yield the claim.

\tmtextbf{Step 1.} We first show that for any $\psi\in C^{\infty}_{c}(\mathbb{R}^{2})$,
\begin{equation}
	\begin{aligned}
		- \log \int \exp (- \langle \varphi, \psi \rangle) \nu_{\tmop{SG}} (\mathd
		\varphi) = & - \frac{1}{2} \| (m^2 - \Delta)^{-1 / 2} \psi \|_{L^2}^2 \\
		           & + \lim_{\rho\to 1}
		[\mathcal{V}^{\rho} (- (m^2 - \Delta)^{-1} \psi) - \mathcal{V}^{\rho} (0)],
		\label{eq:MGF-SG}
	\end{aligned}
\end{equation}
where $\mathcal{V}^{\rho}$ is the value function as defined in
\eqref{eq:def-value-func}.  By Theorem \ref{thm:T,rho-convergence} the measure
$\nu_{\tmop{SG}}$ on $\mathcal{H}$ can be obtained as a weak limit of
measures $\nu_{\tmop{SG}}^{\rho, T}$ that are absolutely continuous with
respect to the Gaussian measure $\mu_T = \mathcal{N} (0, G_T)$.
Combined with the Gaussian tails from Corollary \ref{cor:exp-integrability} this implies convergence of $-\log\nu_{\tmop{sG}}^{\rho,T}(\exp(-\langle\psi,\cdot\rangle))\to -\log \nu_{\tmop{sG}}(\exp(-\langle\psi,\cdot\rangle))$ by uniform integrability for the functionals $\exp(-\langle \psi, \cdot\rangle)$.
Indeed, for every $L>0$,
\begin{align*}
	          & \left|\int \exp(-\langle\psi, \varphi\rangle )\nu^{\rho,T}_{\tmop{sG}}(\mathd \varphi) - \int \exp(-\langle\psi, \varphi\rangle )\nu_{\tmop{sG}}(\mathd \varphi)\right| \\
	\leqslant & \left|\int_{\{\left\| \varphi \right\|_{H^{-\varepsilon, -n}}\leqslant L \}}\exp(-\langle\psi, \varphi\rangle )\nu^{\rho,T}_{\tmop{sG}}(\mathd \varphi) -
	\int_{\{\left\| \varphi \right\|_{H^{-\varepsilon, -n}}\leqslant L \}} \exp(-\langle\psi, \varphi\rangle )\nu_{\tmop{sG}}(\mathd \varphi)\right|                                    \\
	          & + \left|\int_{\{\left\| \varphi \right\|_{H^{-\varepsilon, -n}}\geq L \}} \exp(-\langle\psi, \varphi\rangle )\nu^{\rho,T}_{\tmop{sG}}(\mathd \varphi)\right|
	+ \left|\int_{\{\left\| \varphi \right\|_{H^{-\varepsilon, -n}}\geq L \}} \exp(-\langle\psi, \varphi\rangle )\nu_{\tmop{sG}}(\mathd \varphi)\right|.
\end{align*}
Regarding the two tail integrals, note that for every $\gamma>0$, there is a $K_{\gamma}>0$ such that
\[|-\langle \psi, \varphi\rangle|
	\leqslant\left\| \psi \right\|_{H^{\varepsilon,n}} \left\| \varphi \right\|_{H^{-\varepsilon, -n}}
	\leqslant \gamma\left\| \varphi \right\|^2_{H^{-\varepsilon,-n}}
	+ K_{\gamma} \left\| \psi \right\|^2_{H^{\varepsilon,n}}.
\]
Hence, using Corollary \ref{cor:exp-integrability}, for any $\kappa>0$, we can choose $L>0$ sufficiently large so that
\[
	\sup_{\rho\prec 1,T\leqslant\infty}\left|\int_{\{\left\| \varphi \right\|_{H^{-\varepsilon, -n}}\geq L \}} \exp(-\langle\psi, \varphi\rangle )\nu^{\rho,T}_{\tmop{sG}}(\mathd \varphi)\right| \leqslant \frac{1}{3} \kappa.
\]
The first difference converges by weak convergence of $\nu_{\tmop{sG}}^{\rho,T}$ to $\nu_{\tmop{sG}}$ and the boundedness of $\exp(-\langle \psi, \varphi\rangle)$ on $\{\left\| \varphi \right\|_{H^{-\varepsilon, -n}}\leqslant L \}$. Thus, for $L$ fixed as above, we may choose $\rho,T$ sufficiently close to $1,\infty$ so that also
\[
	\left|\int_{\{\left\| \varphi \right\|_{H^{-\varepsilon, -n}}\leqslant L \}}\exp(-\langle\psi, \varphi\rangle )\nu^{\rho,T}_{\tmop{sG}}(\mathd \varphi) -
	\int_{\{\left\| \varphi \right\|_{H^{-\varepsilon, -n}}\leqslant L \}} \exp(-\langle\psi, \varphi\rangle )\nu_{\tmop{sG}}(\mathd \varphi)\right| \leqslant \frac{\kappa}{3}.
\]
Finally, it remains only to show that also the $\log$ converges. To this end, note that by Lemma \ref{lem:B-to-B}, there are universal constants $K, K_Z>0$ such that uniformly in $\rho\preceq1$, $T\in (0,\infty]$
\begin{align*}
	\nu_{\tmop{sG}}^{{\rho,T}}(\exp(-\langle \psi,\cdot\rangle))
	=    & \mathbb{E} [\exp (-\langle \psi, Z_T^{{\rho,T}}\rangle) \exp(-\langle \psi, W_T \rangle) ]                                                    \\
	\geq & \mathe^{-K_{Z}\left\| \psi \right\|_{H^{\varepsilon,n}}^{2} } \mathbb{E}[\exp(-\gamma \left\| W_\infty \right\|_{H^{-\varepsilon, -n}}^{2} )]
	\geq K>0.
\end{align*}
Therefore, the convergence $-\log \nu_{\tmop{sG}}^{\rho,T}(\exp (-\langle \psi, \cdot\rangle))\to -\log \nu_{\tmop{sG}}^{\rho,T}(\exp (-\langle \psi, \cdot\rangle)) $ also follows immediately.
Combined, this allows us to write
\begin{equation*}
	\begin{aligned}
		 & - \log \int \exp (- \langle \varphi, \psi \rangle) \nu_{\tmop{SG}} (\mathd \varphi)     \\
		 & = \lim_{\tmscript{\begin{array}{c}
					                     T \rightarrow \infty \\
					                     \rho \rightarrow 1
				                     \end{array}}} - \log \left\{ \Xi_{T, \rho}^{- 1} \int \exp (- \langle
		\varphi, \psi \rangle - V_T^{T, \rho} (\varphi)) \mu_T (\mathd \varphi)
		\right\}.
	\end{aligned}
\end{equation*}
For any $h \in H_{\tmop{CM}} (\mu) = H^1$, by the Cameron--Martin theorem \
(see e.g. {\cite[Corollary 2.4.3]{bogachevGaussianMeasures1998}}),
\[ \mu_T (\mathd (\varphi - h)) = \exp \left( \langle \varphi, G_T^{-1} h \rangle -
	\frac{1}{2} \langle h, G_T^{-1} h \rangle \right) \mu_T (\mathd \varphi) . \]
Thus,
\begin{equation*}
	\begin{aligned}
		 & -\log \left\{ \Xi_{T, \rho}^{- 1} \int \exp (- \langle \varphi, \psi
		\rangle - V_T^{T, \rho} (\varphi)) \mu_T (\mathd \varphi) \right\}              \\
		 & = - \log \left\{ \Xi_{T, \rho}^{- 1} \int \exp \left( - \langle \varphi,
		\psi \rangle - V_T^{T, \rho} (\varphi) + \frac{1}{2} \langle h, G_T^{-1}h
		\rangle - \langle \varphi, G_T^{-1}h \rangle \right) \mu_T (\mathd (\varphi -
		h)) \right\}                                                                    \\
		 & = - \log \left\{ \Xi_{T, \rho}^{- 1} \int \exp \left( - \langle (\varphi-h),
		\psi + G_T^{-1}h \rangle - V_T^{T, \rho}
		(\varphi + h) + \frac{1}{2} \langle h, G_T^{-1}h \rangle \right) \mu_T (\mathd
		\varphi) \right\}.
	\end{aligned}
\end{equation*}
Since $\psi\in H^{-1}$ and $G_\infty H^{-1}=H^{1}=H_{\tmop{CM}} (\mu)$,
we may perform a shift along $h_T = - G_T \psi$ to obtain
\begin{equation*}
	\begin{aligned}
		- & \log \int \exp (- \langle \varphi, \psi \rangle) \nu_{\tmop{SG}} (\mathd
		\varphi)                                                                                                                                                         \\
		  & = \lim_{\tmscript{\begin{array}{c}
					                      T \rightarrow \infty \\
					                      \rho \rightarrow 1
				                      \end{array}}} - \log \left\{ \exp \left( \frac{1}{2} \langle \psi, G_T \psi \rangle \right) \Xi_{T, \rho}^{- 1} \int \exp (- V_T^{T, \rho}
		(\varphi - G_T \psi)) \mu_T (\mathd \varphi) \right\}                                                                                                            \\
		  & = - \frac{1}{2} \| (m^2-\Delta)^{\frac{1}{2}}h\|_{L^2}^2
		- \lim_{\tmscript{\begin{array}{c}
					                  T \rightarrow \infty \\
					                  \rho \rightarrow 1
				                  \end{array}}}  \log \left\{
		\Xi_{T, \rho}^{- 1} \int \exp (- V_T^{T, \rho} (\varphi + h_T))
		\mu_T (\mathd \varphi) \right\}.
	\end{aligned}
\end{equation*}
For the latter, Theorem~\ref{thm:infinite-vol-var} with $h=-G_\infty\psi$ implies,
\[ \lim_{\tmscript{\begin{array}{c}
				T \rightarrow \infty
			\end{array}}} - \log \left\{ \Xi_{T, \rho}^{- 1} \int \exp (- V_T^{T, \rho}
	(\varphi + h_T)) \mu_T (\mathd \varphi) \right\} =
	\mathcal{V}^{\rho} (h) - \mathcal{V}^{\rho} (0), \]
which yields \eqref{eq:MGF-SG}.

\tmtextbf{Step 2.}
We now show that \eqref{eq:MGF-Gaussian} implies that $C\psi\in H_{\tmop{CM}}(\nu)$
whenever $\psi\in C_c^\infty$.
To this end it is sufficient to show that $\|\psi\|_{H_{\tmop{CM}}(\nu_{\tmop{SG}})}
	^2\leqslant\infty$ whenever $\psi\in C^\infty_c$.
From Theorem \ref{thm:T,rho-convergence}, we know that $\tmop{supp}(\nu_{\tmop{SG}})
	\subset H^{-1,-n}$. Therefore, applying \eqref{eq:MGF-Gaussian} for
$\mathcal{H}=H^{-1,-n}$ and inserting the estimate from Step 1,
\[
	\begin{aligned}
		\| & C\psi\|_{H_{\tmop{CM}}(\nu_{\tmop{SG}})}^2                                     \\
		   & \leqslant |\langle\psi, b\rangle| + |\log\int\exp(-\langle\varphi,\psi\rangle)
		\nu_{\tmop{SG}}(\mathd\varphi)|                                                     \\
		   & \leqslant \|\psi\|_{H^{-1,-n}}\|b\|_{H^{1,n}}
		+ \frac{1}{2} \|(m^2-\Delta)^{-\frac{1}{2}}\psi\|^2_{L^2}
		+ \sup_{\rho\leqslant 1} |\mathcal{V}^{\rho} (-G_\infty\psi) -
		\mathcal{V}^{\rho} (0)|.
	\end{aligned}
\]
All the terms except the last one are finite for $\psi\in C^\infty_c$.
For the difference of the value functions, we use \eqref{eq:gradV=F+R}
combined with Lemma~\ref{lem:F2-W1-bounds} and Lemma \ref{lem:B-to-B},
\[ \sup_{\rho\leqslant 1}\|\nabla \mathcal{V}^{\rho} (h)\|_{L^\infty} \leqslant
	\sup_{\rho\leqslant 1}\|F_0^{\rho} (h)\|_{L^\infty}
	+ \sup_{\rho\leqslant 1}\|R_0^{\rho} (h)\|_{L^\infty}\lesssim 1, \qquad h
	\in C_{c}^{\infty} (\mathbb{R}^2).
\]
Then, a simple Taylor expansion combined with the estimates on $F$ and $R$
\[
	\sup_{\rho\leqslant 1} |\mathcal{V}^{\rho} (-G_\infty\psi) -
	\mathcal{V}^{\rho} (0)|
	\leqslant
	\sup_{\theta \in [0, 1]}\sup_{\rho\leqslant 1}
	\| \nabla \mathcal{V}^{\rho, T} (\theta (m^2 - \Delta)^{-1} \psi)
	\|_{L^{\infty}} \| (m^2 - \Delta)^{-1} \psi \|_{L^1}<\infty.
\]

\tmtextbf{Step 3.} We conclude by showing that \eqref{eq:MGF-SG} is not
quadratic for $\psi\in C^\infty_c$, which is valid by Step 2.
It will be more convenient to show that its
gradient is non-linear.
We know from Lemma~\ref{lem:F2-W1-bounds} that
\[
	\| F_0^{\rho} (h) + \rho\lambda\sin(\beta h)\|_{L^{\infty}}
	\lesssim \bar{\lambda}^2 \| h \|_{W^{1, \infty}}.
\]
Combined with $\|
	R_0 \|_{L^{\infty}} \lesssim \bar{\lambda}^4$ from Lemma \ref{lem:B-to-B}, we can
gather all contributions $\ell > 1$ in a uniformly bounded function $c^{\rho}$
satisfying
\[ \sup_{\rho \preceq  1} \sup_{\| h \|_{W^{1, \infty}} \leqslant C}
	\| c^{\rho} (h) \|_{L^{\infty}} \lesssim 1, \]
so that for any fixed $C > 0$ and $\| h \|_{W^{1, \infty}} \leqslant C$,
\[ \nabla \mathcal{V}^{\rho} (h) = - \rho \beta \lambda \sin (\beta h) +
	\lambda^2 c^{\rho} (h) . \]
From here, we can see that $\nabla \mathcal{V}^{\rho}$ is not additive. For
example, we may choose $\psi, \tilde{\psi} \in C^{\infty}_c \subset
	H_{\tmop{CM}} (\nu_{\tmop{SG}})$ such that $\mathbbm{1}_{\{ | x | \leqslant 1
		\}} \psi (x) = \frac{\pi}{2 \beta}$ and $\mathbbm{1}_{\{ | x | \leqslant 1 \}}
	\tilde{\psi} = \frac{\pi}{4 \beta}$. If $\rho \equiv 1$ on
$B_1 (0)$, then we verify that on $B_1 (0)$ for $K > 0$ and $\lambda>0$,
\begin{align}
	          & \nabla \mathcal{V}^{\rho} (\psi + \tilde{\psi}) + \nabla
	\mathcal{V}^{\rho} (\psi - \tilde{\psi}) - 2 \nabla \mathcal{V}^{\rho}
	(\psi)                                                                  \\
	=         & -\beta \lambda \left[ \sin \left( \frac{3 \pi}{4} \right) +
		\sin \left( \frac{\pi}{4} \right) - 2 \sin \left( \frac{\pi}{2} \right)
	\right] + O (\lambda^2)                                                 \\
	\geqslant & \lambda \left( 2 - \sqrt{2} \right) - K \lambda^2 .
\end{align}
For $\bar{\lambda}$ sufficiently small, uniformly in $\rho \preceq  1$ it holds
that $\lambda \left(2 - \sqrt{2} \right) - K \lambda^2 \geqslant \tilde{K} >
	0$ and we conclude that \ $\nabla \mathcal{V}^{\rho}$ is non-linear for any
$\rho \preceq  1$. For $\lambda<0$ the the claim follows in the same way.

\appendix\section{Heat kernel estimates}\label{app:hk}

This Appendix contains some basic estimates on the heat kernel which we use
throughout as well as some technical proofs which have been postponed.

\subsection{General estimates}

\begin{lemma}
	\label{lem:hk-basic-bounds}With $G$ as defined in~\eqref{eq:def-G}, there
	are uniformly bounded functions $g_1, g_2$ and constants $C \geqslant 0$
	such that for any $t \in \mathbb{R}_+$ and $x \in \mathbb{R}^2$,
	\begin{align}
		G_t (0)                & = \frac{1}{4 \pi} \log (t \vee 1) + g_1 (t),
		\label{eq:cov-0-bound}                                                 \\
		(G_{\infty} - G_t) (x) & = \frac{1}{4 \pi} \log (| x |^{- 2} t^{- 1}
		\vee 1) + g_2 (t, x) \label{eq:cov-x-s-bound}                          \\
		\nabla \dot{G}_t  (x)  & = C | x | \mathe^{- m^2 / t - \frac{t}{4} | x
			|^2} \label{eq:grad-g},
	\end{align}
	and consequently
	\begin{equation}
		| \dot{G_t} (0) - \dot{G}_t (x) | \lesssim | x |  \langle t \rangle^{- 1 /
			2} \label{eq:hk-g0-gx} .
	\end{equation}
\end{lemma}

\begin{proof}
	The estimate~\eqref{eq:cov-0-bound} follows immediately from the
	$L^2$-kernel representation \eqref{eq:G-kernel}, noting that
	\[ G_t (0) = \frac{1}{4\pi}\int_0^t \frac{\mathd s}{s} \mathe^{- m^2 / s} = \int_0^{1
			\wedge t} \frac{\mathd s}{4 \pi s} \mathe^{- m^2 / s} + \int_{t \wedge
			1}^{t \vee 1} \frac{\mathd s}{4 \pi s} \mathe^{- m^2 / s} = g_1 (t) +
		\frac{1}{4 \pi} \log (t \vee 1) . \]
	Regarding~\eqref{eq:cov-x-s-bound}, we obtain after a substitution with $u
		= s^{- 1} | x |^{- 2}$,
	\[ \begin{aligned}
			(G_{\infty} - G_t) (x) & = \int_t^{\infty}  \frac{\mathd s}{4 \pi s}
			\mathe^{- m^2 / s} \mathe^{- \frac{s}{4} | x |^2}                                           \\
			                       & = \int_0^{t^{- 1} | x |^{- 2}}  \frac{\mathd s}{4 \pi s} \mathe^{-
			m^2 | x |^2 s} \mathe^{- \frac{1}{4 s}}                                                     \\
			                       & = \int_0^{1 \wedge t^{- 1} | x |^{- 2}} \frac{\mathd s}{4 \pi s}
			\mathe^{- m^2 | x |^2 s} \mathe^{- \frac{1}{4 s}} + \int_{1 \wedge t^{-
						1} | x |^{- 2}}^{t^{- 1} | x |^2 \vee 1} \frac{\mathd s}{4 \pi s}
			\mathe^{- m^2 | x |^2 s} \mathe^{- \frac{1}{4 s}}                                           \\
			                       & = g_2 (t, x) + \frac{1}{4 \pi} \log (t^{- 1} | x |^{- 2} \vee 1) .
		\end{aligned} \]
	Finally, \eqref{eq:grad-g} is a direct computation and~\eqref{eq:hk-g0-gx}
	follows from
	\[ \frac{| \dot{G_t} (0) - \dot{G}_t (x) |}{| x |} \lesssim \sup_y  | \nabla
		\dot{G}_t (y) | . \]
	Maximising the right-hand side, we see that the maximum is attained at $y =
		Cs^{- 1 / 2}$ for some constant $C$ which gives the claim.
\end{proof}

\begin{lemma}
	\label{lem:exp-integrability}For $\gamma^2 < 4 c$, it holds that
	\[ \int_{\mathbb{R}^2} \mathd x \mathe^{- ct | x |^2 + m \gamma | x | - m^2
			/ s} | x |^{2 k} \lesssim \langle t \rangle^{- 1 - k} . \]
\end{lemma}

\begin{proof}
	We treat the small and large scales separately. For $t > 1$,
	\[ \mathe^{- \frac{t}{4} | x |^2 + \gamma m | x |} \lesssim \mathe^{-
			\frac{t}{8} | x |^2}, \]
	so that the estimate follows as in the unweighed case. To deal with the
	large scales, note that for any $c \geqslant 0$ the polynomial
	\[ p (x) \assign ct | x |^2 - m (\gamma - \varepsilon) | x | +
		\frac{m^2}{t}, \]
	attains its minimum at $x^{\pm} = \pm \frac{m (\gamma - \varepsilon)}{2
			ct}$ and $p (x) \geqslant p (x^{\pm}) \geqslant - \varepsilon^2 / t$
	provided $(\gamma - \varepsilon)^2 < 4 c$. Therefore, choosing $\varepsilon
		> 0$ sufficiently small depending on $\gamma^2 < 4 c$, we have \[\mathe^{- ct
			| x |^2 + \gamma | x |} \mathe^{- m^2 / t} \leqslant \mathe^{- \varepsilon |
			x |} \mathe^{- \varepsilon^2 / t},\] and thus
	\[ \begin{aligned}
			\sup_{t \in [0, 1]}  \int_{\mathbb{R}^2} \mathd y \frac{1}{4 \pi s}
			\mathe^{- \frac{t}{4} | x |^2 + \gamma | x |} \mathe^{- m^2 / t}
			\lesssim \sup_{t \in [0, 1]}  \int_{\mathbb{R}^2} \mathd y \frac{1}{4
				\pi} t^{- 1} \mathe^{- \varepsilon | x |} \mathe^{- \varepsilon^2 / t}
			< \infty .
		\end{aligned} \]
\end{proof}

\begin{lemma}
	\label{lem:hk-weighted-Lipschitz}For any $t \in \mathbb{R}_+, x \in
		\mathbb{R}^2$ and $c \in (0,1)$, we have
	\[ | \dot{G}_t (x) - \dot{G}_t (y) | \mathe^{- c \frac{t}{4}| x - y |^2} \lesssim | x -
		y | (| x | + | x - y |) \mathe^{- \frac{c}{2} \frac{t}{4}| x |^2} \mathe^{- m^2 / t}
		. \]
	The same estimate holds for $\dot{G}_t$ replaced by $t^{- 1} \mathe^{- m^2 /
			2 t} Q_t$.
\end{lemma}

\begin{proof}
	We start by rewriting the difference as
	\[ \dot{G}_t (x) - \dot{G_t} (y) = (y - x)  \int_0^1 \mathd \vartheta \nabla
		\dot{G}_t (x - \vartheta (x - y)) . \]
	For any $z = x - \vartheta (x - y)$ and $\vartheta \in [0, 1]$, we have
	$\frac{1}{2} | x |^2 \leqslant | x - z |^2 + | z |^2 \leqslant \vartheta^2 |
		x - y |^2 + | z |^2$. Combined with~\eqref{eq:grad-g} this means
	\[ | \nabla \dot{G}_t (z) | \leqslant C | z | \mathe^{- \frac{t}{4}| z |^2} \mathe^{-
			m^2 / t} \leqslant C | z | \mathe^{- \frac{c}{2}  \frac{t}{4}| x |^2} \mathe^{c \frac{t}{4}| x -
			y |^2} \mathe^{- m^2 / t}, \]
	and consequently,
	\[ | \nabla \dot{G}_t (z) | \mathe^{- c \frac{t}{4} | x - y |^2} \leqslant C | z |
		\mathe^{- \frac{c}{2} | x |^2} \mathe^{- m^2 / t} \leqslant C (| x | + |
		x - y |) \mathe^{- \frac{c}{2} \frac{t}{4}| x |^2} \mathe^{- m^2 / t} . \]
	The estimate on $Q$ follows in the exact same way, only replacing the
	estimate on the gradient by
	\[ | \nabla Q_t (x) | \lesssim t | x | \mathe^{- 2 t | x |^2 - m^2 / 2 t} .
	\]
\end{proof}

\begin{lemma}
	\label{lem:hk-exp} \label{lem:hk-scaling}For $k \in \mathbb{R}$, $\alpha >
		0$, and $\gamma \in (- 1, 1)$, consider the weight $w (x) = \langle x \rangle^k$ or $w (x) =
		\exp (\gamma m | x |)$. Then, for any $t \in \mathbb{R}_+$ and $u \in L^p
		(w)$,
	\begin{equation}
		\begin{aligned}
			\| | x |^{2 \alpha / p} Q_t \|_{L^p (w)} = \langle t \rangle^{- 1 / p -
			\alpha / p}, \quad           & \quad \| Q_t u \|_{L^p (w)} \lesssim \langle t \rangle^{- 1
			/ p} \| u \|_{L^1 (w)},                                                                    \\
			\| | x |^{2 \alpha / p}  \dot{G}_t \|_{L^p (w)} = \langle t \rangle^{- 1
			- 1 / p - \alpha / p}, \quad & \quad \| \dot{G}_t u \|_{L^p (w)} \lesssim \langle
			t \rangle^{- 1 - 1 / p} \| u \|_{L^1 (w)}.
		\end{aligned}\label{eq:QLp}
	\end{equation}
	Moreover, with $C_t \assign (G_{\infty} - G_t)$ it holds for any $s
		\geqslant t$ and $c > 0$ sufficiently small, and $\alpha > (1 - 2 \delta)
		\vee 0 = \left( \frac{\beta^2}{4 \pi} - 1 \right) \vee 0$,
	\begin{equation}
		\int_{\mathbb{R}^2} \dot{G}_s (x) \mathe^{\beta^2 C_s (x)} | x |^{2
				\alpha} \mathe^{c t | x |^2} \mathd x \lesssim \langle s \rangle^{- 2 -
			\alpha} . \label{eq:hk-scaling}
	\end{equation}
\end{lemma}

\begin{proof}
	For the first estimate we simply compute from \eqref{eq:G-kernel} and Lemma
	\ref{lem:exp-integrability},
	\[ \| Q_t \|_{L^p (w)}^p \lesssim \int_{\mathbb{R}^2} \mathe^{- pm^2 / 2 s}
		\mathe^{- 2 pt | x |^2} | x |^{2 \alpha} w (x) \mathd x \lesssim \langle
		t \rangle^{- 1 - \alpha} . \]
	In the polynomial case, the second bound is now a simple consequence of
	Young's convolution inequality and \eqref{eq:jpb-triangle-ineq}. For the
	case of the exponential weights, observe that by the triangle inequality,
	\[ \| Q_s u_s \|_{L^p (w_{\gamma})}^p \leqslant \int_{\mathbb{R}^2} \mathd x
		\left( \int_{\mathbb{R}^2} \mathd y \mathe^{\gamma m | x - y |} Q_s (x -
		y) \mathe^{\gamma m | y |} u_s (y) \right)^p \leqslant \| Q_s \|^p_{L^1
				(w)} \| u_s \|_{L^p (w)}^p, \]
	which again implies the claim with Lemma \ref{lem:exp-integrability}. The
	estimates on $\dot{G}_t$ follow from the estimates on $Q_t$ the convolution
	inequalities, since
	\[ \| \dot{G}_t \|_{L^p (w)} = \| Q_t \ast Q_t \|_{L^p (w)} \leqslant \|
		Q_t \|_{L^1} \| Q_t \|_{L^p (w)} \lesssim \langle t \rangle^{- 1 - 1 / p}
		. \]
	For the estimate~\eqref{eq:hk-scaling}, we start from
	Lemma~\ref{lem:hk-basic-bounds} and the definition of $\dot{G}_s$, to
	estimate
	\[ \dot{G}_s (x) \mathe^{\beta^2 C_s (x)} \lesssim s^{- 1} \mathe^{- m^2 /
			s} \mathe^{- \frac{s}{4} | x |^2} | x |^{- \beta^2 / 2 \pi} \langle s
		\rangle^{- \beta^2 / 4 \pi} . \]
	Moreover, \ for $s \geqslant t$ and $c, \tilde{c} > 0$ sufficiently small
	(more precisely, $c \in \left( 0, \frac{1}{4} \right)$ and $\tilde{c} \in
		\left( 0, \frac{1}{4} - c \right)$), we have $\mathe^{- \frac{s}{4} | x |^2}
		\mathe^{c t | x |^2} \leqslant \mathe^{- \tilde{c} s | x |^2}$. Combining
	both observations we see that
	\[ \begin{aligned}
			\int_{\mathbb{R}^2} \dot{G}_s (x) \mathe^{\beta^2 C_s (x)} | x |^{2
			\alpha} \mathe^{c t | x |^2} \mathd x & \lesssim  \int | x |^{2
					(\alpha - \beta^2 / 4 \pi)} \mathe^{- \tilde{c} s | x |^2} \mathe^{-
			m^2 / s} s^{- 1 - \beta^2 / 4 \pi} \mathd x                                                                 \\
			                                      & \lesssim  s^{- \beta^2 / 4 \pi} \mathe^{- m^2 / s}  \int_0^{\infty}
			r^{2 (\alpha - \beta^2 / 4 \pi) + 1} \mathe^{- \tilde{c} s r^2} \mathd
			r                                                                                                           \\
			                                      & \lesssim  \langle s \rangle^{- 2 - \alpha},
		\end{aligned} \]
	provided that \ $r \mapsto r^{2 (\alpha - \beta^2 / 4 \pi) + 1} \mathe^{-
			\tilde{c} s r^2}$ is integrable over $\mathbb{R}_+$, which is exactly the
	condition $\alpha > \frac{\beta^2}{4 \pi} - 1$.
\end{proof}

\begin{lemma}
	\label{lem:I(u)-regularity}For any $\alpha \in [0, 1]$, $k \in \mathbb{R}$
	and $p \in [1, \infty]$, we have
	\begin{equation}
		\left\| \int_0^T Q_s u_s \mathd s \right\|_{B_{p, p}^{\alpha, k}} \lesssim
		\sup_{s \in [0, T]} \| \langle s \rangle^{\alpha / 2 + \varepsilon} u_s
		\|_{L^{p, k}} . \label{eq:Iu-Bp}
	\end{equation}
	Moreover, in $L^2$ the improved bound
	\begin{equation}
		\left\| \int_0^T Q_s u_s \mathd s_{\nosymbol} \right\|_{H^{1, k}}^2
		\lesssim \int_0^T \| u_s \|_{L^{2, k}}^2 \mathd s. \label{eq:Iu-H1}
	\end{equation}
	holds.
\end{lemma}

\begin{proof}
	For any $\bar{\varepsilon} > 0$ and $\bar{p} > p$, it holds
	\begin{align*}
		\left\| \int_0^t Q_s u_s \mathd s \right\|_{B_{p, p}^{\alpha, k}}
		 & \leqslant \left\| \int_0^t Q_s u_s \mathd s \right\|_{B_{\bar{p},
		\bar{p}}^{\alpha, k}}                                                \\
		 & \leqslant \int_0^t \mathd s \| Q_s \|_{B_{1,
		p}^{\alpha, n}} \| u_s \|_{L^{p, k}}                                 \\
		 & \leqslant \sup_s \| \langle s
		\rangle^{\alpha / 2 + \varepsilon} u_s \|_{L^{p, k}}  \int_0^{\infty}
		\mathd s \langle s \rangle^{- \alpha / 2 - \varepsilon} \| Q_s \|_{B_{1,
						p}^{\alpha, n}} .
	\end{align*}
	Moreover, by the interpolation of Besov spaces, for any $p > p (\alpha) = (1
		- \alpha)^{- 1}$ and $\tilde{\varepsilon} > 0$ sufficiently small,
	\[ \| Q_s \|_{B_{1, p}^{\alpha, n}} \leqslant \| Q_s \|_{B_{1, p
						(\alpha)}^{\alpha, n}} \lesssim \| Q_s \|_{L^{1, n}}^{1 - \alpha} \| Q_s
		\|^{\alpha}_{B_{1, \infty}^{1 - \tilde{\varepsilon}, n}} \lesssim \langle
		s \rangle^{- 1 + \alpha} \| Q_s \|_{B^{1 - \tilde{\varepsilon}, n}_{1,
						\infty}}^{\alpha}, \]
	so that the claim will follow once we compute
	\[ \| Q_s \|_{B^{1 - \tilde{\varepsilon}, n}_{1, \infty}} \lesssim \sup_{| y
			| \leqslant 1}  \int \mathd x \frac{| Q_s (x - y) - Q_s (x) |}{| y |}
		\langle x \rangle^n \lesssim \langle s \rangle^{- 1 / 2} . \]
	But this follows from a simple Taylor expansion,
	\[ \begin{aligned}
			\int \mathd x | Q_s (x - y) - Q_s (x) | & = | y | s \mathe^{- m^2 / 2
				s}  \int_0^1 \mathd \vartheta \int \mathd x \langle x \rangle^n | x -
			\vartheta y | \mathe^{- 2 s | x - \vartheta y |^2}                                                          \\
			                                        & = | y | s \mathe^{- m^2 / 2 s}  \int \mathd x | x | \mathe^{- s |
			x |^2}                                                                                                      \\
			                                        & \lesssim | y | s^{- 1 / 2} \mathe^{- m^2 / 2 s} .
		\end{aligned} \]
	To remove the $\varepsilon$ in the $L^2$ estimates, we pass to the Fourier
	transform and use the fact that $Q_s$ is diagonal in Fourier space. Since $w
		(x) \assign \langle x \rangle^k$ grows at most polynomially, we have $w \in
		\mathcal{S}' (\mathbb{R}^2)$. For this computation only, we denote the
	Fourier transform of a distribution $f \in \mathcal{S}' (\mathbb{R}^2)$ by
	$\hat{f} = \mathcal{F} (f)$. Repeatedly applying H{\"o}lder's inequality and
	Parseval's identity yields after some manipulation,
	{\small\begin{align*}
				         & \left\| \int_0^t \mathd s Q_s u_s \right\|_{H^{\alpha} (w)}^2      \\
				=        & C \left\| \int_0^t \mathd s \mathcal{F} (w (1 - \Delta)^{\alpha /
				2} Q_s u_s) \right\|^2_{L^2}                                                  \\
				=        & C \left\| \int_0^t \mathd s \int \mathd k \hat{w} (\xi - k) (1 + |
				k |^2)^{\alpha / 2} s \mathe^{- \frac{m^2 + | k |^2}{2 s}}  \hat{u} (k)
				\right\|_{L^2 (\mathd \xi)}^2                                                 \\
				\lesssim & \left\| \int_{\mathbb{R}^2} \mathd k (1 + | k |^2)^{\alpha /
						2}  \hat{w} (\xi - k) \left( \int_0^t \mathd s \frac{1}{s^{1 + \alpha}}
				\mathe^{- \frac{m^2 + | k |^2}{2 s}} \right)^{1 / 2} \times\right.            \\&\qquad\times \left. \left( \int_0^t
				\mathd s \mathe^{- m^2 / 2 s} s^{- 1 + \alpha} \hat{u}_s^2 (k)
				\right)^{1 / 2} \right\|_{L^2 (\mathd \xi)}^2                                 \\
				\lesssim & \left\| \int_{\mathbb{R}^2} \mathd k \frac{(1 + | k
					|^2)^{\alpha / 2}}{(| k |^2 + m^2)^{\alpha / 2}} \mathe^{- \frac{m^2 +
						| k |^2}{2 t}}  \hat{w} (\xi - k) \left( \int_0^t \mathd s \langle s
				\rangle^{- 1 + \alpha}  \hat{u}_s^2 (k) \right)^{1 / 2} \right\|^2_{L^2
				(\mathd \xi)}                                                                 \\
				\lesssim & \sup_k  \frac{(1 + | k |^2)^{\alpha}}{(| k |^2 +
					m^2)^{\alpha}} \left\| \int_{\mathbb{R}^2} \mathd k \hat{w} (\xi - k)
				\left( \int_0^t \mathd s \langle s \rangle^{- 1 + \alpha}  \hat{u}_s^2
				(k) \right)^{1 / 2} \right\|^2_{L^2 (\mathd \xi)}                             \\
				=        & \int_0^t \mathd s \langle s \rangle^{- 1 + \alpha} \| \hat{w} \ast
				\hat{u} \|_{L^2}^2                                                            \\
				\lesssim & \int_0^t \mathd s \langle s \rangle^{- 1 + \alpha} \| u_s
				\|_{L^2 (w)}^2 .
			\end{align*}}

\end{proof}

\begin{remark}
	Lemma \ref{lem:I(u)-regularity} takes advantage of the concrete choice for
	the scale interpolation to get the optimal regularity estimates
	\eqref{eq:Iu-H1} in $L^2$. For a general scale interpolation, not
	necessarily diagonal in Fourier space, we have to use \eqref{eq:Iu-Bp} and
	give up an arbitrarily small $\varepsilon > 0$ in regularity. This is not
	crucial to the analysis, but would in general lead to slightly worse
	results, e.g. replacing $\mathbb{D}= L^{2, n}$ by $\mathbb{D}=
		H^{\varepsilon, n}$ in the infinite volume variational problem in Theorem
	\ref{thm:infinite-vol-var}.
\end{remark}

\subsection{Proof of
	Lemma~\ref{lem:GFF-convergence}}\label{app:GFF-convergence}

We first restrict ourselves to the case $p \in [1, \infty)$. To this end, we
use the translation invariance of the Law of $W_t$, and hypercontractivity to
estimate
\begin{equation}
	\mathbb{E} \| W_t \|_{B^{- \varepsilon, - n}_{p, p}}^p = \sum_{i \geqslant -
		1} 2^{- i \varepsilon p} \int \mathbb{E} [| \Delta_i W_t (x) |^p] \langle x
	\rangle^{- pn} \mathd x \lesssim \sum_{i \geqslant - 1} 2^{- i \varepsilon
			p} \mathbb{E} [| \Delta_i W_t (0) |^2]^{p / 2} . \label{eq:W-Besov}
\end{equation}
Since $\Delta_i W_t (0) = \langle W_t, K_i \rangle$ and $\tmop{Cov} (W_t) =
	G_t = (m^2 - \Delta)^{- 1} \mathe^{- (m^2 - \Delta) / t}$, we can compute the
expectation on the right hand side as
\[ \begin{aligned}
		\mathbb{E} [| \Delta_i W_t (0) |^2] & = \mathbb{E} [\langle W_t, K_i
		\rangle \langle W_t, K_i \rangle]                                                                         \\
		                                    & = \int \mathd \xi \frac{| \varphi_i (\xi) |^2}{m^2 + | \xi |^2}
		\mathe^{- (m^2 - \Delta) / t}                                                                             \\
		                                    & \leqslant  \int_{R_1  2^i}^{R_2 2^i} \frac{r}{(m^2 + r^2)} \mathd r \\
		                                    & \lesssim  \log (2^i) .
	\end{aligned} \]
Here we used the fact that $\varphi_i$ is supported on an annulus with radii
$R_1 2^i$, $R_2 2^i$ in the second to last estimate. Inserting this bound in
\eqref{eq:W-Besov} yields the claim for $p \in [1, \infty) $. For $p =
	\infty$, we use the Besov embedding $\| \cdot \|_{B^{- \alpha, - n}_{\infty,
					\infty}} \lesssim \| \cdot \|_{B^{- \alpha + \varepsilon}_{p, p}}$ for $p > 2
	/ \varepsilon$.

Finally, applying exactly the same reasoning to the increment $W_{\infty} -
	W_t$ instead shows the convergence in $L^p (\mathd \mathbb{P}, B^{- \varepsilon, -
			n}_{p, p})$ for any $p \in [1, \infty)$.

\subsection{Proof of
	Lemma~\ref{lem:charged-cov-bound}}\label{app:charged-cov-bound}

Suppose for concreteness that $q (\xi_{1 : \ell}) > 0$ and recall that we want
to show
\[ \Gamma_{t, s} (\xi_1, \ldots, \xi_{\ell}) \leqslant \frac{\beta^2}{2} (G_t
	(0) - G_s (0)) + C. \]
We assume that (possibly after relabelling),
\[ \sigma_k = \begin{cases}
		+ 1,    & k \leqslant q, \\
		(- 1)^k & k > q,
	\end{cases} \]
and that \(\xi_1\) is the most isolated charge, that is \(\min_{k>q}|x_1-x_k| = \max_{j\leqslant q}\min_{k>q} |x_j - x_k|\). 
We split the matrix into the 3 components
\begin{equation}
	\Gamma_{t, s} (\xi_1, \ldots, \xi_n) = \Gamma_{t, s} (\xi_1) + \Gamma_{t, s} (\xi_2,
	\ldots, \xi_n) + \sigma_1  \sum_{i > 1} \sigma_i (G_t - G_s) (x_1 - x_i) .
	\label{eq:W-charged-split}
\end{equation}
By the definition of $\Gamma$ and the basic heat kernel estimates
\eqref{eq:cov-0-bound}, the first summand is
\[ \frac{\beta^2}{2} (G_t - G_s) (0) \leqslant \frac{\beta^2}{8 \pi} (\log (t \vee
	1) - \log (s \vee 1)) + C. \]
The second summand is bounded from above by \eqref{eq:W-leq-0}.
So~\eqref{eq:charged-cov-bound} will follow once we establish an upper bound
for the last term in~\eqref{eq:W-charged-split}. Towards this goal, we start
by extracting the charged and neutral part,
\[ \sum_{i > 1} \sigma_1 \sigma_i (G_t - G_s) (x_1 - x_i) = \sum_{i = 2}^q
	\sigma_1 \sigma_i (G_t - G_s) (x_1 - x_i) + \sum_{i = q + 1}^{\ell}
	\sigma_1 \sigma_i (G_t - G_s) (x_1 - x_i) . \]
The first (charged) part satisfies $\sigma_1 \sigma_i = 1$ and we can use the
same reasoning as in \eqref{eq:W-leq-0} to conclude boundedness from the
positivity of $G$. The second (neutral) part also contains contributions with
the ``bad'' signs $\sigma_1 \sigma_i = - 1$ and requires special attention.
Since this part is neutral, we know that the sum contains an even number of
points and we can proceed by considering the neutral pairs $(x_{q + 2 i}, x_{q
			+ 2 i + 1})$, $i = 0, 1, \ldots$ one at a time. In other words, the claim will
follow if there is a constant $C > 0$ such that for any $(y, z) \in
	\mathbb{R}^2 \times \mathbb{R}^2$,
\begin{equation}
	(G_t - G_s) (x_1 - y) - (G_t - G_s) (x_1 - z) \leqslant C.
	\label{eq:3pt-bound}
\end{equation}
By construction, one of the terms in these pairings comes with the ``good''
sign, which we are going to use to bound the neutral contribution. We start by
rewriting the covariance using the kernel representation,
\[ \begin{aligned}
		 & (G_t - G_s) (x_1 - y) - (G_t - G_s) (x_1 - z)
		= - \int_t^s \mathd
		\tau [\dot{G}_{\tau} (x_1 - y) - \dot{G}_{\tau} (x_1 - z)]          \\
		 & = - \int_t^s \mathd \tau \tau^{- 1} \mathe^{- m^2 / \tau} \left(
		\mathe^{- \frac{\tau}{4} | x_1 - y |^2} - \mathe^{- \frac{\tau}{4} | x_1 - z
			|^2} \right) .
	\end{aligned} \]
If the charged edge is the shortest edge in the triangle connecting $x_1, y,
	z$, that is $| x_1 - y | \leqslant | x_1 - z |$, then
\[ \mathe^{- \frac{\tau}{4} | x_1 - y |^2} - \mathe^{- \frac{\tau}{4} | x_1 -
		z |^2} \geqslant 0, \]
and we can bound \eqref{eq:3pt-bound} with $C = 0$. Otherwise, one of the
neutral edges $| x_1 - z |$ or $| z - y |$ is the shortest edge. Thanks to the choice of \(x_1\) as the most isolated charge,  $| y - z | \leqslant | x_1 - z |$. If $| y - z | = 0$, then $(G_t - G_s) (x_1 - y) -
	(G_t - G_s) (x_1 - z) = 0$ and \eqref{eq:3pt-bound} is trivially true. Thus,
we may assume that all edges have positive lengths. On $\tau > | y - z |^{-
			2}$, we directly compute
\[ \int^s_{| z - y |^{- 2}} \mathd \tau \tau^{- 1} \mathe^{- m^2 / r} \left|
	\mathe^{- \frac{\tau}{4} | x_1 - y |^2} - \mathe^{- \frac{\tau}{4} | x_1 -
		z |^2} \right| \leqslant 2 \int^s_{| z - y |^{- 2}} \mathd \tau \tau^{- 1}
	\mathe^{- m^2 / \tau} \mathe^{- \frac{\tau}{4} | z - y |^2} \lesssim 1. \]
On $\tau \leqslant | y - z |^{- 2}$, we use \eqref{eq:hk-g0-gx} combined with
the translation invariance of $\dot{G}$ to conclude
\begin{align*}
	\int_t^{| z - y |^{- 2}} \mathd \tau [\dot{G}_{\tau} (x_1 - y) -
		\dot{G}_{\tau} (x_1 - z)]
	 & \lesssim |y-z| \int_t^{| y - z |^{- 2}} \mathd \tau
	|\nabla\dot{G_{\tau}} (x_1 - y + \tau(y-z))| \\
	 & \lesssim | y - z | \int_t^{| y - z |^{-
				2}} \mathd \tau  \langle \tau \rangle^{- 1 / 2} \lesssim 1,
\end{align*}
which completes the proof of Lemma~\ref{lem:charged-cov-bound}.

\subsection{Proof of Lemma~\ref{lem:W-Halpha}}

\label{app:pf-lem-W-Halpha}We first show the claim for $p < \infty$. To this
end, we start by rewriting $t^{- \alpha / 2} W_t$ for $t \geqslant 1$ using
Ito's formula
\[ t^{- \alpha / 2} W_t = W_1 - \frac{\alpha}{2}\int_1^t s^{- 1 - \alpha / 2} W_s \mathd s +
	\int_1^t s^{- \alpha / 2} \mathd W_s, \]
so that
\begin{align}
	\begin{split}
		\sup_t \| t^{- \alpha / 2} W_t \|_{B_{p, p}^{\alpha - \varepsilon, - n}}
		\lesssim & \| W_1 \|_{B_{p, p}^{\alpha - \varepsilon, - n}} + \int_1^{\infty}
		s^{- 1 - \alpha / 2} \| W_s \|_{B_{p, p}^{\alpha - \varepsilon, - n}} \mathd
		s                                                                             \\ &+ \sup_t \left\| \int_1^t s^{- \alpha / 2} \mathd W_s \right\|_{B_{p,
						p}^{\alpha - \varepsilon, - n}} . \label{eq:DDC-W}
	\end{split}
\end{align}
For the bounded variation part, a similar computation to the proof of Lemma
\ref{lem:GFF-convergence} shows that for any $\tilde{\varepsilon} > 0$ we have
\[ \mathbb{E} \| W_s \|_{B^{\alpha - \varepsilon}_{p, p}} \lesssim \langle s
	\rangle^{\alpha / 2 - (\varepsilon - \tilde{\varepsilon}) / 2} . \]
Therefore, choosing $\tilde{\varepsilon} \in (0, \varepsilon)$,
\begin{align*}
	\mathbb{E} \int_1^{\infty} s^{- 1 - \alpha / 2} \| W_s \|_{B_{p,
					p}^{\alpha, - n}} \mathd s
	 & \lesssim \int_1^{\infty} s^{- 1 -
			\tilde{\varepsilon} / 2} \mathbb{E} [\langle s \rangle^{\alpha / 2 -
			(\varepsilon - \tilde{\varepsilon}) / 2} \| W_s \|_{B^{\alpha -
	\varepsilon}_{p, p}}] \mathd s       \\
	 & \lesssim \int_1^{\infty} s^{- 1 -
			\tilde{\varepsilon} / 2} < \infty .
\end{align*}
Regarding the martingale $M_t = \int_1^t s^{- \varepsilon / 2} \mathd W_s$, we
compute by translation invariance, the maximal inequalities and Gaussian
hypercontractivity,
\[ \begin{aligned}
		\mathbb{E} [\sup_t \| M_t \|^p_{B^{\alpha, - n}_{p, p}}] & = \mathbb{E}
		\left[ \sup_t \sum_{i \geqslant - 1} 2^{i (\alpha - \varepsilon) p}  \int
		| \Delta_i M_t (x) |^p \langle x \rangle^{- pn} \mathd x \right]                                                           \\
		                                                         & \lesssim  \sum_{i \geqslant - 1} 2^{i (\alpha - \varepsilon) p}
		\mathbb{E} [\sup_t  | \Delta_i M_t (0) |^p]                                                                                \\
		                                                         & \lesssim  \sum_{i \geqslant - 1} 2^{i (\alpha - \varepsilon) p}
		\mathbb{E} [| \Delta_i M_{\infty} (0) |^p]                                                                                 \\
		                                                         & \lesssim  \sum_{i \geqslant - 1} 2^{i (\alpha - \varepsilon) p}
		\mathbb{E} [| \Delta_i M_{\infty} (0) |^2]^{p / 2} .
	\end{aligned} \]
The covariance of $M_{\infty}$ can be computed directly, as for some constant
$C$,
\[ \begin{aligned}
		\mathbb{E} [\langle M_{\infty}, f \rangle \langle M_{\infty}, g \rangle]
		 & = \int_0^{\infty} \mathd s \int \mathd x f (x) \int \mathd y s^{-
		\alpha} \dot{G}_s (x - y) g (y)                                         \\
		 & = C \int \mathd x \int \mathd y f (x) ((m^2 - \Delta)^{- 1 - \alpha}
		g) (x)                                                                  \\
		 & = C \langle f, (m^2 - \Delta)^{- 1 - \alpha} g \rangle .
	\end{aligned} \]
Since $\Delta_i M_{\infty} (0) = \langle M_{\infty}, K_i \rangle$, we have
\[ \mathbb{E} [| \Delta_i M_{\infty} (0) |^2] =\mathbb{E} [| \langle
		M_{\infty}, K_i \rangle |^2] = C \langle K_i, (m^2 - \Delta)^{- 1 - \alpha}
	K_i \rangle = \int \mathd \xi \frac{| \varphi_i (\xi) |^2}{(m^2 + | \xi
		|^2)^{1 + \alpha}}, \]
where we used $K_i = \mathcal{F}^{- 1} (\varphi_i)$. Since $\varphi$ is
radially symmetric and supported on an annulus with radii ${R_1}_{}$, $R_2$,
we see passing to spherical coordinates using $\varphi_i = \varphi (2^{- i}
	\cdot)$
\[ \int \mathd \xi \frac{| \varphi_i (\xi) |^2}{(m^2 + | \xi |^2)^{1 +
				\alpha}} \lesssim \int_{R_1  2^i}^{R_2 2^i} \frac{r^{d - 1}}{(m^2 + r^2)^{1
				+ \alpha}} \mathd r \overset{d = 2}{\lesssim} (R_1 2^i)^{- 2 \alpha}
	\lesssim 2^{- 2 \alpha i} . \]
Therefore, for $0 < \alpha < \varepsilon$,
\[ \mathbb{E} [\sup_t \| M_t \|^p_{B^{(\alpha - \varepsilon), - n}_{p, p}}]
	\lesssim \sum_{i \geqslant - 1} 2^{i (\alpha - \varepsilon) p} \mathbb{E}
	[| \Delta_i M_{\infty} (0) |^2]^{p / 2} \lesssim \sum_{i \geqslant - 1}
	2^{- pi \varepsilon} < \infty, \]
and inserting the bounds in the \eqref{eq:DDC-W}
{\small\[ \mathbb{E} [\sup_t  \| t^{- \alpha / 2} W_t \|_{B_{p, p}^{\alpha -
							\varepsilon, - n}}] \lesssim 1 +\mathbb{E} \int_1^{\infty} s^{- 1 - \alpha
				/ 2} \| W_s \|_{B_{p, p}^{\alpha - \varepsilon, - n}} \mathd s +\mathbb{E}
		[\sup_{t \geqslant 1} \| M_t \|_{B_{p, p}^{\alpha - \varepsilon, - n}}] < \infty . \]}
Finally, for the case $p = \infty$, we use the Besov embedding $\| \cdot
	\|_{B^{\alpha - \gamma, - n}_{\infty, \infty}} \lesssim \| \cdot \|_{B_{p,
					p}^{\alpha, - n}}$ for $\alpha - \gamma > 0$ and $p > 2 / \gamma$. Then
choosing $\gamma \in (0, \varepsilon)$ and $p$ sufficiently large it holds
that
\[ \mathbb{E} [\sup_{t \geqslant 1} \| t^{- \alpha / 2} W_t
		\|_{B_{\infty, \infty}^{\alpha - \varepsilon, - n}}] \lesssim
	\mathbb{E} [\sup_{t \geqslant 1}  \| t^{- \alpha / 2} W_t \|_{B_{p,
						p}^{\alpha - (\varepsilon - \gamma), - n}}^p]^{1 / p} \lesssim 1. \]

\section{Auxiliary estimates on the Fourier coefficients}
We collect some additional estimates on the kernels $f$ defined
in~\eqref{eq:fe-v-coeffients}. No new ideas are needed for these estimates and
we only want to briefly illustrate how the proofs in the previous section can
be modified to obtain the additional results.

\subsection{Dependence on the terminal condition}

The following Lemma quantifies the dependence on the terminal
condition~\eqref{eq:fe-terminal-conditions} and is used to show convergence as
the small-scale cut-off is removed in Proposition \ref{prop:force-estimates}.

\begin{lemma}
	\label{lem:coeff-T-dependence}Let $T_1,T_2>0$. For $\varepsilon>0$ sufficiently small
	depending only on $\delta$, it holds for all $t < \infty$,
	\begin{equation}
		\interleave f^{[\ell], T_1}_t - f_t^{[\ell], T_2} \interleave_t \lesssim
		\bar{\lambda}_t^{\ell} \langle t\rangle^{- (\ell - 1) + 2 \varepsilon} \langle T_1 \wedge T_2
		\rangle^{- \varepsilon} . \label{eq:coeff-T-dependence}
	\end{equation}
\end{lemma}

\begin{proof}
	We proceed inductively (mod the procedure required to remove the
	regularising kernel for $\ell \geqslant 3$) starting from $\ell = 1$. For
	simplicity, let us assume that $T_2 \leqslant T_1$, then
	\[ |f^{[1], T_1}_t - f_t^{[1], T_2}|
		=|\mathbbm{1}_{\{ t \geqslant T_2 \}}
		(f_t^{[\ell], T_1} - f_{T_2}^{[\ell], T_2})|
		=\mathbbm{1}_{\{ t \geqslant
			T_2 \}} \lambda_t \left( 1 - \frac{\lambda_{T_2}}{\lambda_t} \right) . \]
	And using that $G_t - G_{T_2} \geqslant 0$ on $\{ t \geqslant T_2 \}$,
	\[ 0 \leqslant 1 - \frac{\lambda_{T_2}}{\lambda_t} = 1 - \mathe^{- \beta^2
			(G_t - G_{T_2}) (0)} \leqslant \beta^2 (G_t - G_{T_2}) (0) \leqslant
		\frac{\beta^2}{4 \pi} \log (t \vee 1) + C. \]
	Therefore,
	\[ \interleave f^{[1], T_1}_t - f_t^{[1], T_2} \interleave_t = | f^{[1],
				T_1}_t - f_t^{[1], T_2} | \lesssim \mathbbm{1}_{\{ t \geqslant T_2 \}}
		\bar{\lambda}_t \log (t \vee 1) \lesssim \bar{\lambda}_t t^{2 \varepsilon} \langle
		T_2 \rangle^{- \varepsilon}, \]
	as required.
	Moving on now to $\ell \geqslant 2$, assuming that
	\eqref{eq:coeff-T-dependence} holds for $\ell' < \ell$, we have
	\begin{equation*}
		\begin{aligned}
			\delta_T f^{[\ell]}_t = & \delta_T f_{T_2}^{[\ell]} -
			\sum_{\tmscript{\begin{array}{c}
						                I_1 \dot{\cup} I_2 = [\ell] \\
						                k \in \{ 1, 2 \}
					                \end{array}}}  \sum_{\tmscript{\begin{array}{c}
						                                               i \in I_1 \\
						                                               j \in I_2
					                                               \end{array}}} C (| I_1 |, | I_2 |)\sigma_i \sigma_j \frac{\beta^2}{2}\times \\
			                        & \times\int_{t \vee T_2}^{T_2}
			\delta_T f_s (\xi_{I_1}) f_s^{[T_k]} (\xi_{I_2}) \mathe^{\Gamma_{t, s}
				(\xi_I)} \dot{G}_s (x_i - x_j) \mathd s.
		\end{aligned}
	\end{equation*}
	\[ \]
	Inserting the bounds on $\delta_T f^{[\ell']}$ for $\ell' < \ell$, and the
	bounds on $\interleave f_t^{[\ell']} \interleave_t$, we find in the same way
	as in section \ref{sec:flow-eq},
	\[ \interleave \int_t^{T_2} \delta_T f_s (\xi_{I_1}) f_s (\xi_{I_2})
		\mathe^{\Gamma_{t, s} (\xi_I)} \dot{G}_s (x_i - x_j) \mathd s \interleave_t
		\lesssim \bar{\lambda}_t^{\ell} t^{- (\ell - 1) + 2 \varepsilon} \langle T_2
		\rangle^{- \varepsilon} . \]
	Regarding the remaining term, for $\ell > 1$, it holds for $t \leqslant T_2$,
	\begin{equation*}
		\begin{aligned}
			\interleave & \delta_T f_{T_2}^{[\ell]} \interleave_t                                                                                          \\
			            & = \interleave
			f_{T_2}^{[\ell], T_1} \interleave_t                                                                                                            \\
			            & \lesssim \sum_{I_1 \dot{\cup} I_2 = [\ell]} \int_{T_2}^{T_1} \interleave
			f_s^{T_1} (\xi_{I_1}) \interleave_s \interleave f_s^{T_1} (\xi_{I_2})
			\interleave_s \times                                                                                                                           \\
			            & \phantom{\lesssim}\times\interleave \sum_{\tmscript{\begin{array}{c}
						                                                                  i \in I_1 \\
						                                                                  j \in I_2
					                                                                  \end{array}}} \mathe^{\Gamma_{t, s} (\xi_I)} \frac{\kappa_t (\xi_I) \sigma_i
				\sigma_j}{\kappa_s (\xi_{I_1}) \kappa_s (\xi_{I_2})} \dot{G}_s (x_i - x_j)
			\interleave \mathd s,                                                                                                                          \\
			            & \lesssim \left\{ \begin{array}{cl}
				                               t^{1 / 2} \bar{\lambda}_{T_2}^2 \langle T_2 \rangle^{- 3 / 2}, & \ell = 2, q
				                               (\xi_I) = 0                                                                       \\
				                               \bar{\lambda}_t^1 \bar{\lambda}_{T_2}^{\ell - 1} \langle T_2 \rangle^{- (\ell -
				                               1)},                                                           & \text{otherwise}
			                               \end{array} \right\}                               \\
			            & \lesssim \bar{\lambda}_t^{\ell} \langle t \rangle^{- (\ell
				- 1) + 2 \varepsilon} \langle T_2 \rangle^{- \varepsilon},
		\end{aligned}
	\end{equation*}
	as required.
\end{proof}

\subsection{Dependence on the mollification}

Suppose that $(\eta^{\varepsilon})_{\varepsilon > 0}$ is an approximation to
the identity on $\mathbb{R}^2$ such that for $(t, x) \in \mathbb{R}_+ \times
	\mathbb{R}^2$,
\[ \lim_{\varepsilon \rightarrow 0} G^{\varepsilon}_t (x) \assign
	\lim_{\varepsilon \rightarrow 0}  \int \mathd y \eta^{\varepsilon} (x - y)
	G_t (y) = G_t (x) . \]
Define the truncated solution to the flow equation $F^{\varepsilon}$, its
Fourier coefficients $f^{\varepsilon}$ and the renormalisation constants
$\lambda^{\varepsilon}$ in the same way as before, with $G$ replaced by
$G^{\varepsilon}$. To prove reflection positivity in
Section~\ref{sec:OS-Axioms}, we have to understand the dependency of the flow
equation on this mollification.
The required argument is essentially the same as for the dependence on $T$ in
the previous section, with the main difference being that we do not need to
distinguish cases $t\geq T_1\wedge T_2$ and $t\leqslant T_1\wedge T_2$.

\begin{lemma}
	\label{lem:f-eps-convergence}If $\bar{\lambda}_t^{\varepsilon} \leqslant C
		\bar{\lambda}_t$ for some constant $C > 0$, then there is a subsequence
	$\varepsilon_N \rightarrow 0$ such that
	\[ \interleave f^{[\ell]}_t - f_t^{[\ell], \varepsilon_N} \interleave
		\lesssim N^{- 1} \interleave f_t^{[\ell]} \interleave . \]
\end{lemma}

\begin{proof}
	Let us first derive the dependency of the renormalisation constant
	$\lambda^{\varepsilon}$ on $\varepsilon$. Since $\bar{\lambda}^{\varepsilon}_t
		\lesssim \bar{\lambda}_t$ uniformly in $\varepsilon > 0$,
	\[ | \lambda_t^{\varepsilon} - \lambda_t | = \left|
		\mathe^{\frac{\beta^2}{2} G_t^{\varepsilon} (0)} -
		\mathe^{\frac{\beta^2}{2} G_t (0)} \right| \lesssim
		\mathe^{\frac{\beta^2}{2} G_t (0)} | G_t^{\varepsilon} (0) - G_t (0) | .
	\]
	Choosing $(\varepsilon_N)_{N \in \mathbb{N}}$ such that $|
		G^{\varepsilon_N}_t (0) - G_t (0) | \leqslant N^{- 1}$, this implies with
	the definition of $\lambda_t$,
	\[ | \lambda_t^{\varepsilon} - \lambda_t | \lesssim N^{- 1} \bar{\lambda}_t, \]
	As a by-product, since $f^{[1]}_t = - \frac{\beta \lambda_t}{2 i}$, this
	shows the claim for $\ell = 1$. For $\ell = 2, 3$ we proceed as in the
	bounds derived in Lemma \ref{lem:f2-kernel-scaling} and
	\ref{lem:f3-kernel-scaling}. For example, for the charged case
	\begin{equation*}
		\begin{aligned}
			 & \interleave f^{[2] (\pm 2)}_t - f_t^{[2], (\pm 2), \varepsilon}
			\interleave                                                          \\
			 & =  C \sup_{x_1}  \left| \int_t^T \mathd s \int \mathd
			x_2 [(\lambda_s^{\varepsilon})^2 \dot{G^{\varepsilon}_s} (x_1 - x_2)
			\mathe^{\Gamma_{t, s}^{\varepsilon} (\xi_1, \xi_2)} - (\lambda_s)^2
			\dot{G}_s (x_1 - x_2) \mathe^{\Gamma_{t, s} (\xi_1, \xi_2)}] \right| \\
			 & =  C \left| \int_t^T \mathd s \int \mathd x_2
			[\lambda_t^{\varepsilon} \lambda_s^{\varepsilon}
			\dot{G^{\varepsilon}_s} (x_2) - \lambda_t \lambda_s  \dot{G}_s (x_2)]
			\right|                                                              \\
			 & =  C \left| \int_t^T \mathd s \left[ \lambda_t^{\varepsilon}
				\lambda_s^{\varepsilon} \int \mathd x_2 \left( \dot{G^{\varepsilon}_s}
				(x_2) - \dot{G_s} \left( x_2 \right) \right) - (\lambda_t \lambda_s -
				\lambda_t^{\varepsilon} \lambda_s^{\varepsilon}) \int \mathd x_2
				\dot{G}_s (x_2) \right] \right| .
		\end{aligned}
	\end{equation*}
	But thanks to the translation invariance,
	\[ \int \mathd x_2 (\dot{G^{\varepsilon}_s} (x_2) - \dot{G_s} (x_2)) = \int
		\mathd x_1 \eta^{\varepsilon} (x_1)  \int \mathd x_2 (\dot{G}_s (x_1 -
		x_2) - \dot{G}_s (x_2)) = 0. \]
	Thus, with the estimates on $| \lambda_s^{\varepsilon} - \lambda_s |
		\lesssim N^{- 1} \bar{\lambda}_s$ and $| \lambda_t \lambda_s -
		\lambda_t^{\varepsilon} \lambda_s^{\varepsilon} | \lesssim \bar{\lambda}_t
		\bar{\lambda}_s N^{- 1}$
	\[ \interleave f^{[2] (\pm 2)}_t - f_t^{[2] (\pm 2), \varepsilon_N}
		\interleave \lesssim \int_t^T \mathd s | \lambda_t \lambda_s -
		\lambda_t^{\varepsilon} \lambda_s^{\varepsilon} |  \int \mathd x_2
		\dot{G}_s (x_2) \lesssim \bar{\lambda}_t^2 \langle t \rangle^{- 1} N^{- 1} . \]
	The remaining bounds on the neutral part $f^{[2] (0)}$ and $f^{[3]}$ follow
	in the same way.
\end{proof}

\section{Wick-ordered cosine}\label{app:wick-cos}

For the large deviations principle in Section~\ref{sec:LDP}, we rely on the
convergence of the Wick-ordered sine and cosine in the first region $\beta^2 <
	4 \pi$.

\begin{lemma}
	\label{lem:wick-conv}Let $\beta^2 \in [0, 4 \pi)$. For any $p \in [1,
		\infty)$ and $\alpha > \beta^2 / 4 \pi$, it holds that
	\[ \sup_T \mathbb{E} \| \llbracket \cos (\beta W_T) \rrbracket - 1 \|_{B_{p,
						p}^{- \alpha} (\langle x \rangle^{- n})}^p \lesssim \beta^2 . \]
	Moreover, as $T \rightarrow \infty$, the martingale $(\llbracket \cos (\beta
		W_T) \rrbracket)$ converges in $L^p (\mathd \mathbb{P} ; B_{p, p}^{- \alpha} (\langle
		x \rangle^{- n}) \nobracket$ and almost surely. We denote the unique limit
	by $\llbracket \cos (\beta W_{\infty}) \rrbracket$. An analogous statement
	holds for the Wick-ordered sine.
\end{lemma}

The main ingredient in for the proof of Lemma~\ref{lem:wick-conv} is the
following point-wise estimate on the quadratic variation.

\begin{lemma}
	\label{lem:variation-bound}Let $N_t = \llbracket \cos (\beta W_t)
		\rrbracket$ and $\beta^2 \in [0, 4 \pi)$. Then, its quadratic variation
	satisfies for any $\varepsilon > 0$,
	\[ | \langle \Delta_i N \rangle_t (x) | \lesssim \beta^2 2^{2 i (\beta^2 / 4
				\pi + \varepsilon)} . \]
	The analogous statement holds for the cosine replaced by the sine.
\end{lemma}

\begin{proof}
	Expanding the Wick-ordered cosine with Ito's formula we find,
	\[ \llbracket \cos (\beta W_t) \rrbracket = 1 - \beta \int_0^t \llbracket
		\sin (\beta W_s) \rrbracket \mathd W_s = 1 - \beta \int_0^t
		\mathe^{\frac{\beta^2}{2} G_s (0)} \sin (\beta W_s) \mathd W_s . \]
	Therefore, using $\tmop{Cov} (W_s) = G_s$ and applying Young's inequality
	repeatedly,
	\begin{equation}
		\begin{aligned}
			          & | \langle \Delta_i N \rangle_t (x) |                               \\
			\leqslant & \beta^2  \int^t_0 \mathd s \mathe^{\beta^2 G_s (0)}  \int
			\mathd y_1 K_i (x - y_1)  \int \mathd y_2 K_i (x - y_2) \times                 \\
			          & \qquad \times \sin (\beta W_s (y_1)) \sin (\beta W_s (y_2)) \mathd
			\langle W (y_1), W (y_2) \rangle_s                                             \\
			\leqslant & \beta^2 \int_0^t \mathd s \mathe^{\beta^2 G_s (0)} \sup_x
			\int \mathd y_1 K_i (x - y_1)  \int \mathd y_2 K_i (x - y_2)  \dot{G}_s
			(y_1 - y_2)                                                                    \\
			\leqslant & \beta^2 \| K_i \|_{L^1} \| K_i \|_{L^p} \sup_{y_1}  \int_0^t
			\mathd s \mathe^{\beta^2 G_s (0)} \| \dot{G}_s (y_1 - y_2) \|_{L^q
			(\mathd y_2)}                                                                  \\
			\leqslant & \beta^2 \| K_i \|_{L^1} \| K_i \|_{L^p}  \int_0^t \mathd s
			\mathe^{\beta^2 G_s (0)} \| \dot{G}_s \|_{L^q},
		\end{aligned} \label{eq:N-q-variation}
	\end{equation}
	where $\frac{1}{p} + \frac{1}{q} = 1$ are to be determined later. So using
	the estimates on the heat kernel
	\[ \| \dot{G_s} \|_{L^q} \lesssim \langle s \rangle^{- 1} \| Q_s \|_{L^q}
		\lesssim \langle s \rangle^{- 1 - 1 / q}, \]
	combined with the estimates on the Littlewood--Paley kernels
	\eqref{eq:LP-kernels}, the previous computation \eqref{eq:N-q-variation} and
	$\mathe^{\beta^2 G_s (0)} \lesssim \langle s \rangle^{\beta^2 / 4 \pi}$ from
	Lemma~\ref{lem:hk-basic-bounds} gives
	\[ | \langle \Delta_i N \rangle_t (x) | \lesssim \beta^2 2^{2 i \frac{p -
					1}{p}}  \int_0^t \mathd s \langle s \rangle^{\beta^2 / 4 \pi}  \langle s
		\rangle^{- 1 - 1 / q} . \]
	For the integral to be bounded uniformly in $t$, we need $\frac{1}{q} >
		\beta^2 / 4 \pi$ and since $p, q$ are H{\"o}lder conjugates this means
	\[ \frac{p - 1}{p} = \frac{1}{q} > \beta^2 / 4 \pi . \]
	Consequently, we can choose $p, q \in [1, \infty)$ if and only if $\beta^2 <
		4 \pi$ which gives the claim.
\end{proof}

\begin{proof}[Proof of Lemma \ref{lem:wick-conv}]
	Recall the definition of the Besov norms,
	\begin{equation}
		\mathbb{E} [\| N_t \|_{B_{p, p}^{\alpha} (\langle x \rangle^{- n})}^p] =
		\sum_{i \geqslant - 1} 2^{- ip \alpha} \mathbb{E} \| \Delta_i N_t
		\|_{L^{p, - n}}^p . \label{eq:N-norm}
	\end{equation}
	We compute by Lemma~\ref{lem:variation-bound} and the
	Burkholder--Davis--Gundy's inequalities, for any $\varepsilon > 0$,
	\[ \mathbb{E} \| \Delta_i N_t \|_{L^p (\langle x \rangle^{- n})}^p \lesssim
		\int \mathd x \langle x \rangle^{- pn} \mathbb{E} [| \langle \Delta_i
			\llbracket \cos (\beta W) \rrbracket (x) \rangle_t |]^{p / 2} \lesssim
		\beta^2 2^{pi (\beta^2 / 4 \pi + \varepsilon)} . \]
	Therefore, \eqref{eq:N-norm} is finite provided $\beta^2 / 4 \pi < \alpha$
	and the convergence now follows from the martingale convergence theorem.
\end{proof}

We also owe the proof of Lemma \ref{lem:sing-wick-estimates}.

\begin{proof}[Proof of Lemma \ref{lem:sing-wick-estimates}]
	\label{proof-lemma-NC}\tmtextbf{(N)} Let
	\[ M_t (x) = \llbracket \cos (\beta (W_t (z - x) - W_t (x))) \rrbracket
		\assign \llbracket \cos (\beta (\delta_z W_t (x))) \rrbracket, \]
	which by Ito's formula can be written as
	\[ \begin{aligned}
			M_t (x) & = \int_0^t - \beta \llbracket \sin (\beta \delta_z W_s (x))
			\rrbracket \mathd (\delta_z W_s (x))                                  \\
			        & = \int_0^t - \beta \llbracket \sin (\beta \delta_z W_s (x))
			\rrbracket \int \mathd y (Q_s (x - z - y) - Q_s (x - y)) \mathd B_s
			(y),
		\end{aligned} \]
	where we recall that $(B_t)_t$ is a cylindrical Brownian motion on $L^2
		(\mathbb{R}^2)$. Now
	\[ Q_s (x - z - y) - Q_s (x - y) = | z | \int_0^1 \mathd \vartheta \nabla
		Q_s (x - y - \vartheta z), \]
	so that by translation invariance and since $\rho$ has compact support, we conclude  $\mathbb{E} \| M_t \|_{L^1} \lesssim \mathbb{E} [| M_t (0) |^2]^{1 / 2}$.
	The latter can be estimates as follows
	\[ \begin{aligned}
			          & \mathbb{E} | M_t (0) |^2                                          \\
			=         & \mathbb{E} \langle M (0), M (0) \rangle_t                         \\
			=         & \beta^2 | z |^2 \int_0^t \llbracket \sin (\beta \delta_z W_s (0))
			\rrbracket \llbracket \sin (\beta \delta_z W_s (0)) \rrbracket \times         \\
			          & \qquad \times \int \mathd y \int_0^1 \mathd \vartheta_1 \int_0^1
			\mathd \vartheta_2 \nabla Q_s (y - \vartheta_1 z) \nabla Q_s (y -
			\vartheta_2 z) \mathd s                                                       \\
			\leqslant & \beta^2 | z |^2 \int_0^t \mathd s \mathe^{2 \beta^2 G_s (0)
				- 2 \beta^2 G_s (z)} \int_0^1 \mathd \vartheta_1 \int_0^1 \mathd
			\vartheta_2 \int \mathd y \nabla Q_s (y - (\vartheta_1 - \vartheta_2)
			z) \nabla Q_s (y) .
		\end{aligned} \]
	Using the estimates from Lemma~\ref{lem:hk-exp}, $\| \nabla^{\alpha} Q_s
		\|_{L^p} \lesssim \langle s \rangle^{- 1 / p + \alpha / 2}$ so that
	\[ \begin{aligned}
			\mathbb{E} | \mathe^{\beta^2 G_t (z)} M_t (0) |^2 & \leqslant  \beta^2
			| z |^2 \int_0^t \mathd s \mathe^{2 \beta^2 G_s (0)} \mathe^{2 \beta^2
			(G_t - G_s) (z)} \times                                                                                                \\
			                                                  & \qquad \times \int_0^1 \mathd \vartheta_1 \int_0^1 \mathd
			\vartheta_2 \int \mathd y \nabla Q_s (y - (\vartheta_1 - \vartheta_2)
			z) \nabla Q_s (y)                                                                                                      \\
			                                                  & \lesssim  | z |^{2 - 8 (1 - \delta)} \int_0^t \mathd s \int_0^1
			\mathd \vartheta_1 \int_0^1 \mathd \vartheta_2 \int \mathd y \nabla Q_s
			(y - (\vartheta_1 - \vartheta_2) z) \nabla Q_s (y)                                                                     \\
			                                                  & \leqslant  | z |^{2 - 8 (1 - \delta)}  \int_0^t \mathd s \| \nabla
			Q_s \|_{L^1} \| \nabla Q_s \|_{L^{\infty}}                                                                             \\
			                                                  & \lesssim  | z |^{2 - 8 (1 - \delta)}  \int_0^t \mathd s \langle s
			\rangle^{- 1 / 2} \langle s \rangle^{1 / 2}                                                                            \\
			                                                  & \lesssim  | z |^{2 - 8 (1 - \delta)} t.
		\end{aligned} \]
	Using the same argument as in the proof of Lemma \ref{lem:W-Halpha} in the
	scales as well as the usual Kolmogorov argument for the $L^{\infty}$-norm in
	$z$ we obtain for any $\gamma_1 > 1 / 2$, $\gamma_2 > - 1 + 4 (1 - \delta)$
	\[ \mathbb{E} [\sup_{t, z} t^{- \gamma_1} | z |^{\gamma_2} \mathe^{\beta^2
				G_t (z)} \| M_t \|_{L^1} ] \lesssim 1, \]
	and thus
	\[ \sup_{t, z} t^{- \gamma_1} | z |^{\gamma_2} \mathe^{\beta^2 G_t (z)} \|
		M_t \|_{L^1}  < \infty, \quad a.s. \]
	\tmtextbf{(C)} We start by estimating
	\[ \begin{aligned}
			 & \mathbb{E} \| \llbracket \cos (\beta (W_t (\cdot - z)) + W_t
			(\cdot)) \rrbracket \rho (\cdot) \|_{B^{- s}_{p, p} (\mathd x)}    \\
			 & = \sum_{i \geqslant - 1} 2^{- isp} \mathbb{E} \| \Delta_i
			\llbracket \cos (\beta (W_t (\cdot - z) + W_t (\cdot))) \rrbracket
			\|^p_{L^p (\mathd x)}                                              \\
			 & \lesssim_{\rho}  \sum_{i \geqslant - 1} 2^{- isp} \mathbb{E} [|
				\Delta_i \llbracket \cos (\beta (W_t (\cdot - z) + W_t (\cdot)))
				\rrbracket (0) |^2]^{p / 2},
		\end{aligned} \]
	where we again used that $\rho$ is smooth and compactly supported and that
	the law of $W$ is translation invariant. Here, the Littlewood-Paley blocks
	act only in $x$, that is
	\[ \Delta_i \llbracket \cos (\beta (W_t (\cdot - z) + W_t (\cdot)))
		\rrbracket (x) = \int \mathd y K_i (x - y) \llbracket \cos (\beta (W_t (y
		- z) + W_t (y))) \rrbracket . \]
	Developing the martingale $\llbracket \cos (\beta (W_t (y - z) + W_t (y)))
		\rrbracket$ along the scales with Ito's formula, we obtain,
	\[ \begin{aligned}
			          & \mathbb{E} [| \Delta_i \llbracket \cos (\beta (W_t (z) + W_t (0)))
			\rrbracket |^2]                                                                \\
			\leqslant & \beta^2 \int_0^t \int \mathd y_1 \int \mathd y_2 K_i (y_1)
			K_i (y_2) \mathe^{\frac{\beta^2}{2} \mathbb{E} [| W_s (y_1 - z) + W_s
					(y_1) |^2]} \mathe^{\frac{\beta^2}{2} \mathbb{E} [| W_s (y_2 - z) + W_s
			(y_2) |^2]}_s                                                                  \\
			          & \times \mathd \langle W (y_1 - z) + W (y_1), W (y_2 - z) + W (y_2)
			\rangle_s                                                                      \\
			=         & \int_0^t \int \mathd y_1 \int \mathd y_2 K_i (y_1) K_i (y_2)
			\mathe^{2 \beta^2 G_s (z)} \mathe^{2 \beta^2 G_s (0)} [\dot{G}_s (y_1 -
				y_2) + \dot{G}_s (y_1 - y_2 - z)] \mathd s.
		\end{aligned} \]
	Thanks to the positivity of $G$, we have the estimate $\mathe^{2 \beta^2
			(G_s - G_t) (z)} \leqslant 1$ for $t \geqslant s$ so that for any $1 / r + 1
		/ q = 1$,
	{\footnotesize\[ \begin{aligned}
					          & \int_0^t \int \mathd y_1 \int \mathd y_2 K_i (y_1) K_i (y_2)
					\mathe^{2 \beta^2 (G_s - G_t) (z)} \mathe^{2 \beta^2 G_s (0)}
					[\dot{G}_s (y_1 - y_2) + \dot{G}_s (y_1 - y_2 - z)] \mathd s             \\
					\leqslant & \int_0^t \int \mathd y_1 \int \mathd y_2 K_i (y_1) K_i
					(y_2) \mathe^{2 \beta^2 G_s (0)} [\dot{G}_s (y_1 - y_2) + \dot{G}_s
					(y_1 - y_2 - z)] \mathd s                                                \\
					\leqslant & \| K_i \|_{L^1}  \| K_i \|_{L^r} \int_0^t \langle s
					\rangle^{4 (1 - \delta)} \| \dot{G_s}  \|_{L^q}                          \\
					\lesssim  & 2^{2 i / q}  \int_0^t \mathd s \langle s \rangle^{- 1 - 1 /
					q + 4 (1 - \delta)}                                                      \\
					\lesssim  & (t^{- 1 / q + 4 (1 - \delta)} \vee 1) 2^{2 i / q} .
				\end{aligned} \]}
	Therefore,
	{\small \[ \mathbb{E} \| \mathe^{- \beta^2 G_t (z)} \llbracket \cos (\beta (W_t
				(\cdot - z) + W_t (\cdot))) \rrbracket \rho (\cdot) \|^p_{B^{- s}_{p, p}
						(\mathd x)} \lesssim (t^{- 1 / 2 q + 2 (1 - \delta)} \vee 1)^p \sum_{i
					\geqslant - 1} 2^{- ip \left( s - \frac{1}{q} \right)} . \]}Using the same argument as in the proof of Lemma \ref{lem:W-Halpha}, we can
	choose $1 / 2 q$ sufficiently close to $s \in (0, 2 \delta)$ sufficiently
	large to conclude for any $\gamma_1 > 0$, $\gamma_2 > 2 - 3 \delta$,
	\[ \sup_{t, z} \| \mathe^{- \beta^2 G_t (z)} | z |^{\gamma_1 / 2} t^{-
				\gamma_2} \llbracket \cos (\beta (W_t (\cdot - z) + W_t (\cdot)))
		\rrbracket \rho (\cdot) \|_{B^{- s}_{p, p} (\mathd x)} < \infty, \quad
		a.s. \]
\end{proof}
\bibliography{local-bib.bib}
\end{document}